\documentclass[12pt,twoside]{report}
\usepackage{epsfig}

\catcode`\@=11

\font\dunhxxv=cmdunh10 scaled \magstep 5
\font\dunhxxi=cmdunh10 scaled \magstep 4
\font\dunhxvii=cmdunh10 scaled \magstep 3
\font\dunhxv=cmdunh10 at 15pt
\font\dunhxiv=cmdunh10 at 14pt
\font\dunhxiii=cmdunh10 scaled \magstep 2
\font\dunhxii=cmdunh10 scaled \magstep 1
\font\dunhx=cmdunh10

\ifcase \@ptsize
\def\@makechapterhead#1{ \vspace*{20pt} { \hrule height .2pt\vspace*{5pt}%
\parindent 0pt \raggedleft
\ifnum \c@secnumdepth >\m@ne \dunhxvii\baselineskip=23pt \@chapapp{} \thechapter \par
\vspace*{5pt}\hrule height .4pt \fi
\vspace*{15pt}\raggedright \dunhxxi\baselineskip=28pt #1\par
\vspace*{5pt}\hrule height .8pt%
\nobreak \vskip 60pt } }

\def\@makeschapterhead#1{ \vspace*{20pt} { \hrule height .2pt\vspace*{15pt}%
\parindent 0pt \raggedright
\dunhxxi\baselineskip=28pt #1\par
\vspace*{5pt}\hrule height .8pt%
\nobreak \vskip 60pt } }

\def\section{\@startsection {section}{1}{\z@}{-8ex plus -1ex minus
-.2ex}{2.3ex plus .2ex}{\cmbx10\baselineskip=19pt}}

\def\subsection{\@startsection{subsection}{2}{\z@}{-6ex plus -1.5ex minus
-.3ex}{1.5ex plus .2ex}{\dunhxiv\baselineskip=19pt}}

\def\l@chapter#1#2{\addpenalty{-\@highpenalty}
\vskip 1.0em plus 1pt \@tempdima 1.5em \begingroup
\parindent \z@ \rightskip \@pnumwidth
\parfillskip -\@pnumwidth
\leavevmode \advance\leftskip\@tempdima \hskip -\leftskip #1\nobreak\hfil
\nobreak\hbox to\@pnumwidth{\hss #2}\par
\penalty\@highpenalty \endgroup}

\def\@part[#1]#2{\ifnum \c@secnumdepth >-2\relax
\refstepcounter{part}%
\addcontentsline{toc}{part}{\thepart \hspace{1em}#1}%
\else
\addcontentsline{toc}{part}{#1}%
\fi
\markboth{}{}%
{\centering
\ifnum \c@secnumdepth >-2\relax
\dunhxiv\baselineskip=19pt \partname{} \thepart \par \vskip 20pt
\fi
\dunhxvii\baselineskip=23pt #2\par}%
\@endpart}

\def\@spart#1{{\centering \dunhxvii\baselineskip=23pt #1\par}\@endpart}

\def\l@part#1#2{\addpenalty{-\@highpenalty}
\addvspace{2.25em plus 1pt} \begingroup
\@tempdima 3em \parindent \z@ \rightskip \@pnumwidth \parfillskip
-\@pnumwidth
{\dunhxii\baselineskip=16pt
\leavevmode #1\hfil \hbox to\@pnumwidth{\hss #2}}\par
\nobreak \global\@nobreaktrue \everypar{\global\@nobreakfalse\everypar{}}
\endgroup}
\or
\or
\def\@makechapterhead#1{ \vspace*{20pt} { \hrule height .2pt\vspace*{5pt}%
\parindent 0pt \raggedleft
\ifnum \c@secnumdepth >\m@ne \dunhxxi\baselineskip=28pt \@chapapp{} \thechapter \par
\vspace*{5pt}\hrule height .4pt \fi
\vspace*{15pt}\raggedright \dunhxxv\baselineskip=33pt #1\par
\vspace*{5pt}\hrule height .8pt%
\nobreak \vskip 60pt } }

\def\@makeschapterhead#1{ \vspace*{20pt} { \hrule height .2pt\vspace*{15pt}%
\parindent 0pt \raggedright
\dunhxxv\baselineskip=33pt #1\par
\vspace*{5pt}\hrule height .8pt%
\nobreak \vskip 60pt } }

\def\section{\@startsection {section}{1}{\z@}{-8ex plus -1ex minus
-.2ex}{2.3ex plus .2ex}{\dunhxvii\baselineskip=23pt}}

\def\subsection{\@startsection{subsection}{2}{\z@}{-6ex plus -1.5ex minus
-.3ex}{1.5ex plus .2ex}{\dunhxvii\baselineskip=23pt}}

\def\l@chapter#1#2{\addpenalty{-\@highpenalty}
\vskip 1.0em plus 1pt \@tempdima 1.5em \begingroup
\parindent \z@ \rightskip \@pnumwidth
\parfillskip -\@pnumwidth
\dunhxii\baselineskip=16pt % Put your favorite font here
\leavevmode \advance\leftskip\@tempdima \hskip -\leftskip #1\nobreak\hfil
\nobreak\hbox to\@pnumwidth{\hss #2}\par
\penalty\@highpenalty \endgroup}

\def\@part[#1]#2{\ifnum \c@secnumdepth >-2\relax
\refstepcounter{part}%
\addcontentsline{toc}{part}{\thepart \hspace{1em}#1}%
\else
\addcontentsline{toc}{part}{#1}%
\fi
\markboth{}{}%
{\centering
\ifnum \c@secnumdepth >-2\relax
\dunhxvii\baselineskip=23pt \partname{} \thepart \par \vskip 20pt
\fi
\dunhxxi\baselineskip=28pt #2\par}%
\@endpart}

\def\@spart#1{{\centering \dunhxxi\baselineskip=28pt #1\par}\@endpart}

\def\l@part#1#2{\addpenalty{-\@highpenalty}
\addvspace{2.25em plus 1pt} \begingroup
\@tempdima 3em \parindent \z@ \rightskip \@pnumwidth \parfillskip
-\@pnumwidth
{\dunhxiv\baselineskip=19pt
\leavevmode #1\hfil \hbox to\@pnumwidth{\hss #2}}\par
\nobreak \global\@nobreaktrue \everypar{\global\@nobreakfalse\everypar{}}
\endgroup}
\fi

\if@twoside
\def\ps@headings{%
\let\@oddfoot\@empty\let\@evenfoot\@empty
\def\@evenhead{\thepage\hfil\slshape\leftmark}%
\def\@oddhead{{\slshape\rightmark}\hfil\thepage}%
\let\@mkboth\markboth
\def\chaptermark##1{%
\markboth {{ \dunhxii  %
\underline{\ifnum \c@secnumdepth >\m@ne
\@chapapp\ \thechapter. \  %
\fi
##1}}}{}}%
\def\sectionmark##1{%
\markright {{\dunhxii %
\underline{\ifnum \c@secnumdepth >\z@
\thesection. \ %
\fi
##1}}}}}
\else
\def\ps@headings{%
\let\@oddfoot\@empty
\def\@oddhead{{\slshape\rightmark}\hfil\thepage}%
\let\@mkboth\markboth
\def\chaptermark##1{%
\markright {{\dunhxii %
\underline{\ifnum \c@secnumdepth >\m@ne
\@chapapp\ \thechapter. \ %
\fi
##1}}}}}
\fi

\newdimen\bibindent
\def\mylarge{\fontsize {17.28}{18}\dunhx}

\renewcommand\tableofcontents{%
\if@twocolumn
\@restonecoltrue\onecolumn
\else
\@restonecolfalse
\fi
\chapter*{\contentsname
\@mkboth{%
\dunhxii\underline{\contentsname}}
{\dunhxii\underline{\contentsname}}}%
\@starttoc{toc}%
\if@restonecol\twocolumn\fi
}

\if@titlepage
\renewcommand\maketitle{\begin{titlepage}%
\let\footnotesize\small
\let\footnoterule\relax
\null\vfil
\vskip 60\p@
\begin{center}%
{\LARGE \dunhxxv \baselineskip=50pt \@title \par}%
\vskip 3em%
{\large
\lineskip .75em%
\begin{tabular}[t]{c}%
\dunhxxi \@author
\end{tabular}\par}%
\vskip 1.5em%
{\large \@date \par}%       % Set date in \large size.
\end{center}\par
\@thanks
\vfil\null
\end{titlepage}%
\setcounter{footnote}{0}%
\let\thanks\relax\let\maketitle\relax
\gdef\@thanks{}\gdef\@author{}\gdef\@title{}}
\else
\renewcommand\maketitle{\par
\begingroup
\renewcommand\thefootnote{\@fnsymbol\c@footnote}%
\def\@makefnmark{\rlap{\@textsuperscript{\normalfont\@thefnmark}}}%
\long\def\@makefntext##1{\parindent 1em\noindent
\hb@xt@1.8em{%
\hss\@textsuperscript{\normalfont\@thefnmark}}##1}%
\if@twocolumn
\ifnum \col@number=\@ne
\@maketitle
\else
\twocolumn[\@maketitle]%
\fi
\else
\newpage
\global\@topnum\z@   % Prevents figures from going at top of page.
\@maketitle
\fi
\thispagestyle{plain}\@thanks
\endgroup
\setcounter{footnote}{0}%
\let\thanks\relax
\let\maketitle\relax\let\@maketitle\relax
\gdef\@thanks{}\gdef\@author{}\gdef\@title{}}
\def\@maketitle{%
\newpage
\null
\vskip 2em%
\begin{center}%
{\LARGE \@title \par}%
\vskip 1.5em%
{\large
\lineskip .5em%
\begin{tabular}[t]{c}%
\@author
\end{tabular}\par}%
\vskip 1em%
{\large \@date}%
\end{center}%
\par
\vskip 1.5em}
\fi

\newskip\humongous \humongous=0pt plus 1000pt minus 1000pt
\def\caja{\mathsurround=0pt}
\def\eqalign#1{\,\vcenter{\openup1\jot \caja
\ialign{\strut $\displaystyle{##}$&$
\displaystyle{{}##}$\hfil\crcr#1\crcr}}\,}
\newif\ifdtup

\def\tr{\mathop{\rm tr}}
\def\Tr{\mathop{\rm Tr}}

\def\Re{\mathop{\rm Re}}

\def\VEV#1{\left\langle #1\right\rangle}

\def\pr#1{#1^\prime}

\def\beq{\begin{equation}}
\def\eeq{\end{equation}}

\def\beqn{\begin{eqnarray}}
\def\eeqn{\end{eqnarray}}

\relax

\def\Qb{\overline{Q}}
\def\qb{\overline{q}}
\def\bb{\overline{b}}
\def\Dh{\hat{D}}
\def\asb{{}\ifmmode \bar{\alpha}_s \else $\bar{\alpha}_s$\fi}
\def\t{\,\log\frac{\mu^2}{\mu_0^2}\,}

\def\xb{\overline{x}}
\def\Bb{\overline{B}}

\def\e{\epsilon}
\catcode`\@=11

\def\@normalsize{\@setsize\normalsize{15pt}\xiipt\@xiipt
\abovedisplayskip 14pt plus3pt minus3pt%
\belowdisplayskip \abovedisplayskip
\abovedisplayshortskip \z@ plus3pt%
\belowdisplayshortskip 7pt plus3.5pt minus0pt}

\def\small{\@setsize\small{13.6pt}\xipt\@xipt
\abovedisplayskip 13pt plus3pt minus3pt%
\belowdisplayskip \abovedisplayskip
\abovedisplayshortskip \z@ plus3pt%
\belowdisplayshortskip 7pt plus3.5pt minus0pt
\def\@listi{\parsep 4.5pt plus 2pt minus 1pt
\itemsep \parsep
\topsep 9pt plus 3pt minus 3pt}}

\@twosidetrue
\relax

\catcode`@=12

\catcode`\@=11

\def\@chapter[#1]#2{ \cleardoublepage\ifnum \c@secnumdepth >\m@ne
\refstepcounter{chapter}%
\typeout{\@chapapp\space\thechapter.}%
\addtocontents{toc}
{\protect\renewcommand\mathbf{}}%
\addcontentsline{toc}{chapter}%
{\@chapapp\
\protect\numberline{\thechapter.} #1}%
\else
\addcontentsline{toc}{chapter}{#1}%
\fi
\chaptermark{#1}%
\addtocontents{lof}{\protect\addvspace{10\p@}}%
\addtocontents{lot}{\protect\addvspace{10\p@}}%
\if@twocolumn
\@topnewpage[\@makechapterhead{#2}]%
\else
\@makechapterhead{#2}%
\@afterheading
\fi}

\catcode`\@=11

\def\section{\@startsection {section}{1}{\z@}{-8ex plus -1ex minus
-.2ex}{2.3ex plus .2ex}{\dunhxv\baselineskip=23pt}}
\def\subsection{\@startsection{subsection}{2}{\z@}{-6ex plus -1.5ex minus
-.3ex}{1.5ex plus .2ex}{\dunhxiv\baselineskip=23pt}}

\def\appendix{\par
\setcounter{chapter}{0}%
\setcounter{section}{0}%
\renewcommand\@chapapp{\appendixname}%
\renewcommand\thechapter{\@Alph\c@chapter}

\catcode`\@=11
}

\catcode`\@=12

\def\figcap{\section*{Figure Captions\markboth
{FIGURECAPTIONS}{FIGURECAPTIONS}}\list
{Fig. \arabic{enumi}:\hfill}{\settowidth\labelwidth{Fig. 999:}
\leftmargin\labelwidth
\advance\leftmargin\labelsep\usecounter{enumi}}}
 \relax
\def\tablecap{\section*{Table Captions\markboth
{TABLECAPTIONS}{TABLECAPTIONS}}\list
{Table \arabic{enumi}:\hfill}{\settowidth\labelwidth{Table 999:}
\leftmargin\labelwidth
\advance\leftmargin\labelsep\usecounter{enumi}}}
 \relax
\def\reflist{\section*{References\markboth
{REFLIST}{REFLIST}}\list
{[\arabic{enumi}]\hfill}{\settowidth\labelwidth{[999]}
\leftmargin\labelwidth
\advance\leftmargin\labelsep\usecounter{enumi}}}
 \relax

\catcode`\@=11

\relax

\catcode `@ 11
\def\biblabel#1{\if@filesw\immediate
\write\@auxout{\string\bibcite{#1}{\the\value{\@listctr }}}\fi}
\catcode `@ 12

\newcommand{\ccaption}[2]{
\begin{center}
\parbox{0.85\textwidth}{
\caption[#1]{\small\it {#2}}}
\end{center}    }
\def    \be             {\begin{equation}}
\def    \ee             {\end{equation}}
\def    \ba             {\begin{eqnarray}}
\def    \ea             {\end{eqnarray}}
\def    \nn             {\nonumber}
\def    \=              {\;=\;}

\def \ep{\epsilon}
\def \as   {\ifmmode \alpha_s \else $\alpha_s$ \fi}

\def\b0{b_0}

\def \mt   {\ifmmode m_{\rm t} \else $m_{\rm t}$ \fi}

\def \to   {\mbox{$\rightarrow$}}
\newcommand     \MSB            {\ifmmode {\overline{\rm MS}} \else
$\overline{\rm MS}$\fi}

\newcommand\hepph[1]{{\tt hep-ph/#1}}

\def\dl#1{$$\displaylines{\quad#1}$$}
\def\nl{\hfill\cr\hfill}
\def\lq{\left[}
\def\rq{\right]}
\def\rg{\right\}}
\def\lg{\left\{}

\def\e{\epsilon}
\def\a{\alpha}
\def\b{\beta}

\def\d{\Delta}
\def\g{\gamma}
\def\dpr{\Delta '}
\def\su{\sigma_1}
\def\sd{\sigma_2}
\def\st{\sigma_3}
\def\lp{\lambda_{+}}
\def\lm{\lambda_{-}}
\def\lpm{\lambda_{\pm}}
\def\taup{\tau_{+}}
\def\taum{\tau_{-}}
\def\taupm{\tau_{\pm}}
\def\th{\theta}
\def\t{s_3}
\def\zp{\xi_{+}}
\def\zm{\xi_{-}}
\def\zpm{\xi_{\pm}}
\def\cp{\xi_{+}}
\def\cm{\xi_{-}}

\def\rop{\rho_{+}}
\def\rom{\rho_{-}}
\def\ropm{\rho_{\pm}}
\def\di{\eta_-}
\def\so{\eta_+}
\def\tol{\longrightarrow}
\def\li#1{\,{\rm Li}_2\left(#1 \right)}
\def\dl#1{$$\displaylines{\quad#1}$$}
\def\nl{\hfill\cr\hfill}
\def\tr{\mathop{\rm Tr}}
\def\ds#1{#1\kern-1ex\hbox{/}}
\def\dsh{h\kern-1.2ex /}
\def\mod#1{|{\bf #1}|}
\def\Nep{{\rm N}(\e)}
\newcommand\nlf{\ifmmode {n_{\rm lf}} \else $n_{\rm lf}$  \fi}
\newlength{\largfig}
\newlength{\Largfig}
\newcommand\sq{{\rm \scriptscriptstyle Q}}
\newcommand\sz{{\rm \scriptscriptstyle Z}}
\newcommand\sB{{\rm \scriptscriptstyle B}}
\newcommand\sBb{{\rm \scriptscriptstyle \overline{B}}}

\def\Mz{M_{\sz}}

\def\figura#1#2#3
{\begin{figure}
\begin{center}
\leavevmode\protect\epsfxsize=#1\protect\epsffile{#2.eps}
\protect\caption{#3}
\label{#2}
\end{center}
\end{figure}
}

\def\({\left(}
\def\){\right)}
\def\ga{\left\{}
\def\gc{\right\}}

\def\PSq{d\Phi_4}
\def\PSt{d\Phi_3}
\def\PSd{d\Phi_2}
\def\vet#1{{\bf #1}}
\def\s{\sigma}
\def\kperp{k_\perp}

\def\gperp{g_\perp}
\def\sp{\hspace{0cm}}
\begin{document}
%%%%%%%%%%%%%%%%%%%%%%%%%%%%%%%%%%%%%%%%%%%%%%%%%%%%%%%%%%%%%%%%%
%\title{ Next-to-Leading-Order Corrections to\\
% the Production of Heavy-Flavour\\
%     Jets  in {\largetitle $e^+e^-$} Collisions \\ \ }
%\author{Carlo Oleari}
%\date{}
%\maketitle
%\input{phd_title_eng}
\def\vsmalltitlefont{\dunhxii}
\def\smalltitlefont{\dunhxiii}
\def\midtitlefont{\dunhxvii}
\def\bigtitlefont{\dunhxxi}

\thispagestyle{empty}
\mbox{}
\vspace{-3cm}
\begin{center}
\smalltitlefont \baselineskip 0.8cm
Universit\`a degli Studi di Milano\\
Facolt\`a di Scienze Matematiche, Fisiche e Naturali\\
Dipartimento di Fisica
\end{center}

\vspace{4cm}

\def\mytitle{\fontsize {20.74}{18}\normalfont}

\centerline{\smalltitlefont Ph.D. Thesis}
\begin{center}
{\bigtitlefont \baselineskip 1.2cm
Next-to-Leading-Order Corrections to\\
the Production of Heavy-Flavour\\
Jets  in {\mytitle $e^+e^-$} Collisions\\
}
\end{center}
\vspace{0.3cm}
\centerline{\smalltitlefont }
\begin{center}
\midtitlefont
Carlo Oleari
\end{center}

%\flushleft

\vspace{4cm}
\begin{minipage}[t]{3cm}
\baselineskip 0.7cm
\vsmalltitlefont
%Il Tutore\\
%Prof.\ P. Nason
\end{minipage}
%\hspace{3cm}
\hfill
\begin{minipage}[t]{6cm}
\baselineskip 0.7cm
\vsmalltitlefont
Tutor: Prof.\ P. Nason
\end{minipage}

\vspace{2cm}
\vfill

\centerline{\vsmalltitlefont
1994--1997
}

\newpage
\thispagestyle{empty}
\mbox{}
%\markboth{\mbox{}}{\mbox{}}
\newpage
\setcounter{page}{1}
%\newpage
%\mbox{}
%\newpage

\pagenumbering{roman}
\tableofcontents
% le righe che seguono, impongono una pagina bianca senza intestazione
\newpage
\thispagestyle{plain}
\mbox{}
%\markboth{\mbox{}}{\mbox{}}
\newpage
%%%%%%%%%%%%%%%%%%%%%%%%%%%%%%%
\pagenumbering{arabic}
%\input{phd_intro}
%\input{oc.sty}
%\genericchapter{Introduction}{}
%\markboth{\dunhxii \underline{Introduction}}
%{\dunhxii \underline{Introduction}}
\def\intro{Introduction}
\chapter*{\intro}
%\addtocounter{chapter}{-1}
\addcontentsline{toc}{chapter}{\intro}
\markboth{\dunhxii \underline{\intro}}
{\dunhxii \underline{\intro}}

Radiative corrections to jet production in $e^+e^-$ annihilation at
order $\as^2$ were computed a long time ago \cite{ERT,VGO,FKSS}. These
calculations were, however, performed for massless quarks.

This is sufficient in most practical applications: in fact, if we
consider the heaviest quark that can be produced nowadays at LEP, the
$b$ quark, we can say that, at relatively low energy, the $b$ fraction
is strongly suppressed, and, at high energy (i.e. on the $Z$ peak and
beyond), mass effects can be neglected for many observables, because
they appear as powers of the ratio of the mass over the total energy.

Nevertheless there are several reasons why a massive next-to-leading-order
calculation is desirable.
\begin{itemize}
\item[-]
First of all, there are quantities, such as the {\it heavy-flavour
momentum correlation}~\cite{NO,NO1}, that, although well defined in
the massless limit, cannot be computed using the massless result.

\item[-]
Secondly, a massive calculation  finds application in the area of
{\it fragmentation functions}, where it can be used
to improve the resummed cross section.

\item[-]
Then, this type of calculation helps to understand the relevance of
{\it mass corrections}  in the
determination of $\as$ from event shape variables and jets, and may be
used to measure {\it quark masses}~\cite{Rodrigo,Bil_Rod_Santa}.

\item[-]
Finally, in the future accelerators, where high energies will be
reached (Next Linear Collider), $t\overline{t}$ pairs will be produced
and mass effects are very likely to be important in those
measurements.
\end{itemize}

In this thesis, we describe a Next-to-Leading-Order (NLO) calculation of the
heavy-flavour production cross section in $e^+e^-$ collisions,
including quark mass effects, for unoriented quantities.

At NLO, we get contributions both from two-, three- and from four-particle
final states.  The amplitudes describing these contributions contain
infrared and ultraviolet divergences. Therefore, some regularization
procedure is needed.  We use the {\it dimensional regularization},
that consists in changing the space dimensions $d$ from $4$ to
$4-2\e$.  In this way, soft and ultraviolet divergences appear as
poles in $\e$.  At the end of the calculation, the renormalization
procedure takes care of the ultraviolet poles, while the infrared
poles must cancel for infrared-safe physical quantities.

Very recently, two calculations have appeared in the literature
dealing with the same problem~\cite{Rodrigo}--\cite{Brandenburg}.
They both use a slicing method, in order to deal with infrared
divergences.
This method consists to exclude, from numerical integration, the
region of soft and collinear singularities.
For this reason, a cut-off parameter is introduced in the energy
region of soft gluons and in the collinear limit.
In the excluded phase space regions, the transition amplitudes are
replaced by their limiting values and the integration is done
analytically, in order to obtain the coefficients of the infrared
poles $1/\e$ and $1/\e^2$, that must cancel the corresponding terms in
the virtual diagrams.
Together with the poles, there is a  finite part, which is
integrated numerically over the remaining three-body phase space.
Finally, the full transitions are integrated numerically in $d=4$
dimensions over the  four-body phase space, above
the energy cut-off.
Both results, the semi-analytical and the full-numerical one, depend
on the choice for the value of the cut-off. Since this parameter
was introduced
arbitrarily, the final result must be independent from this value.\\
The smaller this cut-off is, the better result is obtained: in fact,
this is an approximate method, because it substitutes, in the singular
region, the exact transition amplitudes with their limiting values.
On the other hand, for small cut-off parameters, the numerical
integration start to fail since the integration region gets closer and
closer to the singular points.  The best solution is found in the
region of the cut-off values where the computed physical quantities
become nearly independent from these values.

In our work, we preferred to use  a subtraction method,
since, in this way, we do not need to worry about taking the
limit for small cut-off parameters. Subtraction methods for
the calculation of radiative corrections to $e^+e^-\,\to\,\mbox{jets}$
have been used in Refs.~\cite{ERT,YellowBook}, and
they have also been successfully employed in the calculation
of hadronic production processes.

%We were able to perform a partial comparison of our result with that of
%Ref.~\cite{Rodrigo}, and found satisfactory agreement.

Together with the quoted massless calculations, parts of the massive
calculation were computed by other groups: in the work of
Ref.~\cite{Ballestrero}, a calculation of the process $e^+e^-\,\to\,
Q\overline{Q}gg$ is given, but virtual corrections to the process
$e^+e^-\,\to\, Q\overline{Q}g$ are not included.  In the same work,
the amplitudes for two $Q\Qb$ pairs in the final state are calculated.
In Ref.~\cite{Magnea}, the NLO corrections to the
production of a heavy-quark pair plus a photon are given, including
both real and virtual contributions, while,
in Ref.~\cite{Hagiwara}, the virtual contribution with a three-fermion
loop  is computed.

In this thesis we deal with virtual and real contributions to the
$e^+e^-$ annihilation process, that is with terms coming from the
interference between one-loop diagrams with the tree-level terms and
with final states characterized by a heavy couple of quarks plus a
pair of gluons or light quarks. The computation of these contributions
is the really hard part of the full calculation.

This work is organized in a quite reverse order: in fact, we first
give two physical applications of our calculation, leaving to the
other chapters the detailed description of the full computation.

In Chapter~\ref{chap:bbcorr}, we introduce the heavy-flavour momentum
correlation, its relationship with the measurement of $R_b$, and the
results we have obtained at leading and next-to-leading order.  The
computation of this quantity was the initial reason that conducted us to
perform this calculation.

In Chapter~\ref{chap:fragment}, we revisit the fragmentation-function
formalism for heavy quarks and the connection with our fixed order
calculation. In fact, we check the validity of the initial conditions
for the fragmentation functions, the NLO splitting functions in the
time-like region and we derive an improved cross section, valid at NLO,
for energies of order of the mass, but valid at Next-to-Leading-Log for
energies much larger than the mass.

The complete calculation is given in the following chapters: in
Chapter~\ref{chap:kinematics}, we introduce the kinematical variables
necessary to describe unoriented quantities, and we find the
expression of the phase space  in $d=4-2\e$ dimensions for the
three-body final state ($Q\Qb g$) and for the four-body final states
($Q\Qb g g$, $Q\Qb q \qb$ and $Q\Qb Q\Qb$).

In Chapter~\ref{chap:NLO_amplitudes}, we introduce the notation used
during the calculation and we present the Feynman diagrams and the
cut-diagrams which contribute to the differential cross section.
All the results are in an analytic form and are implemented in a
FORTRAN program. This program can compute each sort of unoriented
infrared-safe shape variables and jet-clustering
algorithms.
This chapter is the hard part of the thesis, and some of the
calculations involved, together with other details, are given in the
appendixes.

The check of the cancellation of the infrared poles is performed in
Chapter~\ref{chap:IR_canc}, where we apply the subtraction method.
We first give a pedagogical introduction of this method before using
it on the full cross section.
In this chapter, we show how we can avoid to compute two-loop
contributions and how we obtain the limits of the amplitudes in the
soft and collinear regime.
At the end, we list some controls we have done: both internal
consistency and external comparisons with the results of other groups.
In fact, we found satisfactory agreement between our jet-clustering
results and those of Ref.~\cite{Rodrigo}.

In Chapter~\ref{chap:results} we present some tables computed
with different values of the ratio $m/E$, and we compare these results
with the massless case.
For this kind of calculations, it would be difficult to perform
analytical comparisons: for example, at the moment, this is
impossible, since the other two
groups~\cite{Rodrigo}--\cite{Brandenburg} have used a different method
of computation.
For this reason, we have chosen a few shape variables, and we have
computed some moments, which can be obtained with a great precision.
In addition, we present the results for the three-jet decay rate,
computed according to four jet-clustering algorithms, for different
values of the cut parameter.

Finally, we summarize our work in the concluding chapter.

\chapter{Momentum correlation}
\label{chap:bbcorr}
\thispagestyle{plain}
In the early of  1996, the discrepancy between the measured
and the theoretical  value of $R_b$ was drawing considerable
attention in the physics community.
We then began investigating on a possible systematic error, of
dynamical origin, in the measurement of $R_b$.

Very soon we realized that our first order result in $\as$ was to be
complemented by a second order calculation, so that we started the
computation of the differential cross section for the process
$e^+e^-\,\to\, Z/\gamma\,\to\, Q\Qb + X$, where $Q$ is a massive
quark, $X$ is anything else, at order $\as^2$.

\section[Momentum correlation in $Z/\gamma \,\to\, b \bar{b}$]{Momentum
correlation in {\mylarge $Z/\gamma \,\to\, b \bar{b}$}}
$R_b$ is defined by the ratio
\beq
R_b = \frac{\displaystyle\Gamma_{b\bar b}}
{\displaystyle\Gamma_{\rm had}}\;,
\eeq
where $\Gamma$ is the decay width of the $Z/\gamma$ into the specified
final state.  In several experimental techniques for the measurement
of $R_b$, the tagging efficiency is extracted from the data by comparing
the sample of events in which only one $b$ has been tagged, with the
one in which both $b$'s are observed. If the production
characteristics of the $b$ and the $\bar b$ were completely
uncorrelated, this method would yield the exact answer, without need of
corrections. Of course, other correlations of experimental nature
should be properly accounted for, but their discussion is outside the
scope of the present theoretical section.
Here we deal with the standard QCD gluon emission, that  generates a
correlation of the quark-antiquark momenta of order $\as$.  Other
dynamical effects, like the production of heavy-quark pairs via a
gluon-splitting mechanism, may affect the measurement.  However, they
are, to some extent, better understood: in fact, they tend to give
soft heavy quarks, and they are, therefore, easily eliminated.

We begin by considering the simple case in which
the efficiency for tagging a $B$ meson is a linear function
of its momentum.
This simplifying assumption allows us to make very precise statements
about the correlation. Furthermore, it is not extremely far from
reality, in the sense that the experimental tagging efficiency is often
a growing function of the $B$ momentum.
For this reason, we introduce two kinematical variables to describe
the momentum carried by the couple quark-antiquark
\beq
\xb_{1(2)}=\frac{\mod{p}_{{\sB}(\sBb)}}{\mod{p}^{(\rm max)}_{\sB}} \;,
\eeq
where $\vet{p}$ denotes the  meson three-momentum.
In terms of the usual notation, where
\beq
x_{1(2)}=\frac{2\,E_{\sB(\sBb)}}{E} \;,
\eeq
with $E$ the total centre-of-mass energy and $E_{\sB(\sBb)}$ the
energy of the $B(\Bb)$ meson, we have
\beq
\xb_{1(2)}^2 = \frac{x_{1(2)}^2-\rho}{1-\rho} \;, \quad\quad\quad\quad
\rho  =\frac{4\,m_{\sB}^2}{E^2}\;.
\eeq
We then assume that the tagging efficiency $\e$ is given by
\beq
\e_1=C \, \xb_1\;, \quad\quad\quad\quad  \e_2=C \, \xb_2\;,
\eeq
and that, in our ideal detector, the detection efficiency is not
influenced by the presence of another tag.
We  separate the events when one single $B$ or $\Bb$ is tagged,
from the events where both $B$ and $\Bb$ are tagged.
Denoting with $N_1$ the number of single tags, with $N_2$ the
number of double tags and with $N$  the total number of events, we have
\beqn
\frac{N_1}{N} &\equiv&  R_b\, \VEV{\e_1 + \e_2} =  R_b \,
C \left(\VEV{\xb_1}+\VEV{\xb_2}\right) = R_b \,
2\,C\,\VEV{\xb}\nn\\[-3mm]
&& \label{eq:system_N1_N2}\\[-3mm]
\frac{N_2}{N} &\equiv&  R_b\,\VEV{\e_1\,\e_2}   = R_b\, C^2 \,
\VEV{\xb_1 \,\xb_2} =R_b\, C^2 \VEV{\xb}^2 \times(1+r) \;,\nn
\eeqn
where
\beq
\VEV{\xb} = \VEV{\xb_1} = \VEV{\xb_2}\;,
\eeq
and $r$ is the momentum correlation
\beq
\label{eq:def_r}
r \equiv \frac{\VEV{\xb_1 \, \xb_2}-\VEV{\xb}^2}{\VEV{\xb}^2}\;.
\eeq
Solving the system~(\ref{eq:system_N1_N2}), we obtain
\beq
R_b=\frac{N_1^2}{4 N N_2}\times(1+r)\;.
\eeq
The quantity $r$ cannot be measured, and therefore one has to compute
it in order to determine $R_b$.
It is convenient to rewrite $r$ in the following way
\beq
\label{eq:betterr}
r=\frac{\VEV{\(1-\xb_1\) \(1-\xb_2\)}-\VEV{1-\xb}^2}{\VEV{\xb}^2}\;,
\eeq
so that we immediately see that terms, to whatever order in
$\as$, with only a $b\bb$ couple in the final state give zero
contributions, so that we do not need to compute two-loop diagrams at
order $\as^2$ (see Sec.~\ref{sec:two_body} for a more detailed discussion).
We can then compute this infrared-safe quantity with our program.
Introducing the differential cross section, normalized to one, we
define
\beqn
\mbox{}\!\!\!\!\!&&\!\!\!
\VEV{1-\xb} =  \int dx_1\, dx_2 \, \frac{d\s}{dx_1\,dx_2} \(1-\xb\) \equiv
\frac{\as}{2\pi}\, a\,+\,{\cal O}\(\as^2\)
\nn\\[-3mm] \mbox{}\!\!\!\!\!&&\\[-3mm]
\mbox{}\!\!\!\!\!&&\!\!\!
\VEV{\(1-\xb_1\) \(1-\xb_2\)} = \int dx_1\, dx_2 \,
\frac{d\s}{dx_1\,dx_2} \(1-\xb_1\) \(1-\xb_2\) \equiv
\frac{\as}{2\pi}\,b + \left(\frac{\as}{2\pi}\right)^2 \!c
+ {\cal O}\(\as^3\)\nn
\eeqn
where the integral is extended to the appropriate phase space region.
Using eq.~(\ref{eq:betterr}), we obtain, for the expansion in $\as$ of
$r$,
\beq
r=\frac{\as}{2\pi}\, b + \left(\frac{\as}{2\pi}\right)^2
\left(c+2ab-a^2\right)+ {\cal O}\(\as^3\)\;.
\eeq
Our results for $r$ are displayed in Tab.~\ref{tab:logdep}, where we have
taken the total energy $E=\Mz$.
\begin{table}[htbp]
\begin{center}
\leavevmode
\begin{tabular}{||c|c|c|c|c||}
\hline\hline
$m$    & $a$  & $b$  & $c$ & $r$ \\
\hline\hline
1 GeV  &$12.79(1)$&$0.6628(1)$&$155.5(3)$&$0.1055\,\as+0.23(1)\,\as^2$   \\
\hline
5 GeV  &$7.170(2)$&$0.6182(1)$&$50.94(5)$& $0.0984\,\as + 0.213(1)\,\as^2$ \\
\hline
10 GeV &$4.858(1)$&$0.5432(1)$&$26.23(2)$&$0.0865\,\as+0.2004(6)\,\as^2$ \\
\hline\hline
\end{tabular}
\caption{Mass dependence of the coefficients $a$, $b$, $c$ and $r$.
The number in parenthesis (to be taken as zero if not given)
represents the accuracy of the last digit. The total
energy $E$ is taken equal to $\Mz$.}
\label{tab:logdep}
\end{center}
\end{table}
From this table, we see that the coefficients $a$ and $c$ are both
plagued by collinear divergences, because of the presence of large
logarithms of the ratio  $E/m$, but the coefficients of the
expansion of $r$ itself do instead converge in the limit of small
mass.  This is due to cancellation of these logarithms in the
definition of $r$.

It is easy to prove that this cancellation must occur to all orders in
perturbation theory.  In fact, according to the
factorization theorem, we can write the double inclusive cross
section for $b \bb$ production, in the limit of $E\gg m$, as
\beq
\label{eq:doubleinc}
\frac{d\sigma}{dx_1\,dx_2}=\int dy_1\,dz_1\,dy_2\,dz_2
\;\Dh(z_1)\;\Dh(z_2)\; \delta(y_1\,z_1-x_1)\;\delta(y_2\,z_2-x_2)\;
\frac{d\hat\sigma}{dy_1\,dy_2}\;,
\eeq
where $d\hat\sigma$ is the short-distance cross section, which, in the
limit $E/m\,\to\,\infty$, has a perturbative expansion in $\as$ with
finite coefficients (observe that the distinction between the
momentum- or energy-defined Feynman $x$ becomes irrelevant in the
limit we are considering here), and $\Dh(z)$ is the fragmentation
function, that absorbs all the divergent terms.  We then have
\beqn
\label{eq:fullavx}
\mbox{}\!\!\!\!\!&&\!\!\!
\VEV{x}=\int dx_1\,  dx_2\; x_1 \frac{d\sigma}{dx_1 dx_2}
=\left(\int dz\,z \,\Dh(z)\right)
\int dx_1\, dx_2\; x_1 \frac{d\hat\sigma}{dx_1dx_2} \nn\\[-3mm]
\mbox{}\!\!\!\!\!&&\\[-3mm]
\mbox{}\!\!\!\!\!&&\!\!\!
\VEV{x_1\, x_2}=\int dx_1 \, dx_2\; x_1\, x_2\; \frac{d\sigma}{dx_1 dx_2}
=\left(\int dz\,z \,\Dh(z)\right)^2
\int dx_1 \, dx_2\; x_1\, x_2\;\frac{d\hat\sigma}{dx_1 dx_2}\;.\nn
\eeqn
In the ratio $\VEV{x_1\,x_2}/\VEV{x}^2$ (and therefore in $r$) the
integral containing $\Dh(z)$ cancels.  Thus, the perturbative
coefficients of $r$ are finite in the limit $E/m\,\to\,
\infty$. Observe that, in the derivation of eqs.~(\ref{eq:fullavx}), we
have assumed the relation
\beq
\int \Dh(z)\,dz=1\;,
\eeq
which is appropriate when we can neglect the secondary production
of $b\bb$ pairs via gluon splitting.

We also define the quantity $r^\prime$
\beq
r^\prime=\frac{\VEV{\xb_1 \, \xb_2\;{\rm cut}(1,2)}-
\VEV{\xb}^2}{\VEV{\xb}^2}\;,
\eeq
where the function ${\rm cut}(1,2)$ assumes value one if the quark and
the antiquark are in opposite hemispheres with respect to the thrust
axis, and zero otherwise.  We define
\beq
\VEV{1-\xb_1-\xb_2+\xb_1\, \xb_2\,{\rm cut}(1,2)}=
\frac{\as}{2\pi}\,b^\prime +
\left(\frac{\as}{2\pi}\right)^2\! c^\prime\,+\,{\cal O}\(\as^3\)\;,
\eeq
and obtain
\beq
r^\prime=\frac{\as}{2\pi}\, b^\prime + \left(\frac{\as}{2\pi}\right)^2
\left(c^\prime+2ab^\prime-a^2\right)+ {\cal O}\(\as^3\)\;.
\eeq
The quantities $b^\prime$, $c^\prime$ and $r^\prime$ are given in
Tab.~\ref{tab:logdepcut}.
\begin{table}[htbp]
\begin{center}
\leavevmode
\begin{tabular}{||c|c|c|c||}
\hline\hline
$m$    & $b^\prime$  & $c^\prime$ & $r^\prime$ \\
\hline\hline
1 GeV  & $0.3800(7)$  & $158.4(3)$ & $0.0605(1)\,\as + 0.12(1)\,\as^2$ \\
\hline
5 GeV  & $0.3653(6)$  & $50.9(1)$ & $0.0581(1)\,\as + 0.119(2)\,\as^2$ \\
\hline
10 GeV & $0.3423(5)$  & $25.51(6)$ & $0.0545(1)\,\as + 0.133(1)\,\as^2$ \\
\hline\hline
\end{tabular}
\caption{Mass dependence of the coefficients
$b^\prime$, $c^\prime$ and $r^\prime$ at $E=\Mz$.}
\label{tab:logdepcut}
\end{center}
\end{table}

\section[Double inclusive cross section at order $\alpha_s$]
{Double inclusive cross section at order {\mylarge $\alpha_s$}}
Up to now, we have assumed that the efficiency is linear in the $B$
momentum. Even in the more realistic case in which the efficiency is a
more complicated function of the kinematical variables, it is possible
to compute the inclusive cross section for the production of a $b\bb$
pair, provided one also knows the $B$ fragmentation function, which is,
to some extent, measured at LEP. The appropriate formula is given in
eq.~(\ref{eq:doubleinc}).  We limit our considerations up to order
$\as$.
The expression for the short distance cross section is
\beq
\label{eq:doubleinchat}
\frac{d\hat\sigma}{dy_1\,dy_2}=\delta(1-y_1)\,\delta(1-y_2)
+\frac{2\as}{3\pi}\left.\frac{y_1^2+y_2^2}{(1-y_1)(1-y_2)}\right|_+\;,
\eeq
where the ``$+$'' distribution sign specifies the way the singularities
at $y_1=1$ and $y_2=1$ should be treated: for any smooth function
of $y_1$ and $y_2$, we define
\beqn
&& \!\!\!\! \int_{y_1+y_2 > 1} dy_1\,dy_2\, \left.
\frac{y_1^2+y_2^2}{(1-y_1)(1-y_2)}\right|_+ G(y_1,y_2) =
\nonumber \\ \label{eq:doubleplus}
&& \!\!\!\!
\int_{y_1+y_2 > 1} dy_1\,dy_2\, \frac{y_1^2+y_2^2}{(1-y_1)(1-y_2)}
\Big[ G(y_1,y_2)-G(1,y_2)-G(y_1,1)+G(1,1) \Big] \;.\phantom{aaaaa}
\eeqn
It is easy now to compute the value of $r$ in the massless limit. Using
eqs.~(\ref{eq:def_r}) and~(\ref{eq:fullavx}), we obtain
\beq
\label{eq:r_value}
r = \frac{\as}{3\,\pi} = 0.1061 \times \as\;,
\eeq
that is consistent with the massive results of Tab.~\ref{tab:logdep}.
In addition, eqs.~(\ref{eq:doubleinc}) and~(\ref{eq:doubleinchat})
give
\beq
\label{eq:scheme}
\frac{d\sigma}{dx_1}=\int dx_2 \,\frac{d\sigma}{dx_1\,dx_2}=\hat{D}(x_1)\;.
\eeq
The above equation fixes the factorization scheme to be the
{\it annihilation scheme} as defined in Ref.~\cite{NasonWebber}. The choice
of the scheme of eq.~(\ref{eq:scheme}) defines unambiguously the result,
without the need of computing explicitly the virtual corrections.  In
fact, the most general formula for the short distance cross section at
order $\as$ is obtained by adding to eq.~(\ref{eq:doubleinchat}) terms
of the form
\beq
\as \lq f(y_1)\,\delta(1-y_2) + f(y_2)\,\delta(1-y_1) + g\,
\delta(1-y_1)\,\delta(1-y_2)  \rq \;,
\eeq
where $f$ is a generic distribution and $g$ is a
constant. However, in order for eq.~(\ref{eq:scheme}) to be respected,
these terms must be absent.

\begin{figure}
\centerline{\epsfig{figure=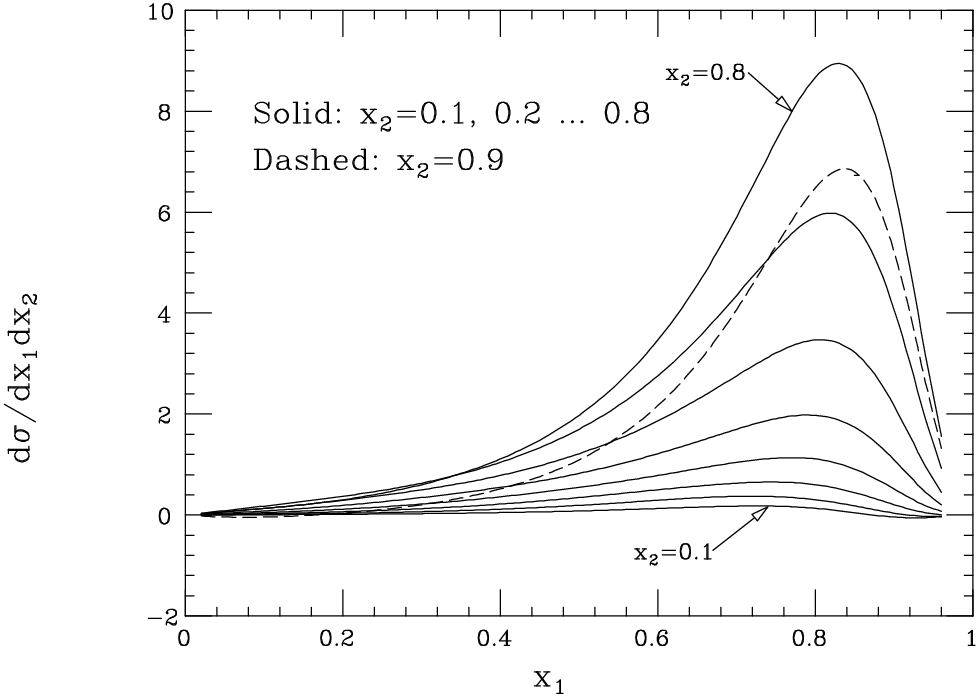,
width=0.7\textwidth,clip=}} \ccaption{}{ \label{sigdx1x2}
Double inclusive cross section $d \sigma/dx_1\,dx_2$, plotted as
a function of $x_1$ for several values of $x_2$. }
\end{figure}

From eqs.~(\ref{eq:doubleinc}), (\ref{eq:doubleinchat})
and~(\ref{eq:doubleplus}), we immediately derive the ${\cal O}\(\as\)$
formula
\beqn
\frac{d\sigma}{dx_1 d x_2}\!\!\! &=&\!\!\!
\Dh(x_1) \, \Dh(x_2)+\frac{2\as}{3\pi} \int_0^1 dy_1\,dy_2
\,\Theta(y_1+y_2-1)\,\frac{y_1^2+y_2^2}{(1-y_1)(1-y_2)} \nn \\
&&\!\!\!\times \Bigg[ \Dh\left(\frac{x_1}{y_1}\right)
\Dh\left(\frac{x_2}{y_2}\right) \frac{1}{y_1 \, y_2}
\,\Theta(y_1-x_1)\,\Theta(y_2-x_2) +\Dh(x_1)\, \Dh(x_2) \nonumber \\
&& \!\!\! - \Dh\left(\frac{x_1}{y_1}\right)\Dh(x_2)\,
\frac{1}{y_1}\,\Theta(y_1-x_1)
- \Dh(x_1)\, \Dh\left(\frac{x_2}{y_2}\right)\, \frac{1}{y_2}\,
\Theta(y_2-x_2)\Bigg]\;, \phantom{aaaaa}
\eeqn
where $\Theta$ is the Heaviside function.
As an illustration, we plot in Fig.~\ref{sigdx1x2} the double
inclusive cross section $d \sigma/dx_1\,dx_2$ as a function of $x_1$,
for several values of $x_2$.  We use the Peterson parametrization of
the fragmentation function
\beq
\Dh(z)=N^\prime_\ep \,\frac{z\,(1-z)^2}{\lq (1-z)^2+z\,\ep\rq^2} \;,
\eeq
where $N^\prime_\ep$ is fixed by the condition $\int dz\,\Dh(z) = 1$.
We took the values $\ep=0.04$, which gives $\VEV{x}=0.70$,
and $\as=0.12$.
%%%%%%%%%%%%%%%%%%%%%%%%%%%%%%%%%%%%%%%%%%%
%%%%%%%%%%%%%%%%%%%%%%%%%%%%%%%%%%%%%%%%%%%
The positive momentum correlation is quite visible in the figure.
As $x_2$ increases, the peak of the distribution in $x_1$ also
moves towards larger values.
Using the above formula, we can  compute $r$ again. The result
should not depend upon the choice of the fragmentation
function. In this case, the quantity $\VEV{x}$ does not receive
corrections at order $\as$, and we  get
\beq
\VEV{x}=0.7036\;,\quad\quad \VEV{x_1\, x_2}=0.4950+0.0525\times\as\;,
\quad\quad r=0.1061\times\as
\eeq
which is consistent with  the value of $r$ previously obtained (see
eq.~(\ref{eq:r_value})).

\section{Final results}
Assuming $\as(\Mz)=0.118$ (corresponding to the Particle Data Group
average~\cite{PDG}) and $m=5$~GeV, we have, at leading order,
$r=0.0984\times\as=0.0116$, and at next-to-leading order
$r=0.0984\times\as+0.213\times\as^2=0.0146$.
With the assumption of a rough geometric growth of the expansion, we
can give an estimate of the theoretical error due to higher orders:
$r=0.0146\pm 0.0007$.
In addition, we get $r^\prime=0.0064$ at leading order, and
$r^\prime=0.0083$ at next-to-leading order.

As a conclusion, we can say that, as far as its perturbative expansion
in powers of $\as$ is concerned, the average momentum correlation is a
quantity that is well behaved in perturbation theory, and it is also
quite small. Since its effect is typically of the order of 1\%, one
may worry that non-perturbative effects, of order $\Lambda/E$ (where
$\Lambda$ is a typical hadronic scale) may compete with the
perturbative result.  This is a very delicate problem, since we know
very little about the hadronization mechanism in QCD. In
Ref.~\cite{NasonWebber_PL}, this problem was addressed in the context of
the renormalon approach to power corrections. It was shown there that
corrections to the momentum correlation
are at least of order $(\Lambda/E)^2$, and thus negligible at LEP
energies. Although this result cannot be considered as a definitive
answer to the problem, it is at least an indication that power
corrections to this quantity are small.

\chapter{Fragmentation functions}
\label{chap:fragment}
\thispagestyle{plain}
In this chapter, we introduce the notion of fragmentation functions for
massive quarks.
Using our ${\cal O}\(\as^2\)$ calculation of the
differential cross section for the production of heavy quarks in
$e^+e^-$ annihilation, we  verify that the Leading (LL) and
Next-to-Leading Logarithmic (NLL) terms in this cross section are correctly
given by the standard NLO fragmentation-function formalism for
heavy-quark production, in the limit $m/E\,\to\,0$.

\section[Fragmentation functions for heavy quarks in $e^+e^-$ collisions]
{Fragmentation functions for heavy quarks in {\mylarge $e^+e^-$} collisions}
The inclusive heavy-quark production is a calculable process
in perturbative QCD, since the heavy-quark mass acts as a cut-off for the
final state collinear singularities.
Thus, the process
\beq
e^+ e^- \,\to\, Z/\gamma\,\to\, Q+ X\;,
\eeq
where $Q$ is the heavy quark and $X$ is anything else, is calculable.
Its cross section can be expressed as a power expansion in the strong
coupling constant
\beq
\label{eq:sigma}
\frac{d\sigma}{dx}(x,E,m)
= \sum_{n=0}^{\infty} a^{(n)}(x,E,m,\mu)\, \asb^n(\mu)\;,
\eeq
where $E$ is the centre-of-mass energy, $m$ is the mass of the heavy quark,
$\mu$ is the renormalization scale, and
\beq
\asb(\mu) =  \frac{\as(\mu)}{2\pi}\;.
\eeq
As usual we define
\beq
x = \frac{2\, p\cdot q}{q^2}\;,
\eeq
where $q$ and $p$ are the four-momenta of the intermediate virtual
boson and of the final heavy quark $Q$.
We define {\it the heavy-quark fragmentation function} in $e^+e^-$
annihilation as
\beq
\Dh(x,E,m) \equiv \frac{1}{\s_{\rm tot}}\, \frac{d\s}{dx}(x,E,m)\;.
\eeq
Since each heavy quark in the final state contributes to the
fragmentation function, its integral with respect to $x$ gives the
average multiplicity of these quarks
\beq
\label{eq:multiplicity}
\VEV{n(E,m)} = \int_0^1 dx\, \Dh(x,E,m)\;.
\eeq

When $E/m$ is not too large, the truncation of eq.~(\ref{eq:sigma}) at
some fixed order in the coupling $\asb$ can be used to compute the cross
section.  On the other hand, if $E\gg m$, since  the $n^{\rm th}$ order
coefficient of the expansion  has the form
\beq
a^{(n)} \sim \( \log\frac{E}{m} \)^n\;,
\eeq
we cannot truncate the series.
In fact, if
\beq
\asb \log\frac{E}{m} \approx 1\;,
\eeq
each term of the series~(\ref{eq:sigma}) is of the same order of
magnitude of the first one, so that we cannot trust a fixed order
calculation, that computes only a finite number of terms.
We must then try to resum these large logarithms. The resummation
procedure is described in Ref.~\cite{MeleNason}, and the resummed
differential cross section that is obtained with this procedure
has the form
\beqn
\label{eq:resummed_sigma}
\left. \frac{d\sigma}{dx}(x,E,m) \right|_{\rm res} \!&=&\!
\sum_{n=0}^{\infty} \beta^{(n)}(x) \,\( \asb(E) \log\frac{E}{m}
\)^n + \asb(E)\sum_{n=0}^{\infty} \gamma^{(n)}(x) \,\( \asb(E)
\log\frac{E}{m}\)^n
\nn \\
&& {} + \asb^2 (E)\sum_{n=0}^{\infty} \delta^{(n)}(x) \,
\( \asb(E) \log\frac{E}{m}\)^n +\ldots + {\cal O}\(\frac{m^2}{E^2}\)\;,
\eeqn
where terms that are suppressed by powers of $m^2/E^2$ are not
included, because they can be neglected if compared with the large
$\log\(E/m\)$, and where we have taken $\mu=E$, for simplicity of
notation.  If we compute, with the resummation procedure, the first
series of eq.~\ref{eq:resummed_sigma}, we talk of Leading-Order (LO)
resummed cross section; if we include the next series, we refer to it
as Next-to-Leading-Order (NLO) resummed cross section, and so on.

In the approximation when you can neglect powers of $m^2/E^2$, it can
be observed that the inclusive heavy-quark cross section must satisfy
the factorization theorem formula
\beqn
\label{eq:factorization}
\frac{d\s}{dx}(x,E,m) &=&
\sum_i \int_0^1 dy\,dz\, \delta(x-y\,z)\,
\frac{d\hat\s_i}{dz}\(z,E,\mu\) \,
\Dh_i\(y,\mu,m\) \nn\\
&=& \sum_i \int_x^1 \frac{dz}{z}\,
\frac{d\hat\s_i}{dz}\(z,E,\mu\) \,
\Dh_i\(\frac{x}{z},\mu,m\)\;,
\eeqn
where $d\hat\s_i(z,E,\mu)/dz$ are the \MSB-subtracted partonic
cross sections for producing the parton $i$, and $\Dh_i(y,\mu,m)$
are the \MSB\ fragmentation functions for the parton $i$ into the
heavy quark $Q$.  In order for eq.~(\ref{eq:factorization}) to hold,
it is essential that you use a renormalization scheme where the heavy
flavour is treated as a light one, like the pure \MSB\ scheme. Thus
$d\hat\s_i(z,E,\mu)/dz$ has a perturbative expansion in terms of $\asb$
with $n_f$ flavours, where $n_f$ includes the heavy one.

In order to compute the differential short-distance cross section that
describes the process $Z/\gamma\,\to\, i + X$, we must introduce a
regulator, because we now deal with a massless quark (no mass
dependence in the expression of $d\hat{\s}_i$), so that we have to face
the presence of collinear singularities.
Choosing the dimensional regularization ($d=4-2\e$), the \MSB\
prescription amounts  to throw away all the terms that contain
poles in $\e$ in the expression of $d\hat{\s}_i$.
After that, the form of $\Dh_i$ is fixed, because its
convolution with  $d\hat{\s}_i$ must give the physical, finite,
differential cross section $d\s$.

With the factorization theorem, we succeed in separating the two scales
of energy $E$ and $m$ into two different terms, through the
introduction of a factorization scale $\mu$, that we take equal to the
renormalization scale, for simplicity of notation.
The scale $\mu$ should be chosen of the order of $E$, in order to
avoid the appearance of large logarithms of $E/\mu$ in the partonic
cross section.

The \MSB\ fragmentation functions $\Dh_i$ obey the
Altarelli-Parisi evolution equations
\beq
\label{eq:AP}
\frac{d \hat D_i}{d\log\mu^2} (x,\mu,m) = \sum_j\int^1_x \frac{dz}{z}\,
P_{ij}\(\frac{x}{z},\asb(\mu)\) \;\Dh_j(z,\mu,m)\;,
\eeq
that resum correctly all the large logarithms.

The Altarelli-Parisi splitting functions $P_{ij}$ have the perturbative
expansion
\beq
\label{eq:APfunctions}
P_{ij}\Bigl(x,\asb(\mu)\Bigr) = \asb(\mu) P^{(0)}_{ij}(x)
+ \asb^2(\mu) P^{(1)}_{ij}(x) + {\cal O}\(\asb^3\)\;,
\eeq
where $P_{ij}^{(0)}$ are given in Ref.~\cite{AltarelliParisi}
and $P_{ij}^{(1)}$ have been computed in
Refs.~\cite{CurciFurmanskiPetronzio}-\cite{Konishi}.
The only missing ingredients for the calculation of the inclusive cross
section are the initial conditions for the \MSB\ fragmentation functions.
These were obtained at NLO level in Ref.~\cite{MeleNason} by matching
the ${\cal O}\(\asb\)$ direct calculation of the process (i.e.
formula~(\ref{eq:sigma})) with the expansion of
eq.~(\ref{eq:factorization}) at order $\asb$. They have the form
\beq
\label{eq:ini}
\eqalign{
\Dh_Q(x,\mu_0,m) = \delta(1-x)+\asb(\mu_0)\,
d^{(1)}_Q(x,\mu_0,m)+{\cal O}\(\asb^2\)
\cr
\Dh_g(x,\mu_0,m) = \asb(\mu_0)\, d^{(1)}_g(x,\mu_0,m) +
{\cal O}\(\asb^2\)\;, }
\eeq
all the other components being of order $\asb^2$.
Thus, to compute the NLO resummed expansion, one takes the
initial conditions~(\ref{eq:ini}), at a value of $\mu_0$ of order
$m$ (so that no large logarithms appear),
evolves them at the scale $\mu$ (taken to be of order $E$), and then
applies formula~(\ref{eq:factorization}), using a NLO expression for
the partonic cross section
\beq\label{eq:sigmahat}
\frac{d\hat\s_i}{dx}(x,E,\mu) =\hat{a}^{(0)}_i(x) +
\hat{a}^{(1)}_i(x,E,\mu) \,\asb(\mu) + {\cal O}\(\asb^2\)\;.
\eeq
For example, if the parton $i$ is the heavy quark itself, one gets
\beq
\label{eq:sigmahatQ}
\frac{d\hat\s_Q}{dx}(x,E,\mu) =   \delta(1-x) + \hat{a}_Q^{(1)}(x,E,\mu)
\,\asb(\mu) + {\cal O}\(\asb^2\) \;,
\eeq
and if  the parton $i$ is a gluon
\beq
\frac{d\hat\s_g}{dx}(x,E,\mu) =    \hat{a}_g^{(1)}(x,E,\mu)
\,\asb(\mu) + {\cal O}\(\asb^2\) \;,
\eeq
where we have  normalized the cross section to one at zeroth order in
the strong coupling constant.

The procedure outlined above guarantees that all terms of the form
$(\asb L)^n$ (leading order) and $\asb (\asb L)^n$ (next-to-leading order),
where $L$ is the large logarithm,
are included correctly in the resummed formula.
Notice that, at NLO level, the scale that appears in $\asb$, in
eqs.~(\ref{eq:ini}) and~(\ref{eq:sigmahat}), could be changed by factors
of order 1, since this amounts to a correction of order $\asb^2$.
However, one cannot set $\mu_0=\mu$ in eqs.~(\ref{eq:ini})
(or $\mu=m$ in formula~(\ref{eq:sigmahat})), since this amounts to
a correction of order $\asb^2 L$, and thus it would spoil the
validity of the resummation formula at NLO level.

The validity of this procedure has however been questioned by the
authors of Ref.~\cite{Spira}.  In their procedure, the heavy-quark
short-distance cross section is replaced by
\beq
\label{eq:sigmahatprime}
\frac{d\hat\s^\prime_Q}{dx}(x,E,\mu) = \delta(1-x) +
\asb(E)\,\left[ \hat{a}^{(1)}_Q(x,E,\mu) + d^{(1)}_Q(x,\mu_0,m) \right]
+ {\cal O}\(\asb^2\)\;,
\eeq
and the initial condition by
\beq\label{eq:iniprime}
\hat{D}^\prime_Q(x)=\delta(1-x)\;,
\eeq
which is to be evolved from the scale $\mu_0$ to the scale $\mu$
using the NLO \MSB\ evolution equations. This procedure differs
at the NLO level from the standard procedure advocated in
Ref.~\cite{MeleNason}. The difference starts to show up in the
terms of order $\asb^2 L$.

We have used our calculation of the differential cross section at
fixed order $\asb^2$ to check the standard formalism and thereby to
dismiss the approach of Ref.~\cite{Spira}.
By the way, the full consistency of the two results gives support to
the correctness of our computation.

\section{Calculation}
Instead of dealing with the realistic case of $Z/\gamma$ decay,
we perform the calculation for a hypothetical vector boson $V$ that
couples only to the heavy quark with vectorial coupling.

We introduce the following notation for the Mellin transform of a
generic function $f(x)$:
\beq
f(N) \equiv \int_0^1 dx \, x^{N-1} f(x)\;.
\eeq
We adopt the convention that, when $N$ appears, instead of $x$, as the
argument of a function, we are actually referring to the Mellin transform
of the function. This notation is somewhat improper, but it should
not generate confusion in the following, since we will work
only with Mellin transforms.

The Mellin transform of the factorization
theorem~(\ref{eq:factorization}) is given by
\beq\label{eq:sigmaN}
\sigma(N,E,m) =\sum_i \hat\sigma_i(N,E,\mu) \; \Dh_i(N,\mu,m)\;,
\eeq
where
\beq
\sigma(N,E,m) = \int_0^1 dx \, x^{N-1} \frac{d\s}{dx}(x,E,m)\;,
\eeq
and a similar one for $\hat{\s}_i(N,E,\mu)$;  the Mellin transform
of the Altarelli-Parisi evolution equations~(\ref{eq:AP}) is
\beq
\label{eq:APN}
\frac{d\Dh_i(N,\mu,m)}{d\log\mu^2} = \sum_j
\asb(\mu) \left[ P_{ij}^{(0)}(N) + P_{ij}^{(1)}(N)\,
\asb(\mu) +   {\cal O}\(\asb^2\)\right] \Dh_j(N,\mu,m)\;.
\eeq
In order to make a comparison with our fixed order calculation,
we need  an expression for $\sigma(N,E,m)$ valid at the second
order in $\asb$.
Thus, we solve eq.~(\ref{eq:APN}), with initial condition
at $\mu=\mu_0$, accurate at order $\asb^2$. This is easily done by rewriting
eq.~(\ref{eq:APN}) as an integral equation
\beqn
\Dh_i(N,\mu,m)\!\!\! &=& \!\!\!\Dh_i(N,\mu_0,m) \nn \\
\label{eq:integAPN0}
\!\!\! &+& \!\!\! \sum_j \int_{\mu_0}^\mu d\log{\mu^\prime}^2\;
\asb(\mu^\prime) \left[ P_{ij}^{(0)}(N) + P_{ij}^{(1)}(N)\,
\asb(\mu^\prime) \right]  \Dh_j(N,\mu^\prime,m)
\;. \phantom{aaaaa}
\eeqn
The terms proportional to $\asb^2$ can be evaluated at any scale
($\mu$ or $\mu_0$),
the difference being of order $\asb^3$. Factors involving a single
power of $\asb$ can instead be expressed in terms of $\asb(\mu_0)$
using the renormalization group equation
\beq
\eqalign{
\label{eq:RGE}
\asb(\mu^\prime) = \asb(\mu_0)-2\,\pi\,b_0\,\asb^2(\mu_0)\,
\log\frac{{\mu^\prime}^2}{\mu_0^2}+{\cal O}\(\asb^3\(\mu_0\)\)
\cr
b_0 = \frac{11\,C_A - 4\, n_f\,T_F}{12\,\pi}\;,}
\eeq
with $n_f$ the number of flavours, including the heavy one.
Equation~(\ref{eq:integAPN0}) then becomes
\beqn
\Dh_i(N,\mu,m) &=& \Dh_i(N,\mu_0,m)
+ \sum_j \int_{\mu_0}^\mu d\log{\mu^\prime}^2\;
\asb(\mu_0) P_{ij}^{(0)}(N) \,
\Dh_j(N,\mu^\prime,m)
\nonumber \\
&&{}+ \sum_j  \asb^2(\mu_0)\, P_{ij}^{(1)}(N) \,
\Dh_j(N,\mu_0,m)\,\log\frac{\mu^2}{\mu_0^2}  \nonumber \\
\label{eq:integAPN1}
&& {} -  2\,\pi\,b_0\,\sum_j
\asb^2(\mu_0)\, P_{ij}^{(0)}(N)\,
\Dh_j(N,\mu_0,m)\,\frac{1}{2}\log^2\frac{\mu^2}{\mu_0^2}  \;.
\eeqn
Now, we  need to express $\Dh_j(N,\mu^\prime,m)$ on the right-hand side
of the above equation as a function of the initial condition,
with an accuracy of order $\asb$. This is simply done by iterating
the above equation once, keeping only the first two terms on the
right-hand side. Our final result is then
\beqn
\Dh_i(N,\mu,m) &=& \Dh_i(N,\mu_0,m)
+ \sum_j
\asb(\mu_0) P_{ij}^{(0)}(N) \,
\Dh_j(N,\mu_0,m) \, \log\frac{\mu^2}{\mu_0^2}
\nonumber \\
&&{} +  \sum_{kj}
\asb^2(\mu_0) P_{ik}^{(0)}(N) \, P_{kj}^{(0)}(N)\,
\Dh_j(N,\mu_0,m) \, \frac{1}{2}\log^2\frac{\mu^2}{\mu_0^2}
\nonumber \\
&&{} +  \sum_j \asb^2(\mu_0) P_{ij}^{(1)}(N) \,
\Dh_j(N,\mu_0,m) \, \log\frac{\mu^2}{\mu_0^2}
\nonumber \\ \label{eq:integAPN2}
&&{} -  \pi\,b_0\,\sum_j
\asb^2(\mu_0) P_{ij}^{(0)}(N) \,
\Dh_j(N,\mu_0,m) \, \log^2\frac{\mu^2}{\mu_0^2} \;.
\eeqn
Since the initial condition is
\beq
\Dh_j(N,\mu_0,m)=\delta_{jQ}+\asb(\mu_0)\,d^{(1)}_j(N,\mu_0,m)
+{\cal O}\(\asb^2\(\mu_0\)\)\;,
\eeq
eq.~(\ref{eq:integAPN2}) becomes, with the required accuracy,
\beqn
\Dh_i(N,\mu,m) &=& \delta_{iQ}+\asb(\mu_0)\,d^{(1)}_i(N,\mu_0,m)
+  \asb(\mu_0) P_{iQ}^{(0)}(N)
\, \log\frac{\mu^2}{\mu_0^2}    \nonumber \\
&& {} +  \sum_j \asb^2(\mu_0) P_{ij}^{(0)}(N) \,
d^{(1)}_j(N,\mu_0,m)\, \log\frac{\mu^2}{\mu_0^2}
\nonumber \\
&&{} +  \sum_{k} \asb^2(\mu_0) P_{ik}^{(0)}(N)\, P_{kQ}^{(0)}(N)
\, \frac{1}{2}\log^2\frac{\mu^2}{\mu_0^2}
\nonumber \\
\label{eq:integAPN3}
&& {} +  \asb^2(\mu_0) P_{iQ}^{(1)}(N)\, \log\frac{\mu^2}{\mu_0^2}
- \pi\,b_0\,
\asb^2(\mu_0) P_{iQ}^{(0)}(N)
\,\log^2\frac{\mu^2}{\mu_0^2} \;.\phantom{aaaaa}
\eeqn
Re-expressing $\asb\(\mu_0\)$ in terms of $\asb\(\mu\)$, using
eqs.~(\ref{eq:RGE}), we get
\beqn
\Dh_i(N,\mu,m) &=& \delta_{iQ}+\asb(\mu)\,d^{(1)}_i(N,\mu_0,m)
+ 2\, \pi\,b_0\,\asb^2(\mu)\,d^{(1)}_i(N,\mu_0,m)\,\log\frac{\mu^2}{\mu_0^2}
\nonumber \\
&&{} + \asb(\mu) P_{iQ}^{(0)}(N) \, \log\frac{\mu^2}{\mu_0^2}
+ \sum_j \asb^2(\mu) P_{ij}^{(0)}(N) \,
d^{(1)}_j(N,\mu_0,m) \, \log\frac{\mu^2}{\mu_0^2}
\nonumber \\
&& {} +  \sum_{k} \asb^2(\mu) P_{ik}^{(0)}(N) \, P_{kQ}^{(0)}(N)
\, \frac{1}{2}\log^2\frac{\mu^2}{\mu_0^2}
+ \asb^2(\mu) P_{iQ}^{(1)}(N)\, \log\frac{\mu^2}{\mu_0^2}
\nonumber \\
\label{eq:integAPN}
&&{} +  \pi\,b_0\, \asb^2(\mu) P_{iQ}^{(0)}(N)
\, \log^2\frac{\mu^2}{\mu_0^2} \;.
\eeqn
The partonic cross sections are given by
\beq\label{eq:partN}
\hat\sigma_i(N,E,\mu)=\delta_{iQ}+\delta_{i\Qb}+
\asb(\mu)\,\hat{a}^{(1)}_i(N,E,\mu)
+{\cal O}\(\asb^2(\mu)\)\;,
\eeq
where $\hat{a}^{(1)}_i$ vanishes unless $i$ is either $Q$, $\overline{Q}$
or $g$.
Thus, combining eq.~(\ref{eq:partN}) with eq.~(\ref{eq:integAPN}),
according to eq.~(\ref{eq:sigmaN}), we obtain
\beqn
\sigma(N,E,m) &=& 1+
\asb(\mu)\left[ \hat{a}^{(1)}_Q(N,E,\mu)+d^{(1)}_Q(N,\mu_0,m)
+P_{QQ}^{(0)}(N)\, \log\frac{\mu^2}{\mu_0^2} \right]
\nonumber \\
&&{}+ \asb^2(\mu)\Bigg\{\sum_i \hat{a}^{(1)}_i(N,E,\mu) P^{(0)}_{iQ}\,
\log\frac{\mu^2}{\mu_0^2}
+ 2\,\pi\,b_0\,d^{(1)}_Q(N,\mu_0,m)\, \log\frac{\mu^2}{\mu_0^2}
\nonumber \\
&&{} + \sum_j \left[ P_{Qj}^{(0)}(N) +  P_{\Qb j}^{(0)}(N)\right]
d^{(1)}_j(N,\mu_0,m)\, \log\frac{\mu^2}{\mu_0^2}
\nonumber \\
&& {} + \sum_{k} \left[ P_{Qk}^{(0)}(N) + P_{\Qb k}^{(0)}(N)\right]
P_{kQ}^{(0)}(N) \, \frac{1}{2}\log^2\frac{\mu^2}{\mu_0^2}
\nonumber \\
\label{eq:sigmaNall}
&& {} +  \left[ P_{QQ}^{(1)}(N)+ P_{\Qb Q}^{(1)}(N)\right]
\log\frac{\mu^2}{\mu_0^2} + \pi\,b_0\, P_{QQ}^{(0)}(N)
\, \log^2\frac{\mu^2}{\mu_0^2} \Bigg\} \;.
\eeqn
The above formula should accurately describe the terms of order
$\asb L$, $\asb$, $\asb^2 L$ and  $\asb^2 L^2$. Terms of order
$\asb^2$, without logarithmic enhancement, are not accurately given
by the fragmentation formalism at NLO level, and have consistently been
neglected.

The lowest order splitting functions are given by
\beqn
P^{(0)}_{QQ}(N) &=&
C_F\left[\frac{3}{2}+\frac{1}{N(N+1)}-2\,S_1(N)\right]\,,
\nonumber \\
P^{(0)}_{Qg}(N) &=& P^{(0)}_{\Qb g}(N) =
C_F\left[\frac{2+N+N^2}{N(N^2-1)}\right]\,,
\nonumber \\
P^{(0)}_{gQ}(N) &=& T_F\left[ \frac{2+N+N^2}{N(N+1)(N+2)}\right]\,,
\eeqn
where, restricting ourselves to integer values of $N$,
\beq
S_1(N)=\sum_{j=1}^N\frac{1}{j}\;.
\eeq
The splitting functions $P_{QQ}^{(1)}(N)$ and $P_{\Qb Q}^{(1)}(N)$ are
given by
\beq\label{QQandQBARQ}
P_{QQ}^{(1)}(N) = P^{\rm NS}_{QQ}(N) + P_{q^\prime q}(N)\;,\quad\quad\quad
P_{\Qb Q}^{(1)}(N) = P^{\rm NS}_{\Qb Q}(N) + P_{q^\prime q}(N)\;,
\eeq
where the non-singlet components are
\beq
\begin{array}{ll}
P_{QQ}^{\rm NS}(N) = P^{C_F}_{QQ}(N)+ P^{C_A}_{QQ}(N)+P^{n_f}_{QQ}(N)
\quad\quad
&  P^{C_F}_{QQ}(N)=C_F^2\left[P_F(N)+\Delta(N)\right]  \\[.3cm]
&  P^{C_A}_{QQ}(N)=\frac{1}{2} C_F C_AP_G(N) \\[.3cm]
&  P^{n_f}_{QQ}(N)=n_f C_F T_F P_{NF}(N) \\[.5cm]
P_{\Qb Q}^{\rm NS}(N) = P^{C_F}_{\Qb Q}(N)+ P^{C_A}_{\Qb Q}(N)
\quad\quad
&  P^{C_F}_{\Qb Q}(N)=C_F^2 P_A(N)  \\[.3cm]
&  P^{C_A}_{\Qb Q}(N)=-\frac{1}{2} C_F C_A P_A(N)\;,
\end{array}
\eeq
and
\beq
P_{q^\prime q}(N)
= -C_F T_F \frac{8 + 44 N + 46 N^2 + 21 N^3 + 14 N^4 + 15 N^5 +
10 N^6 + 2 N^7}{N^3 (N+1)^3(N+2)^2 (N-1)}\;.
\eeq
$P_F(N)$, $\Delta(N)$, $P_G(N)$ and $P_{NF}(N)$ were taken
from the appendix of Ref.~\cite{MeleNason} and $P_A(N)$ is given
in eq.~(5.39) of Ref.~\cite{CurciFurmanskiPetronzio}.
We have obtained our explicit expression for $P_{q^\prime q}(N)$
using the relation
\beq \label{pqpq}
P_{q^\prime q}=\frac{P^{\rm S}_{QQ}-P^{\rm NS}_{QQ}
-P^{\rm NS}_{\Qb Q}}{2\,n_f}\;,
\eeq
where $P^{\rm S}_{QQ}$ is the singlet component\footnote{We warn the reader
that, sometimes, in the literature, the notation $P^{\rm S}$ is used for
the ``sea'' component, and $P_{QQ}$ is used for the singlet one.
Here we use $P_{QQ}$ for the full $QQ$ splitting function.}, calculated
in Ref.~\cite{FurmanskiPetronzio}.
Equation~(\ref{pqpq}) is easily seen to follow from
eqs.~(\ref{QQandQBARQ}) and from eqs.~(2.42) of Ref.~\cite{NasonWebber}.

The expressions for $\hat{a}_{Q}^{(1)}$ and $d^{(1)}_{Q}$ are
respectively given in eqs.~(A.12) and~(A.13) of Ref.~\cite{MeleNason}.
The coefficient $\hat{a}_{g}^{(1)}$ can be obtained by performing the
Mellin transform of the expression $c_{\rm T,g}+c_{\rm L,g}$, where
$c_{\rm T,g}$ and $c_{\rm L,g}$ are given in eq.~(2.16) of
Ref.~\cite{NasonWebber}. Thus
\beqn
\hat{a}_{g}^{(1)}(N,E,\mu) &=& C_F \Biggl\{
\frac{2(2+N+N^2)}{N(N^2-1)} \log\frac{E^2}{\mu^2}
+ 4 \left[ -\frac{2}{(N-1)^2} +
\frac{2}{N^2} -\frac{1}{(N+1)^2} \right]
\nonumber\\
&&{} - 2 \left[ \frac{2}{N-1} S_1(N-1) -\frac{2}{N} S_1(N) +
\frac{1}{N+1} S_1(N+1)  \right]
\Biggr\}
\nonumber \\
d^{(1)}_{g}(N,\mu_0,m)&=&P^{(0)}_{gQ}(N)\log\frac{\mu_0^2}{m^2}\;.
\eeqn
In order to make a more detailed comparison with our fixed order
calculation, we separate the ${\cal O}\(\asb^2\)$ contributions to
$\sigma(N,E,m)$ according to their colour factors. Choosing for
simplicity $\mu=E$ and $\mu_0=m$, and using the notation
\beq
\hat{a}^{(1)}_{Q/g}(N)=\hat{a}^{(1)}_{Q/g}(N,E,\mu)\vert_{\mu=E}\;,
\quad\quad\quad
d^{(1)}_{Q/g}(N)=d^{(1)}_{Q/g}(N,\mu_0,m)\vert_{\mu_0=m}\;,
\eeq
we write
\newcommand\tsq{\log^2\frac{E^2}{m^2}}%{L^2}
\renewcommand\t{\log\frac{E^2}{m^2}}%{L}
\beq
\sigma(N,E,m) = 1+ \asb(E)\, A(N,E,m) + \asb^2(E)\, B(N,E,m)\;,
\eeq
with
\beqn
A(N,E,m) &=&  \hat{a}^{(1)}_Q(N)+d^{(1)}_Q(N) + P_{QQ}^{(0)}(N) \t
\nn\\[-3mm]
&& \label{eq:A_B_coef} \\[-3mm]
B(N,E,m) &=&  B_{C_F}(N,E,m)+B_{C_A}(N,E,m)
+B_{n_f}(N,E,m)+B_{T_F}(N,E,m) \;, \nn
\eeqn
and
\beqn
B_{C_F}(N,E,m) &=& \left\{ P^{(0)}_{QQ}(N) \left[ d^{(1)}_{Q}(N) +
\hat{a}^{(1)}_{Q}(N)\right] +
P^{C_F}_{QQ}(N)+ P^{C_F}_{\Qb Q}(N) \right\} \t
\nonumber \\ &&{}+ \frac{1}{2} \left[P^{(0)}_{QQ}(N) \right]^2 \tsq
\nonumber \\
B_{C_A}(N,E,m)&=&\left[P^{C_A}_{QQ}(N) + P^{C_A}_{\Qb Q}(N)+
\frac{11}{6}C_A\,
d^{(1)}_{Q}(N)\right] \t
\nonumber \\ &&{}+ \frac{11}{12} C_A \, P^{(0)}_{QQ}(N) \tsq
\nonumber \\
B_{n_f}(N,E,m)&=&\left[ P^{n_f}_{QQ}(N) -\frac{2}{3} n_f T_F \,
d^{(1)}_{Q}(N) \right] \t
\nonumber \\      &&{}-\frac{1}{3}n_f T_F P^{(0)}_{QQ}(N) \tsq
\nonumber \\
B_{T_F}(N,E,m)&=&\left[\hat{a}_g^{(1)}(N)\, P_{gQ}^{(0)}
+2\,P_{Qg}^{(0)}(N)\,d^{(1)}_g(N) + 2\, P_{q^\prime q}(N)\right]\t
\nonumber \\ &&{} +P_{Qg}^{(0)}(N)\, P_{gQ}^{(0)}(N)\tsq\;,
\label{eq:colcoef}
\eeqn
where the subscripts $C_F$, $C_A$, $n_f$ and $T_F$ denote the
$C_F^2$, $C_F C_A$, $n_f C_F T_F\,$ and $C_F T_F$ colour components.

\begin{figure}[htb]
\centerline{\epsfig{figure=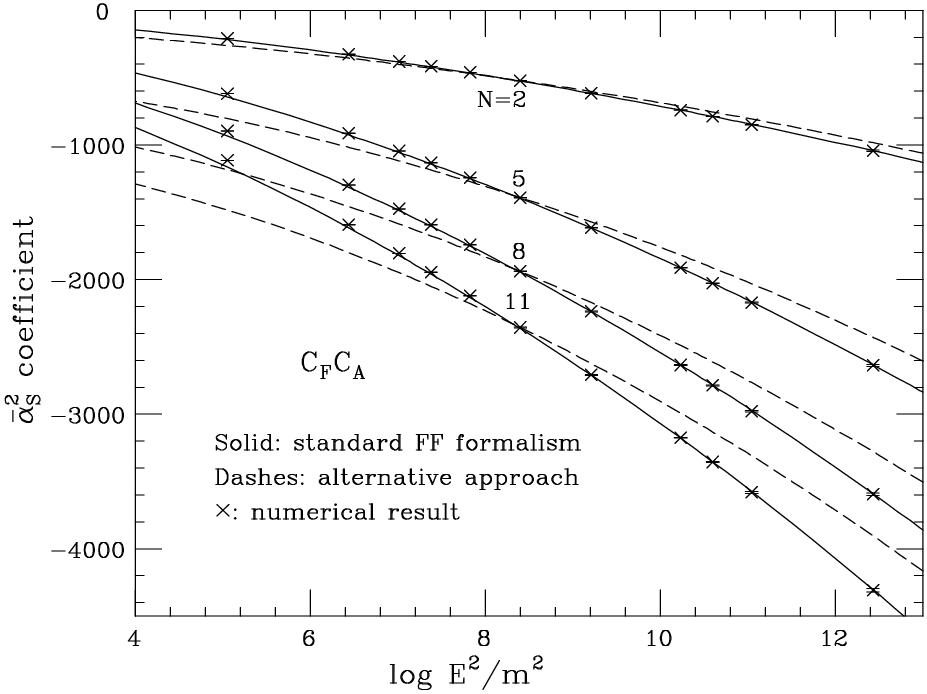,width=0.7
\textwidth,clip=}}
\ccaption{}{ \label{fig:ca}
$C_FC_A$ component of the $\asb^2$ coefficient in $\sigma(N,E,m)$,
as a function of $\log E^2/m^2$, for $N=$2, 5, 8 and 11.
The dashed lines correspond to the alternative approach proposed by
the authors of Ref.~\cite{Spira}.}
\end{figure}

\begin{figure}[htb]
\centerline{\epsfig{figure=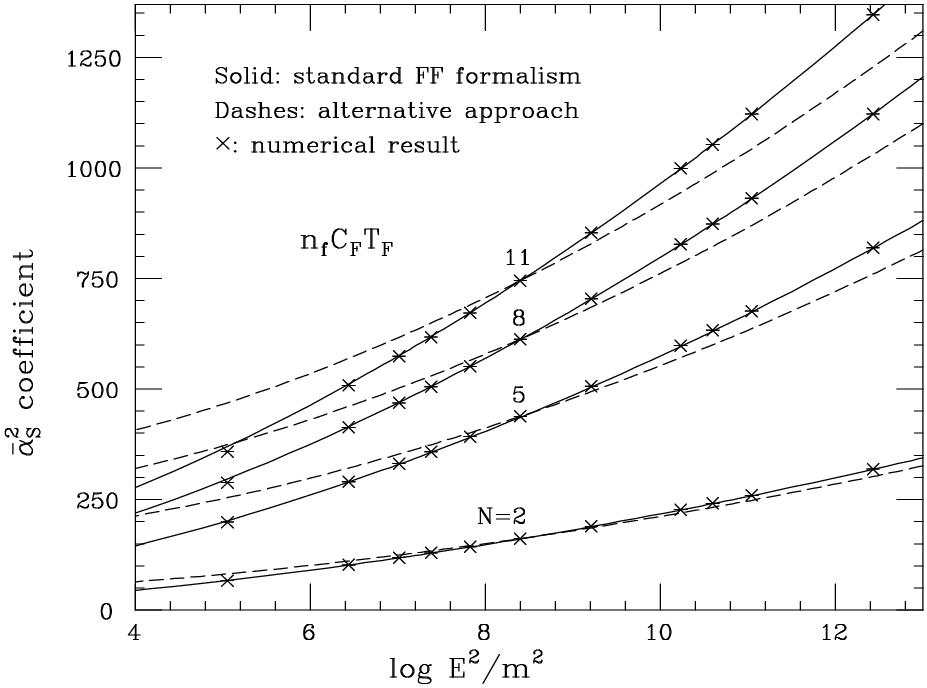,
width=0.7\textwidth,clip=}}
\ccaption{}{ \label{fig:nf}
Same as in Fig.~\protect{\ref{fig:ca}}, for the $ n_f C_F T_F$ component.}
\end{figure}

\begin{figure}[htb]
\centerline{\epsfig{figure=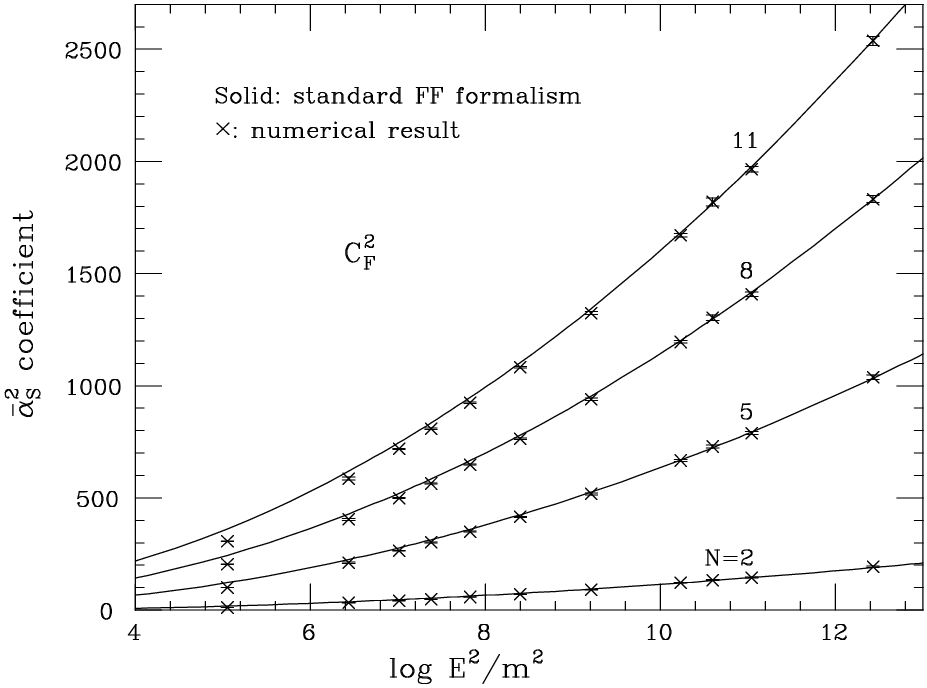,
width=0.7\textwidth,clip=}}
\ccaption{}{ \label{fig:cf}
Same as in Fig.~\protect{\ref{fig:ca}}, for the $C_F^2$ component.}
\end{figure}

\begin{figure}[htb]
\centerline{\epsfig{figure=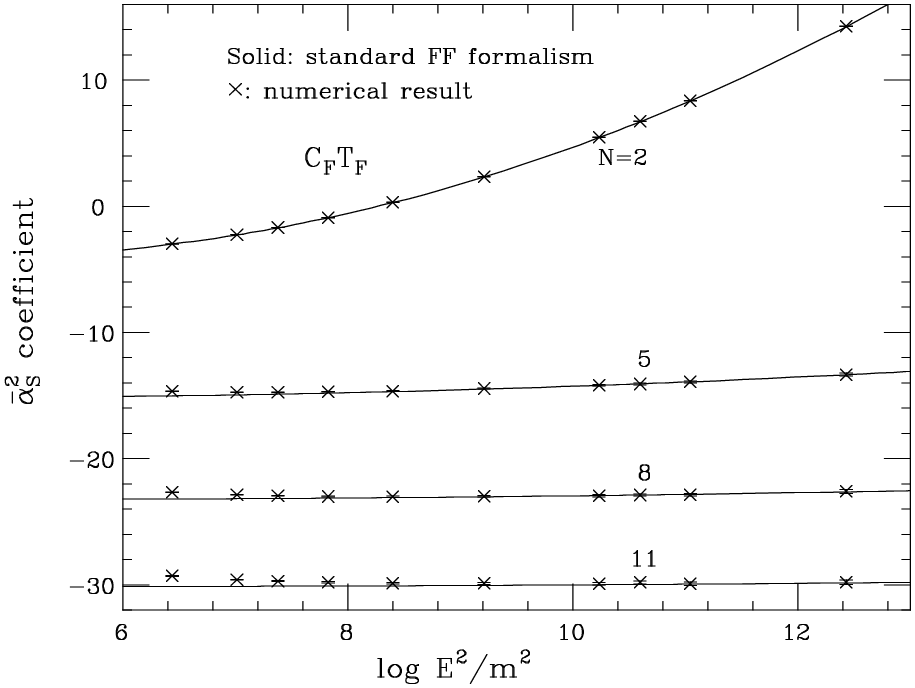,
width=0.7\textwidth,clip=}}
\ccaption{}{ \label{fig:tf}
Same as in Fig.~\protect{\ref{fig:ca}}, for the $C_F T_F$ component.}
\end{figure}

Our fixed order calculation can be used to compute the cross section
for the production of a heavy-quark pair plus one or two more partons,
at order $\asb^2$.  We separate contributions in which four heavy
quarks are present in the final state, from those where a single
$Q\Qb$ pair is present together with one or two light partons.  These
last contributions are computed only in a three-jet configuration,
and they are singular in the two-jet limit, that is to say, when $x\to
1$.  Furthermore, the virtual corrections to the two-body process
$V\,\to\, Q+\Qb$ are not included in our calculation.  In order to
remedy for these problems, we proceed in the following way (see
Sec.~\ref{sec:two_body} for a detailed description of the method).
The ${\cal O}\(\asb^2\)$ inclusive cross section for $V\,\to\, Q+\Qb+X$,
can be written, symbolically, in the following form
\beq
\frac{d\s}{dx} =
a^{(0)}\delta(1-x) +  \asb\int dY \,a^{(1)}(x,Y)
+  \asb^2 \left[\int dY\, a^{(2)}_l(x,Y)\,
+ 2 \int dY\, a^{(2)}_h(x,Y) \right]
\eeq
where $Y$ denotes all the other kinematical variables, besides $x$,
upon which the final state may depend (see Chapter~\ref{chap:kinematics}).
We assume $\mu=E$, and we do
not indicate, for ease of notation, the dependence upon $E$ and $m$ of
the various quantities. The term $a^{(2)}_l$ arises from final states
with a single $Q\Qb$ pair plus at most two light partons, while
$a^{(2)}_h$ arises from final states with two $Q\Qb$ pairs. The factor
of 2 in front of the $a^{(2)}_h$ contribution takes account for the fact
that we may detect either one of the two heavy quarks, as is
illustrated in eq.~(\ref{eq:multiplicity}).
The moments of the inclusive cross section can be written in the
following way
\beqn
\s(N)\!\!\! &\equiv& \!\!\! \int dx\, x^{N-1}\,\frac{d\s}{dx} \; =\;
\sigma\, + \, \asb\int dx\,dY \,\(x^{N-1}-1\)\,a^{(1)}(x,Y)
\nn \\
&&\!\!\! {}+\asb^2 \left[\int dx\,dY \,\(x^{N-1}-1\)\,a^{(2)}_l(x,Y)\,
+ 2\int dx\,dY \,\(x^{N-1}-  \frac{1}{2}\)\,
a^{(2)}_h(x,Y)\right]\nn\\[-2mm] &&\label{eq:sig_N}
\eeqn
where
\beq
\sigma =    a^{(0)} +  \asb\int dx\,dY \,a^{(1)}(x,Y)
+  \asb^2 \left[\int dx\, dY\, a^{(2)}_l(x,Y)\,
+\int dx\, dY\, a^{(2)}_h(x,Y) \right]\;.
\eeq
The expression for $\s(N)$ can now be easily computed with our
program, since the \mbox{$\(x^{N-1}-1\)$} factors regularize the
singularities in the two-jet limit, and suppress the two-body
$V\,\to\, Q+\Qb$ virtual terms.  Furthermore, in the massless limit,
\beq
\label{eq:massless_Xsec}
\sigma = 1 + 2\,\asb(E) + c\,\asb^2(E)\,
+ {\cal O}\left(\frac{m^2}{E^2}\right) + {\cal O}\(\asb^3\) \;,
\eeq
where $c$ is a constant. In fact, the ${\cal O}\(\asb^2\)$ term does not
contain any large logarithm, as long as $\asb$ is the coupling with
$n_f$ flavours, including the heavy one.  If, instead, the cross
section formulae are expressed in terms of $\asb^{(n_f-1)}$, we have to
take account of the different number of flavours.
From the renormalization group equation~(\ref{eq:RGE}), that we
rewrite remarking the number $n_f$ of flavours,
\beq
\asb^{(n_f)}(\mu) = \asb^{(n_f)}(\mu_0)-\frac{11\,C_A - 4\,
n_f\,T_F}{6} \,\asb^2(\mu_0)\,
\log\frac{{\mu}^2}{\mu_0^2}+{\cal O}\(\asb^3\(\mu_0\)\) \;,
\eeq
and the matching condition
\beq
\asb^{(n_f)}(\mu_0) = \asb^{(n_f-1)}(\mu_0) \;,
\eeq
we derive
\beq
\asb^{(n_f)}(\mu) - \asb^{(n_f-1)}(\mu) = \frac{2}{3}\, T_F \,\asb^2\,
\log\frac{{\mu}^2}{\mu_0^2}+{\cal O}\(\asb^3\(\mu_0\)\) \;.
\eeq
In this way, the total  cross section of
eq.~(\ref{eq:massless_Xsec}) becomes
\beq
\sigma = 1 + 2\,\asb^{(n_f-1)}(E) + \frac{4}{3} \, T_F \, \asb^2(E) \,
\log\frac{E^2}{m^2} + {\cal O}\(\asb^2\) \;.
\eeq
Reintroducing the energy and mass dependence in eq.~(\ref{eq:sig_N})
and using eq.~(\ref{eq:massless_Xsec}), we have
\beqn
\s(N,E,m) &=& 1 +
\asb(E)\, L(N,E,m)+\asb^2(E)\, M(N,E,m)  +
c\,\asb^2(E)  \nonumber \\
&& {}+ {\cal O}\left(\frac{m^2}{E^2}\right) +{\cal O}\(\asb^3\)\;,
\eeqn
where
\beqn
L(N,E,m)&=& 2 + \int dx\, dY \(x^{N-1}-1\)a^{(1)}(x,Y)
\nonumber \\
M(N,E,m)&=& \int dx\,dY\, \(x^{N-1}-1\)
a^{(2)}_l(x,Y) \nonumber \\
&& {}+ 2\,\int dx\, dY\, \(x^{N-1}-\frac{1}{2}\)
a^{(2)}_h(x,Y)\;.
\eeqn
We have calculated $L(N,E,m)$ and $M(N,E,m)$ numerically, using $E =$
100~GeV and $m=$ 8, 4, 3, 2.5, 2, 1.5, 1, 0.6, 0.5, 0.4, 0.2~GeV, for
a vector current coupled to the heavy quark.
We expect that, for small masses, $A(N,E,m)$ of eqs.~(\ref{eq:A_B_coef})
should coincide with
$L(N,E,m)$, and $M(N,E,m)$ should differ from $B(N,E,m)$ by a
mass and energy independent quantity, since such term is actually
beyond the next-to-leading logarithmic approximation. We find
very good agreement between $A(N,E,m)$ and $L(N,E,m)$.
We present the results for $M(N,E,m)$ separated into the different
colour components
\beq
M(N,E,m)=M_{C_F}(N,E,m)+M_{C_A}(N,E,m)
+M_{n_f}(N,E,m)+M_{T_F}(N,E,m)\;.
\eeq
In Fig.~\ref{fig:ca} we have plotted our results for $M_{C_A}$ (crosses
with error bars) and for $B_{C_A}$ (solid lines).
%%%%%%%%%%%%%%%%%%%%%%%%%%%%%%%%%%%%%%%%%%%
%%%%%%%%%%%%%%%%%%%%%%%%%%%%%%%%%%%%%%%%%%%
An arbitrary ($N$-dependent) constant has been added to the curves for
$B_{C_A}$, in order to make them coincide with the numerical result
for $m/E=0.015$.
We find, as the mass gets smaller, satisfactory agreement for all
moments.  Notice that for
higher moments we need smaller masses to approach the massless limit.
In Figs.~\ref{fig:nf}, \ref{fig:cf} and~\ref{fig:tf} we report the
analogous results for the remaining colour combinations.
%%%%%%%%%%%%%%%%%%%%%%%%%%%%%%%%%%%%%%%%%%%
%%%%%%%%%%%%%%%%%%%%%%%%%%%%%%%%%%%%%%%%%%%
%%%%%%%%%%%%%%%%%%%%%%%%%%%%%%%%%%%%%%%%%%%
%%%%%%%%%%%%%%%%%%%%%%%%%%%%%%%%%%%%%%%%%%%
%%%%%%%%%%%%%%%%%%%%%%%%%%%%%%%%%%%%%%%%%%%
%%%%%%%%%%%%%%%%%%%%%%%%%%%%%%%%%%%%%%%%%%%
Again, we find satisfactory agreement.

If one follows the procedure proposed by the authors of
Ref.~\cite{Spira}, eqs.~(\ref{eq:colcoef}) are modified in the $C_F
C_A$ and in the $n_f C_F T_F$ coefficients. More specifically, the
terms proportional to $d^{(1)}_Q$ all disappear from the expressions
of the $C_F C_A$ and of the $n_F C_F T_F$ coefficients. In fact, by
inspecting formulae~(\ref{eq:sigmahatprime}) and~(\ref{eq:iniprime}),
and the derivation of eq.~(\ref{eq:sigmaNall}), we see that the only
difference between the two approaches is that the term
$\asb\(\mu_0\)\,d^{(1)}_Q$ is replaced by $\asb\(\mu\)\,d^{(1)}_Q$,
which, using the renormalization group equation~(\ref{eq:RGE}), amounts
to a difference of $-2\,\pi\, b_0\, \asb^2
\,d^{(1)}_Q\,\log\mu^2/\mu_0^2 $, precisely what is needed to cancel
the term of the same form appearing in eq.~(\ref{eq:sigmaNall}). The
modified result  is also shown in Figs.~\ref{fig:ca}
and~\ref{fig:nf} (dashed lines).  It is quite clear that the
approach proposed by these authors does not work.

\section{Improved cross section}
We can get an improved cross section by merging the fixed order
calculation with the NLO resummed cross section, to obtain a formula
that, for $E\approx m$, is accurate to order $\as^2$, and, for $E\gg
m$, is accurate at NLO level.  This can be accomplished with the
following considerations: we have computed the differential cross
section till order $\as^2$ so that we know the first three
coefficients of the series~(\ref{eq:sigma}), that is
\beq
\left. \frac{d\sigma}{dx}(x,E,m) \right|_{\rm fix} =
a^{(0)}(x,E,m) + a^{(1)}(x,E,m)\,\asb(E) +  a^{(2)}(x,E,m)\, \asb^2(E)\;,
\eeq
where we have taken $\mu=E$, for simplicity of notation. In this
equation, we have not computed the contribution to  $a^{(2)}$ in the
two-jet region. This is not a limiting point since an approximate
expression for $a^{(2)}$ till order ${\cal O}\(\(m^2/E^2\)^6\)$ is
now available~\cite{Harlander}.

The NLO resummed cross section is given by (see
eq.~(\ref{eq:resummed_sigma}))
\beq
\label{eq:NLO_resum}
\left. \frac{d\sigma}{dx}(x,E,m) \right|_{\rm res} =
\sum_{n=0}^{\infty} \beta^{(n)}(x) \,\( \asb(E) \log\frac{E}{m} \)^n +
\asb(E)\sum_{n=0}^{\infty}\gamma^{(n)}(x) \,\( \asb(E)
\log\frac{E}{m}\)^n\;,
\eeq
so that, the improved formula reads
\beqn
\label{eq:improved}
\left. \frac{d\sigma}{dx}(x,E,m) \right|_{\rm imp} &= &
a^{(0)}(x,E,m) + a^{(1)}(x,E,m)\,\asb(E) +  a^{(2)}(x,E,m)\, \asb^2(E)
\nn \\
&& {} +\sum_{n=3}^{\infty} \beta^{(n)}(x) \,\( \asb(E) \log\frac{E}{m}
\)^n
\nn\\
&&{}+\asb(E)\sum_{n=2}^{\infty} \gamma^{(n)}(x) \,\( \asb(E)
\log\frac{E}{m}\)^n \;.
\eeqn
The actual way this improved formula is obtained is a bit more
complicated, because we do not have eq.~(\ref{eq:NLO_resum}) in
this form, but we have a numerical function of $\asb$.
We start from the following considerations
\beqn
\lim_{m\to 0}  a^{(0)}(x,E,m) &\approx& \beta^{(0)}(x) \nn\\
\lim_{m\to 0}  a^{(1)}(x,E,m) &\approx& \beta^{(1)}(x)
\log\frac{E}{m} + \gamma^{(0)}(x)\nn\\
\lim_{m\to 0}  a^{(2)}(x,E,m) &\approx& \beta^{(2)}(x)
\log^2\frac{E}{m} + \gamma^{(1)}(x)
\log\frac{E}{m} + \delta^{(0)}(x)\;,\label{eq:delta_0}
\eeqn
which allow us to write
\beqn
\left. \frac{d\sigma}{dx}(x,E,m) \right|_{\rm fix} \!\!\!&-&
\!\!\!  \lim_{m\to 0} \lg\left. \frac{d\sigma}{dx}(x,E,m) \right|_{\rm
fix}\rg + \left. \frac{d\sigma}{dx}(x,E,m) \right|_{\rm res} =\nn \\
&=& \!\! a^{(0)}(x,E,m) + a^{(1)}(x,E,m)\,\asb(E) +
a^{(2)}(x,E,m)\, \asb^2(E)
\nn \\
&&  \!\!\! {} +\sum_{n=3}^{\infty} b^{(n)}(x) \,\( \asb(E) \log\frac{E}{m}
\)^n
\nn\\
&& \!\!\!{}+\asb(E)\sum_{n=2}^{\infty} c^{(n)}(x) \,\( \asb(E)
\log\frac{E}{m}\)^n  - \delta^{(0)}(x)\,\asb^2(E) \;.
\label{eq:numerical_improved}
\eeqn
By comparing eq.~(\ref{eq:improved}) with
eq.~(\ref{eq:numerical_improved}) we obtain
\beqn
\left. \frac{d\sigma}{dx}(x,E,m) \right|_{\rm imp} &=&
\left. \frac{d\sigma}{dx}(x,E,m) \right|_{\rm fix} -
\lim_{m\to 0} \lg\left. \frac{d\sigma}{dx}(x,E,m) \right|_{\rm
fix}\rg + \left. \frac{d\sigma}{dx}(x,E,m) \right|_{\rm res} \nn\\
&&{} +\delta^{(0)}(x)\,\asb^2(E) \;,
\eeqn
where $\delta^{(0)}(x)$ is defined by the limiting procedure of
eq.~(\ref{eq:delta_0}).

\chapter{Kinematics}
\label{chap:kinematics}
\thispagestyle{plain}
\section[Kinematics and four-body phase space with two massless
particles]{Kinematics and four-body phase space with two massless\\
particles}
\label{sec:kinem_2m}
%%%%%%%%%%%%%%%%%%%%%%%%%%%%%%%%%%%%%%%%%%%
\begin{figure}[htb]
\centerline{\epsfig{figure=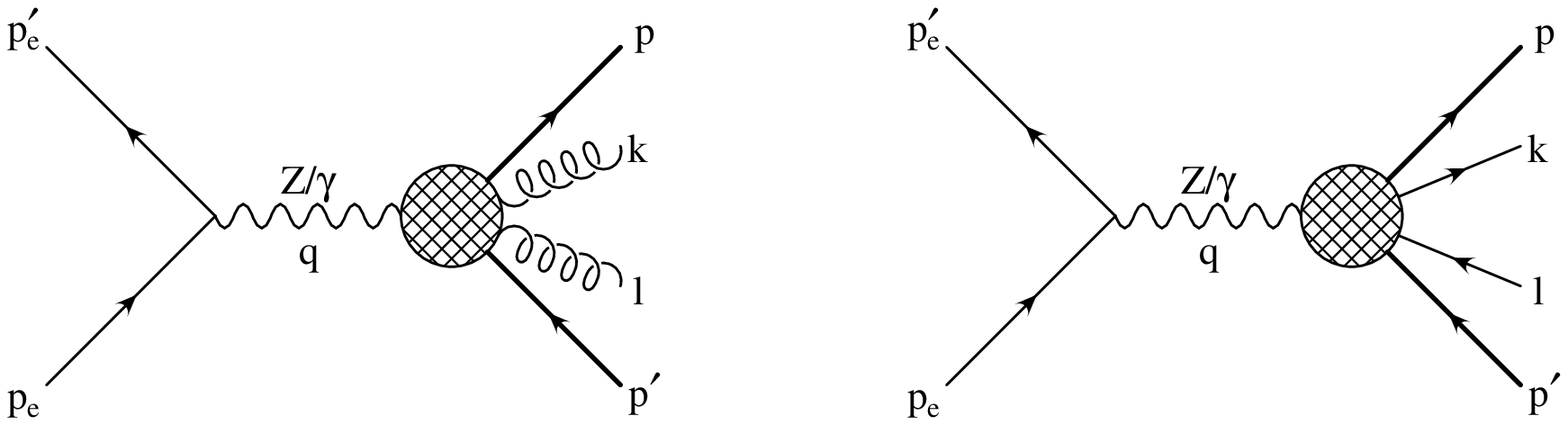,
width=0.9\textwidth,clip=}}\ccaption{}{ \label{fig:4kin}
Four-body kinematics.}
\end{figure}
%%%%%%%%%%%%%%%%%%%%%%%%%%%%%%%%%%%%%%%%%%%
The four-body processes we are considering  are illustrated in
Fig.~\ref{fig:4kin}, and summarized by
\beqn
e^+\(p_e'\)+ e^-\(p_e\) \ \, \to\ \,
Z/\gamma\,(q) \!\!&\to&\!\! Q(p) + \overline{Q}(p') + g(k) + g(l) \nn \\
e^+\(p_e'\)+ e^-\(p_e\) \ \,\to\ \,
Z/\gamma\,(q) \!\!&\to&\!\! Q(p) + \overline{Q}(p') + q(k) + \bar{q}(l)\;,
\nonumber
\eeqn
where $Q$ is the massive quark, $q$ is the massless quark and
the momenta satisfy
\beq
l^2 = k^2 = 0\;,  \quad\quad \quad  p^2 = p'^2 = m^2\;,  \quad \quad\quad
q^2=\(p_e'+p_e\)^2 \;.
\eeq
We hope that no confusion arises between the total momentum $q$ and
the light quark $q$.
In the centre-of-mass system of the two massless particles,
we can express the $d$-momenta in the following way
\beqn
\label{eq:def_l}
l &=& l_0 \, \bigl( 1,\ldots,\sin\th\sin\phi,
\sin\th \cos\phi,\cos\th \bigr)   \\
k &=& k_0 \, \bigl( 1,\ldots,-\sin\th\sin\phi,
-\sin\th \cos\phi,-\cos\th \bigr) \\
p &=& p_0 \left( 1,\ldots,0,0,\sqrt{1-\frac{m^2}{p_0^2}}\right)  \\
p' &=& p'_0 \left(1,\ldots,0,\sqrt{1-\frac{m^2}{{p'_0}^2}}\sin\a,
\sqrt{1-\frac{m^2}{{p'_0}^2}}\cos\a\right) \;,
\label{eq:def_p'}
\eeqn
where the dots indicate $d-3$ equal and opposite components in the
expression for $l$ and $k$, and $d-3$ zeros in the expression for $p$
and $p'$.

To describe the unoriented four-body phase space, we need five independent
variables, which we choose to be
\beq
\label{eq:x1x2y}
x_1=\frac{2q\cdot p}{q^2}\;,\quad\quad x_2=\frac{2q\cdot p^\prime}{q^2}\;,
\quad\quad y=\frac{(k+l)^2}{q^2}\;,\quad\quad \th, \quad\quad\phi\;.
\eeq
In the centre-of-mass system of the light particles
\beq
\label{eq:l0k0_def}
l_0 = k_0 = \sqrt{q^2}\, \frac{\sqrt{y}}{2}\;,
\eeq
where we have used the definition of $y$ of eq.~(\ref{eq:x1x2y}).
From momentum conservation, written in the following way,
\beq
\eqalign{
(q-p)^2 = (p'+k+l)^2 \cr
(q-p')^2 =  (p+k+l)^2 \;, \cr}
\eeq
we have
\beq\label{eq:p0pp0}
p_0 = \sqrt{q^2}\,\frac{1-x_2-y}{2\sqrt{y}}\;, \quad\quad\quad
p'_0 =\sqrt{q^2}\,\frac{1-x_1-y}{2\sqrt{y}}\;,
\eeq
and from
\beq
(p+p'+k+l)^2 = q^2 \;,
\eeq
we obtain
\[
\(p_0 + p_0' + \sqrt{q^2}
\sqrt{y}\)^2 - \(p_0'^2-m^2\) \sin^2\alpha - \( \sqrt{p_0^2-m^2} +
\sqrt{p_0'^2-m^2} \cos\alpha\)^2 = q^2\;.
\]
We can solve this expression to give
\beq
\label{eq:cosalfa}
\cos\a = \frac{ y\, (\rho-x_1-x_2) + (1-x_1)(1-x_2)-y^2}{
\sqrt{\( 1-x_1-y \)^2-4 \rho\, y}\;
\sqrt{\( 1-x_2-y \)^2-4 \rho\, y}}\;,
\eeq
where
\beq
\label{eq:def_rho}
\rho = \frac{4 m^2}{q^2}\;.
\eeq
Starting from the four-particle phase space  in $d$ dimensions
\beq
\label{eq:PS4_def}
\PSq = \int \frac{d^{d-1}p}{2\,p_0(2\pi)^{d-1}}\,
\frac{d^{d-1}p'}{2\,p'_0(2\pi)^{d-1}}
\,\frac{d^{d-1}l}{2\,l_0(2\pi)^{d-1}}
\,\frac{d^{d-1}k}{2\,k_0(2\pi)^{d-1}} \,(2\pi)^d\, \delta^d(q-p-p'-l-k)\;,
\eeq
we introduce the following two identities
\beq
\int\frac{d^dt}{(2\pi)^d}\, (2\pi)^d\, \delta^d(t-l-k) = 1\;,
\quad\quad\quad
q^2\,\int\frac{dy}{2\pi} \, 2\pi \,\delta(t^2 - q^2\,y) = 1\;,
\eeq
which allow us to integrate eq.~(\ref{eq:PS4_def}) in $t_0$, and to obtain
\beqn
\PSq \!&=&\! q^2 \int\frac{dy}{2\pi}\,\Theta(y)\,
\underbrace{
\frac{d^{d-1}t}{2\,t_0(2\pi)^{d-1}}
\,\frac{d^{d-1}p}{2\,p_0(2\pi)^{d-1}}
\,\frac{d^{d-1}p'}{2\,p'_0(2\pi)^{d-1}}\,
(2\pi)^d \, \delta^d(q-t-p-p') }_{\PSt}  \nn\\
&& \phantom{\int\frac{dy}{2\pi}\,\Theta(y)\, }\times
\underbrace{ \frac{d^{d-1}l}{2\,l_0(2\pi)^{d-1}} \,
\frac{d^{d-1}k}{2\,k_0(2\pi)^{d-1}} \, (2\pi)^d \,\delta^d(t-k-l)}_{\PSd}\;,
\label{eq:PS4_separated}
\eeqn
where
\beq
\label{eq:t_0_gen}
t_0 = \sqrt{\mod{t}^2 + q^2 y}\;,
\eeq
and where $\Theta(y)$ is the Heaviside function.
In this way we succeed in dividing the four-body phase space into two
simpler Lorentz-invariant scalars, that we can evaluate in the most
appropriate reference system.

We compute  $\PSd$ in  the centre-of-mass system of the two massless
particles.  In this system
\beq
\label{eq:t_0_cms}
\vet{t} = \vet{k}+\vet{l} = \vet{0}\;, \quad\quad\quad t_0 = k_0+l_0 =
\sqrt{q^2}\sqrt{y} \;,
\eeq
where we have used eq.~(\ref{eq:l0k0_def}).  Integrating in $k$, we get
\beqn
\PSd &=& \int \frac{d^{d-1}l}{2\,l_0(2\pi)^{d-1}} \,
\frac{1}{2\,k_0} \, 2\pi \,\delta(t_0-k_0-l_0) =
\int \frac{d^{d-1}l}{4\,l_0^2\,(2\pi)^{d-2}} \,\delta(t_0-2\,l_0)\nn\\
&=& \int \frac{dl_y\, dl_z\, dl^{d-3}_\perp}
{4\,l_0^2\,(2\pi)^{d-2}} \,\delta(t_0-2\,l_0) =
\int \frac{dl_y\, dl_z\, \mod{l_\perp}^{d-4} d\mod{l_\perp}
\, d\Omega^{d-3}}
{4\,l_0^2\,(2\pi)^{d-2}} \,\delta(t_0-2\,l_0)\;,\nn
\eeqn
where we have introduced the integration over the solid angle $\Omega$
in $d-3$ dimensions:
\beq
\Omega^{d} \equiv  \int d\Omega^{d} =
\frac{2\,\pi^{\frac{d}{2}}}{\Gamma\(\frac{d}{2}\)}\;.
\eeq
We perform the following change of variables
\beqn
l_y &=& l_0 \sin\th\cos\phi\nn\\
l_z &=& l_0 \cos\th\\
\mod{l_\perp} &=& \sqrt{\mod{l}^2 - l_y^2 - l_z^2} = l_0
\sin\th\sin\phi\;,
\eeqn
with Jacobian
\beq
dl_y\, dl_z\, d\mod{l_\perp} = l_0^2\, \sin\th \, d\th\, d\phi \, dl_0\;,
\eeq
and considering that $\mod{l}=l_0$, we can write
\beq
\PSd =\int \frac{d\th \, d\phi\,  dl_0}
{4\,l_0^2\,(2\pi)^{d-2}}\,  l_0^{d-2}\, \sin\th \,
\(\sin\th\sin\phi\)^{d-4} \,
\frac{2\,\pi^{\frac{d-3}{2}}}{\Gamma\(\frac{d-3}{2}\)}\,\delta(t_0-2\,l_0)
\;.
\eeq
The integration over $l_0$ is straightforward and, with the last
change of variable
\beq
\label{eq:v}
v = \frac{1}{2}(1-\cos\th) \;,
\eeq
we have, in $d=4-2\e$ dimensions,
\beq
\label{eq:PS2_final}
\PSd = \frac{1}{8\pi}\,
\frac{\(4\pi\)^\e}{\Gamma(1-\e)}   \(q^2 y \)^{-\e}
\int_0^1 dv \,[v(1-v)]^{-\e} \frac{1}{N_{\phi}}
\int_0^\pi d\phi\,(\sin\phi)^{-2\e} \;,
\eeq
where
\beq
N_{\phi} = \int_0^\pi d\phi \,(\sin\phi)^{-2\e} =  4^\e \pi
\frac{\Gamma(1-2\e)}{\Gamma^2(1-\e)}
\eeq
and where we have used eq.~(\ref{eq:t_0_cms}).

We can immediately integrate in $t$ the three-body phase space of
eq.~(\ref{eq:PS4_separated}) to obtain
\beq
\label{eq:PS3_1}
\PSt = \int \frac{2\pi}{2\,t_0} \,\frac{d^{d-1}p}{2\,p_0(2\pi)^{d-1}}
\,\frac{d^{d-1}p'}{2\,p'_0(2\pi)^{d-1}}\, \delta(q_0-t_0-p_0-p_0')
\eeq
with the condition
\beq
\vet{q}= \vet{t}+\vet{p}+\vet{p'}\;.
\eeq
We evaluate this integral in the laboratory frame, where
\beq
\label{eq:t_vet}
q = (q_0,\vet{0})\;,\quad \quad \quad  \vet{t}=-\vet{p}-\vet{p'}\;.
\eeq
We orient our reference axes in such a way that $\vet{p}$ is along the
$z$ axis and $\vet{p'}$ belongs to the $yz$ plane, forming an angle
$\beta$ with $\vet{p}$.
In this system
\beqn
d^{d-1}p &=& \mod{p}^{d-2} \, d\mod{p}\, \Omega^{d-1}\nn\\
d^{d-1}p' &=& dp_z' \, d^{d-2}p'_\perp = dp'_z \, \mod{p'_\perp}^{d-3}
d\mod{p'_\perp} \, \Omega^{d-2} = \mod{p'} \, d\mod{p'} \, d\beta
\mod{p'_\perp}^{d-3}  \, \Omega^{d-2}\nn\\
&=& \mod{p'}^{d-2} \, d\mod{p'} \, \(\sin\beta\)^{d-3}\, d\beta
\, \Omega^{d-2}\;,
\eeqn
where we have used
\beqn
p'_z &=& \mod{p'} \cos\beta \nn\\
\mod{p'_\perp} &=& \mod{p'} \sin\beta \;,\nn
\eeqn
so that eq.~(\ref{eq:PS3_1}) can be written
\beq
\PSt = \int \frac{2\pi}{2\,t_0}
\,\frac{\mod{p}^{d-2} \, d\mod{p}\, \Omega^{d-1}}{2\,p_0(2\pi)^{d-1}}
\,\frac{\mod{p'}^{d-2}\, d\mod{p'}\, \Omega^{d-2}
\(\sin\beta\)^{d-4} d\cos\beta}
{2\,p'_0(2\pi)^{d-1}}\,  \delta(q_0-t_0-p_0-p_0')\;.
\eeq
From eqs.~(\ref{eq:t_0_gen}) and~(\ref{eq:t_vet}) we have
\beq
t_0 = \sqrt{\mod{p}^2+\mod{p}^2+2\,\mod{p}\mod{p'}\cos\beta + q^2 y}\;,
\quad\quad\quad\quad
\frac{\partial t_0}{\partial\cos\beta} = \frac{\mod{p}\mod{p'}}{t_0}\;,
\eeq
and  we can integrate in $\beta$ to obtain
\beq
\PSt = \int \pi
\,\frac{\mod{p}^{d-3} \, d\mod{p}\, \Omega^{d-1}}{2\,p_0(2\pi)^{d-1}}
\,\frac{\mod{p'}^{d-3}\, d\mod{p'}\, \Omega^{d-2}
\(\sin\beta\)^{d-4}} {2\,p'_0(2\pi)^{d-1}}\;,
\eeq
with $\beta$ determined by the $\delta$-function argument
\beq
\label{eq:beta}
q_0-p_0-p'_0 =  \sqrt{\mod{p}^2+\mod{p}^2+2\,\mod{p}\mod{p'}\cos\beta
+ q^2 y} \;.
\eeq
If we evaluate $x_1$ and $x_2$ of eq.~(\ref{eq:x1x2y}) in the
laboratory system, we get
\beq
x_1=\frac{2\, p_0}{q_0}\;,\quad\quad\quad\quad x_2=\frac{2\, p'_0}{q_0}\;,
\eeq
and considering that
\beq
\mod{p}^2 = p_0^2 -m^2 = \frac{q^2}{4} \(x_1^2-\rho\)\;, \quad\quad\quad
\mod{p'}^2=  p_0'^2 -m^2 = \frac{q^2}{4} \(x_2^2-\rho\)\;,
\eeq
we can write $\PSt$ as
\beqn
\PSt &=& \frac{1}{\Gamma(2-2\e)}\,
\frac{(8 \pi)^{2\e}}{2\, (4\pi)^3}\,(q^2)^{1-2\e}\!\!\int  dx_1\, dx_2 \,
\(\sin \beta\)^{-2\e}\(x_1^2-\rho\)^{-\e}\(x_2^2-\rho\)^{-\e}\nn\\
\label{eq:PS3_final}
&=& \frac{1}{\Gamma(2-2\e)}\,
\frac{(8 \pi)^{2\e}}{2\, (4\pi)^3}\,(q^2)^{1-2\e}\!\!\int  dx_1\, dx_2 \,
\(1-\cos^2\beta\)^{-\e}\(x_1^2-\rho\)^{-\e}\(x_2^2-\rho\)^{-\e}\;.\nn\\[-3mm]
\eeqn
From eq.~(\ref{eq:beta}), we have
\beq
\label{eq:cos_beta}
\cos\beta = \frac{\(x_g^2-4\,y\)-\(x_1^2-\rho\)-\(x_2^2-\rho\)}
{2\sqrt{x_1^2-\rho}\sqrt{x_2^2-\rho}}\;,
\eeq
where we  define
\beq
\label{eq:xg_def}
x_g \equiv 2 - x_1 - x_2\;,
\eeq
and  the integration range for $x_1$ and $x_2$ is determined by the reality
condition for $\cos\beta$
\beq
\label{eq:reality_cond}
-1 \leq \cos\beta \leq 1\;.
\eeq
Inserting the expressions of eqs.~(\ref{eq:PS2_final})
and~(\ref{eq:PS3_final}) into eq.~(\ref{eq:PS4_separated}), we obtain
\beqn
\PSq &=& \frac{q^2} {(4\pi)^2 \,
\Gamma(1-\e)}\( \frac{4\pi}{q^2} \)^{\e} H
\int dx_1 \, dx_2 \, dy \, y^{-\e}\,\Theta(y) \nn \\
&&\times \ga 4\(x_1^2-\rho\) \(x_2^2-\rho\)-\left[
(x_g^2-4\, y)-(x_1^2-\rho)-(x_2^2-\rho) \right]^2\gc ^{-\e} \nn \\
\label{eq:ps4_inter}
&&\times \int_0^1 dv \,[v(1-v)]^{-\e} \,\frac{1}{N_{\phi}}
\int_0^\pi d\phi\,(\sin\phi)^{-2\e}\;,
\eeqn
with
\beq
H \equiv \frac{1}{\Gamma(2-2\e)}\,
\frac{q^2}{2\, (4\pi)^3}\, \( \frac{16 \pi}{q^2} \)^{2\e}\;.
\eeq

\begin{figure}[htb]
\centerline{\epsfig{figure=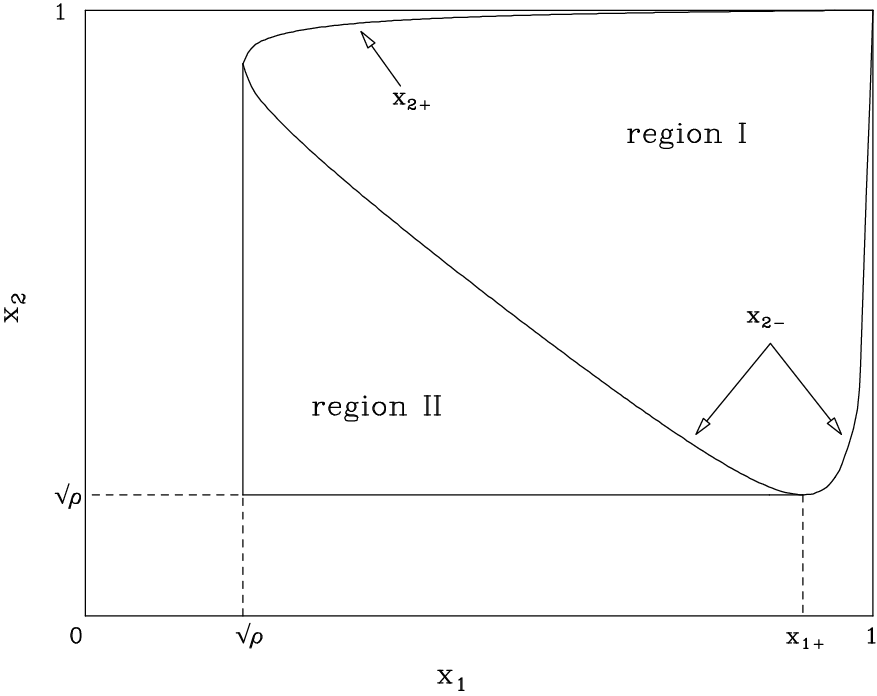,
width=0.7\textwidth,clip=}}\ccaption{}{ \label{phd_phs4}
The two different areas in the $x_1$-$x_2$ plane  correspond to the
{\rm region~I} and to the {\rm region~II}
of eq.~(\protect{\ref{eq:ps4}}). }
\end{figure}

The last step to perform is the determination of the integration
region. First of all, the physical boundaries for the integration
variables are
\beq
\sqrt{\rho} \leq x_1,x_2 \leq 1\;, \quad\quad\quad  0 \leq y \leq
\(1-\sqrt{\rho}\)^2 \;.
\eeq
If  we choose to integrate first in $y$, we solve
eq.~(\ref{eq:reality_cond}) with respect to $y$, and we get
\beq
y_{-} \leq  y \leq y_{+}     \;,
\eeq
with
\beq
y_{\pm} \equiv \frac{1}{4} \left[
x_g^2 -(x_1^2-\rho)-(x_2^2-\rho) \pm 2\sqrt{x_1^2-\rho}
\sqrt{x_2^2-\rho}  \right] \;.
\eeq
The Heaviside function $\Theta(y)$ of eq.~(\ref{eq:ps4_inter}) imposes
the condition
\beq
y_{+} \geq 0\;,
\eeq
which forces  the allowed regions for $x_1$ and $x_2$ integration
to be  ``region I'' and ``region II'' of Fig.~\ref{phd_phs4}.
%%%%%%%%%%%%%%%%%%%%%%%%%%%%%%%%%%%%%%%%%%%
%%%%%%%%%%%%%%%%%%%%%%%%%%%%%%%%%%%%%%%%%%%
In these regions, the relation
\beq
y \leq  \(1-\sqrt{\rho}\)^2
\eeq
is always satisfied.
In addition, we have
\beq
y_{-} \geq 0  {\rm \ \ \ if\ \ \ } x_1,x_2 \in {\rm region\ II}\;.
\eeq
Following this procedure, we succeed in dividing the phase space into
two regions:
\begin{enumerate}
\item
{\bf region I}: $y$ can reach zero, where  collinear and soft
divergences arise. The integration range is (see
Fig.~\ref{phd_phs4})
\beq
\int_{\sqrt{\rho}}^1 dx_1  \int_{x_{2-}}^{x_{2+}} dx_2
\int_0^{y_{+}} dy \;,
\eeq
where
\beq
\label{eq:x2_limits}
x_{2\pm} = \frac{1}{4(1-x_1)+\rho} \left[ 2\(1-x_1\) \(2-x_1\) +
\rho\(2-x_1\)  \pm
2\(1-x_1\)\sqrt{x_1^2-\rho}   \right]
\eeq

\item
{\bf region II}: $y$ cannot reach $0$ and  the region is free from
infrared divergences. The integration range is
\beq
\int_{\sqrt{\rho}}^{x_{1+}} dx_1  \int_{\sqrt{\rho}}^{x_{2-}} dx_2
\int_{y_{-}}^{y_{+}} dy \;,
\eeq
where
\beq
x_{1+} = 2-\frac{2-\rho}{2-\sqrt{\rho}}\;.
\eeq
\end{enumerate}
We can then summarize the full four-body phase space
\beqn
\PSq &=& \frac{q^2} {(4\pi)^2 \,
\Gamma(1-\e)}\( \frac{4\pi}{q^2} \)^{\e} H  \nonumber \\
&&\times \Biggl\{
\underbrace{\int_{\sqrt{\rho}}^1 dx_1  \int_{x_{2-}}^{x_{2+}} dx_2
\int_0^{y_{+}} dy}_{\rm{region\ I}}\,
+ \underbrace{
\int_{\sqrt{\rho}}^{x_{1+}} dx_1  \int_{\sqrt{\rho}}^{x_{2-}} dx_2
\int_{y_{-}}^{y_{+}} dy }_{\rm{region\ II}}
\Biggr\} \, y^{-\e} \nn \\
&&\times \ga 4\(x_1^2-\rho\) \(x_2^2-\rho\)-\left[
(x_g^2-4 y)-(x_1^2-\rho)-(x_2^2-\rho) \right]^2\gc ^{-\e} \nn \\
\label{eq:ps4}
&&\times \int_0^1 dv \,[v(1-v)]^{-\e} \,\frac{1}{N_{\phi}}
\int_0^\pi d\phi\,(\sin\phi)^{-2\e}\;.
\eeqn
A statistical factor $1/2!$ must be supplied if we consider the final
state with the two identical gluons.

Sometimes we will need an analogous set of final-state variables,
in which the role of $p$ and $p'$ is interchanged.
The variable $y$ remains the same, $x_1$ and $x_2$ are exchanged, and
the other two variables, denoted by $v'$ and $\phi'$,
are related to $v$ and $\phi$ by the equations
\beq
\label{eq:ppexch}
\eqalign{
v' = \frac{1}{2}\(1-\cos\th'\)= \frac{1}{2}\lq 1-\cos\alpha
-\(2\,\sqrt{v\,(1-v)}\cos\phi\sin\alpha-2\,v\cos\alpha\)\rq
\cr
\cos\phi' = \frac{1-\cos\alpha-2\,
(v-v'\cos\alpha)}{2\sin\alpha\,\sqrt{v'(1-v')}} \;.}
\eeq
These relations can be obtained considering the definition of the
involved angles
\beq
\eqalign{
\vet{p'} \cdot \vet{k} = \mod{p'}\mod{k} \( \sin\th\cos\phi\sin\alpha +
\cos\th\cos\alpha \) \equiv \mod{p'}\mod{k}  \cos\th'  \cr
\vet{p} \cdot \vet{k} = \mod{p}\mod{k} \cos\th \equiv
\mod{p}\mod{k}\( \sin\alpha\sin\th'\cos\phi'+\cos\th'\cos\alpha\)\;,}
\eeq
where the same scalar products have been computed in the reference system
of eqs.~(\ref{eq:def_l})--(\ref{eq:def_p'}) and in the reference
system where $p$ and $p'$ have been interchanged.

Exchanging  the roles of $l$ and $k$ brings about the following
transformations
\beq
\label{eq:klexch}
v\,\to\,1-v \;,\quad\quad\quad \phi\,\to\, \pi+\phi\;, \quad\quad\quad
v'\,\to\, 1-v'\;,  \quad\quad\quad  \phi'\,\to\, \pi+\phi'\;.
\eeq

\section{Kinematics and three-body phase space with one massless
particle}
%%%%%%%%%%%%%%%%%%%%%%%%%%%%%%%%%%%%%%%%%%%
\begin{figure}[htb]
\centerline{\epsfig{figure=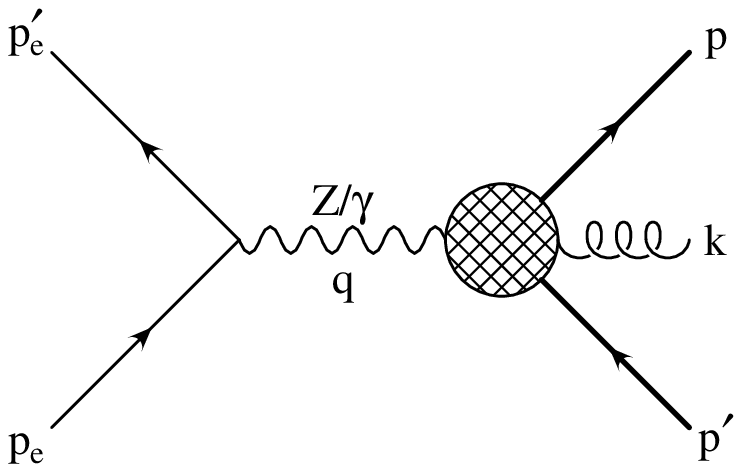,
width=0.43\textwidth,clip=}}\ccaption{}{ \label{fig:3kin}
Three-body kinematics.}
\end{figure}
%%%%%%%%%%%%%%%%%%%%%%%%%%%%%%%%%%%%%%%%%%%
We consider the three-body process depicted in Fig.~\ref{fig:3kin}
\beqn
e^+\(p_e'\)+ e^-\(p_e\) \;\to\;
Z/\gamma\,(q) \;\to\; Q(p) + \overline{Q}(p') + g(k)\;,
\eeqn
where the momenta of the particles satisfy
\beq
p^2=p'^2 = m^2\;,  \quad\quad\quad   k^2 = 0\;.
\eeq
For unoriented shape variables, we can express the three-body phase
space in terms of two variables, that we choose to be $x_1$ and $x_2$
of eq.~(\ref{eq:x1x2y}).
Using eqs.~(\ref{eq:PS3_final}) and~(\ref{eq:cos_beta}) we have
\beqn
\PSt &=&  H \int_{\sqrt{\rho}}^{1}  dx_1 \int_{x_{2-}}^{x_{2+}}
dx_2 \nn \\
&& \label{eq:ps3}
\times\ga 4\(x_1^2-\rho\) \(x_2^2-\rho\)-\left[
x_g^2-(x_1^2-\rho)-(x_2^2-\rho) \right]^2\gc ^{-\e}\;,
\eeqn
where we put $y=0$, because the mass of the light system is zero
(final gluon on-shell).  The integration range can be found
by imposing the reality
condition~(\ref{eq:reality_cond}), with $y=0$.

\section[Kinematics and four-body phase space with four massive
particles]{Kinematics and four-body phase space with four massive\\
particles}
%%%%%%%%%%%%%%%%%%%%%%%%%%%%%%%%%%%%%%%%%%%
\begin{figure}[htb]
\centerline{\epsfig{figure=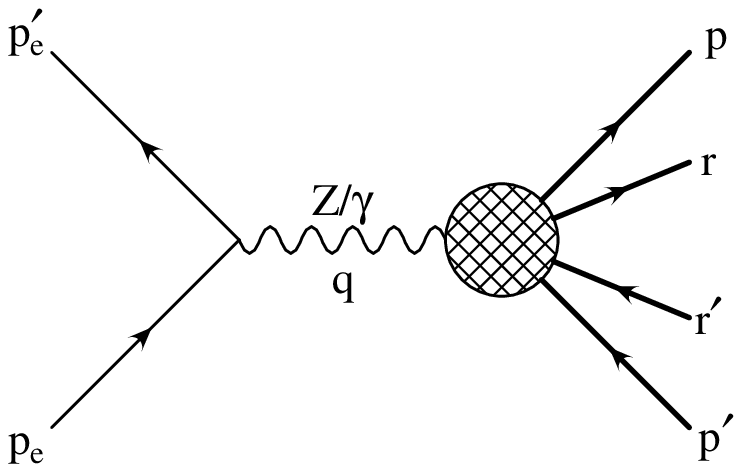,
width=0.43\textwidth,clip=}}\ccaption{}{ \label{fig:4m_kin}
Four massive body kinematics.}
\end{figure}
%%%%%%%%%%%%%%%%%%%%%%%%%%%%%%%%%%%%%%%%%%%
The four-body process describing the production of four massive quarks
is illustrated in Fig.~\ref{fig:4m_kin} and summarized by
\beq
e^+\(p_e'\)+ e^-\(p_e\) \;\to\;
Z/\gamma\,(q) \;\to\; Q(p) + \overline{Q}(p') + Q(r) + \overline{Q}(r')\;,
\eeq
where
\[     r^2=r'^2=p^2=p'^2=m^2\;.
\]
The four-body phase space is obtained with a procedure similar to the
one given in Sec.~\ref{sec:kinem_2m}, with the simplification that now
the entire cross section has no soft or collinear divergences, so
that we can put ourselves directly in $d=4$ dimensions and we do not
need to divide the allowed phase-space  into two different regions.  In
the centre-of-mass frame of one heavy quark-antiquark pair we have
\beqn
r &=& (r_0,\mod{r}\sin\th\sin\phi,
\mod{r}\sin\th \cos\phi,\mod{r}\cos\th)  \nn\\
r' &=& (r_0,-\mod{r}\sin\th\sin\phi,
-\mod{r}\sin\th \cos\phi,-\mod{r}\cos\th)   \nn\\
p &=& p_0 \( 1,0,0,\sqrt{1-\frac{m^2}{p_0^2}}\) \nn\\
p' &=& p'_0 \(1,0,\sqrt{1-\frac{m^2}{{p'_0}^2}}\sin\a,
\sqrt{1-\frac{m^2}{{p'_0}^2}}\cos\a\) \;,
\eeqn
where
\[
y =\frac{(r+r')^2}{q^2} \; \Longrightarrow \;
r_0 = \sqrt{q^2}\,\frac{\sqrt{y}}{2} \;,
\]
and $p_0$, $p'_0$ and $\cos\a$ are given by~(\ref{eq:p0pp0})
and~(\ref{eq:cosalfa}), while
\beq
\label{eq:mod_r}
\mod{r}=\sqrt{r_0^2-m^2}\;.
\eeq
Starting from eq.~(\ref{eq:PS4_separated}), we need to re-compute the
two-body phase space in $d=4$ dimensions for massive particles
\beqn
\PSd\!\! &=&\!\! \int \frac{d^{3}r}{2\,r_0(2\pi)^{3}} \,
\frac{d^{3}r'}{2\,r_0'(2\pi)^{3}} \, (2\pi)^4 \,\delta^4(t-r-r') =\int
\frac{d^3 r}{4\, r_0^2 \,(2\pi)^2} \, \delta\(t_0-2r_0\) \nn\\
&=& \!\!
\int\frac{\mod{r}^2 d\mod{r}\, \sin\theta\,d\theta\,d\phi}{4\,r_0^2\,
(2\pi)^2} \, \delta\(t_0-2r_0\) = \frac{1}{4} \frac{1}{(2\pi)^2}
\sqrt{1-\frac{\rho}{y}} \int_0^1 dv \int_0^{2\pi}d\phi \;,\phantom{aaaaaa}
\label{eq:ps2_4m}
\eeqn
where we have used eqs.~(\ref{eq:mod_r}) and~(\ref{eq:v}).
The four-body phase space, according to eqs.~(\ref{eq:PS4_separated}),
(\ref{eq:PS3_final}) and~(\ref{eq:ps2_4m}) becomes
\beq
\PSq =  \frac{q^4}{\(4\pi\)^6} \int dy\, dx_1 \,dx_2
\sqrt{1-\frac{\rho}{y}} \int_0^1 dv \int_0^{2\pi} d\phi\;.
\ee
The physical boundaries for the kinematical variables are
\beq
\label{eq:4m_reality}
\rho \leq y \leq \(1-\sqrt{\rho}\)^2 \;,\quad\quad\quad
\sqrt{\rho} \leq x_1,x_2 \leq 1-2\rho\;,
\eeq
and from the reality condition~(\ref{eq:reality_cond}), we have
\beq
\bar{x}_{2-} \leq x_2 \leq \bar{x}_{2+}\;,
\eeq
where
\beq
\bar{x}_{2\pm} = \frac{1}{4(1-x_1)+\rho} \lq (2-x_1)(2+\rho-2y-2x_1)
\pm 2 \sqrt{\(x_1^2-\rho\) \left[(x_1-1+y)^2-\rho\,y \right]}\rq
\eeq
According to eq.~(\ref{eq:4m_reality}), $\bar{x}_{2+}$ must satisfy
\beq
\bar{x}_{2+} \geq \sqrt{\rho}\;,
\eeq
which implies
\beq
x_1 \leq 1-y-\sqrt{\rho\,y}\;,
\eeq
so  we have
\beq
\label{eq:ps4m}
\PSq =  \frac{q^4}{\(4\pi\)^6}
\int_{\rho}^{\bar{y}_{+}} dy \sqrt{1-\frac{\rho}{y}}
\int_{\sqrt{\rho}}^{\bar{x}_{1+}}  dx_1
\int_{\bar{x}_{2-}}^{\bar{x}_{2+}} dx_2  \int_0^1 dv \int_0^{2\pi} d\phi\;,
\eeq
where
\beq
\eqalign{
\bar{y}_{+} = \(1-\sqrt{\rho}\)^2  \cr
\bar{x}_{1+} \!\!= 1-y-\sqrt{\rho\,y} \;.
}
\eeq
A statistical factor $1/(2!2!)=1/4$ must be supplied
to eq.~(\ref{eq:ps4m}), because of the presence of two pairs of identical
particles in the final state.

\chapter{Next-to-leading-order amplitudes}
\label{chap:NLO_amplitudes}
\thispagestyle{plain}
\section{Introduction and notation}
\label{sec:outlinecalc}
In writing the  amplitude for the process we are investigating, we
disregard, at present, the  contributions coming from the decay of the
$Z/\gamma$ boson into a light couple of quarks.
We postpone all the comments on this subject to
Secs.~\ref{sec:hagiwara} and~\ref{sec:QQqq}.
We can then write the amplitude, up to an irrelevant phase, as
\beqn
{\cal A}&=& \bar{v}(p_e') \Bigg[
g_\sz^2 \,\frac{-g_{\mu\nu}}{q^2-M_{\rm \sz}^2+i\Gamma_{\rm \sz}
M_{\rm \sz}}
\left(v_e\gamma^\mu-a_e\gamma^\mu\gamma^5\right)\langle 0|
J^\nu_V(0) v_\sq  -J^\nu_A(0) a_\sq |f \rangle \nonumber \\
&& \mbox{}+g^2\frac{-g_{\mu\nu}}{q^2}(c_e\gamma^\mu)\langle 0|
J^\nu_V(0) c_\sq |f \rangle \Bigg] u(p_e) \label{eq:amp_gener}\;,
\eeqn
where $|f \rangle$ refers to states with four-momentum $q$ and
$J_V(0)$, $J_A(0)$ are the currents that describe the decay of the
vector boson into the final state $|f \rangle$.
In this equation we see the contributions coming from the $\gamma$
propagator, second term,  and the $Z$ propagator, first term,
where we have neglected terms
proportional to $q_\mu q_\nu$ because of leptonic  current
conservation, always verified for the vectorial part but verified  in
the massless limit for the axial part
\beq
\bar{v}(p_e') \, q_\mu \,  \Gamma_{V/A}^\mu \, u(p_e) = 0
\quad\quad {\rm with} \quad\quad
\Gamma_{V}^\mu = \gamma^\mu \;, \ \
\Gamma_{A}^\mu = \gamma^\mu\gamma^5 {\rm \ \ and \ \ } m_e =0\;.
\eeq
The notation used in eq.~(\ref{eq:amp_gener}) is
\beqn
g_{\rm \sz} &\equiv& \frac{g}{2\sin \theta_{\rm W} \cos \theta_{\rm W} }
\nn\\
\label{eq:coupling_constants}
v_i &\equiv& T_{3i}-2c_i\sin^2\theta_{\rm W}  \\
a_i &\equiv& T_{3i}  \nn
\eeqn
where $g$ is the electromagnetic coupling, $T_{3i}$ is the third
component of the (left) isospin of fermion $i$,
$c_i$ is its electric charge in units of the positron charge and
$\theta_{\rm W}$ is the Weinberg angle.
$M_{\rm \sz}$ and $\Gamma_{\rm \sz}$ are the $Z$ mass and total decay
width.

We are interested in describing only  unoriented events: for this
reason, we can neglect the axial-vector interference term in the
square of the amplitude.
In fact, for the three-parton final state of Fig.~\ref{fig:3kin},
there are not enough momenta to construct an invariant with
an $\epsilon$ symbol.
For the four-parton final state of Fig.~\ref{fig:4kin}, one could in
principle build such an invariant, but the cross section must be
symmetric in the light parton momenta, so that such an invariant
cannot survive. This is strictly true for the $Q\Qb gg$ final state,
and for the state $Q\Qb q \qb$, if the weak current is coupled to the
heavy quark. We refer to Sec.~\ref{sec:QQqq} for further details.

In addition, we have no problems between the use of the dimensional
regularization procedure and the axial coupling.  In fact, we can
circumvent the presence of $\gamma_5$ considering the case of a
generic vector current coupled to two fermions with different masses
$m_1$ and $m_2$. One can easily convince oneself that the case of the
axial coupling can be obtained by setting $m_1=m$ and $m_2=-m$, since
one can turn $-m$ into $m$ by a chiral rotation.  This procedure is
bound to work if there are no anomalies involved in the calculation,
and this is certainly the case for the graphs we have chosen to
compute (see Secs.~\ref{sec:virtual}--\ref{sec:real}).

When squaring the amplitude to compute the differential cross section,
we have two different tensorial structures which describe the leptonic
and hadronic part of the process.
The leptonic tensor, obtained by averaging over the initial
polarization, is given by
\beq
L^{\mu\nu} = \frac{1}{4}\Tr\(\ds{p}_e'\,\Gamma^\mu_{V/A}\, \ds{p}_e \,
\Gamma^\nu_{V/A} \)=
p_e^\mu \, {p'}_e^{\nu} + p_e^\nu \,{p'}_e^{\mu} - \(p_e\!\cdot
p'_e\) \, g^{\mu\nu} \; ,
\eeq
and, if we further average over the incoming electron beam direction, we
obtain
\beq
\overline{L^{\mu\nu}} = \frac{q^2}{3} \(-g^{\mu\nu} +
\frac{q^\mu q^\nu}{q^2} \)\;.
\eeq
We can then write  the differential cross section as
\beqn
d\sigma &=& \frac{4\,\pi}{3\, q^2}\, \alpha^2 N_c
\bigg\{dT_V
\left[\rho_2(q^2)\,(v_e^2+a_e^2)\,v_\sq^2+c_e^2 \,c_\sq^2-2\,\rho_1(q^2)\,
v_e\, v_\sq \,c_e \, c_\sq  \right]\nn \\
&& {}+  dT_A  \left[\rho_2(q^2)\,(v_e^2+a_e^2)\,a_\sq^2\right] \bigg\}
\; ,\nn
\eeqn
where $\alpha=g^2/(4\pi)$ is the electromagnetic coupling
constant, $N_c=3$ is the number of colours and
\beqn
\rho_1(q^2) &=& \frac{1}{4\sin^2\theta_{\rm W}\cos^2\theta_{\rm W}}\;
\frac{q^2\,(M_\sz^2-q^2)}{(M_\sz^2-q^2)^2+M_\sz^2\,\Gamma_\sz^2}
\nonumber \\
\rho_2(q^2) &=& \left(\frac{1}{4\sin^2\theta_{\rm W}
\cos^2\theta_{\rm W}}\right)^2
\frac{q^4}{(M_\sz^2-q^2)^2+M_\sz^2\,\Gamma_\sz^2} \;. \nn
\eeqn
We have also defined
\beqn
dT_{V/A}&=&\sum_n {\cal M}^{(f_n)}_{V/A} d\Phi_n \nonumber \\
\label{calMdef}
{\cal M}^{(f_n)}_{V/A}&=&\frac{2\pi}{N_c\, q^2}
\left(-g_{\mu\nu}+\frac{q_\mu q_\nu}{q^2}\right)
\langle 0| J^\mu_{V/A}(0)|f_n\rangle\langle f_n|J^\nu_{V/A}(0)|0\rangle\;,
\eeqn
where $d\Phi_n$ represents the $n$-body phase space, and $|f_n\rangle$
represents an $n$-body final state.  The $q^\mu q^\nu$ term in the
projector in eq.~(\ref{calMdef}) is, of course, irrelevant for the
vector current component, but it should be kept for the axial current
when the quark mass is non-zero.

In the following we will be interested in strong corrections up to the
second order, and in the final states:
$Q\Qb$, $Q\Qb g$, $Q\Qb gg$, $Q\Qb q\bar{q}$ and $Q\Qb Q \Qb$.
We will use the following simplified notation:
\begin{itemize}
\item[-]
${\cal M}^{(2)}_{V/A}$ for the $Q\Qb $ tree-level term
\item[-]
${\cal M}^{(b)}_{V/A}$ or ${\cal M}_b$
to indicate the three-body Born $Q\Qb g$, ${\cal O}\(\as\)$ term
\item[-]
${\cal M}^{(v)}_{V/A}$ or ${\cal M}_v$
to indicate the three-body virtual $Q\Qb g$, ${\cal O}\(\as^2\)$ term
\item[-]
${\cal M}^{(gg)}_{V/A}$ or ${\cal M}_{gg}$
for the four-body $Q\Qb gg$, ${\cal O}\(\as^2\)$ term
\item[-]
${\cal M}^{(q\qb )}_{V/A}$ or ${\cal M}_{q\qb }$
for the four-body $Q\Qb q\qb $, ${\cal O}\(\as^2\)$ term,
\end{itemize}
and equivalent ones for the $dT_{V/A}$ terms.

We will drop the ${V/A}$ suffix when not referring
specifically to the axial or vector contribution.

\section[$Q\Qb$ cross section]
{{\mylarge $Q\Qb$} cross section}
%%%%%%%%%%%%%%%%%%%%%%%%%%%%%%%%%%%%%%%%%%%
\begin{figure}[htb]
\centerline{\epsfig{figure=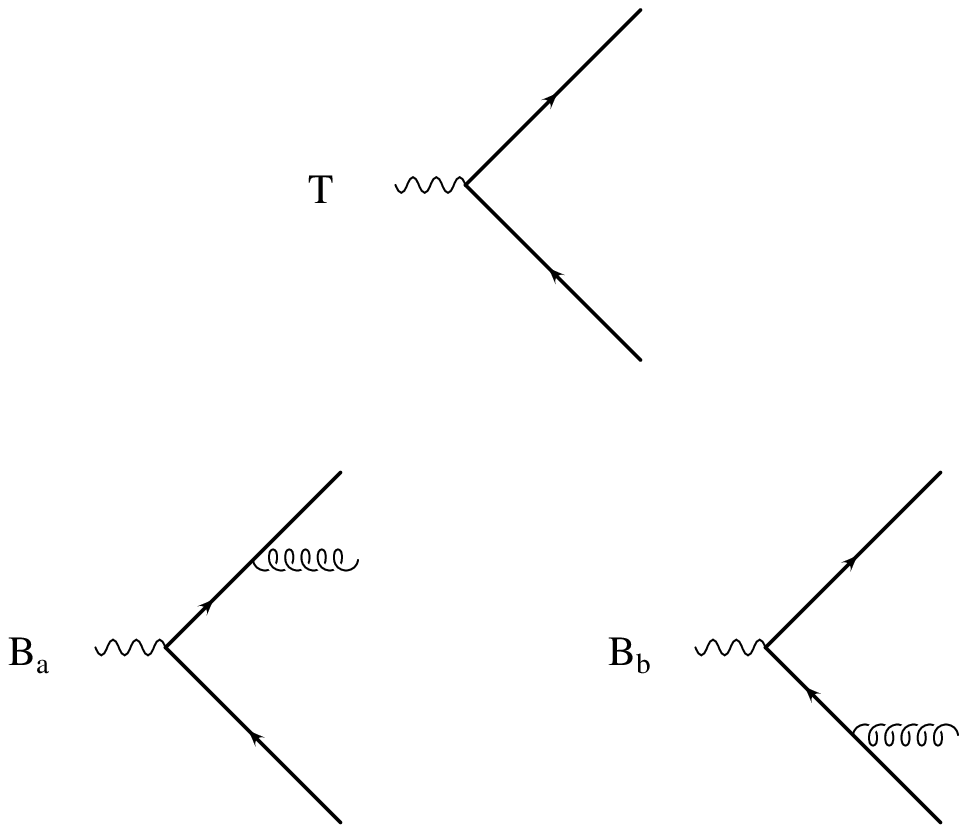,
width=0.6\textwidth,clip=}}\ccaption{}{ \label{fig:Born}
Tree {\rm (T)} and Born {\rm (B)} Feynman diagrams.}
\end{figure}
%%%%%%%%%%%%%%%%%%%%%%%%%%%%%%%%%%%%%%%%%%%
In the higher part of Fig.~\ref{fig:Born} we have depicted the Feynman
diagram representing the tree-level amplitude
\beq
{\cal A}^{\mu}_{V/A} = \overline {u}(p) \Gamma^\mu_{V/A}  v (p')\;,
\eeq
that, once squared, gives, for the two-body contribution to
eq.~(\ref{calMdef}), at zeroth order in $\as$,
\beq
{\cal M}_V^{(2)}
= \frac{2\pi}{N_c\, q^2} N_c\, 4\,q^2\,\left(1+\frac{\rho}{2}\right)\;,
\quad\quad\quad  {\cal M}_A^{(2)} =
\frac{2\pi}{N_c\, q^2} N_c\, 4\,q^2 \beta^2 \;,\label{ttwobody}
\eeq
where $\rho$ is defined in eq.~(\ref{eq:def_rho}) and
\beq
\beta=\sqrt{1-\rho}\;.
\eeq
Multiplying eq.~(\ref{ttwobody}) by the 2-body phase space $\beta/(8\pi)$,
we get the zeroth-order contributions to the total cross section
\beq
T_V^{(2)}=\beta\,\left(1+\frac{\rho}{2}\right)\;, \quad\quad\quad
T_A^{(2)}=\beta^3\;.
\eeq
This justifies our choice for the normalization factor in
eq.~(\ref{calMdef}): in the massless limit, $T_V^{(2)}=T_A^{(2)}=1$.

\section[$Q\overline{Q}g$ cross section at order
$\alpha_s$]
{{\mylarge $Q\overline{Q}g$} cross section at order
{\mylarge $\alpha_s$}}
The amplitude for the Born term represented in the lower part of
Fig.~\ref{fig:Born} is
\beqn
{\cal A}^{\mu\sigma}_{V/A}=\overline{u}(p) \Bigg[
\(-i g_s t^a_{ij}\g^\s\)
\frac{\ds{p} +\ds{k}+m}{(p+k)^2-m^2} \Gamma^\mu_{V/A} &&\nn\\
&&\hspace{-2.8cm}{}+ \Gamma^\mu_{V/A}
\frac{\ds{p} -\ds{q}+m}{(p-q)^2-m^2}\(-i g_s t^a_{ij}\g^\s\)
\Bigg]  v(p') \;, \phantom{aaaaaaa}
\eeqn
where $t_{ij}^a$ are the generators of SU($N_c$) gauge symmetry,
$i$ and $j$ refer to the colour indexes of quarks, while $a$ to
the colour index of the external gluon.
We remind here that
\beq
\sum_{a=1}^{N_c^2-1} \sum_{k=1}^{N_c} t^a_{ik}\, t^a_{kj} =
C_F \,\delta_{ij}  \quad\quad\quad   i,j=1\ldots N_c \;,
\eeq
where
\beq
C_F = \frac{N_c^2 -1}{2N_c} = \frac{4}{3} \quad{\rm for}\quad N_c=3\;.
\eeq
We define
\beq
\label{eq:M_Born_}
M^{\sigma\sigma'}_{V/A} \equiv \frac{1}{ g_s^2 \, C_F N_c}
\left(-g_{\mu\nu}+\frac{q_\mu q_\nu}{q^2}\right) \sum
{\cal A}^{\mu\sigma}_{V/A}
{\cal A}^{*\nu\sigma'}_{V/A}
\eeq
where the sum refers to the spin and colour of the quarks
and to the colour of the gluon in the final state.
In the Feynman gauge, the sum over the polarization of the final
gluon gives
\beq
\label{eq:sum_over_pol}
\sum_{\rm pol} \e^\s(k)\,\e^{\s'}(k) = -g^{\s\s'}
\eeq
so that we can define
\beq
M_{V/A}=-g_{\sigma\sigma'}\,M_{V/A}^{\sigma\sigma'}\;.
\eeq
We need the expressions for $M_{V/A}$ in $d=4-2\e$
dimensions. Computing the trace in eq.~(\ref{eq:M_Born_}),
\beqn
M_V &=&
8 \frac{x_1^2+x_2^2}{(1-x_1)(1-x_2)} +\frac{16}{(1-x_1)^2(1-x_2)^2}
\(\frac{m^2}{q^2}\) \Bigl[ 2\,x_1x_2(x_1+x_2) \nonumber\\&&
{}-3\,(x_1^2+x_2^2)
-8\,(1-x_1)(1-x_2)+2  \Bigr] - \frac{32}{(1-x_1)^2(1-x_2)^2}
\(\frac{m^2}{q^2}\)^2 x_g^2 \nonumber \\&&
{}-\frac{16\e}{(1-x_1)(1-x_2)}\Biggl[
x_1^2+x_2^2+(1-x_1)(1-x_2)+x_g-1 \nonumber \\&&
{} - \(\frac{m^2}{q^2}\) \frac{x_g^2}
{(1-x_1)(1-x_2)}
\Biggr]+ \frac{8\e^2}{(1-x_1)(1-x_2)} x_g^2\;,
\eeqn
and
\beqn
M_A &=&
8 \frac{x_1^2+x_2^2}{(1-x_1)(1-x_2)} +\frac{16}{(1-x_1)^2(1-x_2)^2}
\(\frac{m^2}{q^2}\) \Bigl[-12 (x_1+x_2-2x_1x_2) \nonumber \\&&
{}-11 x_1 x_2 (x_1+x_2) + 8(x_1^2+x_2^2)
+ x_1^3(x_2-1)+ x_2^3(x_1-1)\nonumber \\&&
{}+2 x_1^2 x_2^2  + 4\Bigr] (1-\e)
+ \frac{64}{(1-x_1)^2(1-x_2)^2}
\(\frac{m^2}{q^2}\)^2 x_g^2 \,(1-\e)\nonumber \\&&
{}+  \frac{8\e^2}{(1-x_1)(1-x_2)} x_g^2\;,
\eeqn
where $x_1,x_2$ and $x_g$ are defined by~(\ref{eq:x1x2y})
and~(\ref{eq:xg_def}).
According to eq.~(\ref{calMdef}) we have
\beq
{\cal M}^{(b)}_{V/A}=\frac{2\pi}{q^2} \; C_F\, g_s^2 \mu^{2\e}
\, M_{V/A}\;,
\eeq
where $\mu$ is the mass  parameter of dimensional regularization,
introduced in order to keep $g_s$ dimensionless.

For a future use, we introduce now a unit three-vector $\vet{j}$,
belonging to the event plane (i.e. the plane defined by $\vet{p}$,
$\vet{p^\prime}$ and $\vet{k}$) and perpendicular to $\vet{k}$. In the
centre-of-mass system of the process, we can then write
\beq
\label{eq:j_def}
\vet{j}^2 = 1\;,  \hspace{1cm} \vet{k}\cdot \vet{j} = 0\;,
\hspace{1cm}  \vet{j} = a\,\vet{p}+ b\, \vet{k}\;.
\eeq
The generalization of these equations is easily obtained. From the
fact that $j$ is a unit, purely space-like vector, we have
\beq
\label{eq:j2}
j^2 = j_0^2 -\vet{j}^2 = -1  \quad \Longrightarrow \quad j_0=0\;.
\eeq
Since $q=\(q_0,\vet{0}\)$
\beq
\label{eq:j_conditions}
\begin{array}{l}
q\cdot j = q_0 j_0 - \vet{q}\cdot \vet{j} = 0\\
k\cdot j = k_0 j_0 - \vet{k}\cdot \vet{j} = 0 \;,\\
\end{array}
\eeq
and
\beq
\label{eq:def_j}
j = a\,p+ b\, k + c\, q\;.
\eeq
We can then determine the coefficients $a,b$ and $c$ by imposing the
validity of eqs.~(\ref{eq:j_conditions})
\beqn
q\cdot j &= & 0 \quad\Longrightarrow\quad a\,(p\cdot q) + b \,(k\cdot q) +
c\, q^2 = 0\nn\\
k\cdot j &= & 0 \quad\Longrightarrow\quad a\,(p\cdot k) + c\,(q\cdot k)
= 0  \;,
\eeqn
that, once solved, give
\beq
a = \frac{N}{p\cdot k}\;,  \quad\quad\quad  c =-\frac{N}{q \cdot k} \;,
\quad\quad\quad b = N\, \frac{\displaystyle  \frac{q^2}{q \cdot k} -
\frac{p\cdot q}{p\cdot k}} {q \cdot k }\;,
\eeq
with $N$ determined by the normalization condition~(\ref{eq:j2}).
From the expression of $j$ we can compute
\beqn
\label{eq:pdj}
p\cdot j &=& -\frac{1}{2\,x_g}\sqrt{q^2}
\Big[ x_2^2\(4x_1-\rho-4\) + x_1^2\(4x_2-\rho-4\) -2 x_1 x_2 (\rho+6)
\nn\\
&& {}  + 4\(\rho+2\)\(x_1+x_2\)-4(\rho+1)
\Big]^\frac{1}{2} \;.
\eeqn
We are now in a position to give a decomposition of the tensor
$M^{\sigma\sigma'}$ of eq.~(\ref{eq:M_Born_}) (we disregard from now
on the suffix $V/A$) according to the direction of $j$: in fact, the
Lorentz indexes of this tensor take origin from the composition of the
three independent vectors $q$, $p$ and $k$. Using eq.~(\ref{eq:def_j}), we
substitute $j$ instead of $p$ so that we can write $M^{\sigma\sigma'}$
in a quite general form
\beq
\label{Mform}
M^{\sigma\sigma'} = M^\perp g^{\sigma\sigma'}_\perp +
M^j j^\sigma j^{\sigma'} + \mbox{terms involving $q$ or $k$}\;,
\eeq
where (see eq.~(\ref{eq:g_perp}))
\beq
g_\perp^{\s\s'} \equiv g^{\s\s'} - \frac{\eta^{\s}k^{\s'} +
\eta^{\s'}k^{\s}}{\eta\cdot k}\;,
\eeq
with $\eta=(k_0,-\vet{k})$, so that
\beq
\eta^2 = 0\;, \quad\quad\quad  \eta\cdot j = 0 \;,
\quad\quad\quad q_\s\, g_{\perp}^{\s\s'}=0\;.
\eeq
Contracting eq.~(\ref{Mform}) with $g_{\perp\s\s'}$ and with $j_\s
j_{\s'}$, we obtain, in $d=4$ dimensions
\beq
M^j = M^\s_\s + 2\, M^{\s\s'}j_\s j_{\s'}\;.
\eeq
This expression can easily be computed, starting from the
consideration of the previous paragraph about the dependence of
$M^{\s\s'}$ from $q$, $p$, $k$, and from eqs.~(\ref{eq:j_conditions})
and~(\ref{eq:pdj})
\beqn
M^j_{V/A}&=&  \frac{2\,c_{\scriptscriptstyle V/A}}{(1-x_1)^2(1-x_2)^2}
\Big[ 4x_1x_2(x_1+x_2)
-\rho (x_1+x_2)^2 -4(x_1^2+x_2^2)  \nonumber \\&&
{}-12 x_1 x_2+ 4(\rho+2)(x_1+x_2)-4(\rho+1) \Big]\;,
\eeqn
where
\beq
c_{\scriptscriptstyle V} = \rho + 2 \;, \quad\quad\quad\quad
c_{\scriptscriptstyle A} = -2(\rho-1)\;.
\eeq
We also define, consistently with our previous notation
\beq
\label{eq:Mcalssp}
{\cal M}^{(b)\s\s'}_{V/A} = \frac{2\pi}{q^2} \; C_F\, g_s^2 \mu^{2\e}
\, M^{\s\s'}_{V/A}\;,\quad\quad\quad\quad
{\cal M}^{(b)\perp/j}_{V/A} = \frac{2\pi}{q^2} \; C_F\, g_s^2 \mu^{2\e}
\, M^{\perp/j}_{V/A}\;.
\eeq
In the cases when the $V/A$ suffix needs not be specified,
we will simply write ${\cal M}_b^{\s\s'}$, ${\cal M}_b^\perp$ and
${\cal M}_b^j$.

\section{Virtual contributions}
\label{sec:virtual}
\begin{figure}[htb]
\centerline{\epsfig{figure=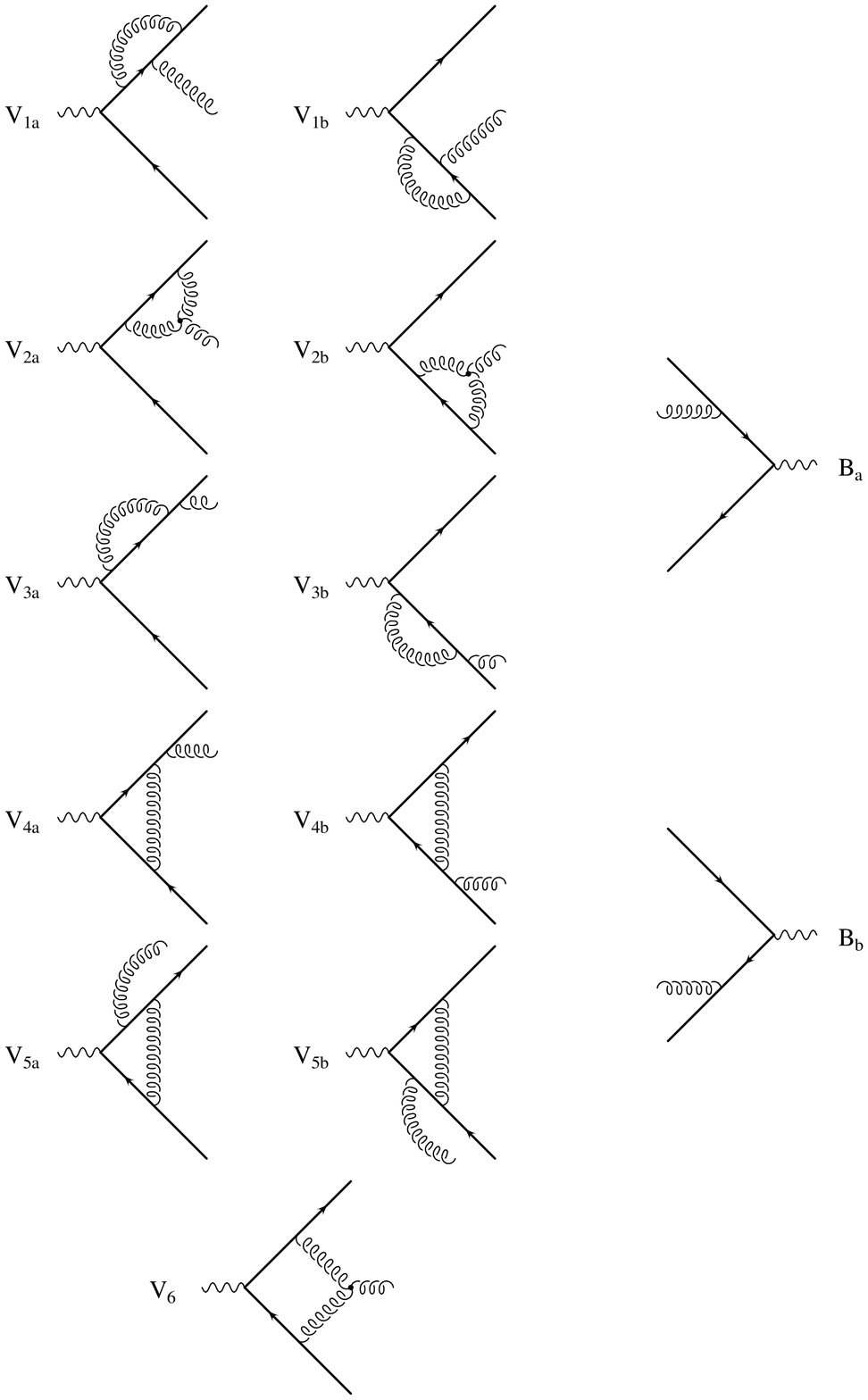,
width=0.787\textwidth,clip=}}\ccaption{}{ \label{fig:virtual}
Radiative corrections to the process $Z/\gamma\,\to\,Q\Qb g$. The
interference of these diagrams (left hand side of the figure) with the
tree-level Born term (right hand side) gives rise to contributions to
order $\as^2$.}
\end{figure}

Corrections to the three-jet decay rate to order $\as^2$
come from the interference of the one-loop graphs
with the tree-level Born ones.
In Fig.~\ref{fig:virtual} we have depicted these contributions.

The algebra to compute these terms in $d=4-2\ep$ dimensions has been
carried out in a straightforward way, using a MACSYMA program, which
reduces the original Feynman graphs to a linear combination of
scalar, one-loop integrals. The scalar integrals have been computed
analytically and their values are listed in Appendix~\ref{app:one_loop}.

Loop corrections to on-shell external lines (not illustrated in
Fig.~\ref{fig:virtual}) require particular attention.

We start from the self-energy corrections to heavy-flavour external
lines. As described in Appendix~\ref{app:mass_cnt}, the effect of the
fermion self-energy correction to an external line, including the
mass counterterm, is equivalent to multiply the external propagator by
the factor $Z_Q$ of eq.~(\ref{eq:Z_Q}), so that we have a contribution
to ${\cal M}_v$ equal to
\beq
z_q\times {\cal M}_b = - N_\ep C_F g_s^2 \left(\frac{\mu^2}{m^2}\right)^\e
\left(\frac{3}{\ep}+4\right)\times {\cal M}_b\;.
\eeq

\begin{figure}[htb]
\centerline{\epsfig{figure=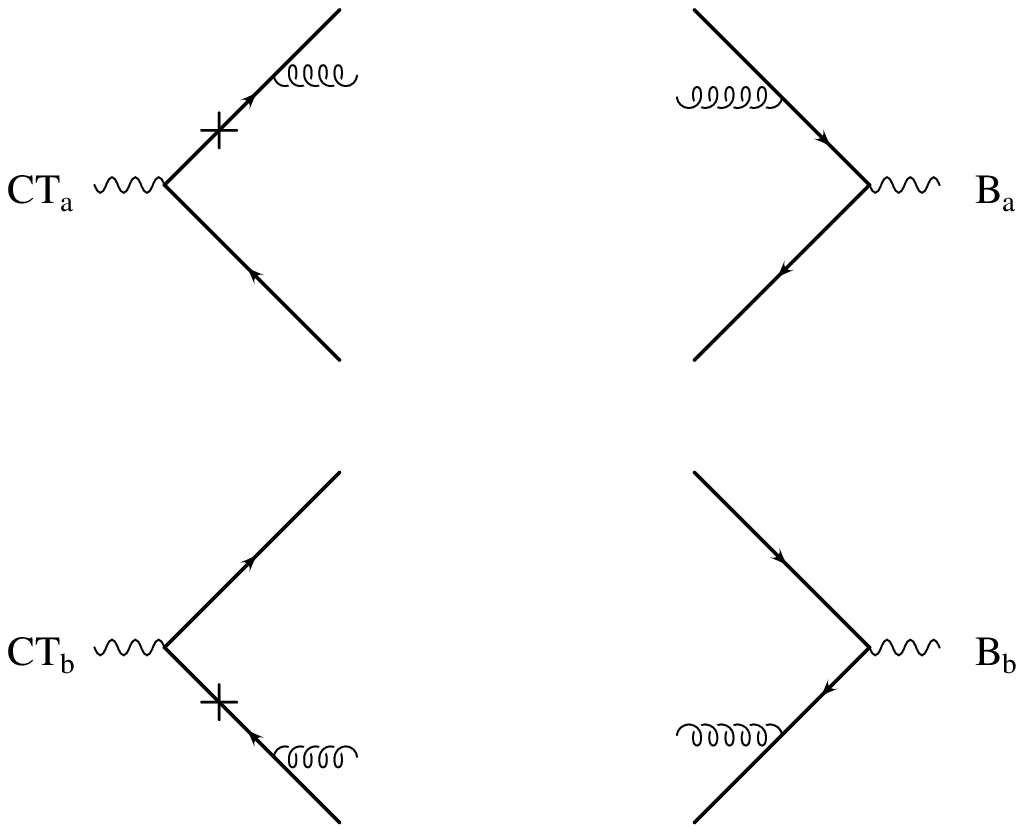,
width=0.65\textwidth,clip=}}\ccaption{}{ \label{fig:counterterm}
Contributions to order $\as^2$ coming from the interference between
diagrams with the mass counterterm insertion
and the tree-level Born term.}
\end{figure}

We have to consider also the diagrams obtained with the mass counterterm
insertion in internal fermion lines. In fact, according to
eq.~(\ref{eq:mass_counterterm}), we have to add a counterterm $m_c$ in
order to keep $m$ as the pole mass, after radiative corrections.
These  diagrams are depicted in Fig.~\ref{fig:counterterm}.

Similar considerations apply to the self-energy corrections to
external gluon lines.
As remarked in Appendix~\ref{app:gluon_renorm},
gluon, ghost and light-fermion self-energy corrections to external
gluon lines vanish in dimensional regularization.
Only the correction coming from a heavy-flavour loop needs be
considered, and it gives a contribution at order $\as^2$ equal to (see
eq.~(\ref{eq:Z_g}))
\beq
z_g  \times {\cal M}_b = - N_\ep T_F g_s^2
\left(\frac{\mu^2}{m^2}\right)^\ep \frac{4}{3\ep} \times {\cal M}_b\;.
\eeq

After that, charge renormalization  is all that is needed,
since we are computing a physical cross section.
Charge renormalization in the mixed scheme of Ref.~\cite{CWZ} is
described in Appendix~\ref{app:charge_renorm}.
From eq.~(\ref{eq:charge_ren}) we can see that the last correction to our
virtual term is equal to
\beq
z_{\as} \times  {\cal M}_b =  g_s^2 N_\e
\left[ \( \frac{4}{3\e}T_F\,n_{\rm lf} -\frac{11}{3\e}C_A \)
+ \(\frac{\mu^2}{m^2}\)^\e
\frac{4}{3\e}T_F \right]\times  {\cal M}_b\;.
\eeq

We can now summarize the combined effect of external line corrections
and renormalization to be included with ${\cal M}_v$
\beq
N_\ep  g_s^2  \left(\frac{\mu^2}{m^2}\right)^\ep \Bigg\{
- 2\, C_F
\left(\frac{3}{\ep}+4\right)
+ \( \frac{4}{3\e}T_F\,n_{\rm lf} -\frac{11}{3\e}C_A \)
\left(\frac{\mu^2}{m^2}\right)^{-\ep}
\Bigg\} \times {\cal M}_b\;.
\eeq
The factor of 2 in front of the fermion external line corrections is
to account for the two fermion lines.
%%%%%%%%%%%%%%%%%%%%%%%%%%%%%%%%%%%%%%%%%%%
%%%%%%%%%%%%%%%%%%%%%%%%%%%%%%%%%%%%%%%%%%%

\clearpage

%%%%%%%%%%%%%%%%%%%%%%%%%%%%%%%%%%%%%%%%%%%
%%%%%%%%%%%%%%%%%%%%%%%%%%%%%%%%%%%%%%%%%%%

\section{Hagiwara contributions}
\label{sec:hagiwara}
%%%%%%%%%%%%%%%%%%%%%%%%%%%%%%%%%%%%%%%%%%%
\begin{figure}[htb]
\centerline{\epsfig{figure=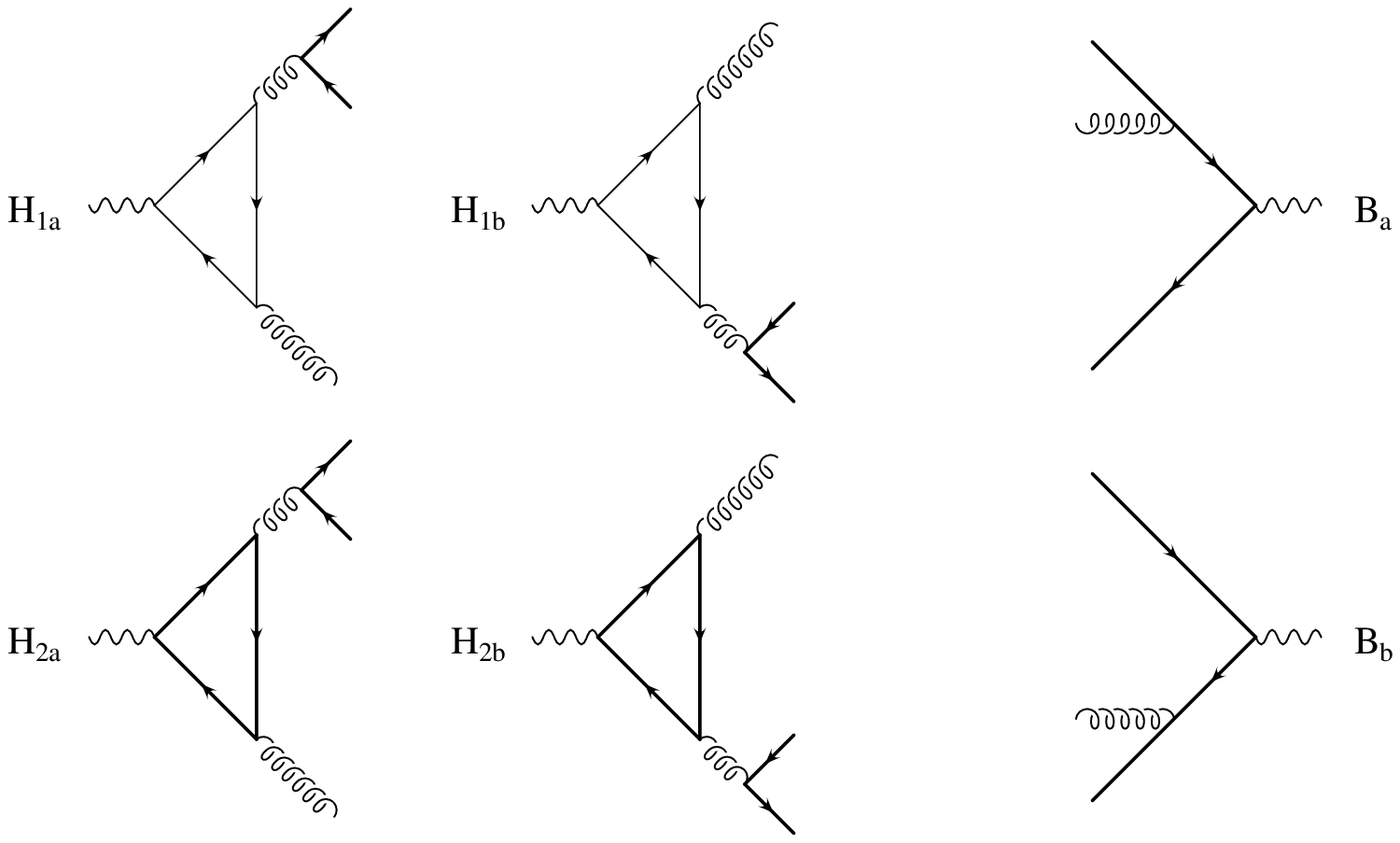,
width=0.85\textwidth,clip=}}\ccaption{}{ \label{fig:hagiwara} Radiative
corrections to the process $Z/\gamma\,\to\,Q\Qb g$ (Hagiwara
contributions).  The interference of these diagrams (left hand side of
the figure) with the tree-level Born term (right hand side) gives rise
to contributions to order $\as^2$.}
\end{figure}
%%%%%%%%%%%%%%%%%%%%%%%%%%%%%%%%%%%%%%%%%%%
At order $\as^2$, we have also contributions coming from the
interference between terms in which the weak current is coupled to the
heavy quarks and to quarks of different flavours.

These diagrams are represented in Fig.~\ref{fig:hagiwara}: the higher part
of the figure describes the contributions coming from light-quark loops,
while the lower part describes the contributions coming from
heavy-quark loops.

By C-invariance (Furry's theorem), these diagrams   vanish for vector
currents.
For axial currents, they cancel in pairs of up-type and down-type
quarks, because they have opposite axial coupling (see
eqs.~(\ref{eq:coupling_constants})), as long as the loop of different
flavour may be regarded as massless.
Thus, the up-quark contribution cancels with the down-quark, and, if
the charm mass is neglected, the charm contribution cancels with the
strange. Only the graph with a top quark loop remains.

Paired to this last type of diagrams are the contributions
where the massive quark in the loop is the same as the heavy quark in
the final state.

We have not included  these diagrams in our calculation,
because they have been computed in Ref.~\cite{Hagiwara}, where it was
shown that their contribution is of order of $1\%$.

%\clearpage

%\input{phd_real}
\section{Real contributions}
\label{sec:real}
The square of the Feynman diagrams with four particles in the final
state gives rise to the real contributions.
These amplitudes are easily obtained, in an analytic form, with a
little ``Diracology''  in $d=4$ dimensions.
The behavior of these amplitudes in the soft and collinear limit is
obtained in the next chapter, where the asymptotic forms of the
amplitudes in derived in $d=4-2\e$ dimensions.

\subsection[Real contributions to the  $Q \Qb gg$
cross section]{Real contributions to the {\mylarge $Q \Qb gg$}
cross section}
\label{sec:QQgg}
\begin{figure}[htb]
\centerline{\epsfig{figure=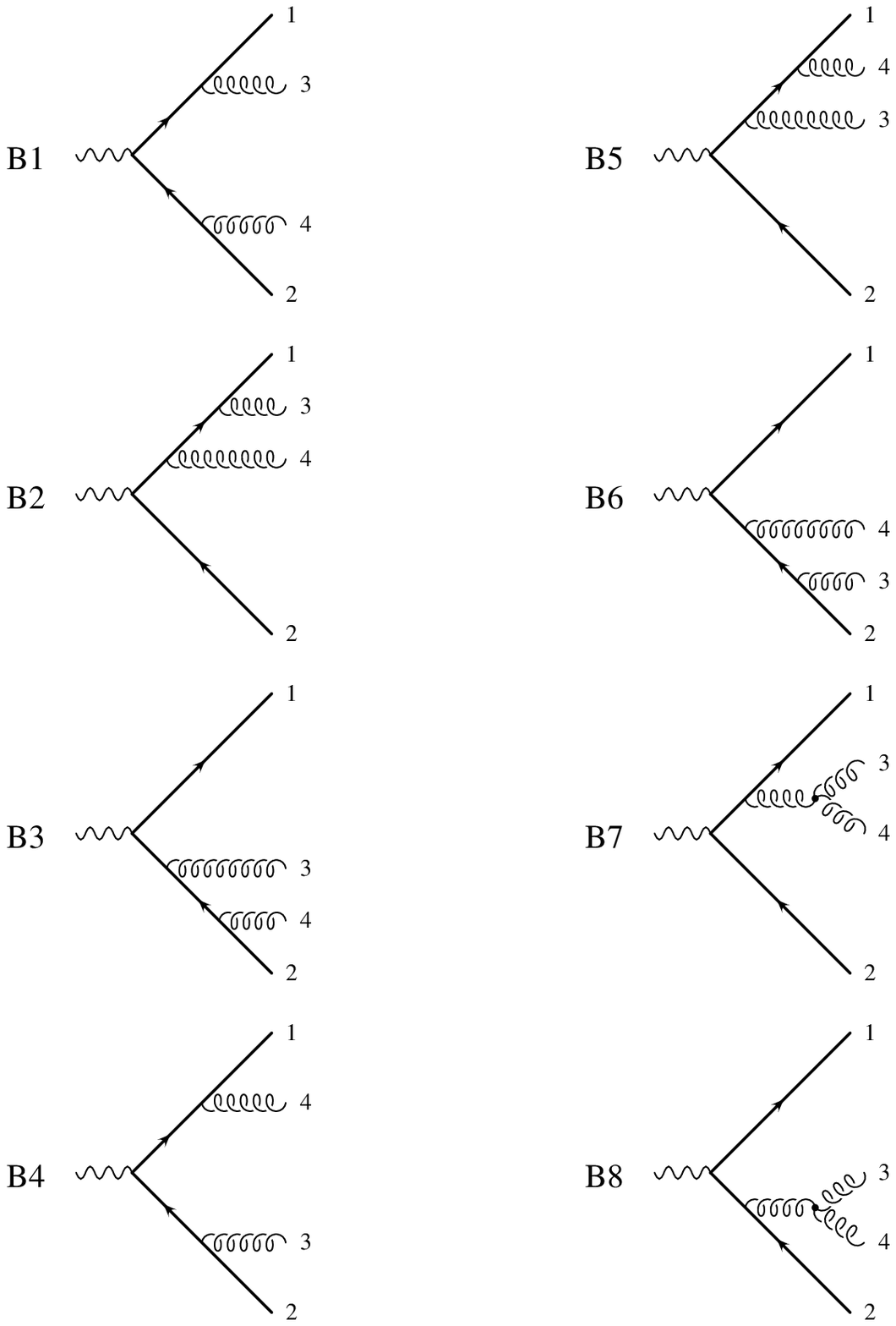,
width=0.7\textwidth,clip=}}\ccaption{}{ \label{fig:phd_gg}
Feynman diagrams contributing to the process
$Z/\gamma\,\to\,Q\Qb g g$.}
\end{figure}

The diagrams contributing to the process
\beq
e^+ e^-\, \to\, Z/\gamma\,\to\, Q \Qb g g
\eeq
are depicted in Fig.~\ref{fig:phd_gg}. From the square of these eight
diagrams, we can obtain thirty-six terms, but most of them are related by
interchange of the momentum labels, so that only thirteen
amplitudes need  be considered.
We follow the notation used in Ref.~\cite{ERT}, and indicate with
Bij the interference of diagram Bi with diagram Bj (i$\geq$ j).

\begin{figure}[htb]
\centerline{\epsfig{figure=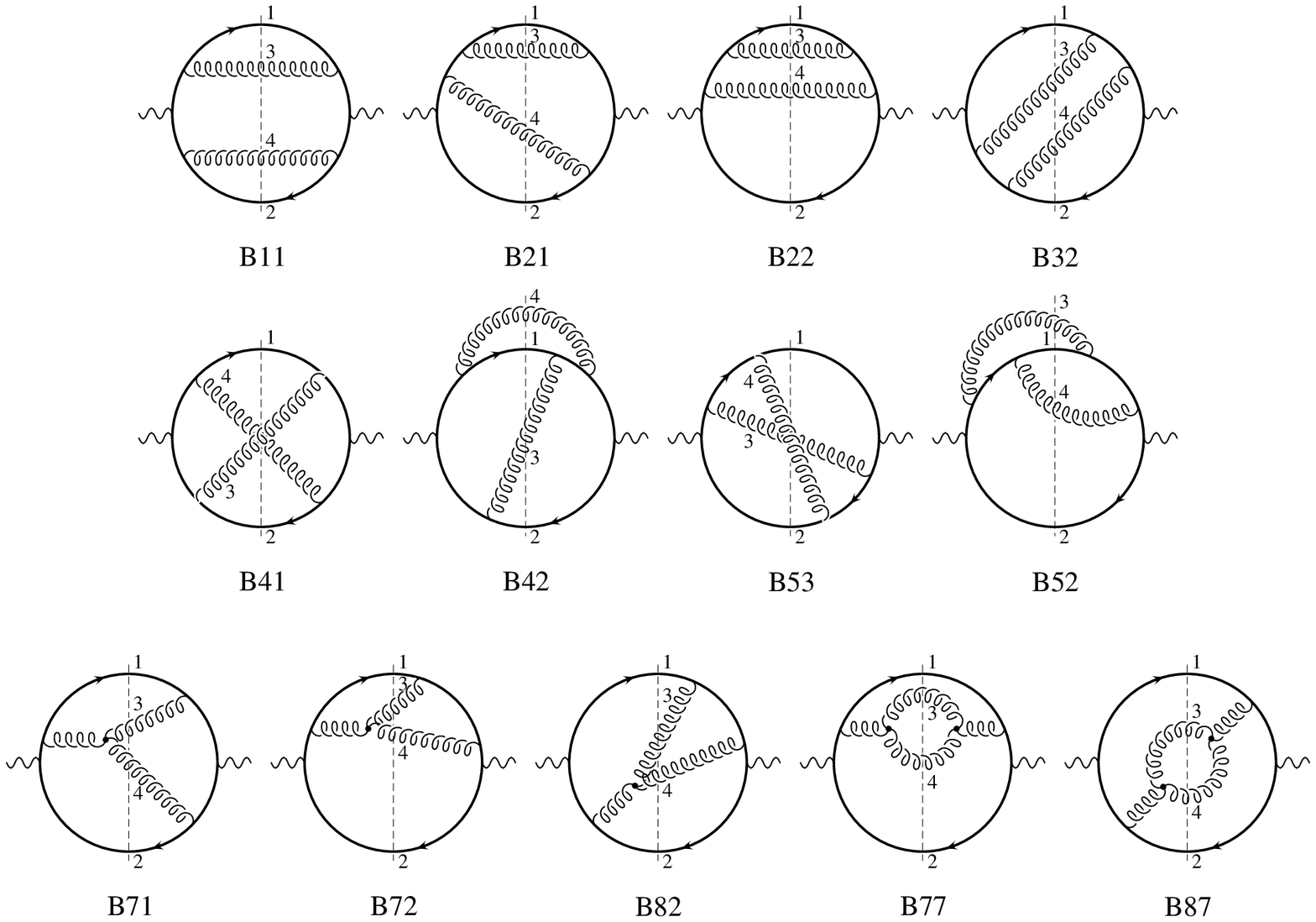,
width=1.1\textwidth,clip=}}\ccaption{}{ \label{fig:phd_ggcut}
The thirteen amplitudes necessary to complete the computation of the cross
section for the process $Z/\gamma\,\to\,Q\Qb g g$.
Cut propagators refer to on-shell particles, while numbers label the
different momenta.}
\end{figure}

The different contributions to the cross section can be classified
into three classes, according to their colour and spatial structure. We
will always factorize out the colour factor common to the Born term,
equal to $C_F\,N_c$, so that we have:
\begin{enumerate}
\item {\bf ${\mathbf C_F}$ class:} planar QED-type diagrams
\item{\bf ${\mathbf C_F-\frac{1}{2}C_A}$ class:} non-planar QED-type
graphs
\item{\bf ${\mathbf C_A}$ class:} QCD graphs, involving the
three-gluon vertex
\end{enumerate}
where $C_A = N_c$.
In Tab.~\ref{tab:gg} we collect the thirty-six terms and the label
interchanges, needed to compute all of them from the thirteen we have
calculated (first row), and that are represented in
Fig.~\ref{fig:phd_ggcut}. In the first row of this figure, there are
the  contributions belonging to the first class, in the second, the
contributions to the second class and in the last row, the
contributions to the third class.
\begin{table}[htbp]
\begin{center}
\leavevmode
\begin{tabular}{||c|c|c|c||}
\hline\hline
permutation & {\bf ${\mathbf C_F}$ class} & {\bf ${\mathbf
C_F-\frac{1}{2}C_A}$ class} & {\bf ${\mathbf C_A}$ class} \\
\hline
& B11 \sp B21 \sp B22 \sp B32
& B41 \sp B42 \sp B53 \sp B52
& B71 \sp B72 \sp B82 \sp B77 \sp B87 \\
$(1 \leftrightarrow 2)$
& \phantom{B11} \sp B64 \sp B66 \phantom{B32}
& \phantom{B41} \sp B61 \sp \phantom{B53} \sp \phantom{B52}
& B84 \sp B86 \sp B76 \sp B88 \sp \phantom{B87} \\
$(3 \leftrightarrow 4)$
& B44 \sp B54 \sp B55 \sp B65
& \phantom{B41} \sp B51 \sp B62 \sp \phantom{B52}
& B74 \sp B75 \sp B85 \sp \phantom{B77} \sp \phantom{B87} \\
$(1 \leftrightarrow 2)\,(3 \leftrightarrow 4)$
& \phantom{B11} \sp B31 \sp B33 \sp \phantom{B32}
& \phantom{B41} \sp B43 \sp \phantom{B53} \sp B63
& B81 \sp B83 \sp B73 \sp \phantom{B77} \sp \phantom{B87} \\
\hline\hline
\end{tabular}
\caption{Label interchanges needed to compute the thirty-six terms
contributing to the process $e^+ e^-\, \to\, Z/\gamma\,\to\,Q\Qb g g$.}
\label{tab:gg}
\end{center}
\end{table}
One final remark is needed, in order to sum over the final gluon
polarizations. In fact, due to non-conservation of gluon current, we
can use eq.~(\ref{eq:sum_over_pol}) to sum over polarization only if
we include B7- and B8-like diagrams with ``external'' ghost.

%%%%%%%%%%%%%%%%%%%%%%%%%%%%%%%%%%%%%%%%%%%
%%%%%%%%%%%%%%%%%%%%%%%%%%%%%%%%%%%%%%%%%%%

\clearpage
%%%%%%%%%%%%%%%%%%%%%%%%%%%%%%%%%%%%%%%%%%%
%%%%%%%%%%%%%%%%%%%%%%%%%%%%%%%%%%%%%%%%%%%

\subsection[Real contributions to $ Q\Qb q\qb$ cross
section]
{Real contributions to the {\mylarge $Q\Qb q\qb$} cross section}
\label{sec:QQqq}
%%%%%%%%%%%%%%%%%%%%%%%%%%%%%%%%%%%%%%%%%%%
\begin{figure}[htb]
\centerline{\epsfig{figure=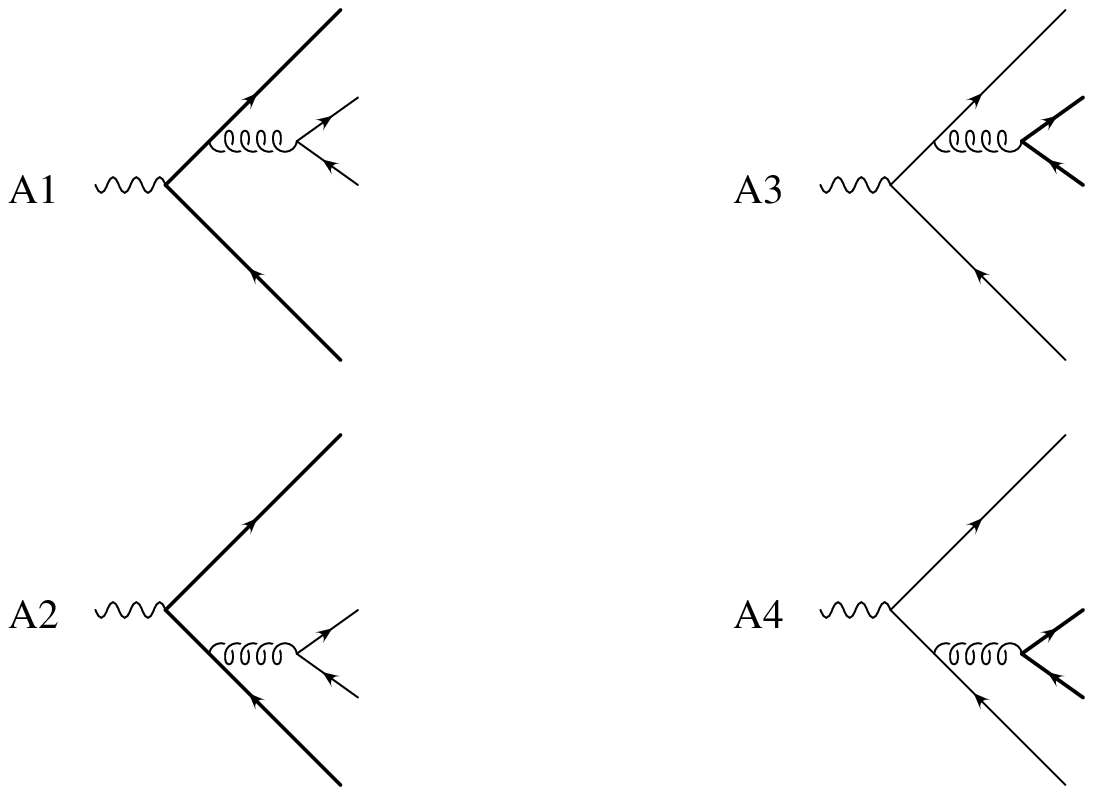,
width=0.67\textwidth,clip=}}\ccaption{}{ \label{fig:phd_2q2m}
Feynman diagrams contributing to the process
$Z/\gamma\,\to\,Q\Qb q \qb$.}
\end{figure}
%%%%%%%%%%%%%%%%%%%%%%%%%%%%%%%%%%%%%%%%%%%
The diagrams contributing to the process
\beq
e^+ e^-\, \to\, Z/\gamma\,\to\, Q \Qb q \qb
\eeq
are illustrated in Fig.~\ref{fig:phd_2q2m}.
%%%%%%%%%%%%%%%%%%%%%%%%%%%%%%%%%%%%%%%%%%%
\begin{figure}[htb]
\centerline{\epsfig{figure=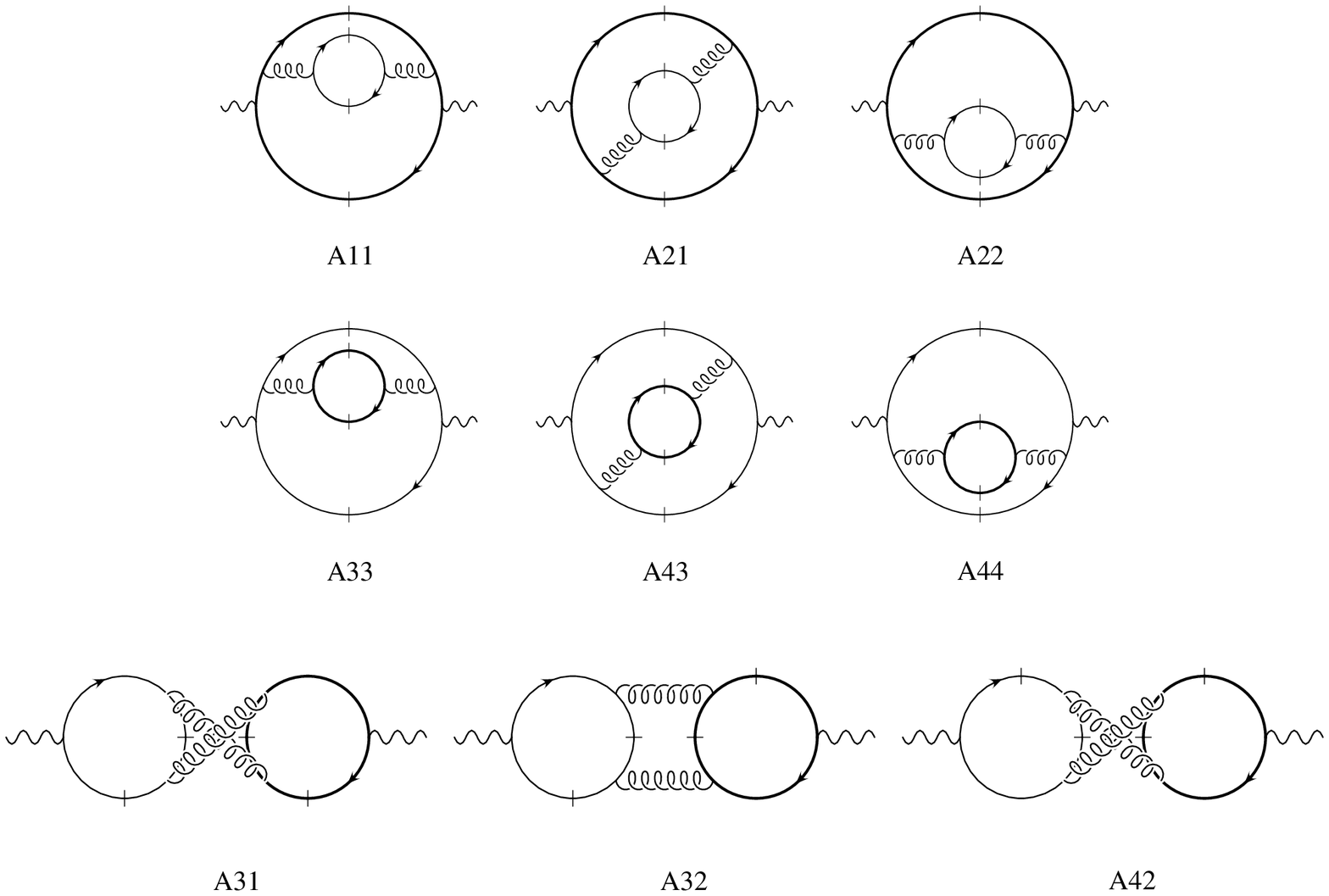,
width=1.1\textwidth,clip=}}\ccaption{}{ \label{fig:phd_2q2m_cut}
The nine amplitudes necessary to complete the computation of the cross
section for the process $Z/\gamma\,\to\,Q\Qb q \qb$.
Cut propagators refer to on-shell particles.}
\end{figure}
%%%%%%%%%%%%%%%%%%%%%%%%%%%%%%%%%%%%%%%%%%%
From the square of these four diagrams, we generate ten terms.
In Fig.~\ref{fig:phd_2q2m_cut} we have depicted nine of them, because
the tenth, A41, can be obtained with the massive loop of diagram
A31 and the massless loop of diagram A42.

Although the colour coefficient is the same for all the ten diagrams
(${\mathbf T_F}$), their infrared structure is different: in fact, only
the diagrams in the first row of Fig.~\ref{fig:phd_2q2m_cut} contain
infrared divergences, due to the massless-quark collinear region, all
the other diagrams being finite.

For this reason, it is mandatory to include the first three diagrams,
to check infrared cancellation, and it is custom to assign the
other diagrams to the light channel: in fact, they are, in general,
characterized by a large invariant mass for the light quarks and a
small invariant mass for the heavy-quark couple. We have not included
these last diagrams in our calculation.

\subsection[Real contributions to the $Q\Qb Q\Qb$ cross
section]
{Real contributions to the {\mylarge $Q\Qb Q\Qb$} cross section}
\label{sec:QQQQ}
%%%%%%%%%%%%%%%%%%%%%%%%%%%%%%%%%%%%%%%%%%%
\begin{figure}[htb]
\centerline{\epsfig{figure=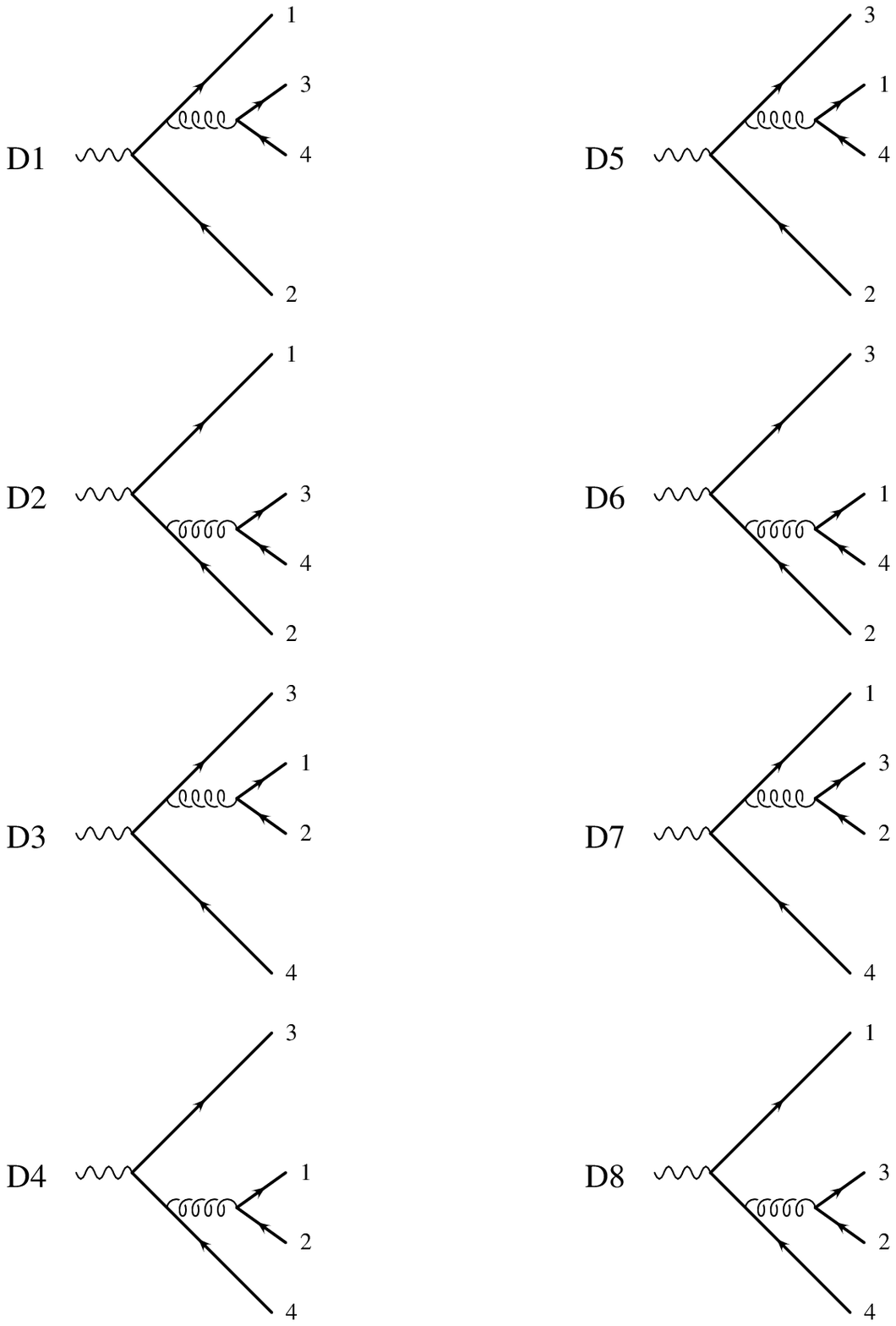,
width=0.63\textwidth,clip=}}\ccaption{}{ \label{fig:phd_4m}
Feynman diagrams contributing to the process
$Z/\gamma\,\to\,Q\Qb Q \Qb$.}
\end{figure}
%%%%%%%%%%%%%%%%%%%%%%%%%%%%%%%%%%%%%%%%%%%
The eight diagrams contributing to the process
\beq
e^+ e^-\, \to\, Z/\gamma\,\to\, Q \Qb Q \Qb
\eeq
are illustrated in Fig.~\ref{fig:phd_4m}.
%%%%%%%%%%%%%%%%%%%%%%%%%%%%%%%%%%%%%%%%%%%
\begin{figure}[htb]
\centerline{\epsfig{figure=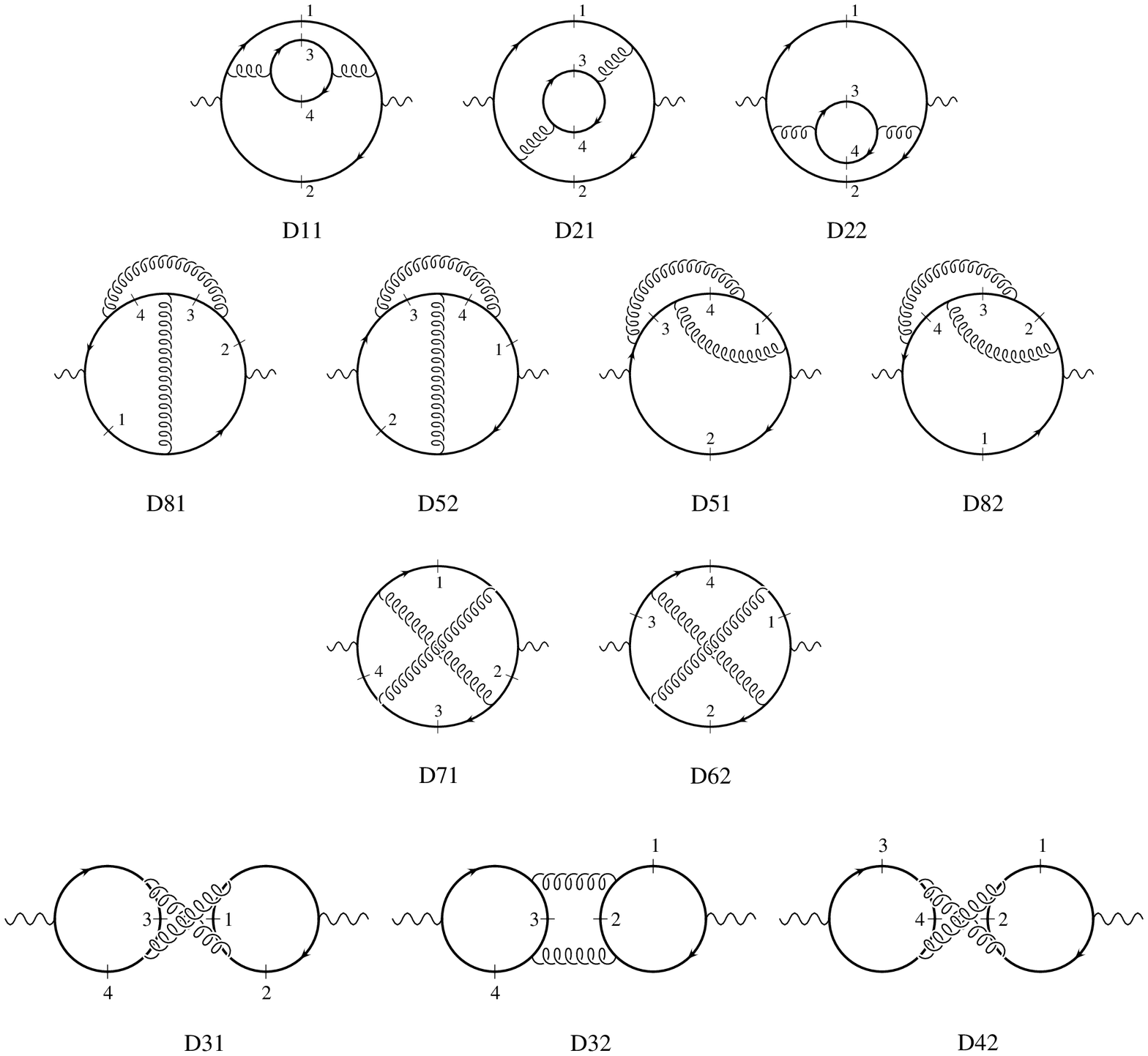,
width=1.1\textwidth,clip=}}\ccaption{}{ \label{fig:phd_4mcut}
The twelve amplitudes necessary to complete the computation of the cross
section for the process $Z/\gamma\,\to\,Q\Qb Q \Qb$.
Cut propagators refer to on-shell particles, while numbers label the
different momenta.}
\end{figure}
%%%%%%%%%%%%%%%%%%%%%%%%%%%%%%%%%%%%%%%%%%%
From the square of these diagrams we have thirty-six infrared-safe
amplitudes. We need to compute only twelve of them, that we have
represented in Fig.~\ref{fig:phd_4mcut}, because the others can be
obtained by permuting the quark momenta.  In Tab.~\ref{tab:QQQQ} we
give the twelve amplitudes we have computed (first row) and the label
interchanges necessary to build all the others.
As previously done for the $Q\Qb g g$ process, we divide these
contributions according to their colour factors and according to their
structure:
\begin{enumerate}
\item {\bf ${\mathbf 1^{\bf st}\ T_F}$ class:} diagrams illustrated in the
first row of Fig.~\ref{fig:phd_4mcut}
\item{\bf ${\mathbf C_F-\frac{1}{2}C_A}$ class:} diagrams from D81 to
D62
\item {\bf ${\mathbf 2^{\bf nd}\ T_F}$ class:} diagrams
composed by two different massive loops, coupled to the weak current, that
can be split apart by cutting the two joining gluon lines (singlet
contributions).
\end{enumerate}

\begin{table}[htbp]
\begin{center}
\leavevmode
\begin{tabular}{||c|c|c|c||}
\hline\hline
permutation & {\bf ${\mathbf 1^{\bf st}\ T_F}$ class} & {\bf ${\mathbf
C_F-\frac{1}{2}C_A}$ class} & {\bf ${\mathbf 2^{\bf nd}\ T_F}$ class} \\
\hline
& D11 \sp D21 \sp D22
& D81 \sp D52 \sp D51 \sp D82 \sp D71 \sp D62
& D31 \sp D32 \sp D42 \\
$(1 \leftrightarrow 3)$
& D55 \sp D65 \sp D66
& D54  \sp D61 \sp \phantom{D51} \sp D64 \sp D53 \sp \phantom{D62}
& D75 \sp D76 \sp D86\\
$(2 \leftrightarrow 4)$
& D77 \sp D87 \sp D88
& D72 \sp D83 \sp D73 \sp \phantom{D82} \sp \phantom{D71} \sp D84
& \phantom{D75} \sp D85 \sp \phantom{D86}\\
$(1 \leftrightarrow 3)\,(2 \leftrightarrow 4)$
& D33 \sp D43 \sp D44
& D63 \sp D74 \sp \phantom{D73} \sp \phantom{D64} \sp \phantom{D53}
\sp \phantom{D84}
& \phantom{D31} \sp D41 \sp \phantom{D42} \\
\hline\hline
\end{tabular}
\caption{Label interchanges needed to compute the thirty-six terms
contributing to the process $e^+ e^-\, \to\, Z/\gamma\,\to\,Q\Qb Q\Qb$.}
\label{tab:QQQQ}
\end{center}
\end{table}

%\clearpage

\chapter{Infrared cancellation}
\label{chap:IR_canc}
\thispagestyle{plain}
\section{The subtraction method}
\label{sec:subtraction}
In this section we want to introduce the method we have used in order
to deal with the infrared divergences.
In fact, as we have seen from the previous sections,
infrared divergences arise both in the virtual  and in the real
terms:
\begin{enumerate}
\item in the virtual terms of Sec.~\ref{sec:virtual}, divergences arise
from loop integration, as you can  see from Appendix~\ref{app:one_loop}.
Since we have chosen to regularize our integrals with a dimensional
method, we obtain poles in $\e$;

\item in the real terms of Secs.~\ref{sec:QQgg} and~\ref{sec:QQqq},
divergences in the differential cross section appear when we
integrate over a particular region of the phase space.  In
the diagrams depicted in these sections, we have propagators
proportional to $1/y$ (according to eq.~(\ref{eq:x1x2y}), $y$ is
proportional to the mass of the light system), so that we have
divergences when integrating over the phase space region of small
$y$. In this  region,  a four-particle final state  masquerades as a
three-particle event, because one of the emitted particles becomes
soft, or two particles become collinear. Notice that
the Born and the virtual graphs always have $y=0$.

\end{enumerate}
According to Bloch-Nordsiek~\cite{BN} and
Kinoshita-Lee-Nauenberg~\cite{KLN}  theorems, infrared divergences
must cancel for sufficiently inclusive physical quantities.

Now, we show  how we have checked this cancellation.
If we compute the  ${\cal O}\(\as^2\)$
differential cross section, in $d=4-2 \epsilon $ dimensions,
we  get a result of the form
\beq
\label{eq:d_sigma_formale}
d\sigma =    d\sigma^{(0)} +
\left(\frac{\as}{2\pi}\right) d\sigma^{(1)}
+ \left(\frac{\as}{2\pi}\right)^2 d\sigma^{(2)}\;,
\eeq
with
\beq
\label{eq:dsigma_all}
d\sigma^{(2)} = \frac{ d\sigma_2^{(2)}}{d\Phi_2} d\Phi_2+
\frac{ d\sigma_3^{(2)}}{d\Phi_3} d\Phi_3
+ \frac{ d\sigma_4^{(2)}}{d\Phi_4} d\Phi_4 \;,
\eeq
where we sum up the contributions  coming from
two-, three- and four-particle final state.
The  two-particle final-state contribution at this order arises from the
interference of two-loop diagrams with the tree-level term.
We have not computed this differential cross section term,
because we want to calculate
only three-jet-related quantities, that give zero contribution in the
two-jet limit.
We give a detailed description of this fact in
Sec.~\ref{sec:two_body}, so that  we  neglect
$d\sigma_2^{(2)}$ in eq.~(\ref{eq:dsigma_all}).

Using eqs.~(\ref{eq:ps3}) and~(\ref{eq:ps4_inter}), we can rewrite the
three- and four-body phase space as
\beqn
d\Phi_3 &=& dx_1\,dx_2\,J_3(x_1,x_2) \nn \\
d\Phi_4 &=& dx_1\,dx_2\,dy\,d\theta\,d\phi\,J_4(x_1,x_2,y,\theta,\phi)\;,
\eeqn
where $J_3$ and $J_4$ represent all the other factors, whose exact
structure we do not need to know in this section.

In order to implement the cancellation of the soft and collinear
singularities, we now imagine to compute some physical quantity $G$
($G$ stands for ``Generic''), which depends on the
final-state variables.
$G$ may be a combination of theta functions that characterize a
histogram bin for some infrared-safe shape variable, like the thrust,
the $c$ parameter and the heavy-jet mass.
In general the definition of $G$ is specified for any number of
particles in the final state.  Since we are  dealing with three-
and four-parton final states, $G$ is
characterized by only two functions, $G_{3}(x_1,x_2)$ and
$G_{4}(x_1,x_2,y,\theta,\phi)$.
Soft and collinear finiteness of $G$ requires that
\beq\label{eq:irsafeness}
\lim_{y\to 0} G_{4}(x_1,x_2,y,\theta,\phi)=G_{3}(x_1,x_2)\;.
\eeq
When computing the contributions to $G$ at second order in $\as$, we have
\beqn
\int  d\sigma^{(2)}  G &=& \int dx_1\,dx_2\,J_3(x_1,x_2)
\frac{ d\sigma_3^{(2)}}{d\Phi_3}
\,G_3(x_1,x_2)\nonumber \\
&& {}+
\int dx_1\,dx_2\,dy\,d\theta\,d\phi\,
J_4(x_1,x_2,y,\theta,\phi)\, \frac{d\sigma_4^{(2)}}{d\Phi_4}\,
G_4(x_1,x_2,y,\theta,\phi) \;, \phantom{aaaaa}
\eeqn
where each term on the right-hand side contains soft and collinear
divergences that cancel in the sum.
The integration of the complete differential cross
section  is too difficult to be performed
analytically, but it is not a limiting point. In fact, we can add and
subtract a suitable quantity, to obtain
\beq
\eqalign{
\int  d\sigma^{(2)} G = \!
\int\! dx_1 dx_2 \,G_3(x_1,x_2) \Bigg\{
\frac{ d\sigma_3^{(2)}}{d\Phi_3} J_3(x_1,x_2)   +  \!
\int\! dy\,d\theta\,d\phi \frac{d\bar{\sigma}_4^{(2)}}{d\Phi_4}
J_4(x_1,x_2,y,\theta,\phi)\! \Bigg\}   \cr
\,\,\,+\int dx_1\,dx_2\,dy\,d\theta\,d\phi\,J_4(x_1,x_2,y,\theta,\phi)
\Bigg\{\frac{ d\sigma_4^{(2)}}{d\Phi_4}
G_4(x_1,x_2,y,\theta,\phi) - \frac{ d\bar{\sigma}_4^{(2)}}{d\Phi_4}
G_3(x_1,x_2) \!\Bigg\}
\label{eq:subtractionmethod} }
\eeq
where $\bar{\sigma}^{(2)}_4$ is chosen  in such a way that it
has the same soft and collinear singular behaviour of $\sigma^{(2)}_4$,
or, mathematically
\beq
\label{eq:samesing}
\lim_{y\to 0} \,{\displaystyle\frac{\displaystyle\frac{
d\bar{\sigma}_4^{(2)}}{d\Phi_4}}
{\displaystyle\frac{ d \sigma_4^{(2)}}{d\Phi_4} }}\,=\,1\;.
\eeq
The aim of this procedure is to obtain an approximation for
$d\sigma^{(2)}_4$ that can be integrated in $dy\,d\theta\,d\phi$, so
that the divergent parts appear as poles in $\e$.
In this way, the first term of eq.~(\ref{eq:subtractionmethod}) can be
computed analytically:  the single and double poles present in
$d\sigma^{(2)}_3/d\Phi_3$ all cancel with the poles arising from the
$dy\,d\theta\,d\phi$ integration of $d\bar{\sigma}_4^{(2)}/d\Phi_4$,
and thus this term is finite.

The second term in eq.~(\ref{eq:subtractionmethod}), because of
eqs.~(\ref{eq:irsafeness}) and (\ref{eq:samesing}), has no soft or
collinear singularities, and thus can be evaluated directly in four
dimensions\footnote{ Observe that both eq.~(\ref{eq:irsafeness}) and
eq.~(\ref{eq:samesing}) must be satisfied in $d$ dimensions in order
for this argument to apply.}.

To implement numerically the computation of this term, we first
generate a four-body configuration $x_1,x_2,y,\theta,\phi$, in the
corresponding four-body phase space, with weight $J_4$.  We associate
to this configuration two events: one four-body event, with kinematics
$x_1,x_2,y,\theta,\phi$ and weight $d\sigma^{(2)}_4/d\Phi_4$, and one
three-body event, with kinematics $x_1$, $x_2$, $y=0$, and weight
$-d\bar{\sigma}_4^{(2)}/d\Phi_4$.  The computation of a shape variable
using the above scheme reproduces exactly the second term of
eq.~(\ref{eq:subtractionmethod}).

\section{Two-loop diagrams}
\label{sec:two_body}
We want now to show why, in at least two cases, we do not need to
compute the two-body  contribution at order $\as^2$, to the differential
cross section.
\begin{enumerate}
\item If we compute the average value of some physical quantity, we
have
\beqn
\overline{G} \equiv \frac{\int d\s \, G}{\int d\s} &=& \frac{1}{\s}
\lg \s^{(0)} G(1,1,0,\ldots) + \left(\frac{\as}{2\pi}\right)
\int d\s^{(1)} G(x_1,x_2,0,\ldots)
\right.\nn\\
&&\left. {}+ \left(\frac{\as}{2\pi}\right)^2
\int d\s^{(2)} G(x_1,x_2,y,\theta,\phi)
+{\cal O}\(\as^3\) \rg
\eeqn
where $\s$ is obtained by integrating $d\s$ of
eq.~(\ref{eq:d_sigma_formale})
\beq
\s = \s^{(0)} + \left(\frac{\as}{2\pi}\right) \sigma^{(1)}
+ \left(\frac{\as}{2\pi}\right)^2 \sigma^{(2)} +{\cal O}\(\as^3\)\;,
\eeq
and where we no longer label $G$ according to the number of particles
in the final state, which is  completely specified by  the arguments
of $G$ itself:
\begin{itemize}
\item[-] for the two-particle final state, $x_1=x_2=1$ and $y=0$, while
$\theta$ and $\phi$ are undefined, so that $G_{2} = G(1,1,0,\ldots)$
\item[-] for the three-particle final state, we must give  $x_1$ and
$x_2$, but $y=0$, while  $\theta$ and $\phi$  are undefined, so that
$G_3(x_1,x_2) = G(x_1,x_2,0,\ldots)$
\item[-] for the four-particle final state, we have to specify all the
five arguments, so that $ G_4(x_1,x_2,y,\theta,\phi) =
G(x_1,x_2,y,\theta,\phi)$.
\end{itemize}
Adding and subtracting the two-body contributions in the integrals,
we obtain
\beqn
\overline{G} &=& \frac{1}{\s}
\lg \s\, G(1,1,0,\ldots) + \left(\frac{\as}{2\pi}\right)
\int d\s^{(1)} \lq G(x_1,x_2,0,\ldots) - G(1,1,0,\ldots) \rq
\right.\nn\\
&& {} + \left(\frac{\as}{2\pi}\right)^2
\int \lq d\sigma_2^{(2)} + d\sigma_3^{(2)} +
d\sigma_4^{(2)} \rq \lq G(x_1,x_2,y,\theta,\phi) - G(1,1,0,\ldots) \rq
\nn\\
&&{}+\left. {\cal O}\(\as^3\) \rg \;,
\label{eq:G_average}
\eeqn
where we have used eq.~(\ref{eq:dsigma_all}) to expand $d\s^{(2)}$.
Since
\beq
d\s_2^{(2)} = \delta\(x_1-1\)\,\delta\(x_2-1\)\,\delta\(y\)\,\s_2^{(2)}
\,dx_1\,dx_2\,dy\;,
\eeq
the contribution of $d\s_2^{(2)}$ to the second integral is zero.  The
same thing happens for the two-body  contribution, at order $\as$,
$d\s_2^{(1)}$: in fact, this term too is proportional to the product
of the three $\delta$'s, and gives zero contribution to the first
integral.

The dependence from $\s_2^{(1)}$ and  $\s_2^{(2)}$ is in the
expression of $\s$. We can then
expand the denominator till order $\as^2$, to obtain
\beqn
\overline{G}  &=&  G(1,1,0,\ldots) + \left(\frac{\as}{2\pi}\right)
\int \frac{d\s^{(1)}}{\s^{(0)}}
\lq G(x_1,x_2,0,\ldots)- G(1,1,0,\ldots) \rq \nn\\
&&{}+\left(\frac{\as}{2\pi}\right)^2  \Bigg\{ \int \Bigg[
\frac{d\sigma_3^{(2)}}{\s^{(0)}} +
\frac{d\sigma_4^{(2)}}{\s^{(0)}} \Bigg] \lq G(x_1,x_2,y,\theta,\phi) -
G(1,1,0,\ldots) \rq  \nn\\
&&{} - \frac{\s^{(1)}}{\s^{(0)}} \int \frac{d\s^{(1)}}{\s^{(0)}}
\lq G(x_1,x_2,0,\ldots)- G(1,1,0,\ldots) \rq  \Bigg\}
+{\cal O}\(\as^3\) \;. \phantom{aaa}
\eeqn
We can see that no dependence from $\s_2^{(2)}$ is left, confirming
that we do not need to compute two-loop diagrams.  We need, however,
the total two-body contribution at first order $\s_2^{(1)}$, but the
analytic expression of this term is known from long time (see, for
example, Ref.~\cite{Reinders}).

\item
The second case, where we do not need to compute the two-loop diagrams,
arises if we deal with physical quantities that assume zero value in
the two-jet region. In fact, starting from eq.~(\ref{eq:G_average}),
not normalized to the total cross section
\beq
\VEV{G} = \int d\s \, G
\eeq
we see that if $G(1,1,0,\ldots)=0$, we accomplish our goal.
This is what we have done in our program: we have computed quantities
like
\begin{itemize}
\item[-] $1-t$, $t$ being the thrust of the process
\item[-] the $c$ parameter
\item[-] $M_h^2 -m^2$, $M_h^2$ being the heavy-jet mass
\end{itemize}
which give zero contribution in the limit $x_1\,\to\, 1$,
$x_2\,\to\,1$, $y\,\to\,0$.
\end{enumerate}
\section[Soft and collinear limit of the $ Q\Qb gg$
cross section]{Soft and collinear limit of the {\mylarge $Q\Qb gg$}
cross section}
\label{sec:QQgg_sc_limit}
In order to apply the subtraction method described in
Sec.~\ref{sec:subtraction}, we need to derive an expression for the
singular part of the four-body cross section, valid in both the
collinear and the soft limit.
These limits are both characterized by $y\,\to\, 0$, except that,
in the soft limit, at the same time, $v\,\to\, 0$ ($l$ soft)
or $v\,\to\, 1$ ($k$ soft).
In fact, from the definition of $y$ given  in eq.~(\ref{eq:x1x2y}),
we see that, if $k$ becomes collinear with $l$ or if
$k$ or $l$ become soft, we have $y\,\to\,0$.
In this limit, from eqs.~(\ref{eq:def_l})--(\ref{eq:cosalfa}), we have
\beqn
&&\cos\a \sim 1 -a\, y \nn\\
&&\sin\a \sim \sqrt{2 a} \sqrt{y} \nn\\
&&\frac{\mod{p}}{p_0} \equiv \sqrt{1-\frac{m^2}{p_0^2}} \sim   1 -f_2\, y
\nn\\
&&\frac{\mod{p'}}{p'_0}\equiv \sqrt{1-\frac{m^2}{{p'_0}^2}}
\sim 1 - f_1\,y  \nn\\
&&\frac{p'_z}{p'_0} \equiv \sqrt{1-\frac{m^2}{{p'_0}^2}} \cos\a
\sim 1 -b\, y  \nn\\
&&\frac{p'_y}{p'_0} \equiv \sqrt{1-\frac{m^2}{{p'_0}^2}} \sin\a
\sim  \sqrt{2a}\sqrt{y}\nn \;,
\eeqn
where
\beqn
&& a= \frac{2}{(1-x_1)(1-x_2)} \ga
x_1+x_2-1-\frac{m^2}{q^2}\left[2+\frac{1-x_2}{1-x_1}+ \frac{1-x_1}{1-x_2}
\right] \gc \nn\\
&& b=\frac{2}{(1-x_1)(1-x_2)} \ga
x_1+x_2-1-\frac{m^2}{q^2}\left[2+ \frac{1-x_1}{1-x_2}
\right] \gc \nn\\
&& f_1 = \frac{2m^2}{q^2}  \frac{1}{(1-x_1)^2} \quad\quad\quad\nn\\
&& f_2 = \frac{2m^2}{q^2}  \frac{1}{(1-x_2)^2} \nn\;.
\eeqn
We can then write, always in the $y\,\to\, 0$ limit,
\beqn
\!\!\!&& p\cdot l \sim \frac{q^2(1-x_2)}{4}\ga 2\,v + y\left[ f_2 (1-2\,v)
- \frac{2\,v}{1-x_2}\right] \gc \nn\\
&& p'\cdot l \sim \frac{q^2(1-x_1)}{4} \ga 2\,v
-2\sqrt{2a}\,\cos\phi \, \sqrt{v(1-v)}\sqrt{y} + y \left[ b\,(1-2\,v)
-\frac{2\,v}{1-x_1} \right]  \gc \nn\\
&& p\cdot k \sim \frac{q^2(1-x_2)}{4}\ga 2(1-v) -y
\left[f_2 (1-2\,v)+\frac{2(1-v)}{1-x_2} \right] \gc\nn
\\[-3mm]
&&\label{eq:prod_approx}\\[-3mm]
&& p'\cdot k \sim \frac{q^2(1-x_1)}{4}\Bigg\{
2(1-v)+2\sqrt{2a}\, \cos\phi\, \sqrt{v(1-v)}\sqrt{y} \nn\\
&& \phantom{ p'\cdot k \sim \frac{q^2(1-x_1)}{4}\Bigg\{}
-y\left[b\,(1-2\,v)
+ \frac{2(1-v)}{1-x_1} \right] \Bigg\} \nn \\
&& p\cdot p' \sim \frac{q^2}{2} \left[ -(1-x_1-x_2)- \frac{2m^2}{q^2}
+y \right] \;,\nn
\eeqn
where we have made use of eq.~(\ref{eq:v}). We can  see that
\begin{itemize}
\item[-] {\bf ${\mathbf l}$ soft}
$$
\left.
\eqalign{
(p+l)^2-m^2 = 2\, p\cdot l \,\to \,0 \cr
(p'+l)^2-m^2 = 2\, p'\cdot l \,\to\, 0
}
\gc  y\,\to\, 0  {\rm\ and\ }  v\,\to\, 0
$$
and
\beqn
p\cdot l &\sim&
m^2\, \frac{1}{2\(1-x_2\)} \left[ y+\frac{q^2}{m^2}\(1-x_2\)^2 v
\right] \nn\\
&=& m^2\, \frac{1}{2\(1-x_2\)} \left[ y+h\, v \right] \nn\\
p'\cdot l &\sim&  q^2\, \frac{b\,(1-x_1) }{4} \lq y
- \frac{2\sqrt{2a}}{b}\, \cos\phi\,\sqrt{v}\sqrt{y} + \frac{2}{b}\,v
\rq\nn\\
&=&  q^2\, \frac{b\,(1-x_1) }{4} \lq y
- c\, \cos\phi\,\sqrt{v}\sqrt{y} + g\,v \rq\nn
\eeqn

\item[-]{\bf ${\mathbf k}$ soft}
$$
\left.
\eqalign{
(p+k)^2-m^2 = 2\, p\cdot k\, \to\,  0 \cr
(p'+k)^2-m^2 = 2\, p'\cdot k\,\to\,  0
}
\gc  y\, \to\,  0  {\rm\ and\ } v\, \to\,  1
$$
and
\beqn
p\cdot k &\sim&
m^2\, \frac{1}{2\(1-x_2\)} \left[ y+\frac{q^2}{m^2}\(1-x_2\)^2 (1-v)
\right] \nn\\
&=& m^2\, \frac{1}{2\(1-x_2\)} \left[ y+ h\,(1-v)
\right] \nn\\
p'\cdot k &\sim&
q^2\, \frac{b\,(1-x_1)}{4} \lq y
+ \frac{2\sqrt{2a}}{b}\, \cos\phi \, \sqrt{1-v}\sqrt{y} +
\frac{2}{b}\,(1-v) \rq \nn\\
&=& q^2\, \frac{b\,(1-x_1)}{4} \lq y
+ c \, \cos\phi \, \sqrt{1-v}\sqrt{y} + g\,(1-v) \rq \nn
\eeqn
\end{itemize}
where
\beqn
\label{eq:parameter1}
&& h = \frac{q^2}{m^2}\(1-x_2\)^2  \\
&& c = \frac{2\sqrt{2a}}{b} \\
\label{eq:parameter2}
&& g = \frac{2}{b} \;.
\eeqn

\subsection{Soft contribution}
We begin with the soft singularities of ${\cal M}_{gg}$ (since the
same formulae apply irrespective of the vector or axial case, we will
always drop the $V/A$ suffix). They are given by eq.~(\ref{eq:lsoft}),
which we now rewrite
\beqn
{\cal M}_{gg}^{\rm soft} &=& g_s^2\,\mu^{2\e}
\Bigg\{
C_A
\left[\frac{p\cdot k} {(p\cdot l) \, (k\cdot l)} +
\frac{p'\cdot k} {(p'\cdot l) \, (k\cdot l)} \right]
+2 \(C_F-\frac{C_A}{2}\) \frac{p\cdot p'}{(p\cdot l)\,(p'\cdot l)}
\nonumber \\
\label{eq:lsoff}
&& \phantom{g_s^2\,\mu^{2\e} \Bigg\{}
{}-C_F \left[ \frac{m^2}{(p\cdot l)^2} + \frac{m^2}{(p'\cdot l)^2} \right]
+(k\leftrightarrow l)
\Bigg\}\times {\cal M}_b\;.
\eeqn
Exploiting the formulae~(\ref{eq:prod_approx}) for the approximated scalar
products in the soft limit, we can write, in the
limit of $l$ soft,
\beqn
&&\frac{p\cdot k}{(p\cdot l)(k\cdot l)} \sim \frac{2h}{q^2}\,
\frac{1}{y\,[y+h\,v]} \equiv E_{p,k;l}(x_1,x_2,y,v)
\nn\\
&& \frac{p\cdot p'}{(p\cdot l)(p'\cdot l)} \sim
\frac{K}{m^2} \, \frac{1}{y+h\,v}\, \frac{1}{y-c
\cos\phi\sqrt{y}\sqrt{v} + g\,v} \equiv E_{p,p';l}(x_1,x_2,y,v,\phi)
\nn\\
&& \frac{m^2}{(p\cdot l)^2} \sim \frac{4h}{q^2}\,
\frac{1}{[y+h\,v]^2} \equiv E_{p,p;l}(x_1,x_2,y,v)\;,
\eeqn
where the first two indexes in $E$ refer to the numerator, while the
third index refers to the soft momentum, and
\beq
K = \frac{1-x_2}{1-x_1}\,\frac{4}{b}\left[ x_1+x_2-1-\frac{2
m^2}{q^2} \right]\;.
\eeq
We will also need analogous formulae  in which the roles
of $p$ and $p'$ are interchanged.
With the help of eqs.~(\ref{eq:ppexch}), we have, in the limit of $l$ soft,
\dl{
\frac{p'\cdot k}{(p'\cdot l)(k\cdot l)} \sim E'_{p',k;l}(x_1,x_2,y,v')
\equiv E_{p,k;l}(x_2,x_1,y,v')
\hfill}
\dl{
\frac{p'\cdot p}{(p\cdot l)(p'\cdot l)} \sim
E'_{p',p;l}(x_1,x_2,y,v',\phi') \equiv E_{p,p';l}(x_2,x_1,y,v',\phi')
\hfill}
\dl{
\frac{m^2}{(p'\cdot l)^2} \sim  E'_{p',p';l}(x_1,x_2,y,v')
\equiv E_{p,p;l}(x_2,x_1,y,v')\;.
\hfill}
The contributions for the case when $k$ is soft are instead
obtained from the above using eqs.~(\ref{eq:klexch}).
For example
\beq
\eqalign{
E_{p,l;k}(x_1,x_2,y,v) = E_{p,k;l}(x_1,x_2,y,1-v) \cr
E_{p,p';k}(x_1,x_2,y,v,\phi) = E_{p,p';l}(x_1,x_2,y,1-v,\phi+\pi)\;.}
\eeq
We can now write down our approximate soft cross section:
\beqn
\label{eq:M_gg^soft}
{\cal M}_{gg}^{\rm soft}&=& g_s^2\,\mu^{2\e}
\Bigg\{C_A\left[E_{p,k;l}+E'_{p',k;l}
+ E_{p,l;k}+E'_{p',l;k}\right] \nonumber \\ &&
+(C_F-\frac{C_A}{2})
\left[E_{p,p';l}+E'_{p',p;l}+E_{p,p';k}+E'_{p',p;k}\right]
\nonumber \\ &&
-C_F\left[E_{p,p;l}+E'_{p',p';l}+E_{p,p;k}+E'_{p',p';k}\right]\Bigg\}\times
{\cal M}_b \;.
\eeqn
The soft cross section written in this way is symmetric under the
interchange of $k$ and $l$, and of $p$ and $p'$.

\subsection{Collinear contribution}
\label{sec:coll_contrib}
The collinear part of the cross section that receives contributions
from the $Q \Qb gg$  final state can be written, according to
eq.~(\ref{eq:ggcollfin}),
\beqn
{\cal M}_{gg}^{\rm coll+soft}\!\! &=&\!\!
g_s^2\,\mu^{2\e}\, \frac{4 \, C_A}{q^2 y} \,
\Bigg\{-\left[-2+\frac{1}{z}+\frac{1}{1-z}+z(1-z)\right]g_{\sigma\sigma'}
\nonumber \\ \label{eq:coll+soft_lim}
&&\!\!
\phantom{g_s^2\,\mu^{2\e}\,\frac{4\, C_A}{q^2 y}\,\Bigg\{ }
{}- 2 z(1-z)(1-\e)\left[\frac{k_{\perp\s}k_{\perp\s'}}{\kperp^2}
-\frac{g_{\perp\s\s'}}{2-2\e}\right]\Bigg\}\times {\cal
M}_b^{\sigma\sigma'} \;, \phantom{aaaaa}
\eeqn
where $z$ is the momentum fraction of $l$ versus $l+k$ in the
collinear limit. It can be chosen to be equal to $v$ or to $v'$.
Notice that, as it is explained in Appendix~\ref{app:gg_splitting},
parts of the collinear singularities are already contained in the
soft-limit expression. In fact, for $y\,\to\, 0$ at $v$ fixed, we have
\beq
E_{p,k;l} \approx E'_{p',k;l} \approx \frac{2}{q^2 y\, v}\;,\quad\quad\quad
E_{p,l;k} \approx E'_{p',l;k} \approx \frac{2}{q^2 y\,(1-v)}\;.
\eeq
Thus, the $1/z$ and $1/(1-z)$ terms in the collinear limit
formula~(\ref{eq:coll+soft_lim}) should not be included, since they are
already present in the soft term, and we get
\beqn
{\cal M}_{gg}^{\rm coll} &=&
g_s^2\,\mu^{2\e}\, \frac{4 \, C_A}{q^2 y} \,
\Bigg\{-\left[-2+v(1-v)\right]g_{\sigma\sigma'}
\nonumber \\
\label{eq:coll_lim} &&
\phantom{g_s^2\,\mu^{2\e}\,\frac{4\, C_A}{q^2 y}\,\Bigg\{ }
{}- 2 v(1-v)(1-\e)\left[\frac{k_{\perp\s}k_{\perp\s'}}{\kperp^2}
-\frac{g_{\perp\s\s'}}{2-2\e}\right]\Bigg\}\times {\cal
M}_b^{\sigma\sigma'} \;. \phantom{aaaaa}
\eeqn
According to eq.~(\ref{eq:kperp_conds}), the perpendicular direction
refers  to a direction orthogonal to $l+k$ in the
centre-of-mass system and in the collinear limit.  For this reason,
$\vet{k_\perp}$ lies in the same perpendicular plane as the vector
$\vet{j}$ defined in eq.~(\ref{eq:j_def}), and forming the angle
$\phi$ with $\vet{j}$, so that
\beq
\frac{\(\vet{k_\perp}\cdot\vet{j}\)^2}{\mod{{k_\perp}}^2} = \cos^2\phi
\quad \Longrightarrow \quad   \frac{\(k_\perp \cdot j\)^2}{k_\perp^2} =
-\cos^2\phi
\eeq
Using eq.~(\ref{Mform}), the azimuth-dependent term
of eq.~(\ref{eq:coll_lim}) becomes
\beqn
\left.{\cal M}_{gg}^{\rm coll}\right|_{\rm az} &\sim&
-2 v(1-v)(1-\e)\left[{\cal M}^\perp_b+{\cal M}^j_b
\frac{(k_\perp\cdot j)^2}{k_\perp^2}
-\frac{{\cal M}^\perp_b(2-2\e)-{\cal M}^j_b}{2-2\e}\right]
\nonumber \\ &=&
-v(1-v)\,{\cal M}^j_b\,\left[\frac{(k_\perp\cdot j)^2}{k_\perp^2}2(1-\e)
+1\right] \nn\\
&=& -v(1-v)\,{\cal M}^j_b\,\left[-2(1-\e)\cos^2\phi
+1\right]
\;.
\eeqn
Considering now the integration over the allowed phase space of
eq.~(\ref{eq:ps4}), we obtain, for the $\phi$ contribution,
\beq
\label{eq:azimuth_integration}
\int_0^\pi d\phi\,(\sin\phi)^{-2\e}
\left.{\cal M}_{gg}^{\rm coll}\right|_{\rm az} \sim \int_0^\pi
d\phi\,(\sin\phi)^{-2\e} \left[-2(1-\e)\cos^2\phi +1\right] = 0 \;.
\eeq
Despite of this fact, this term must be present in the program,
because, for $\e=0$, the corresponding phase space integral becomes
indefinite, since the $y$ integration is divergent, and the azimuthal
term (that gives zero) preserves the numerical integration to fail.

We thus arrive to the following expression for the collinear
term to be added to the soft term
\beqn
{\cal M}_{gg}^{\rm coll}\!\! &=& \!\! g_s^2 \, \mu^{2\e}\,
\frac{4\,  C_A}{q^2 y} \,
\Bigg\{ {\cal M}_b \left[\frac{v(1-v)+v'(1-v')}{2}-2\right] \nn \\
&+& \!\!\frac{{\cal M}_b^j}{2} \left[ v(1-v)\left(2(1-\e) \cos^2\phi -
1\right) + v'(1-v')
\left(2(1-\e)\cos^2\phi'-1\right)\right]\Bigg\}\;,\nn\\[-2mm]&&
\label{eq:M_gg^coll}
\eeqn
where we have symmetrized the expression in $v$, $v'$ and $\phi$, $\phi'$.

\section[Collinear limit of the $Q\Qb q \qb$
cross section]{Collinear limit of the {\mylarge $Q\Qb q \qb$}
cross section}
\label{sec:QQqq_c_limit}
With a procedure analogous to that one used in
Sec.\ref{sec:coll_contrib} for the collinear part of the $Q\Qb gg$
cross section, we can obtain the collinear part of ${\cal
M}_{q\overline{q}}$. From eq.~(\ref{eq:qqcollfin}), we have
\beqn
\label{eq:M_qq^coll}
{\cal M}_{q\bar{q}}^{\rm coll}\!\! &=&\!\! g_s^2\, \mu^{2\e}\,
\frac{4 \, n_{\rm lf} \, T_F}{q^2 y}\, \Bigg\{ {\cal M}_b
\frac{1}{4(1-\e)}\left[v'^2+(1-v')^2+v^2+(1-v)^2-2\e\right] \nn \\
&-& \!\!\frac{{\cal M}_b^j}{2} \left[ v(1-v)\!\left(2\cos^2\phi
-\frac{1}{1-\e}\right)+v'(1-v')\!
\left(2\cos^2\phi'-\frac{1}{1-\e}\right)\right]\!\Bigg\}\;.\phantom{aaaaa}
\eeqn

The expressions ${\cal M}_{gg}^{\rm soft}$ (eq.~(\ref{eq:M_gg^soft})),
${\cal M}_{gg}^{\rm coll}$ (eq.~(\ref{eq:M_gg^coll})) and ${\cal
M}_{q\bar{q}}^{\rm coll}$ (eq.~(\ref{eq:M_qq^coll})) depend upon $x_1$
and $x_2$ via ${\cal M}_b$ and ${\cal M}_b^j$.
These expressions are meaningful only if $x_1$ and $x_2$ belong to the
domain of the three-body phase space.
We thus define
\beq
\label{eq:mtilda}
\eqalign{
\widetilde{\cal M}_{gg} =
\left({\cal M}_{gg}^{\rm soft} + {\cal M}_{gg}^{\rm coll}\right)\;
\Theta_3(x_1,x_2) \cr
\widetilde{\cal M}_{q\bar{q}}=
{\cal M}_{q\bar{q}}^{\rm coll}\;
\Theta_3(x_1,x_2)\;,
}
\eeq
where the $\Theta_3$ function is precisely defined to be zero
when $x_1$ and $x_2$ are outside the three-body phase-space region.
More specifically, using the integration limits of eq.~(\ref{eq:ps3}),
we have
\beq
\Theta_3(x_1,x_2)=\Theta(1-x_1)\,\Theta(x_1-\sqrt{\rho})\,
\Theta(x_{2+}-x_2)\,\Theta(x_2-x_{2-})\;.
\eeq
We are now in a position to specify the subtraction procedure
outlined in Sec.~\ref{sec:subtraction}.
Our expression for the second-order contribution to a soft- and
collinear-safe quantity $G$ is given by
\beqn &&
\frac{1}{2}\int  d\Phi_4 \, {\cal M}_{gg}(x_1,x_2,y,v,\phi)\,
G(x_1,x_2,y,v,\phi)
\nonumber \\ && \hspace{1cm}+
\int  d\Phi_4 \,{\cal M}_{q\bar{q}}(x_1,x_2,y,v,\phi)\,
G(x_1,x_2,y,v,\phi) \nonumber
+\int d\Phi_3\,{\cal M}_v(x_1,x_2) G(x_1,x_2) \;, \nonumber
\eeqn
where all quantities are computed in $d=4-2\e$ dimensions.
The factor  $1/2$ in front of the $gg$ contribution accounts for the
two identical gluons in the final state.
We rewrite the above expression adding and subtracting the same
expression, that is the soft and collinear limit of the cross
sections, to obtain
\beqn
&&\frac{1}{2}\int d\Phi_4 \,
\left({\cal M}_{gg}(x_1,x_2,y,v,\phi) G(x_1,x_2,y,v,\phi)
- \widetilde{\cal M}_{gg}(x_1,x_2,y,v,\phi) G(x_1,x_2)\right)
\nonumber \\
&&+\int  d\Phi_4 \,
\left({\cal M}_{q\bar{q}}(x_1,x_2,y,v,\phi) G(x_1,x_2,y,v,\phi)
- \widetilde{\cal M}_{q\bar{q}}(x_1,x_2,y,v,\phi) G(x_1,x_2)\right)
\nonumber \\
&&+\int d\Phi_3 \, \left({\cal M}_v(x_1,x_2)
+\widetilde{\cal M}_i(x_1,x_2)\right) G(x_1,x_2) \;,
\eeqn
where we have defined
\beq
\widetilde{\cal M}_i(x_1,x_2)=
\frac{1}{2} \int d\Phi_{4/3} \, \widetilde{\cal M}_{gg}(x_1,x_2,y,v,\phi)
+\int d\Phi_{4/3}\, \widetilde{\cal M}_{q\bar{q}}(x_1,x_2,y,v,\phi)\;,
\eeq
and $d\Phi_{4/3}$ is defined by
\beq
d\Phi_4\; \Theta_3(x_1,x_2) = d\Phi_{4/3}\; d\Phi_3\;.
\eeq
The explicit expression for $d\Phi_{4/3}$ can be obtained
from eqs.~(\ref{eq:ps4}) and (\ref{eq:ps3}). We first notice that
the four-body phase space is almost proportional to the three-body
phase space, except for the ratio
\beq
\left(
\frac{ 4\(x_1^2-\rho\) \(x_2^2-\rho\)-\lq
\(x_g^2-4 y\)-\(x_1^2-\rho\)-\(x_2^2-\rho\)\rq^2 }
{ 4\(x_1^2-\rho\) \(x_2^2-\rho\)-\lq
x_g^2-\(x_1^2-\rho\)-\(x_2^2-\rho\) \rq^2 }
\right)^{-\e}\;=\;1+{\cal O}\(y\e\)\;.
\eeq
On the other hand, terms of order $y\e$ can be neglected,
since they cannot generate infrared singularities, because of the $y$
factor, and therefore they can only produce terms of order $\e$.
Thus we can write
\beq
d\Phi_{4/3}=  N_\e \,R_\e\, q^2 \,q^{-2\e}
\int_0^{y_+} dy\, y^{-\e}
\int_0^1 dv \,[v(1-v)]^{-\e} \,\frac{1}{N_{\phi}}
\int_0^\pi d\phi\,(\sin\phi)^{-2\e} \;,
\eeq
or the analogous one in the $v',\phi'$ variables.
The normalization factor $N_\e$ is defined as
\beq
\label{eq:def_Nep}
N_\e = -i\,N(\e) = \frac{1}{16\pi^2} \(4\pi\)^\e \Gamma(1+\e)\;,
\eeq
while
\beq
R_\e = \frac{1}{\Gamma(1+\e)\Gamma(1-\e)}=
1-\frac{\pi^2\e^2}{6}+{\cal O}\(\e^3\)\;.
\eeq
Since we are free to choose the set of variables we prefer in the
$d\Phi_{4/3}$ integration,
it is easy to see that the $\widetilde{\cal M}_i(x_1,x_2)$ term reduces to
\beqn
\widetilde{\cal M}_i(x_1,x_2)&=&g_s^2\, \mu^{2\e} \int d\Phi_{4/3}\,
\Bigg\{
\frac{1}{2}\, \frac{4\, C_A}{q^2 y} \lq v\,(1-v)-2 \rq \nn \\
&&{} + \frac{2\, n_{\rm lf}\, T_F}{q^2 y}
\frac{1}{1-\e}\left[v^2+(1-v)^2 -\e\right] \nn \\
&&{}+ \frac{1}{2} \Big[ 4 \,C_A E_{p,k;l}
+4\,\(C_F-\frac{C_A}{2}\) E_{p,p';l}-4\, C_F E_{p,p;l} \Big] \Bigg\}
\times {\cal M}_b \;, \nn
\eeqn
where the term proportional to ${\cal M}^j$ has been dropped, since it vanishes
in $d=4-2\e$ dimensions, after the azimuthal integration, as shown in
eq.~(\ref{eq:azimuth_integration}).

We can now integrate over $d\Phi_{4/3}$ to obtain the
analytic  expression of $\widetilde{\cal M}_i(x_1,x_2)$.
We define, for the collinear term,
\begin{eqnarray}
I_{gg}^{\rm coll} &=&
\int_0^{y_+} dy\, y^{-\e}
\int_0^1 dv \,[v(1-v)]^{-\e} \,\frac{1}{N_{\phi}}
\int_0^\pi d\phi\,(\sin\phi)^{-2\e}
\,\frac{1}{y}\, \left[v\,(1-v) -2 \right] \nn\\
&=& -\frac{1}{\e}\left[1-\e\log
\(y_{+}\)\right] \(-\frac{11}{6} - \frac{67}{18}\e\) + {\cal O}\(\e\)\;,
\label{eq:I_ggcoll}
\end{eqnarray}
and
\begin{eqnarray}
I_{q\bar{q}}^{\rm coll} &=&
\int_0^{y_+} dy \, y^{-\e}
\int_0^1 dv \,[v(1-v)]^{-\e} \,\frac{1}{N_{\phi}}
\int_0^\pi d\phi\,(\sin\phi)^{-2\e}
\,\frac{1}{y}\, \frac{v^2+(1-v)^2-\e}{1-\e} \nn\\
&=& -\frac{1}{\e}\left[1-\e\log
\(y_{+}\)\right] \(\frac{2}{3}+\frac{10}{9}\e\) + {\cal
O}\(\e\)\;.
\label{eq:I_qqcoll}
\end{eqnarray}
For the integration of the soft term, we define
\beq
I_{p,k;l} = q^2 \,
\int_0^{y_+} dy \, y^{-\e}
\int_0^1 dv \,[v(1-v)]^{-\e} \,\frac{1}{N_{\phi}}
\int_0^\pi d\phi\,(\sin\phi)^{-2\e} E_{p,k;l}\;,
\eeq
and the analogous ones for $I_{p,p;l}$ and $I_{p,p';l}$.
With this notation, we have
\beq
I_{p,k;l} = 2 h \,I_1\;,  \quad\quad\quad
I_{p,p;l} = 4 h \, I_2\;,  \quad\quad\quad
I_{p,p';l} = K\, \frac{q^2}{m^2}\, I_3\;,
\eeq
where the values of $I_1$, $I_2$ and $I_3$ are collected in
Appendix~\ref{app:soft_integrals}, and are given precisely by
eqs.~(\ref{eq:I_1_finale}), (\ref{eq:I_2_finale}) and~(\ref{eq:I_3_finale}).

Our final expression for $\widetilde{\cal M}_i(x_1,x_2)$ is therefore
\beqn
\widetilde{\cal M}_i(x_1,x_2)&=& N_\e \,R_\e\,
g_s^2 \left(\frac{\mu^2}{q^2}\right)^\ep
\bigg\{ 2 \, C_A
\,I_{gg}^{\rm coll}  + 2\, n_{\rm lf}\, T_F\, I_{q\bar{q}}^{\rm coll}
\nn \\
&&{} + \lq 2 \,C_A I_{p,k;l} +2\,\(C_F-\frac{C_A}{2}\) I_{p,p';l}
- 2\, C_F I_{p,p;l} \rq \bigg\} \times{\cal M}_b\; .
\nonumber
\eeqn

\section{Checks of the calculation}
\label{sec:checks}
We have performed several checks  to control the correctness of our
results, both internal and external, by comparing our results with the
known ones.

\begin{enumerate}
\item
The divergences coming from UV and IR poles all cancel. This is surely
one of the most important analytical check, that covers the virtual
terms and the integrals of the soft and collinear limits of the
four-body final states.

\item
The full calculation, as $m\,\to\, 0$, agrees with the massless
results of Ref.~\cite{YellowBook}. In the tables of
Chapter~\ref{chap:results}, this is easily seen. In fact, in the first
column, we report these massless results, and we can see that our
massive calculation reaches the massless limit as the ratio $m/E$ goes
to zero.

\item
Our four-dimensional matrix elements for the processes $e^+e^-\,\to\,
Z/\gamma\,\to\,Q\overline{Q} g g$ and
$e^+e^-\,\to\,Z/\gamma\,\to\,Q\overline{Q} Q\overline{Q}$
agree with Ref.~\cite{Ballestrero}.
We have performed a numerical comparison between their results and ours.
Furthermore, the soft and collinear
limits of the four-body matrix elements for the process
$Z/\gamma\,\to\,Q\overline{Q}$ plus two light partons
are correctly given by formulae~(\ref{eq:mtilda}).
\item
Near the production threshold, we  recover the Coulomb singularity.
If $\beta$ is the velocity of the two massive
quarks in the fermion centre-of-mass system, then (see Ref.~\cite{NDE})
\beq
\label{eq:coulomb}
d\sigma^{(v)}_{V/A}(x_1,x_2)
\stackrel{\beta \to 0}{\longrightarrow}
\frac{\pi^2}{\beta}\(C_F
-\frac{C_A}{2}\) d\sigma^{(b)}_{V/A}(x_1,x_2)\;.
\eeq
By evaluating $(p+p')^2$ in the centre of mass of the two massive
quarks, for small $\beta$, we get
\beq
(p+p')^2 = \left[ 2\(m + \frac{m}{2} \beta^2
+ {\cal O}\(\beta^4\)\) \right]^2
= (q-k)^2 = q^2\, (x_1+x_2 -1)\;.
\eeq
Choosing for example $x_1 = x_2$ we have
\[
x_1=x_2  =\frac{1}{2}\(1+\rho+\rho\,\beta^2\)\;.
\]
By letting $\beta$ get smaller and smaller we have checked that the
behaviour of the virtual differential cross section is in agreement
with eq.~(\ref{eq:coulomb}).

\item
The last check we have made, much more involved than the previous
ones, is described in Chapter~\ref{chap:fragment} and in
Ref.~\cite{no_fragmentation}.
We give here only a brief sketch of it.
We have used the fact that the semi-inclusive
differential cross section for the production of a heavy quark is
calculable in perturbative QCD: in fact, the mass of the final quark
acts as a cut-off for the collinear divergences and logarithms of the
ratio $q^2/m^2$ appears in the final result.

On the other hand, using the factorization theorem and the
Altarelli-Parisi evolution equations, we can obtain an expression of
the differential cross section in which the large logarithms are
correctly resummed, while powers of the ratio $m^2/q^2$ are completely
neglected.

We have made an expansion in $\as$ of the resummed expression till
order $\as^2$, and we have checked that some  moments of the coefficients
of the large logarithms are correctly given by our
program, in the limit of small masses.

\end{enumerate}

\chapter{Numerical results}
\label{chap:results}
\thispagestyle{plain}
%\input{phd_results}
%\section{Results}
We implemented our analytical result in a FORTRAN program, which
behaves like a ``partonic'' Monte Carlo generator, analogous to the
program EVENT~\cite{YellowBook}.  We collect here some results
obtained with our code.  Since for this kind of calculations it would
be difficult to perform analytical comparisons, the only possible
alternative is to choose a few shape variables, and compare numerical
results, in the spirit of what has been done in Ref.~\cite{LEP2}, for
the case of the massless calculation.

We include in these results only the contributions from cut graphs of
Secs.~\ref{sec:virtual}, \ref{sec:QQgg} and~\ref{sec:QQqq}, in which
the weak current couples to the same heavy-flavour loop, and there is
a single $Q\Qb$ pair in the final state, which is the really hard part
of the calculation.

For the contributions involving two heavy-quark pairs in the final
state, it is easier to compare directly the value of the matrix
elements squared (this part of our program was, in fact, checked in this way
with the program of Ref.~\cite{Ballestrero}).

We have chosen a set of shape variables for which it should be easy to
obtain quite accurate numerical results.  We have fixed the
centre-of-mass energy to be 100~GeV, and the mass of the heavy quark
has been taken to be equal to 1, 10 and 30~GeV.  We present
separately the results for a hypothetical vector boson with purely
axial or purely vector couplings, normalized to the massless total
cross section at  zeroth order in $\as$.  We have chosen the
following shape variables: the thrust $t$, the $c$ parameter, the mass
of the heavy jet squared $M_h^2$ (according to the thrust axis), the
energy--energy correlation EEC, the three-jet fractions according to
the E, EM~\cite{Bil_Rod_Santa}, JADE, and DURHAM schemes.

For some shape variables, the presence of massive particles in the
final state may introduce ambiguities in the definition, owing
to the fact that, in the massless case, energy and momentum can be
interchanged.
We thus refer to the exact definitions given in Ref.~\cite{YellowBook}
for $t$, $c$,  $M_h^2$ and in Ref.~\cite{LEP2} for the EEC. We collect
here these definitions.
Thrust is defined as
\beq
t = \max_{\vet{n}}  \frac{\displaystyle\sum_i
\left|\vet{p}_i \cdot \vet{n} \right|}
{\displaystyle\sum_i \left|\vet{p}_i\right|} \;,
\eeq
where $\vet{p}_i$ denotes the three-momentum of the $i^{\rm th}$
particle in the centre-of-mass system,
and the sum extends over all final-state particles. The direction of
$\vet{n}$ that maximizes the above quantity is called thrust axis.

The $c$ parameter is derived from the eigenvalues of the
infrared-safe momentum tensor
\beq
\theta^{ab} = \frac{\displaystyle
\sum_i\frac{p_i^a\,p_i^b}{\left|\vet{p}_i\right|}}
{\displaystyle \sum_i \left|\vet{p}_i\right|}\;,
\eeq
where $p_i^a$ is the $a^{\rm th}$ component of the three-momentum
$\vet{p}_i$. If we denote with $\lambda_1$, $\lambda_2$ and $\lambda_3$
the eigenvalues, we define
\beq
c = 3 \(\lambda_1 \lambda_2 +\lambda_2 \lambda_3+\lambda_1 \lambda_3\)\;.
\eeq

The definition of the heavy-jet mass $M_h^2$, according to the thrust
axis, is obtained with the following procedure: it is the maximum value
between the  invariant masses of the particles belonging to the two
different hemispheres separated by the plane orthogonal to the thrust
axis.

For $t$, $c$, $M_h^2$ and EEC we present moments, instead of distributions,
because they can be obtained with higher precision.
For thrust, for example, we  compute, according to the notation of
Sec.~\ref{sec:outlinecalc}
\beq
\int dT_{V/A} \, (1-t)^n = \left(\frac{\as}{2\pi}\right)   A^t_{V/A}(n) +
\left(\frac{\as}{2\pi}\right)^2 B^t_{V/A}(n)\;.
\eeq
We  further decompose
\beq
B^t_{V/A} = B_{V/A,\,C_A}^{t} + B_{V/A,\,C_F}^{t} +B_{V/A,\,T_F}^{t}\;,
\eeq
where the $C_A$, $C_F$ and $T_F$ subscripts denote the $C_F C_A$,
$C_F^2$ and $n_{\rm f} C_F T_F$ colour components.
For the other quantities, moments are defined as
\beqn
\int dT_{V/A}\; c^n &=& \left(\frac{\as}{2\pi}\right)   A^c_{V/A}(n) +
\left(\frac{\as}{2\pi}\right)^2 B^c_{V/A}(n)\;,
\nonumber \\
\int dT_{V/A} \left(\frac{M_h^2-m^2}{q^2}\right)^n &=&
\left(\frac{\as}{2\pi}\right)   A^{M_h}_{V/A}(n) +
\left(\frac{\as}{2\pi}\right)^2 B^{M_h}_{V/A}(n)\;,
\nonumber \\
\int dT_{V/A} \sum_{ij} \frac{E_i E_j}{q^2} \cos^k\theta_{ij} \,
\sin^{2+n}\theta_{ij} &=&
\left(\frac{\as}{2\pi}\right)   A^{\rm EEC}_{V/A}(n,k) +
\left(\frac{\as}{2\pi}\right)^2 B^{\rm EEC}_{V/A}(n,k)\;,
\nonumber
\eeqn
where the sum runs over all the final particles, and $\theta_{ij}$ is
the angle between the corresponding three-momenta.

Jet clustering algorithms are defined giving two ingredients:
\begin{itemize}
\item[-] the rule to
compute the resolution parameter $y_{ij}$ for each pair of particles $ij$
in the final state
\item[-] the recombination rule for the two particles.
\end{itemize}
Before starting the algorithm,
you  fix the value of a resolution parameter $y_{cut}$ to be used
as a discriminant condition.  Then, according to the given rule,
you  compute the resolution
parameter $y_{ij}$ for each couple of final-state particles.  If the
minimum values of $y_{ij}$'s is less than $y_{cut}$, then the two
particles, for which this value was computed, are recombined into one
pseudo-particle, and you start again, by calculating a new set of
$y_{ij}$'s.  If, instead, no value of $y_{ij}$'s satisfies the condition
$y_{ij}<y_{cut}$, then the algorithm is over, and the final number of
pseudo-particles obtained gives the number of jets.

We have considered four different jet-clustering algorithms: E,
EM, JADE and DURHAM.
Their resolution parameter $y_{ij}$ is defined by
\beqn
\mbox{E}&:& y_{ij} = \frac{(p_i+p_j)^2}{q^2}\;,
\nonumber \\
\mbox{EM}&:&  y_{ij} = 2\; \frac{p_i\cdot p_j}{q^2}\;,
\nonumber \\
\mbox{JADE}&:& y_{ij} =  2\; \frac{E_i E_j}{q^2} (1-\cos\theta_{ij})\;,
\nonumber \\
\mbox{DURHAM}&:&  y_{ij} =  2\min{\(\frac{E_i^2}{q^2},
\frac{E_j^2}{q^2}\)} (1-\cos\theta_{ij})\;,
\eeqn
while their recombination rule is the same for all of them
\beq
p_{ij} = p_i + p_j\;.
\eeq
Observe that the E scheme is not infrared-safe if $y_{cut}<m^2/q^2$.
In fact, in this range, the configuration made up of two heavy quarks
plus a soft gluon cannot be reduced to two pseudo-particles, since the
recombination parameter will fail the cut.
The cancellation of soft divergences cannot therefore
work for these values of the cut parameter.

For the jet clustering algorithms, we have computed
\beq
\int dT_{V/A} \delta_{N_{\rm X}(y_{cut}),\,3}=
\left(\frac{\as}{2\pi}\right)   A^{\rm X}_{V/A}(y_{cut}) +
\left(\frac{\as}{2\pi}\right)^2 B^{\rm X}_{V/A}(y_{cut})\;,
\eeq
where ${\rm X}$ stands for one of the jet-clustering algorithms, and
$N_{\rm X}(y_{cut})$ is the number of pseudo-particles in the final state
after the clustering procedure is over.

We have chosen the renormalization scale $\mu=E$, and $\nlf=5$. The
results are given in Tabs.~\ref{th} to~\ref{DU}. The first column of
each table contains the massless limit, obtained with the program
that has generated the results of Refs.~\cite{YellowBook}
and~\cite{LEP2}, in order to allow a comparison with our  massive
calculation.

Further results for $m/E=0.2$~GeV can be found in Ref.~\cite{NasonOleari}.

 \begin{table}[htbp]
 \begin{center}
 \footnotesize
\begin{tabular}{||c|c|c|c|c||}
 \hline\hline
 $n$&$m/E=0$ & $m/E=0.01$ & $m/E=0.1$                       &$m/E=0.3$\\ \hline
 &\multicolumn{4}{c||}{$\phantom{\Bigl[}B_{V,\,C_A  }^{t            }
 (n)$}
 \\ \hline
 1&
 $71.56 \pm 0.026$ & $71.53 \pm 0.045$ & $57.14 \pm 0.024$ & $21.83 \pm 0.006$ \\                                                                                          
 2&
 $5.249 \pm 9\,10^{-4}$ & $5.303 \pm 0.005$ & $4.47 \pm 0.0034$ & $1.524 \pm 8\,10^{-4}$ \\                                                                                
 3&
 $0.7956 \pm 1.7\,10^{-4}$ & $0.8056 \pm 0.0012$ & $0.6887 \pm 8\,10^{-4}$ & $0.2127 \pm 2\,10^{-4}$ \\                                                                    
 4&
 $0.1582 \pm 4\,10^{-5}$ & $0.1604 \pm 3.4\,10^{-4}$ & $0.1381 \pm 2.4\,10^{-4}$ & $0.04003 \pm 6\,10^{-5}$ \\                                                             
 5&
 $0.03614 \pm 1\,10^{-5}$ & $0.03668 \pm 1\,10^{-4}$ & $0.03174 \pm 7\,10^{-5}$ & $0.008845 \pm 2\,10^{-5}$ \\                                                             
 \hline
 &\multicolumn{4}{c||}{$\phantom{\Bigl[}B_{V,\,C_F  }^{t            }
 (n)$}
 \\
\hline
 \hline
 \end{tabular}
 \caption{The thrust $t$.                                   }
 \label{th}
 \end{center}
 \end{table}
 \begin{table}[htbp]
 \begin{center}
 \footnotesize
 \begin{tabular}{||c|c|c|c|c||}
 \hline\hline
 $n$&$m/E=0$ & $m/E=0.01$ & $m/E=0.1$                       &$m/E=0.3$\\ \hline
 &\multicolumn{4}{c||}{$\phantom{\Bigl[}B_{V,\,C_A  }^{M_h          }
 (n)$}
 \\ \hline
 1&
 $68.14 \pm 0.026$ & $68.32 \pm 0.05$ & $56.71 \pm 0.03$ & $19.46 \pm 0.008$ \\                                                                                            
 2&
 $4.542 \pm 0.001$ & $4.591 \pm 0.006$ & $4.014 \pm 0.004$ & $1.029 \pm 0.0024$ \\                                                                                         
 3&
 $0.6258 \pm 2\,10^{-4}$ & $0.6336 \pm 0.0013$ & $0.5549 \pm 0.001$ & $0.1072 \pm 0.001$ \\                                                                                
 4&
 $0.1141 \pm 5\,10^{-5}$ & $0.1156 \pm 3.6\,10^{-4}$ & $0.1002 \pm 3\,10^{-4}$ & $0.0167 \pm 4.5\,10^{-4}$ \\                                                              
 5&
 $0.0241 \pm 1.4\,10^{-5}$ & $0.02447 \pm 1\,10^{-4}$ & $0.02084 \pm 1\,10^{-4}$ & $0.0037 \pm 2\,10^{-4}$ \\                                                              
 \hline
 &\multicolumn{4}{c||}{$\phantom{\Bigl[}B_{V,\,C_F  }^{M_h          }
 (n)$}
 \\
\hline
 \hline
 \end{tabular}
 \caption{The mass of the heavy jet squared $M_h^2$.        }
 \label{mh}
 \end{center}
 \end{table}
 \begin{table}[htbp]
 \begin{center}
 \footnotesize
 \begin{tabular}{||c|c|c|c|c||}
 \hline\hline
 $n$&$m/E=0$ & $m/E=0.01$ & $m/E=0.1$                       &$m/E=0.3$\\ \hline
 &\multicolumn{4}{c||}{$\phantom{\Bigl[}B_{V,\,C_A  }^{c            }
 (n)$}
 \\ \hline
 1&
 $304.3 \pm 0.1$ & $303.3 \pm 0.18$ & $233.9 \pm 0.09$ & $84.7 \pm 0.023$ \\                                                                                               
 2&
 $70.37 \pm 0.011$ & $70.97 \pm 0.06$ & $58.25 \pm 0.03$ & $19.18 \pm 0.007$ \\                                                                                            
 3&
 $30.21 \pm 0.006$ & $30.55 \pm 0.03$ & $25.56 \pm 0.02$ & $7.789 \pm 0.004$ \\                                                                                            
 4&
 $16.16 \pm 0.0034$ & $16.36 \pm 0.02$ & $13.85 \pm 0.012$ & $4.008 \pm 0.003$ \\                                                                                          
 5&
 $9.658 \pm 0.0022$ & $9.784 \pm 0.012$ & $8.341 \pm 0.008$ & $2.335 \pm 0.002$ \\                                                                                         
 \hline
 &\multicolumn{4}{c||}{$\phantom{\Bigl[}B_{V,\,C_F  }^{c            }
 (n)$}
 \\
\hline
 \hline
 \end{tabular}
 \caption{The $c$ parameter.                                }
 \label{cp}
 \end{center}
 \end{table}
 \begin{table}[htbp]
 \begin{center}
 \footnotesize
 \begin{tabular}{||c|c|c|c|c||}
 \hline\hline
 $n$&$m/E=0$ & $m/E=0.01$ & $m/E=0.1$                       &$m/E=0.3$\\ \hline
 &\multicolumn{4}{c||}{$\phantom{\Bigl[}B_{V,\,C_A  }^{{\rm EEC}    }
 (n,0)$}
 \\ \hline
 0&
 $202.8 \pm 0.07$ & $202.3 \pm 0.12$ & $154.4 \pm 0.06$ & $46.81 \pm 0.013$ \\                                                                                             
 1&
 $142.6 \pm 0.06$ & $143.1 \pm 0.1$ & $117 \pm 0.05$ & $38.2 \pm 0.01$ \\                                                                                                  
 2&
 $115.3 \pm 0.05$ & $115.9 \pm 0.09$ & $97.55 \pm 0.04$ & $33.55 \pm 0.009$ \\                                                                                             
 3&
 $99.28 \pm 0.05$ & $99.79 \pm 0.08$ & $85.27 \pm 0.04$ & $30.38 \pm 0.008$ \\                                                                                             
 4&
 $88.43 \pm 0.05$ & $88.92 \pm 0.08$ & $76.65 \pm 0.04$ & $28.01 \pm 0.007$ \\                                                                                             
 5&
 $80.48 \pm 0.05$ & $80.95 \pm 0.07$ & $70.18 \pm 0.04$ & $26.13 \pm 0.007$ \\
\hline
 \hline
 \end{tabular}
 \caption{The energy--energy correlation EEC, $k=0$.        }
 \label{e0}
 \end{center}
 \end{table}
 \begin{table}[htbp]
 \begin{center}
 \footnotesize
 \begin{tabular}{||c|c|c|c|c||}
 \hline\hline
 $n$&$m/E=0$ & $m/E=0.01$ & $m/E=0.1$                       &$m/E=0.3$\\ \hline
 &\multicolumn{4}{c||}{$\phantom{\Bigl[}B_{V,\,C_A  }^{{\rm EEC}    }
 (n,1)$}
 \\ \hline
 0&
 $-25.89 \pm 0.006$ & $-26.06 \pm 0.025$ & $-21.45 \pm 0.014$ & $-7.006 \pm 0.003$ \\                                                                                      
 1&
 $-9.426 \pm 0.004$ & $-9.56 \pm 0.016$ & $-8.967 \pm 0.01$ & $-3.152 \pm 0.0023$ \\                                                                                       
 2&
 $-4.75 \pm 0.004$ & $-4.827 \pm 0.013$ & $-4.892 \pm 0.008$ & $-1.898 \pm 0.002$ \\                                                                                       
 3&
 $-2.844 \pm 0.004$ & $-2.894 \pm 0.011$ & $-3.086 \pm 0.008$ & $-1.316 \pm 0.0017$ \\                                                                                     
 4&
 $-1.894 \pm 0.0036$ & $-1.93 \pm 0.01$ & $-2.133 \pm 0.007$ & $-0.9876 \pm 0.0015$ \\                                                                                     
 5&
 $-1.354 \pm 0.0035$ & $-1.38 \pm 0.01$ & $-1.57 \pm 0.007$ & $-0.779 \pm 0.0014$ \\
\hline
 \hline
 \end{tabular}
 \caption{The energy--energy correlation EEC, $k=1$.        }
 \label{e1}
 \end{center}
 \end{table}
 \begin{table}[htbp]
 \begin{center}
 \footnotesize
 \begin{tabular}{||c|c|c|c|c||}
 \hline\hline
 $y_{cut}$&$m/E=0$ & $m/E=0.01$ & $m/E=0.1$                 &$m/E=0.3$\\ \hline
 &\multicolumn{4}{c||}{$\phantom{\Bigl[}B_{V,\,C_A  }^{{\rm E}      }(y_{cut})$}\\ \hline
0.01&
 $1343 \pm 1.5$ & $1357 \pm 1.5$ & $-$ & $-$ \\                                                                                                                            
0.05&
 $336.5 \pm 0.34$ & $341.3 \pm 0.5$ & $374 \pm 0.3$ & $-$ \\                                                                                                               
0.10&
 $131.7 \pm 0.15$ & $133.7 \pm 0.3$ & $138.6 \pm 0.17$ & $459.5 \pm 0.17$ \\                                                                                               
0.15&
 $60.06 \pm 0.07$ & $60.75 \pm 0.18$ & $63.1 \pm 0.13$ & $79.06 \pm 0.04$ \\                                                                                               
0.20&
 $26.66 \pm 0.04$ & $27.13 \pm 0.16$ & $29 \pm 0.12$ & $24.54 \pm 0.027$ \\                                                                                                
 \hline
 &\multicolumn{4}{c||}{$\phantom{\Bigl[}B_{V,\,C_F  }^{{\rm E}      }(y_{cut})$}\\
\hline
 \hline
 \end{tabular}
 \caption{The E clustering algorithm.                       }
 \label{EA}
 \end{center}
 \end{table}
 \begin{table}[htbp]
 \begin{center}
 \footnotesize
 \begin{tabular}{||c|c|c|c|c||}
 \hline\hline
 $y_{cut}$&$m/E=0$ & $m/E=0.01$ & $m/E=0.1$                 &$m/E=0.3$\\ \hline
 &\multicolumn{4}{c||}{$\phantom{\Bigl[}B_{V,\,C_A  }^{{\rm EM}     }(y_{cut})$}\\ \hline
0.01&
 $1337 \pm 1.5$ & $1340 \pm 1.7$ & $1135 \pm 0.8$ & $455.7 \pm 0.18$ \\                                                                                                    
0.05&
 $329.4 \pm 0.3$ & $333.2 \pm 0.5$ & $291.2 \pm 0.4$ & $101.6 \pm 0.07$ \\                                                                                                 
0.10&
 $124.7 \pm 0.12$ & $126.3 \pm 0.3$ & $111.8 \pm 0.23$ & $29.24 \pm 0.04$ \\                                                                                               
0.15&
 $53.69 \pm 0.06$ & $54.24 \pm 0.3$ & $48 \pm 0.17$ & $8.176 \pm 0.027$ \\                                                                                                 
0.20&
 $21.68 \pm 0.03$ & $22.02 \pm 0.2$ & $18.85 \pm 0.14$ & $1.52 \pm 0.024$ \\                                                                                               
 \hline
 &\multicolumn{4}{c||}{$\phantom{\Bigl[}B_{V,\,C_F  }^{{\rm EM}     }(y_{cut})$}\\
\hline
 \hline
 \end{tabular}
 \caption{The EM clustering algorithm.                      }
 \label{EM}
 \end{center}
 \end{table}
 \begin{table}[htbp]
 \begin{center}
 \footnotesize
 \begin{tabular}{||c|c|c|c|c||}
 \hline\hline
 $y_{cut}$&$m/E=0$ & $m/E=0.01$ & $m/E=0.1$                 &$m/E=0.3$\\ \hline
 &\multicolumn{4}{c||}{$\phantom{\Bigl[}B_{V,\,C_A  }^{{\rm JADE}   }(y_{cut})$}\\ \hline
0.01&
 $1352 \pm 1.1$ & $1352 \pm 2$ & $1088 \pm 1.6$ & $389.2 \pm 0.26$ \\                                                                                                      
0.05&
 $326 \pm 0.25$ & $328.6 \pm 0.6$ & $275 \pm 0.5$ & $82.34 \pm 0.09$ \\                                                                                                    
0.10&
 $122.7 \pm 0.1$ & $124.2 \pm 0.3$ & $106 \pm 0.26$ & $24.91 \pm 0.05$ \\                                                                                                  
0.15&
 $53.55 \pm 0.05$ & $54.1 \pm 0.3$ & $46.84 \pm 0.26$ & $8.675 \pm 0.03$ \\                                                                                                
0.20&
 $22.46 \pm 0.03$ & $22.86 \pm 0.2$ & $19.84 \pm 0.14$ & $2.98 \pm 0.03$ \\                                                                                                
 \hline
 &\multicolumn{4}{c||}{$\phantom{\Bigl[}B_{V,\,C_F  }^{{\rm JADE}   }(y_{cut})$}\\
\hline
 \hline
 \end{tabular}
 \caption{The JADE clustering algorithm.                    }
 \label{JA}
 \end{center}
 \end{table}
 \begin{table}[htbp]
 \begin{center}
 \footnotesize
 \begin{tabular}{||c|c|c|c|c||}
 \hline\hline
 $y_{cut}$&$m/E=0$ & $m/E=0.01$ & $m/E=0.1$                 &$m/E=0.3$\\ \hline
 &\multicolumn{4}{c||}{$\phantom{\Bigl[}B_{V,\,C_A  }^{{\rm DUR}    }(y_{cut})$}\\ \hline
0.01&
 $442.5 \pm 0.3$ & $443 \pm 1.6$ & $329.9 \pm 0.5$ & $71.09 \pm 0.07$ \\                                                                                                   
0.05&
 $114.7 \pm 0.1$ & $116 \pm 0.3$ & $97.26 \pm 0.22$ & $12.98 \pm 0.05$ \\                                                                                                  
0.10&
 $45.05 \pm 0.06$ & $45.25 \pm 0.2$ & $39.53 \pm 0.2$ & $3.62 \pm 0.03$ \\                                                                                                 
0.15&
 $20.2 \pm 0.03$ & $20.4 \pm 0.18$ & $17.82 \pm 0.13$ & $1.16 \pm 0.026$ \\                                                                                                
0.20&
 $8.593 \pm 0.02$ & $8.7 \pm 0.12$ & $7.61 \pm 0.1$ & $0.41 \pm 0.02$ \\                                                                                                   
 \hline
 &\multicolumn{4}{c||}{$\phantom{\Bigl[}B_{V,\,C_F  }^{{\rm DUR}    }(y_{cut})$}\\
\hline
 \hline
 \end{tabular}
 \caption{The DURHAM clustering algorithm.                  }
 \label{DU}
 \end{center}
 \end{table}

%\input{phd_conclusion}
%\cleardoublepage
\newpage
\thispagestyle{plain}
\mbox{}
%\markboth{\mbox{}}{\mbox{}}
\newpage
\def\conc{Conclusion}
\chapter*{\conc}
%\addtocounter{chapter}{-1}
\addcontentsline{toc}{chapter}{\conc}
\markboth{\dunhxii \underline{\conc}}
{\dunhxii \underline{\conc}}

In this thesis, we have described a next-to-leading-order calculation
of the cross section for the heavy-flavour production in $e^+e^-$
collisions, including quark mass effects.  The computation of the
amplitudes was
performed in a fully analytical way, and the different transition
amplitudes are available as FORTRAN subroutines.

The complete calculation is implemented in a program which behaves like a
``partonic'' Monte Carlo generator, in which pairs of weighted
correlated events are produced, according to the subtraction procedure
described in Sec.~\ref{sec:subtraction}.
The program can compute every kind of differential distribution for
unoriented infrared-safe quantities, and the results can be
plotted in histograms for future analysis.
At the same time, it implements the calculation of the three-jet decay
rate according to the E, EM, JADE
and DURHAM jet-clustering algorithms.

Instead of publishing figures illustrating the differential  behavior of
the shape variables, we have
preferred to compute some moments of these distributions, because these
results can be obtained with high accuracy and the comparison with
other groups is easier.

In this spirit, we have calculated the first five moments of the
thrust, the heavy-jet mass, the $c$ parameter and the energy--energy
correlation, for the ratio $m/E$ equal to 0.01, 0.1 and 0.3.  We have
also computed the three-jet decay rate with a cut parameter ranging
from 0.01 to 0.2.  We have calculated separately these quantities for
a hypothetical vector boson with purely axial or purely vector
couplings, and we have collected our results in the tables of
Chapter~\ref{chap:results}.  In this way, we have succeeded to perform
a partial comparison between our results for the jet-clustering
algorithms with those  of Ref.~\cite{Rodrigo}, and we have found
satisfactory agreement.

Some other applications of our calculation have appeared in the
literature.\\
In Ref.~\cite{NO1}, we have studied the  momentum correlation
in the decay of the $Z/\gamma$ boson into a couple of heavy quarks, at
order $\as^2$. We have found small corrections, of order of 1\%, to this
quantity, so that, non-perturbative  effects, of order $\Lambda/E$ (where
$\Lambda$ is a typical hadronic scale) may compete with the
perturbative result.

In Ref.~\cite{no_fragmentation}, we have given a verification of
several ingredients that are used in the fragmentation-function
formalism for heavy-quark production, such as the initial conditions,
computed in Ref.~\cite{MeleNason}, and the NLO splitting functions in
the time-like region~\cite{CurciFurmanskiPetronzio}.  This check has
been a further test of the validity of our massive
calculation.  In addition, by merging our fixed order cross section
with the NLO resummed one, we have derived an improved differential
cross section, which is accurate at NLO for $E \approx m$, but at
next-to-leading-log for $E \gg m$.

%  start APPENDIX
\appendix
\chapter{Useful integrals}
\thispagestyle{plain}
%\input{phd_useful_int}
%\section{Useful integrals}

In this appendix, we want to introduce some useful functions and
integrals, with their properties, which have been extensively used
throughout our computations.

The Gamma function is defined by
\beq
\Gamma(z) = \int_0^\infty dx\,{\rm e}^{-x} x^{z-1} \hspace{1 cm} \Re
z > 0\;,
\eeq
and, from this definition, you can demonstrate
\beq
\int_0^1 dx\, x^\a (1-x)^\b = 2\int_0^{\frac{\pi}{2}} d\phi\,
\(\sin\phi\)^{2\alpha+1} \(\cos\phi\)^{2\beta+1} =
\frac{\Gamma(\a+1)\,\Gamma(\b+1)}{\Gamma(\a + \b +2)} \;.
\eeq
A useful expansion is given by
\beq
\Gamma(1+ \e) = 1 - \gamma_E\,\e +
\frac{6\,\gamma_E ^2 + \pi^2}{12} \,\e^2 + {\cal O}\(\e^3\)\;,
\eeq
where $\gamma_E=0.5772157\ldots$ is the Euler-Mascheroni constant.

The dilogarithm function is given by
\beq
\li{x} = -\int_0^x dz \, \frac{\log(1-z)}{z} \hspace{1cm} x \leq 1\;,
\eeq
and an immediate consequence of this definition  is the following
expansion in powers of $\e$
\beqn
\int_0^1 dx\, x^{-1-\gamma\,\e} \(1-\a \,x\)^{\beta\, \e} &=&
\int_0^1 dx\, x^{-1-\gamma\,\e} \left[1 + \beta\,\e \log(1-\a \,x) + {\cal
O}\(\e^2\) \right]  \nonumber \\
&=& -\frac{1}{\gamma\,\e} -\beta\,\e \li{\a} + {\cal O}\(\e^2 \)\;.
\eeqn
One of the most used properties is the analytic continuation of the
dilogarithm function
\beq
\label{eq:continuaz_dilog}
\li{x\pm i\eta} = -\li{\frac{1}{x}} -\frac{1}{2}
\log^2 x +\frac{\pi^2}{3} \pm i \pi \log x \hspace{1 cm} x > 1\;,
\eeq
that can be demonstrated with the help of
\beq
\label{eq:continuaz_log}
\log(-x \pm i\eta) = \log x \pm i \pi   \hspace{1cm} x>0\;.
\eeq
In addition
\beq
\label{eq:projective}
\int_0^1 dx \, x^{-1-\e} \frac{1}{(1+\a\,x)^{1+\e}} = (1+\a)^\e
\int_0^1 dx\, x^{-1-\e} \( 1-\frac{\a}{1+\a} x\)^{2\e} \quad\quad
\a > -1  \;, \phantom{aa}
\eeq
where we have used the projective transformation
\beq
x \rightarrow \frac{x}{1 +\a\,(1-x)}\;.
\eeq
When calculating loop integrals, it is often useful to employ the
Feynman parametrization, which, in the most general form, reads
\beq
\label{eq:Feynman_param}
\prod_{i=1}^n \frac{ \displaystyle 1}{ \displaystyle D_i^{c_i}} =
\frac{ \displaystyle\Gamma\(c\)}
{ \displaystyle\prod_{i=1}^n \Gamma\(c_i\)} \int_0^1
\prod_{i=1}^n d\a_i\,\a_i^{c_i-1}\,
\delta\(1-\sum_{j=1}^n \a_j\)\frac{ \displaystyle  1}{ \displaystyle
\(\sum_{k=1}^n \a_k\, D_k \)^c} \;,
\eeq
where $c_i$ are arbitrary complex numbers and
\beq
c = \sum_{i=1}^n c_i  \;.
\eeq
For our future purpose, we need to compute the following Feynman integral
\beqn
I &=& \int \frac{d^dl}{(2\pi)^d}\ \frac{1}{(l+p_1)^2 -m_1^2+i\eta}\,
\frac{1}{(l+p_1+p_2)^2 -m_2^2+i\eta} \cdots\nn \\
\label{eq:def_loop}
&& \phantom{\int \frac{d^dl}{(2\pi)^d}\  }
\cdots\frac{1}{(l+p_1+p_2+\cdots +
p_n)^2 -m_n^2+i\eta}  \;,
\eeqn
where $(+i\eta)$, with $\eta > 0$, gives the prescription of how the
contour integral has to be deformed around the poles, and
\beq
\sum_{i=1}^n p_i = 0\;.
\eeq
We can look at $I$ as being
\beq
I = \int \frac{d^dl}{(2\pi)^d}\ \prod_{i=1}^n \frac{1}{D_i} \;,
\eeq
with $c_i=1\ (i=1,2,\ldots,n)$, $c=n$ and $D_i$ being the single
denominators in eq.~(\ref{eq:def_loop}).
Using the identity~(\ref{eq:Feynman_param}) and integrating over $l$,
we can rewrite $I$ in the following way
\beqn
\label{eq:Feynman_reduction}
\nonumber
I &=& (-1)^n \frac{i}{\(4\pi\)^{\(\frac{d}{2}\)}}\,
\Gamma\(n-\frac{d}{2}\)\int_0^1 \prod_{i=1}^n d\a_i\
\delta\(1-\sum_{j=1}^n \a_j\) \ \frac{1}{D^{n-\frac{d}{2}}} \\
&\equiv& (-1)^n \frac{i}{\(4\pi\)^{\(\frac{d}{2}\)}}\,
\Gamma\(n-\frac{d}{2}\)\int
\frac{[d\a]}{D^{n-\frac{d}{2}}}\;,
\eeqn
where
\beq
\label{eq:Feynman_denominator}
D = -\sum_{i>j} \a_i\,\a_j\, s_{ij} + \sum_i \a_i\, m^2_i -i\eta\;,
\eeq
and $s_{ij}$ is the square of the momentum flowing through the $i$-$j$
cut of the diagram representing $I$.
\chapter{One-loop scalar integrals}
\thispagestyle{plain}
%\input{phd_one_loop}
% \section{One-loop scalar integrals}
\label{app:one_loop}

\section{Kinematical invariants}
We  introduce now the following kinematical invariants (not independent)
\beqn
\label{eq:def_s_1}
s_1 &=& (q-p')^2 = (k+p)^2  \geq m^2 \\
s_2 &=& (q-p)^2  =(k+p')^2  \geq m^2\\
\label{eq:def_s_3}
s_3 &=& (q-k)^2 =  (p+p')^2 \geq 4 m^2 \\
\su &=& (q-p')^2 -m^2 = q^2(1-x_2)  \\
\sd &=& (q-p)^2 -m^2 = q^2(1-x_1) \\
\label{eq:def_sig3}
\st &=& (q-k)^2 = q^2(1-x_g) \;,
\eeqn
where $x_1$ and $x_2$ are defined by~(\ref{eq:x1x2y}) and
$x_g$ by~(\ref{eq:xg_def}).

We can classify the different types of scalar integrals according to
the number of massive propagators in the loop and according to the
``shape'' of the loop: box (B) and triangle (T).

\section[Box with two massive propagators: $B_{2m}$]
{Box with two massive propagators: {\mylarge $B_{2m}$}}
\label{app:b_2m}
%%%%%%%%%%%%%%%%%%%%%%%%%%%%%%%%%%%%%%%%%%%
\begin{figure}[htb]
\centerline{\epsfig{figure=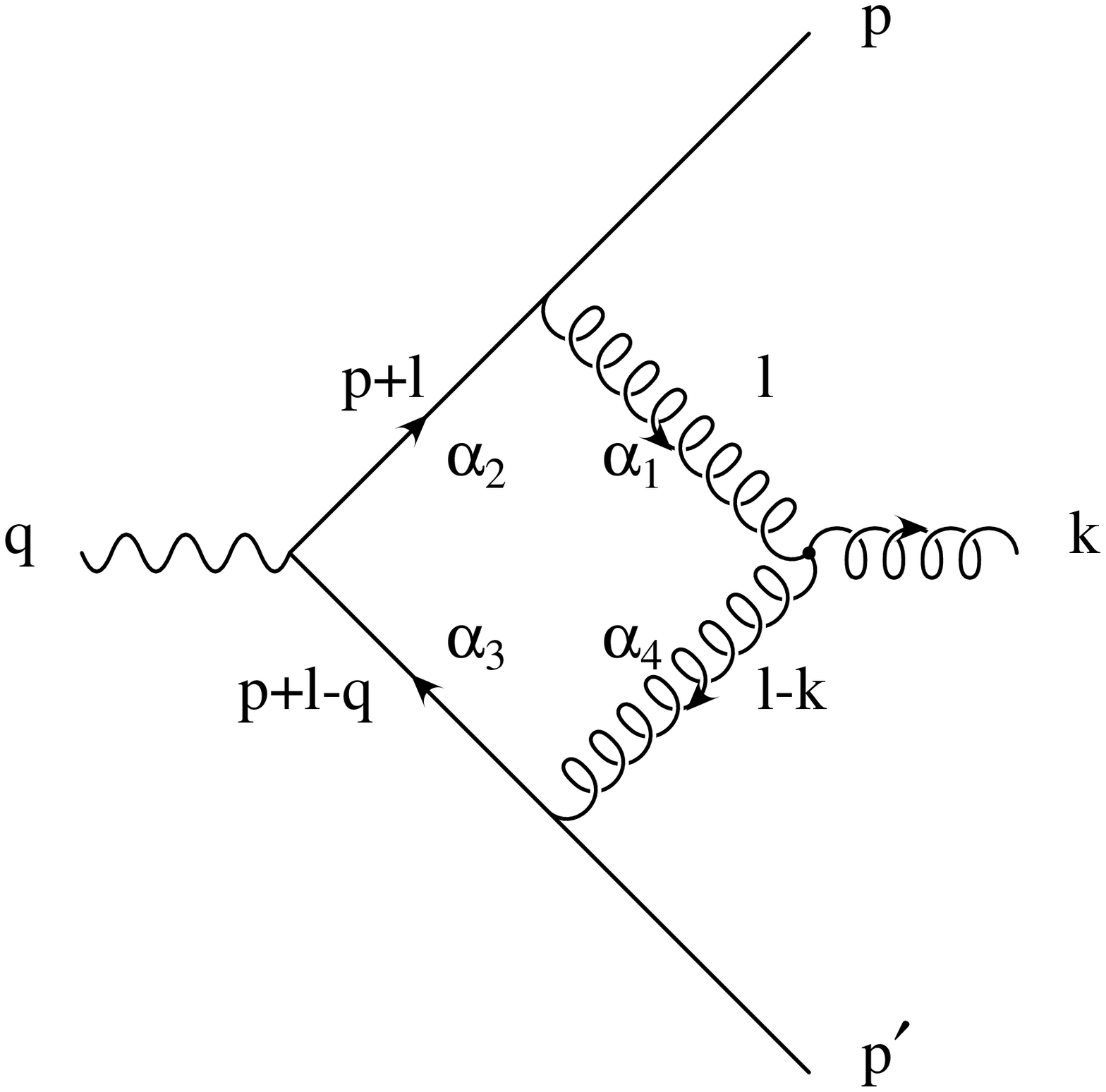,
width=0.43 \textwidth,clip=}}\ccaption{}{ \label{fig:b_2m}
Diagram representing $B_{2m}$.}
\end{figure}
%%%%%%%%%%%%%%%%%%%%%%%%%%%%%%%%%%%%%%%%%%%
We are referring, as far as the denominator structure is concerned, to
a Feynman diagram of the type depicted in Fig.~\ref{fig:b_2m}
\beq
B_{2m}\equiv\int \frac{d^dl}{(2\pi)^d}\ \frac{1}{l^2} \ \frac{1}{(l-k)^2} \
\frac{1}{(l+p-q)^2-m^2} \ \frac{1}{(l+p)^2-m^2} \;.
\eeq
Using eqs.~(\ref{eq:Feynman_reduction})
and~(\ref{eq:Feynman_denominator}), we can rewrite this integral as
\beq
\label{eq:b_2m}
B_{2m}= \frac{i}{16\pi^2}\ (4\pi)^\e \ \Gamma(2+\e) \int
\frac{[d\a]}{D^{2+\e}}\;,\quad\quad\quad d=4 -2\e\;,
\end{equation}
where
\beqn
D &=& {}-\lq\a_1\,\a_2\, m^2 +\a_1\,\a_3\,(q-p)^2
+\a_2\,\a_3\,q^2+\a_2\,
\a_4\, (p+k)^2+ \a_3\,\a_4\, m^2\rq \nn\\
 &&{} +(\a_2+\a_3)\,m^2-i\eta \;.
\eeqn
According to the definitions given in
eqs.~(\ref{eq:def_s_1})--(\ref{eq:def_sig3}), $D$ becomes
\beqn
D \!\!&=& \!\!
-\lq m^2(\a_1\,\a_2+\a_3\,\a_4-\a_2-\a_3)+s_2\,\a_1\,\a_3+q^2\a_2\,\a_3 +
s_1\,\a_2\,\a_4 + i\eta\rq \nn \\
 &=& \!\!-\lq- m^2(\a_2+\a_3)^2 +(s_2-m^2)\a_1\,\a_3
+q^2\a_2\,\a_3+(s_1-m^2)\a_2\,\a_4+i\eta\rq\;.\phantom{aaaa}
\eeqn
In order to perform the integration, it is useful to continue the
denominator in a region where it is a negative-definite form.  For
this reason, we assume $m^2 < 0$, and we perform the integration of a
function that now has no poles on the integration contour.  At the
end, we return to the physical region according to the $i\eta$
prescription
\beq
\label{eq:M2}
   M^2 \equiv  -(m^2-i\eta) > 0 \;.
\eeq
With this definition, we have
\beq
 D=-\lq M^2 (\a_2+\a_3)^2
 +\sd\,\a_1\,\a_3+q^2\a_2\,\a_3+\su\,\a_2\,\a_4 \rq \;,
\eeq
and eq.~(\ref{eq:b_2m}) becomes
\beq
\label{eq:b_2m_I}
  B_{2m} = \frac{i}{16\pi^2}\ (4\pi)^\e \ \Gamma(2+\e) \
\frac{1}{(-1-i\eta)^\e} \int
  \frac{[d\a]}{\left[-D\right]^{2+\e}} = 
 \frac{i}{16\pi^2}\ (4\pi)^\e \ \Gamma(2+\e) \ {\rm e}^{i\pi\e}
 \,I \;. \phantom{aa} 
\eeq
We make the following change of variables
\beq
\label{eq:change_var_b_2m}
\begin{array}{l}
  \a_1 = (1-y)z  \\
  \a_2 = xy      \\
  \a_3 = (1-x) y  \\
  \a_4 = (1-y)(1-z)\;, \\
\end{array}
\eeq
where the Jacobian of the transformation is
\[   \frac{\partial(\a_1\,\a_2\,\a_3)}{\partial(x\,y\,z)} = y(1-y)\;.
\]
In this way, we can write $I$ of eq.~(\ref{eq:b_2m_I}) as
\beqn   
I \!&=& \!\int_0^1dx\,dy\,dz\ y\,(1-y) \nn \\
  &&\! {}\times\frac{1}{\Bigl[ M^2y^2+\sd yz(1-x)(1-y)+q^2
    xy^2(1-x)+\su xy(1-y)(1-z)\Bigr]^{2+\e}} \nn\\
 &=& \!\int_0^1dx\,dy\,dz\ y^{-1-\e}(1-y)\nn\\
 &&  \! {}\times\frac{1}{\Bigl\{ z\Bigl[\sd(1-x)(1-y)-
 \su x(1-y)\Bigr] + M^2y+q^2 x
    y(1-x)+\su x(1-y)\Bigr\}^{2+\e}}\;. \nn
\eeqn
The $z$ integration is easily done
\beqn
 I  &=& \int_0^1dx\,dy \frac{1}{1+\e}\ \frac{y^{-1-\e}}{\sd(1-x)-\su
 x} \left\{ \frac{1}{\Bigl[M^2y + q^2xy(1-x)+\su
      x(1-y)\Bigr]^{1+\e}} \right.  \nn \\
&&
\left. {}- \frac{1}{\Bigl[\sd (1-x)(1-y)+M^2y+q^2 xy(1-x)\Bigr]^{1+\e}}
\right\}\;, 
\eeqn
and, making the change of variable $x \rightarrow 1-x$ in the second
integral, we have
\beqn
I &=& \int_0^1dx\,dy \frac{-1}{1+\e}\ \frac{y^{-1-\e}}{\sd x -
\su(1- x)} \
\frac{1}{\Bigl[\sd x(1-y)+M^2y+q^2 xy(1-x)\Bigr]^{1+\e}}\nn\\
&& {} +  \int_0^1dx\,dy \frac{-1}{1+\e}\ \frac{y^{-1-\e}}{\su
  x-\sd(1-x)} \ \frac{1}{\Bigl[\su x(1-y)+M^2y+q^2
    xy(1-x)\Bigr]^{1+\e}} \;. \nn
\eeqn
We then decompose $I$ as the sum of the two integrals
\[
  I = I_1(\su,\sd)+I_2(\su,\sd)\;,
\]
which have  the symmetry
\[  I_2(\su,\sd) = I_1(\sd,\su)\;,
\]
so that $I$ becomes
\beq
\label{eq:int_totale}
  I = I_1(\su,\sd)+I_1(\sd,\su)\;,
\eeq
and we have to perform only one integration
\[ I_1 = \int_0^1dx\,dy \frac{-1}{1+\e}\ \frac{1}{\sd x-\su(1- x)} \
\frac{1}{(\sd x)^{1+\e}}\ \frac{y^{-1-\e}}{\left\{ 1
    +y\left[-1+\frac{M^2}{\sd x}+ \frac{q^2}{\sd}(1-x)\right]
  \right\}^{1+\e}} \;.
\]
Using the identity~(\ref{eq:projective})
\beqn
I_1 &=& \int_0^1dx\,dy \frac{-1}{1+\e}\ \frac{1}{\sd x-\su(1- x)} \
\frac{1}{(\sd x)^{1+\e}}\ \( \frac{M^2}{\sd x}+ \frac{q^2}{\sd}(1-x)
\)^\e  \nn\\
&& {}\times y^{-1-\e}\,  \left[ 1 -y \frac{-x+\frac{M^2}{\sd}+
     \frac{q^2}{\sd} x (1-x)}{\frac{M^2}{\sd }+ \frac{q^2}{\sd} x
     (1-x)}\right]^{2\e}  \nn \\
 &=& \frac{-1}{1+\e} (M^2)^{-\e} \int_0^1dx\,dy\, \frac{1}{\sd x
  -\su(1-x)} \ \frac{M^{2\e}}{(\sd x)^{1+2\e}} \left[ M^2+q^2 x(1-x)
\right]^\e \nn \\
 && {}\times y^{-1-\e}\,  \Biggl\{ 
    \Biggl(1-y+y\, \frac{x}{\frac{M^2}{\sd}+ \frac{q^2}{\sd}x(1-x)}
    \Biggr) ^{2\e}-(1-y)^{2\e}  {}+(1-y)^{2\e} \Biggr\} \;,\phantom{aaaa}
\label{eq:step1}
\eeqn
where, in the curly braces, we have added and subtracted the same quantity.
If we consider now the first two terms in the curly braces and expand them
in $\e$, we have
\beqn
f(x,y) & \equiv&\Biggl(1-y+y\, \frac{x}{\frac{M^2}{\sd}+ 
\frac{q^2}{\sd}x(1-x)} \Biggr) ^{2\e}-(1-y)^{2\e} \nn\\
&=& 2\,\e \, \Biggl\{ 
 \log \Biggl(1-y+y\, \frac{x}{\frac{M^2}{\sd}+ \frac{q^2}{\sd}x(1-x)}
    \Biggr) - \log (1-y) \Biggr\} + {\cal O}\(\e^2\)\;. \phantom{aaaaa}
\eeqn
Since we are keeping only terms till order ${\cal O}\(\ep^0\)$,
$f(x,y)$ gives contribution to the integral only if it is multiplied
by factors at least of order  $1/\e$. Such factors may arise only at
singular points, which, for eq.~(\ref{eq:step1}), are $x=0$ and $y=0$: in
fact $f(x,y)$ is multiplied by $x^{-1-2\e} y^{-1-\e}$.
In the limit $x\,\to\,0$, $y\,\to\,0$, we have
\beqn
x^{-1-2\e} y^{-1-\e} f(x,y) &=&  2\,\e \,x^{-1-2\e} y^{-1-\e} \Biggl(
-y+y\, \frac{x}{\frac{M^2}{\sd}+ \frac{q^2}{\sd}x(1-x)} +y +
{\cal O}\(xy^2\)\Biggr)\nn \\
 &\approx& 2\,\e\,x^{-2\e} y^{-\e}
\frac{1}{\frac{M^2}{\sd}+ \frac{q^2}{\sd}x(1-x)} \;,
\eeqn
that gives zero contribution in the limit $\e\,\to\,0$. In this way,
$f(x,y)$ gives no contributions and can be disregarded.
By integrating  eq.~(\ref{eq:step1}) in $y$, we obtain
\beqn
I_1 &=& \frac{(M^2)^{-\e}}{1+\e} \ \frac{1}{\e}\
 F_{\e} \nn\\
&&{}\times\int_0^1 dx\ \frac{1}{\sd x-\su(1-x)}\ \frac{1}{\sd x} \ x^{-2\e}
\(\frac{M^2}{\sd}\)^{2\e} \left[1 +\frac{q^2}{M^2}x(1-x)
\right]^\e \label{eq:step2}\\
 & = &\frac{(M^2)^{-\e}}{1+\e} \ \frac{1}{\e}\ \frac{F_{\e}}{\su \sd}
  \nn \\
&& {}\times
\int_0^1 dx\ \left[ \frac{\su+\sd}{\sd x-\su(1-x)}- \frac{1}{x}\right]
\(\frac{M^2}{\sd}\)^{2\e} x^{-2\e} \left[1
  +\frac{q^2}{M^2}x(1-x) \right]^\e  \nn \\
 &=& \frac{(M^2)^{-\e}}{1+\e} \ \frac{1}{\e}\ \frac{F_{\e}}{\su \sd} 
 \left\{
  \int_0^1 dx\ \frac{\su+\sd}{\sd x-\su(1-x)} \( 1 + 2\e
    \log\frac{M^2}{\sd x} + {\cal O}\(\e^2\)
  \)   \right.\nn \\
 && \times \left[ 1 + \e \log\(1+\frac{q^2}{M^2}x(1-x)\) + {\cal O}\(\e^2\)
\right] \nn \\
&&\left. {} -\int_0^1 dx \ \( \frac{M^2}{\sd} \)^{2\e} \, x^{-1-2\e}
  \left[ 1+\e\log\(1+\frac{q^2}{M^2}x(1-x)\) + {\cal O}\(\e^2\)
  \right] \right\} \nn \\
& =& \frac{(M^2)^{-\e}}{1+\e} \ \frac{1}{\e}\ \frac{F_{\e}}{\su \sd}
\ \Biggl\{
  \int_0^1 dx\ \frac{\su+\sd}{\sd x-\su(1-x)} \nn \\
&& \times\Biggl[ \underbrace{ 1+\e
      \log\Biggl( 1+\frac{q^2}{M^2} x(1-x)\Biggr) }_{A_1} +2\e
    \log\frac{M^2}{\sd x}+ {\cal O}\(\e^2\) \Biggr] \nn \\
&&-\left[1+2\e \log\frac{M^2}{\sd}+2\e^2 \log^2 \frac{M^2}{\sd}+
  {\cal O}\(\e^3\)\right] \nn \\
&&\times  \Biggl[ -\frac{1}{2\e}+ \e \underbrace{ \int_0^1 dx \ 
      x^{-1-2\e}\log\Biggl( 1 + \frac{q^2}{M^2}x(1-x) \Biggr) } _{{\rm
        finite\ for\ } \e\,\rightarrow\,0} + {\cal O}\(\e^2\) \Biggr]
 \Biggr\} \;,
\eeqn
where
\beq
F_{\e}= \frac{\Gamma(1-\e) \, \Gamma(1+2\e)}{\Gamma(1+\e)} =
1+\frac{\pi^2}{3} \e^2 +   {\cal O}\(\e^3\)  \;.
\eeq
The contribution given by $A_1$  to the total integral $I$  is zero. 
In fact, using eq.~(\ref{eq:int_totale}), we have
\[  \int_0^1 dx\  \frac{\su+\sd}{\sd x-\su(1-x)} A_1 +
 \int_0^1 dx\  \frac{\sd+\su}{\su x-\sd(1-x)} A_1 = 0\;,
\]
and then
\beqn
I_1 &=& \frac{(M^2)^{-\e}}{1+\e} \ \frac{F_{\e}}{\su \sd} 
\left\{ -2(\su+\sd)   \int_0^1 dx\ \frac{\log \frac{\sd x}{M^2} }
{\sd x -\su(1-x)} \right.\nn \\
&&  {} -\int_0^1 dx
  \ \frac{1}{x}\log\( 1+\frac{q^2}{M^2}x(1-x) \) +
\left.  \frac{1}{2\e^2} +\frac{1}{\e}\log\frac{M^2}{\sd} +
  \log^2\frac{M^2}{\sd} \right\}\;. \phantom{aaaaa}
\eeqn
From eq.~(\ref{eq:int_totale}) we have
\beqn
I &=& \frac{(M^2)^{-\e}}{1+\e} \ \frac{F_{\e}}{\su \sd}\nn \\
&& {}\times\left\{ -2(\su+\sd)
  \int_0^1 dx\ \frac{\log\frac{\sd x}{\su(1-x)}}{\sd x -\su(1-x)}
  -2\int_0^1  \frac{dx}{x}\log\( 1+\frac{q^2}{M^2}x(1-x)
  \)\right.\nn \\
&& \left.{} + \frac{1}{\e^2}
  +\frac{1}{\e}\(\log\frac{M^2}{\su}+\log\frac{M^2}{\sd}\) +
  \log^2\frac{M^2}{\su} + \log^2\frac{M^2}{\sd} \right\}\;.
\label{eq:semifinal}
\eeqn
In this way, we succeed in isolating all the divergent contributions,
which have appeared as poles in $\e$. The remaining integrals are
finite. For the first integral in
eq.~(\ref{eq:semifinal}) we have
\[
\int_0^1 dx\ \frac{\log\frac{\sd x}{\su(1-x)}}{\sd x -\su(1-x)} =
\frac{1}{2\(\su+\sd\)} \(\log^2\frac{\su}{\sd} +\pi^2 \) \;,
\]
that can be demonstrated using the contour integrals around the poles $0,1$
and $\su/\(\su+\sd\)$, and the identity
\beq
  \log\(\frac{\sd}{\su} \frac{x}{1-x}\) = \frac{1}{4\pi\,i} 
 \left[ \log^2  \left.\( -\frac{\sd}{\su} \frac{x}{1-x}\)\right|_{x-i\e} -
  \log^2 \left.\( -\frac{\sd}{\su} \frac{x}{1-x}\)\right|_{x+i\e} \right]\;.
\eeq
The last integral in eq.~(\ref{eq:semifinal}) is solved with
\beqn
\int_0^1 \frac{dx}{x}\log\left[ 1+x(1-x)\frac{q^2}{M^2}\right] &=&
\int_0^1 \frac{dx}{x}\log\( 1-\frac{x}{\lp}\) +
\int_0^1 \frac{dx}{x}\log\( 1-\frac{x}{\lm}\) \nn \\
&=&-\li{\frac{1}{\lp}}-\li{\frac{1}{\lm}} = \frac{1}{2} \log^2 \(
1-\frac{1}{\lm}\) \nn \\
&=& \frac{1}{2}\log^2\frac{-\lm}{\lp}\;,
\eeqn
where we have defined
\beq
\label{eq:lambdapm}
\lpm = \frac{1}{2}\( 1 \pm \sqrt{1+\frac{4M^2}{q^2}}
\) \equiv \frac{1}{2}\( 1\pm \d \)
\eeq
\[
     \lp+\lm= 1 \;.
\]
The integral $I$  becomes
\beqn
I &=& \frac{(M^2)^{-\e}}{1+\e} \ \frac{F_{\e}}{\su \sd} \left\{ - \(
    \log^2\frac{\su}{\sd}+\pi^2\)
    -\log^2 \(1-\frac{1}{\lm}\)  +
  \log^2\frac{M^2}{\su} + \log^2\frac{M^2}{\sd} \right.
\nn \\
&&\left.{} + \frac{1}{\e^2}
  +\frac{1}{\e}\(\log\frac{M^2}{\su}+\log\frac{M^2}{\sd}\)\right\} \;,
\eeqn
and, according to eq.~(\ref{eq:b_2m_I}),
\beqn
B_{2m} &=& \frac{i}{16\pi^2} \
  (4\pi)^\e \ \Gamma(1+\e)  \ {\rm e}^{i\pi\e}\ 
(M^2)^{-\e} \frac{1}{\su \sd}  \left\{ \frac{1}{\e^2}
  +\frac{1}{\e}\(\log\frac{M^2}{\su}  +\log \frac{M^2}{\sd}\)
  \right.\nn \\
&&\left.{} + 2
  \log\(\frac{M^2}{\su} \) \log\(\frac{M^2}{\sd} \) -
  \frac{2}{3}\pi^2 - \log^2\(1-\frac{1}{\lm} \)
\right\} \;.
\eeqn

We return to the physical region with the analytic
continuation of $B_{2m}$, that is, using eq.~(\ref{eq:M2})
\beq
\label{eq:prolung_1}
 (M^2)^{-\e}=(m^2)^{-\e} {\rm e}^{-i\pi \e}
\eeq
\beq
\label{eq:prolung_2} 
\log M^2= \log \(-m^2+i\eta\)= \log m^2 + i\pi
\eeq
\beq  
\label{eq:lambdapm_m}
\lpm = \frac{1}{2}\( 1 \pm \sqrt{1-\frac{4m^2}{q^2}} \) \pm i\eta =
\frac{1}{2}\( 1 \pm \d \) \pm i\eta \;.
\eeq
You can see that
\beq
   0\leq \lm\leq 1/2\;, \quad\quad\quad\quad
   1/2\leq \lp\leq 1 \;,
\eeq
and then
\dl{
1-\frac{1}{\lm} <0 \tol 1-\frac{1}{\lm} -i\eta \;,
\hfill}
so that the logarithm acquires an imaginary part
\dl{
\log\(1-\frac{1}{\lm}\) = -\log\frac{-\lm}{\lp} \tol \log\frac{\lp}{\lm}
-i\pi \;.
\hfill}
The final result is
\beqn
B_{2m} &=& N(\e)\, \(m^2\)^{-\e}
\frac{1}{\su \sd} \left\{ \frac{1}{\e^2}+\frac{1}{\e}\(
    \log\frac{m^2}{\su}+\log\frac{m^2}{\sd}\) + 2
  \log\frac{m^2}{\su}\, \log\frac{m^2}{\sd} \right. \nn \\
&& \left. {} -\frac{5}{3}\pi^2
  -\log^2 \frac{\lp}{\lm} + 2 \,\pi\, i \left[
    \frac{1}{\e}+\log\frac{m^2}{\su}+\log\frac{m^2}{\sd} + \log
    \frac{\lp}{\lm} \right] \right\} \;,
\eeqn
where
\beq
 \Nep = \frac{i}{16\pi^2} \(4\pi\)^\e \Gamma(1+\e) \;.
\eeq
\section[Box with three massive propagators: $B_{3m}$]
{Box with three massive propagators: {\mylarge $B_{3m}$}}
\label{app:b_3m}
%%%%%%%%%%%%%%%%%%%%%%%%%%%%%%%%%%%%%%%%%%%
\begin{figure}[htb]
\centerline{\epsfig{figure=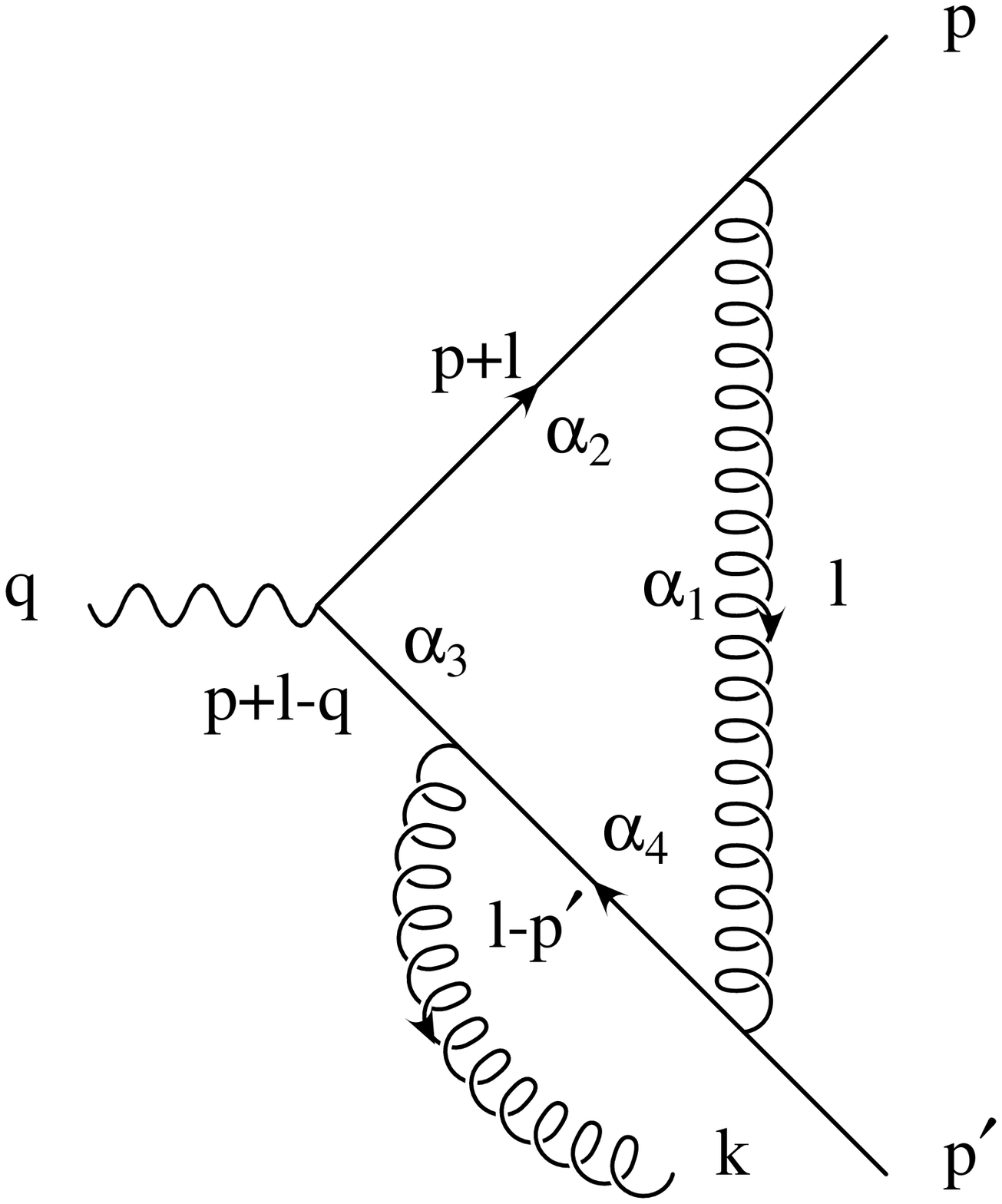,
width=0.36\textwidth,clip=}}\ccaption{}{ \label{fig:b_3m}
Diagram representing $B_{3m}$.}
\end{figure}
%%%%%%%%%%%%%%%%%%%%%%%%%%%%%%%%%%%%%%%%%%%
We are referring, as far as the denominator structure is concerned, to
a Feynman diagram of the type represented in Fig.~\ref{fig:b_3m}
\beq
B_{3m}\equiv \int \frac{d^dl}{(2\pi)^d}\ \frac{1}{l^2} \
\frac{1}{(l+p)^2-m^2} \
\frac{1}{(l+p-q)^2-m^2} \ \frac{1}{(l-p')^2-m^2}\;. \nn 
\eeq
Performing the same steps done for the calculation of $B_{2m}$, we have
\beq
\label{eq:b_3m}
 B_{3m}= \frac{i}{16\pi^2}\ (4\pi)^\e \ \Gamma(2+\e) \int
\frac{[d\a]}{D^{2+\e}}\;, \quad\quad\quad  d=4 -2\e \;,
\eeq
where
\beqn
D &=& -\lq\a_1\,\a_2\, m^2 + \a_1\, \a_3\,(q-p)^2 + \a_1\,\a_4\, m^2 +
\a_2\,\a_3\, q^2 +\a_2\,\a_4\, (q-k)^2\rq \nn \\
&& + (\a_2 + \a_3+\a_4)\, m^2 -i \eta \;.
\eeqn
Using the definitions of the kinematical invariants of
eqs.~(\ref{eq:def_s_1})--(\ref{eq:def_sig3}) and continuing the
integral into the unphysical region of negative mass, according to
eq.~(\ref{eq:M2}), we have
\beqn
D &=& -\lq\a_1(\a_2+\a_4)\,m^2 + \a_1\,\a_3\, s_2 + \a_2\,\a_3\,
q^2+ \a_2\,\a_4\,s_3 -(1-\a_1)\, m^2 + i\eta\rq \nn\\
 &=& -\lq- m^2 (1-\a_1)^2 + \sd \,\a_1\,\a_3 + q^2\, \a_3\,\a_2+
s_3\,\a_2\,\a_4 +i\eta\rq \nn \\
 &=& - \left[ M^2 (1-\a_1)^2 +\sd\,\a_1\,\a_3+ q^2\, \a_2\,\a_3+s_3\,
\a_2\,\a_4 \right] \;.
\eeqn
The integral~(\ref{eq:b_3m}) then becomes
\beq 
 B_{3m} = \frac{i}{16\pi^2}\ (4\pi)^\e \ \Gamma(2+\e) \
{\rm e}^{i\pi\e} \int \frac{[d\a]}{\left[-D\right]^{2+\e}} = (1+\e)\,
\Nep \, {\rm e}^{i\pi\e} \, I \;.
\eeq
By making the following change of variables
\beq
\label{eq:change_var_b_3m}
\begin{array}{l}
  \a_1 = 1-x  \\
  \a_2 = x(1-z)      \\
  \a_3 = xyz  \\
  \a_4 = x(1-y)z\;,  \\
\end{array}
\eeq
and taking into account the Jacobian of the transformation
\[    \frac{\partial(\a_1\,\a_2\,\a_3)}{\partial(x\,y\,z)} = x^2 z \;,
\]
we have
\beqn
 I &=& \int_0^1dx\,dy\,dz\,
  \frac{x^2 z}{\Bigl[ M^2 x^2 + \sd(1-x)xyz
  +s_3 x^2(1-y)z(1-z) + q^2 x^2yz(1-z)\Bigr]^{2+\e}} \nn\\
   &=& \int_0^1dx\,dy\,dz\, \frac{z x^{-\e}}{ \Bigl[ M^2 x + \sd(1-x)yz
  +s_3 x(1-y)z(1-z) + q^2 xyz(1-z)\Bigr]^{2+\e}} \nn \\
 &=& \int_0^1dx\,dz\, \frac{-1}{1+\e}\ \frac{x^{-\e}}{\sd(1-x)-s_3 x(1-z) +
q^2 x(1-z)} \nn \\
 &&\times \left[ \frac{1}{\Bigl[ M^2 x +\sd(1-x)z+q^2 xz(1-z)\Bigr]^{1+\e}}
-\frac{1}{\Bigl[ M^2 x +s_3 xz(1-z)\Bigr]^{1+\e}}
\right] \nn \\
 &=& - \frac{1}{1+\e}\int_0^1dx\,dz\,\frac{1}{\sd(1-x)+(q^2-s_3) x(1-z)
} \nn \\
&&\times \left[ \frac{x^{-\e}}{\Bigl[ M^2 x +\sd(1-x)z+q^2 xz(1-z)
\Bigr]^{1+\e}} -\frac{x^{-1-2\e}}{\Bigl[ M^2  +s_3 z(1-z)\Bigr]^{1+\e}}
\right] \;.
\eeqn
When $x\,\to\, 0$, the second integral is divergent: for this reason, we
add and subtract the asymptotic expression of the divergent function 
when $x$ goes to zero
\beqn
I \!\!&=& \!\! -\frac{1}{1+\e} \lg \int_0^1dx\,dz\, \frac{1}{\sd(1-x) +
(q^2-s_3) x(1-z) }  \phantom{\frac{x^{-\e}}{\Bigl[ M^2\Bigr] ^{1+\e}}} 
\right.\nn \\
&&  \!\!\times \lq \frac{x^{-\e}}{\Bigl[ M^2 x +\sd(1-x)z+q^2
xz(1-z)\Bigr]^{1+\e}}
-\frac{x^{-1-2\e}}{\Bigl[ M^2  +s_3 z(1-z)\Bigr] ^{1+\e}} \rq  \nn \\
&& \!\! \left. +\int_0^1dx\,dz\, \frac{x^{-1-2\e}}{\sd\Bigl[ M^2 +s_3
z(1-z)\Bigr]^{1+\e}} - \int_0^1dx\,dz\, \frac{x^{-1-2\e}}{\sd\Bigl[ M^2 +s_3
z(1-z)\Bigr]^{1+\e}} \rg \phantom{aaaaa}
\label{eq:I_3rd}
\eeqn
We can make the following considerations:
\begin{enumerate}
  \item the first term is finite when  $\e\,\to\,0$
  \item the sum of the second and the third term is finite when  
   $x\,\to\, 0$ by construction, and then
  we can again put $\e = 0$
  \item  the integral of the last term is 
    \beqn
  I_L &\equiv&     \int_0^1dx\,dz\, \frac{x^{-1-2\e}}{\sd\,\Bigl[ M^2 +s_3
        z(1-z)\Bigr]^{1+\e}} \label{eq:I_R}\\
      &=&  -\frac{(M^2)^{-\e}}{2\,\sd\,\e}\int_0^1dz\,\frac{1}{M^2\Bigl[ 1+
        \frac{s_3}{M^2}  z(1-z)\Bigr]^{1+\e}} \nn \\
      &=&  -\frac{(M^2)^{-\e}}{2\,\sd\,\e}\int_0^1dz\,\frac{1}{M^2+s_3
           z(1-z)} +\frac{1}{2\,\sd}\int_0^1dz\,\frac{
        \log\left[1 +\frac{s_3}{M^2} z(1-z)\right]}{M^2+s_3 z(1-z)}\;.
         \nn 
     \eeqn
\end{enumerate}
By collecting a global factor $(M^2)^{-\e}$, eq.~(\ref{eq:I_3rd}) becomes
\beqn
I &=& -\frac{(M^2)^{-\e}}{1+\e}\left\{ \int_0^1dx\,dz\,\left[
\frac{1}{\sd(1-x)+(q^2-s_3) x(1-z)}  
\phantom{\frac{1}{\Bigl[ M^2}}
\right.\right.
\nn\\
&& \times \(\frac{1}{M^2 x +\sd(1-x)z+q^2
xz(1-z)}-\frac{1}{x\,\Bigl[ M^2  +s_3 z(1-z)\Bigr] } \)\nn\\
&&+ \left.\frac{1}{\sd\, x\,\Bigl[ M^2+s_3
z(1-z)\Bigr] } \right] \nn\\
&&  + \left. \frac{1}{2\,\sd\,\e}\int_0^1dz\,\frac{1}{M^2+s_3
  z(1-z)}-
  \frac{1}{2\,\sd}\int_0^1dz\,\frac{\log\left[1 +\frac{s_3}{M^2}
  z(1-z)\right]}{M^2+s_3
  z(1-z)}  \right\} \nn\\
&=& -\frac{(M^2)^{-\e}}{1+\e} \left\{ \int_0^1dx\,dz\,\frac{M^2 -\sd
z + q^2 z(1-z)}{\sd\, \Bigl[ M^2+s_3 z(1-z)\Bigr]\, \Bigl[ M^2 x
+\sd(1-x)z+q^2xz(1-z)\Bigr]} \right.
\nn\\
&& +\left. \frac{1}{2\,\sd\,\e}\int_0^1dz\,\frac{1}{M^2+s_3
  z(1-z)}-
  \frac{1}{2\,\sd}\int_0^1dz\,\frac{\log\left[1 +\frac{s_3}{M^2}
  z(1-z)\right]}{M^2+s_3
  z(1-z)}  \right\} \;.
\eeqn
The integration in  $x$ gives
\beqn
I &=& -\frac{(M^2)^{-\e}}{1+\e}\frac{1}{\sd}\int_0^1 dz \frac{1}{M^2
 +s_3 z(1-z)} 
 \Bigg\{ \log\frac{M^2 + q^2 z(1-z)}{\sd z}\nn\cr
 && + \frac{1}{2\e}-\frac{1}{2}\log\left[1+\frac{s_3}{M^2}
 z(1-z)\right] \Bigg\} \;,
\eeqn
and decomposing the denominator
\[ \frac{1}{M^2 +s_3 z(1-z)} = -\frac{1}{s_3 \dpr}\left[ \frac{1}{z-\zp} -
\frac{1}{z-\zm}\right]\;,
\]
where
\[  \dpr = \sqrt{1+\frac{4M^2}{s_3}} \;,\quad\quad\quad\quad
 \zpm = \frac{1}{2}\left[ 1 \pm \dpr \right]\;,
\]
and
\[  \zp +\zm = 1 \;,
\]
we have
\beqn
I &=& \frac{(M^2)^{-\e}}{1+\e} \frac{1}{\sd s_3 \dpr } \int_0^1 dz\, \left[
\frac{1}{z-\zp} - \frac{1}{z-\zm}\right] \Biggl\{ \( \frac{1}{2\e}+
\log\frac{q^2}{\sd} -\frac{1}{2}\log\frac{s_3}{M^2}\) 
\nn\\
&& - \log z + \log (z-\lm) +\log(\lp-z) -\frac{1}{2}\log(z-\zm)
-\frac{1}{2}\log(\zp-z) \Biggr\}\;,\phantom{aaaaa}
\eeqn
where  $\lpm$ are given by eq.~(\ref{eq:lambdapm}).

In order to compute these integrals, it is worth noticing that
\[  \zm \leq \lm \leq 0 < 1 \leq \zp \leq \lp \;,
\]
where we have used the fact that $q^2 \geq s_3$ (see
eq.~(\ref{eq:def_s_3})). In addition
%integrale
\dl{
\int_0^1 dz\, \left[\frac{1}{z-\zp} - \frac{1}{z-\zm}\right] =
2\log\frac{-\zm}{\zp} \hfill
 }
%integrale
\dl{
\int_0^1 dz\, \frac{\log z}{z-\zpm} = \li{\frac{1}{\zpm}}  \hfill
}
%integrale
\dl{
\int_0^1 dz\, \frac{\log(z-\zm)}{z-\zp} =
-\int_0^1 dz\, \frac{\log(\zp-z)}{z-\zm} \hfill\cr
\quad\phantom{\int_0^1 dz\, \frac{\log(z-\zm)}{z-\zp}}
 = \log \dpr \log\frac{-\zm}{\zp}
- \li{\frac{-\zm}{\dpr}}+ \li{\frac{\zp}{\dpr}} \hfill
}
%integrale
\dl{
\int_0^1 dz\, \frac{\log(z-\zm)}{z-\zm} \stackrel{\rm B.P.}{=}
\frac{1}{2}\log^2(\zp) -\frac{1}{2}\log^2(-\zm)  \hfill
}
%integrale
\dl{
\int_0^1 dz\, \frac{\log(\zp-z)}{z-\zp} \stackrel{\rm B.P.}{=}
\frac{1}{2}\log^2(-\zm) -\frac{1}{2}\log^2(\zp)  \hfill
}
%integrale
\dl{
\int_0^1 dz\, \frac{\log(z-\lm)}{z-\zm}= \log\lp \log\frac{\zp}{\di}
-\log(-\lm)\log\frac{-\zm}{\di}
-\li{\frac{\lm}{\di}}
+\li{\frac{-\lp}{\di}}\hfill
}
%integrale
\dl{
\int_0^1 dz\, \frac{\log(z-\lm)}{z-\zp}=\log\so
\log{\frac{-\zm}{\zp}}+\li{\frac{\zp}{\so}}-\li{\frac{-\zm}{\so}} \hfill
}
%integrale
\dl{
\int_0^1 dz\, \frac{\log(\lp-z)}{z-\zm} \stackrel{\rm B.P.}{=}
\log\zp\log(-\lm) -\log(-\zm) \log\lp -
\int_0^1 dz\, \frac{\log(z-\zm)}{z-\lp} \hfill \cr
\quad\phantom{\int_0^1 dz\,\frac{\log(\lp-z)}{z-\zm} }
\stackrel{\phantom{\rm B.P.}}{=}\log\zp\log(-\lm) -\log(-\zm) \log\lp
\hfill\cr 
\quad\phantom{\int_0^1 dz\, \frac{\log(z-\zm)}{z-\lp}
\stackrel{\rm B.P.}{=}} - \left[
\log\so \log\frac{-\lm}{\lp}+\li{\frac{\lp}{\so}} - \li{\frac{-\lm}{\so}}
\right]\hfill
}
%integrale
\dl{
\int_0^1 dz\, \frac{\log(\lp-z)}{z-\zp} \stackrel{\rm B.P.}{=}
\log(-\zm)\log(-\lm) -\log\zp \log\lp -
\int_0^1 dz\, \frac{\log(\zp-z)}{z-\lp}  \hfill\cr
\quad\phantom{\int_0^1 dz\, \frac{\log(\lp-z)}{z-\zp}}
\stackrel{\phantom{\rm B.P.}}{=}\log(-\zm)\log(-\lm) -\log\zp \log\lp
\hfill\cr
\quad\phantom{\int_0^1 dz\, \frac{\log(\lp-z)}{z-\zp} 
\stackrel{\rm B.P.}{=}}
- \left[
\log\di \log\frac{-\lm}{\lp}+\li{\frac{-\lp}{\di}} - \li{\frac{\lm}{\di}}
\right]\;,\hfill
}
where
\[
  \eta_\pm \equiv \frac{1}{2}(\dpr \pm \d)\;,
\]
and B.P. means integration ``By Parts''.
The integral $I$ then becomes
\beqn
I &=& \frac{(M^2)^{-\e}}{1+\e} \frac{1}{\sd s_3 \dpr } \left\{
\(\frac{1}{\e}+2\log\frac{q^2}{\sd}+\log\frac{M^2}{s_3}
\)  \log\frac{-\zm}{\zp}  \right.\nn\\
&&{} - \li{\frac{1}{\zp}}+\li{\frac{1}{\zm}}+\li{\frac{\zp}{\so}} -
\li{\frac{-\zm}{\so}}\nn\\ 
&& {} + 2\log\lp\log\di-2\log(-\lm)\log\di+2\li{\frac{\lm}{\di}} -
2\li{\frac{-\lp}{\di}} \nn\\
&&{} +2\log(-\zm)\log(-\lm)-2\log\zp\log\lp
-\log\zp\log(-\lm)+\log(-\zm)\log\lp \nn\\
&& {} + \li{\frac{\lp}{\so}}
-\li{\frac{-\lm}{\so}}-\log\dpr\log\frac{-\zm}{\zp}+\li{\frac{-\zm}{\dpr}} -
\li{\frac{\zp}{\dpr}} \nn\\
&& {}+\left.\frac{1}{2}\log^2\zp -\frac{1}{2}\log^2(-\zm) +
\log\so\log\(\frac{-\zm}{\zp}\frac{-\lm}{\lp}\)
 \right\} \;.
\eeqn
We can use the following algebraic relations 
\dl{\frac{\zp}{\so}= \frac{1+\dpr}{\dpr+\d}
=\frac{1+\dpr+\d-\d}{\dpr+\d}= \frac{\lm}{\so}+1
\hfill}
\dl{-\frac{\zm}{\so} = 1-\frac{\lp}{\so}\;,
\hfill}
and some properties of the dilogarithm function
\dl{
\li{\frac{\zp}{\dpr}}=-\li{-\frac{-\zm}{\dpr}}-
\log\frac{\zp}{\dpr}\log\frac{-\zm}{\dpr}+ \frac{\pi^2}{6}
\hfill}

\dl{ \li{\frac{\lp}{\so}}-\li{-\frac{\zm}{\so}}= 2\li{\frac{\lp}{\so}}+
\log\frac{\lp}{\so}\log\frac{-\zm}{\so} -\frac{\pi^2}{6}
\hfill}

\dl{\li{\frac{\zp}{\so}}-\li{-\frac{\lm}{\so}} = -2\li{-\frac{\lm}{\so}}
- \log\frac{-\lm}{\so}\log\frac{\zp}{\so} + \frac{\pi^2}{6}
\hfill}

\dl{ \li{\frac{1}{\zm}}-\li{\frac{1}{\zp}} = 2\li{\frac{1}{\zm}} +
\frac{1}{2}\log^2\frac{-\zm}{\zp}\;,
\hfill}
to obtain
\beqn
I &=& \frac{(M^2)^{-\e}}{1+\e} \frac{1}{\sd s_3 \dpr } \left\{
\frac{1}{\e}\log\frac{-\zm}{\zp} + \(
2\log\frac{M^2}{\sd}+\log\frac{M^2}{s_3}
\) \log\frac{-\zm}{\zp}  \right.
\nn\\
&&{} +2\li{\frac{1}{\zm}} -2\li{-\frac{\lm}{\so}}+ 2\li{\frac{\lp}{\so}} +
2\li{\frac{\lm}{\di}}  \nn\\
&&{}-2\li{-\frac{\lp}{\di}}+2\li{-\frac{\zm}{\dpr}}-\frac{\pi^2}{6} 
-2\log\di \log\frac{-\lm}{\lp}\nn\\
&& \left.{} +2\log\so \log \frac{-\lm}{\lp}+\log^2\zp
-2\log\dpr\log(-\zm)+\log^2\dpr  \right\} \;.
\eeqn
To return to the physical region, we must continue
analytically the solution, using
eqs.~(\ref{eq:prolung_1})--(\ref{eq:lambdapm_m}) and
\beq
\label{eq:csipm}
  \zpm = \frac{1}{2}\( 1 \pm \sqrt{1-\frac{4m^2}{s_3}}
\) \pm i\eta =  \frac{1}{2}\( 1\pm \dpr \) \pm i\eta \;.
\eeq
You can see that
\beq   0\leq \lm,\zm \leq 1/2\;,
\quad\quad\quad\quad  1/2\leq \lp,\zp \leq 1\;,
\eeq
and then
\dl{ \frac{1}{\zm}> 1 \tol \frac{1}{\zm} + i\eta  \hfill
}
\dl{
-\frac{\zm}{\zp}= -\frac{1-\dpr}{1+\dpr}<0 \tol -\frac{1-\dpr-\frac{2}{s_3}
\frac{1}{\dpr} i\eta}{1+\dpr+\frac{2}{s_3} \frac{1}{\dpr} i\eta} 
\hfill \cr
\quad \phantom{-\frac{\zm}{\zp}} =-\frac{1-\dpr}{1+\dpr} +
\underbrace{\frac{1-\dpr}{1+\dpr}\(\frac{1}{1-\dpr}+
\frac{1}{1+\dpr} \)}_{>0} i\eta = -\frac{\zm}{\zp} + i\eta \;.
\hfill
}
In a similar way, you can demonstrate that
\dl{ -\frac{\lm}{\lp} \tol -\frac{\lm}{\lp} + i\eta\;,
\hfill}
and
\dl{\frac{\lp}{\so}=\frac{1+\d}{\dpr+\d} > 1
\tol\frac{1+\d+\frac{2}{q^2}\frac{1}{\d} i\eta}{\dpr + \d +i \eta
\( \frac{2}{q^2}\frac{1}{\d}+\frac{2}{s_3}\frac{1}{\dpr} \)  } 
\hfill\cr
\quad\phantom{\frac{\lp}{\so}} =
\underbrace{\frac{1+\d}{\dpr+\d}}_{>0} \Biggl[1 +
i\eta \underbrace{\Biggl( \frac{1}{1+\d}-\frac{1+\frac{q^2}{s_3}
\frac{\d}{\dpr}}{\dpr+\d} \Biggr)}_{<0} \Biggr]
= \frac{\lp}{\so} -i\eta  \hfill
}
\dl{-\frac{\lp}{\di} \tol  -\frac{\lp}{\di} + i\eta
\hfill}
\dl{\di \tol \di + i\eta \;.
\hfill}
Using eqs.~(\ref{eq:continuaz_dilog}) and~(\ref{eq:continuaz_log})
we can write $B_{3m}$ as
\beqn
B_{3m} &=& N(\e)\ (m^2)^{-\e} \frac{1}{\sd s_3 \dpr }
\left\{\frac{1}{\e}\log\frac{\zm}{\zp} +
\(2\log\frac{m^2}{\sd}+\log\frac{m^2}{s_3}
\)  \log\frac{\zm}{\zp} \right. \nn\\
&&{} -2\li{\zm} -\log^2\zm -2\li{-\frac{\lm}{\so}}-2\li{\frac{\so}{\lp}}
-\log^2\frac{\lp}{\so} \nn\\
&&{} +2\li{\frac{\lm}{\di}}+2\li{-\frac{\di}{\lp}}
+\log^2\(-\frac{\lp}{\di}\) +2\li{-\frac{\zm}{\dpr}}
-\frac{\pi^2}{2} \nn\\
&&{} + 2\log\(-\frac{\so}{\di}\) \log\frac{\lm}{\lp} +\log^2\zp
-2\log\dpr\log\zm +\log^2\dpr \nn\\
&& \left.{} + i\,\pi\left[\frac{1}{\e} +2\log\frac{q^2}{\sd}
 +4\log\so -2\log\dpr +2\log\frac{\zm}{\zp} \right]
\right\} \;.
\eeqn

\section[First triangle with two massive propagators:
        $T_{2m}^q$]
{First triangle with two massive propagators:
        {\mylarge $T_{2m}^q$}}
\label{app:t_2mq}
%%%%%%%%%%%%%%%%%%%%%%%%%%%%%%%%%%%%%%%%%%%
\begin{figure}[htb]
\centerline{\epsfig{figure=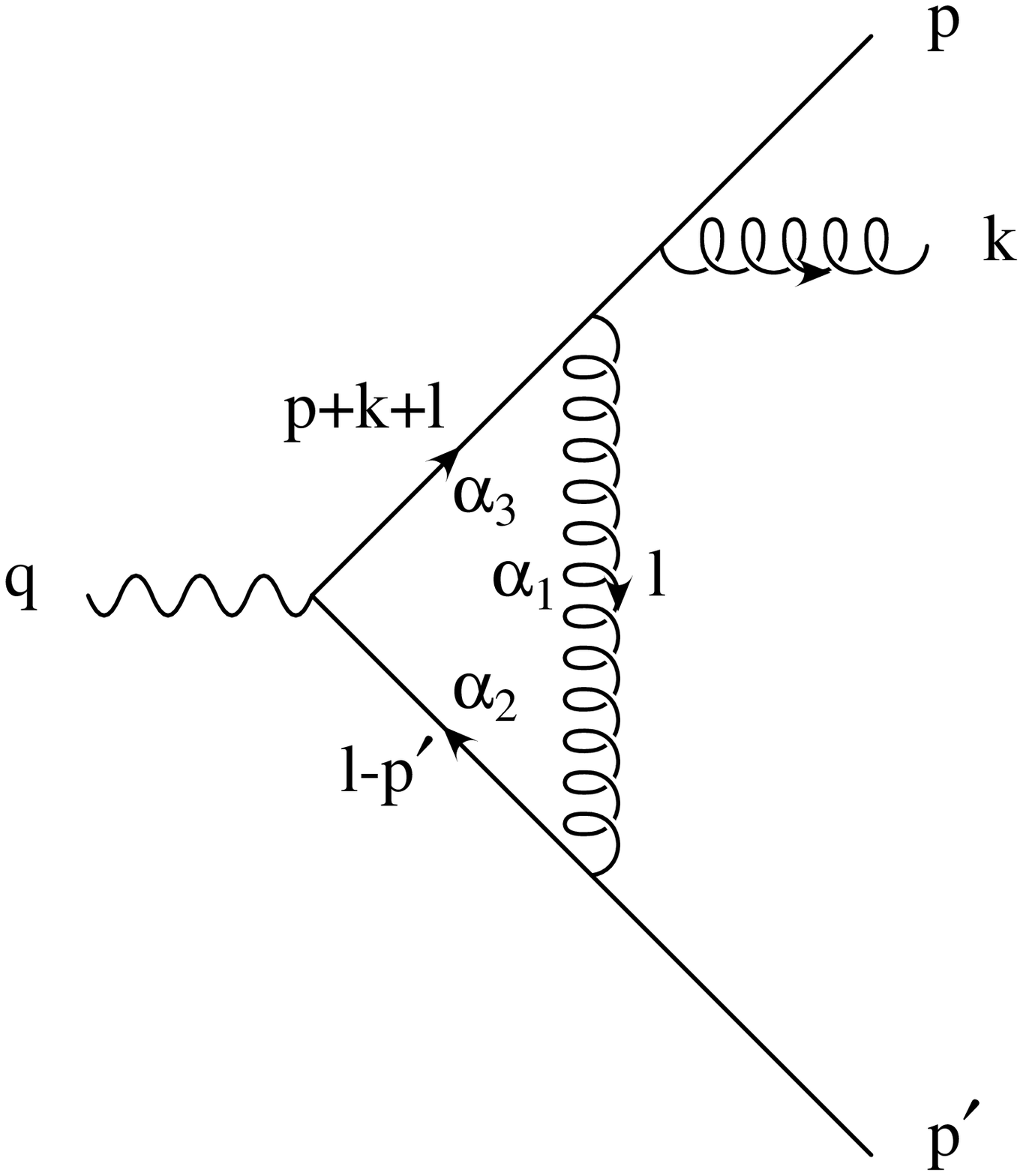,
width=\largfig,clip=}}\ccaption{}{ \label{fig:t_2mq}
Diagram representing $T_{2m}^q$.}
\end{figure}
%%%%%%%%%%%%%%%%%%%%%%%%%%%%%%%%%%%%%%%%%%%
We are referring, as far as the denominator structure is concerned, to
a Feynman diagram of the type illustrated in Fig.~\ref{fig:t_2mq}
\beq
\label{eq:t_2mq_1}
T_{2m}^q\equiv\int \frac{d^dl}{(2\pi)^d}\ \frac{1}{l^2} \
\frac{1}{(l-p')^2-m^2} \
\frac{1}{(l+p+k)^2-m^2}  \;.
\eeq
Using eqs.~(\ref{eq:Feynman_reduction})
and~(\ref{eq:Feynman_denominator}) we can write
\beq
\label{eq:t_2mq}
T_{2m}^q= \frac{i}{16\pi^2}\ (4\pi)^\e \ \Gamma(1+\e) (-1)\int
\frac{[d\a]}{D^{1+\e}}\;,\quad\quad\quad  d=4 -2\e \;,
\eeq
where, with the help of eqs.~(\ref{eq:def_s_1})--(\ref{eq:def_sig3}),
\beqn
D &=& -\lq \a_1\,\a_2\, m^2 +\a_1\, \a_3 (p+k)^2+\a_2\, \a_3
 \,q^2\rq +(\a_2+\a_3)\,  m^2 -i \eta \nn\\
 & =& -\lq \a_1\,\a_2\,  m^2 +\a_1\,\a_3\, s_1 + \a_2\, \a_3 \, q^2
 -(\a_2+\a_3)\, m^2 +  i\eta\rq  \nn\\
 &=& -\lq -m^2(1-\a_1)^2  + \su \,\a_1\,\a_3 +q^2 \a_2\,\a_3 + i\eta\rq \;.
\eeqn
Continuing the integral into the unphysical region, according to
eq.~(\ref{eq:M2}), we have
\beq
D = -\lq M^2 (1-\a_1)^2  + \su \,\a_1\,\a_3 +q^2 \a_2\,\a_3\rq    \;.
\eeq
The integral~(\ref{eq:t_2mq}) then becomes
\beq
  T_{2m}^q = \frac{i}{16\pi^2}\ (4\pi)^\e \ \Gamma(1+\e) \
{\rm e}^{i\pi\e} \int
  \frac{[d\a]}{\left[-D\right]^{1+\e}} = \Nep \, {\rm e}^{i\pi\e} \, I\;.
\eeq
The following change of variables
\[\begin{array}{l}
  \a_1 = 1-x  \\
  \a_2 = x(1-y)      \\
  \a_3 = x y\;,  \\
\end{array}
\]
together with the Jacobian of the transformation 
\[    \frac{\partial(\a_1\,\a_2)}{\partial(x\,y)} = x \;,
\]
gives
\beqn
I &=& \int_0^1 dx\, dy\,
\frac{x}{\Bigl[M^2x^2 +\su x(1-x)y+q^2 x^2 y(1-y)\Bigr]^{1+\e}} \nn\\
&=& \int_0^1 dx\, dy\,
\frac{x^{-\e}}{\Bigl[M^2x +\su (1-x)y+q^2 x y(1-y)\Bigr]^{1+\e}} \;.
\eeqn
The integral $I$ is finite when $\e\,\to\,0$, as can be seen also by
inspecting eq.~(\ref{eq:t_2mq_1}), that has no soft or collinear
divergences. For this reason we put $\e=0$ and integrate in $x$ to obtain
\beq
I = \int_0^1 dy \frac{1}{M^2-\su y +q^2 y(1-y)}\log\frac{M^2 + q^2
y(1-y)}{\su y}\;.
\eeq
By decomposing the denominator in the following way
\beqn
\frac{1}{M^2-\su y +q^2 y(1-y)} &=& \frac{1}{-q^2 \left[y^2
-\(1-\frac{\su}{q^2}\) -\frac{M^2}{q^2}\right]}  \nn\\
&=& -\frac{1}{q^2}\frac{1}{\sqrt{\a_1^2 + \frac{4M^2}{q^2}}} 
\(\frac{1}{y-\rop} -  \frac{1}{y-\rom} \)\;,
\eeqn
where
\[
 \a_1 \equiv 1-\frac{\su}{q^2}\;, \quad\quad\quad\quad 
\sqrt{\frac{4 m^2}{q^2}} \leq \a_1 \leq 1\;,
\]
\[    \ropm = \frac{1}{2}\left[ \a_1 \pm
  \sqrt{\a_1^2+\frac{4M^2}{q^2}}\right] \;,
\]
we have
\beq
I = \int_0^1 dy \frac{-1}{q^2 \sqrt{\a_1^2+\frac{4M^2}{q^2}}}
       \(\frac{1}{y-\rop}- \frac{1}{y-\rom} \)
       \log \left[\frac{-q^2}{\su}\frac{(y-\lp)(y-\lm)}{y}\right]\;.
\eeq
It is easy to see that
\beq
\label{eq:sequenza}
    \rom < \lm < 0 < \rop < 1 < \lp \;,
\eeq
and  $I$ becomes
\beqn
I &=& -\frac{1}{q^2 \sqrt{\a_1^2 +\frac{4M^2}{q^2}}} \left\{
\int_0^1 dy \,\frac{1}{y-\rop}\left[ \log\frac{\lp-y}{\lp-\rop}+
\log\frac{y-\lm}{\rop-\lm} -\log\frac{y}{\rop} \right] \right.
\nn\\
&& \left.{}-\int_0^1 dy\, \frac{1}{y-\rom}\left[ \log\frac{\lp-y}{\lp-\rom}+
\log\frac{y-\lm}{\lm-\rom} -\log\frac{y}{-\rom} \right]  \right\}\;.
\eeqn

We summarize here the results of the integration in $y$
\dl{
\int_0^1 dy \frac{1}{y-\rop} \log\frac{\lp-y}{\lp-\rop}=
\li{-\frac{\rop}{\lp-\rom}} - \li{\frac{1-\rop}{\lp-\rop}}
\hfill}
\dl{
\int_0^1 dy \frac{1}{y-\rop} \log\frac{y-\lm}{\rop-\lm}=
\li{\frac{\rop}{\rop-\lm}} - \li{\frac{1-\rop}{\lm-\rop}}
\hfill}

\dl{
\int_0^1 dy \frac{1}{y-\rop} \log\frac{y}{\rop}=
- \li{1-\frac{1}{\rop}}  +\frac{\pi^2}{6}
\hfill}

\dl{
\int_0^1 dy \frac{1}{y-\rom} \log\frac{\lp-y}{\lp-\rom}=
\li{\frac{-\rom}{\lp-\rom}} - \li{\frac{1-\rom}{\lp-\rom}}
\hfill}

\dl{
\int_0^1 dy \frac{1}{y-\rom} \log\frac{y-\lm}{\lm-\rom}=
\li{-\frac{1-\lm}{\lm-\rom}}-\li{\frac{\lm}{\lm-\rom}} 
\nl
{}+\log\frac{1-\rom}{\lm-\rom}\log\frac{1-\lm}{\lm-\rom} -
\log\frac{-\rom}{\lm-\rom}\log\frac{-\lm}{\lm-\rom}
}

\dl{
\int_0^1 dy \frac{1}{y-\rom} \log\frac{y}{-\rom}=
\li{\frac{1}{\rom}} -\log(-\rom)\log\frac{1-\rom}{-\rom} \;.
\hfill
}
We can then write
\beqn
I &=& -\frac{1}{q^2 \sqrt{\a_1^2 +\frac{4M^2}{q^2}}} \left\{
-\frac{\pi^2}{6}+\li{1-\frac{1}{\rop}}+\li{-\frac{\rop}{\lp-\rop}}\right.
\nn\\
&&- \li{\frac{1-\rop}{\lp-\rop}}
+ \li{\frac{\rop}{\rop-\lm}} -\li{\frac{1-\rop}{\lm-\rop}} +
\li{\frac{1}{\rom}} \nn\\
&&-\log(-\rom)\log\frac{1-\rom}{-\rom} +\li{\frac{1-\rom}{\lp-\rom}}
-\li{-\frac{\rom}{\lp-\rom}} \nn\\
&&+ \li{\frac{\lm}{\lm-\rom}} -\li{-\frac{1-\lm}{\lm-\rom}}-
\log\frac{1-\rom}{\lm-\rom}\log\frac{1-\lm}{\lm-\rom} \nn\\
&&+ \left.\log\frac{-\rom}{\lm-\rom}\log\frac{-\lm}{\lm-\rom}
\right\}\;. 
\eeqn
According to eqs.~(\ref{eq:prolung_1})--(\ref{eq:lambdapm_m}) and
noticing that
\beq
\label{eq:ropm}
  \ropm = \frac{1}{2}\( \a_1 \pm \sqrt{\a_1^2-\frac{4m^2}{q^2}}
\) \pm i\eta \;,
\eeq
we see that the ordered sequence~(\ref{eq:sequenza}) becomes
\[
  0 < \lm < \rom < \rop < \lp < 1 \;.
\]
This fact, together with the values of the signs of the imaginary
parts, helps us to continue the solution into the physical region. In
fact
\dl{
 \frac{1-\rop}{\lp-\rop} > 1 \tol  \frac{1-\rop}{\lp-\rop} -i\eta
\hfill}
\dl{
 \frac{1-\rom}{\lp-\rom} > 1 \tol  \frac{1-\rom}{\lp-\rom} -i\eta
\hfill}
\dl{
 \frac{1-\lm}{\rom-\lm} > 1 \tol  \frac{1-\lm}{\rom-\lm} +i\eta
\hfill}
\dl{
 \frac{\rop}{\rop-\lm} > 1 \tol  \frac{\rop}{\rop-\lm} -i\eta
\hfill}
\dl{
 \frac{1}{\rom} > 1 \tol  \frac{1}{\rom} +i\eta
\hfill}
\dl{
 \frac{1-\rom}{\lm-\rom} <0 \tol  \frac{1-\rom}{\lm-\rom} -i\eta
\hfill}
\dl{
 \frac{1-\lm}{\lm-\rom} <0 \tol  \frac{1-\lm}{\lm-\rom} -i\eta
\hfill}
\dl{
 \frac{1-\rom}{-\rom} <0 \tol  \frac{1-\rom}{-\rom} -i\eta \;.
\hfill}
The final expression is
\beqn
T_{2m}^q &=& N(\e)\ (m^2)^{-\e} \frac{-1}{q^2\sqrt{\a_1^2-\frac{4m^2}{q^2}}}
\left\{\li{1-\frac{1}{\rop}}+\li{-\frac{\rop}{\lp-\rop}} \right.
\nn\\
&&+ \li{\frac{\lp-\rop}{1-\rop}}-\li{\frac{\rop-\lm}{\rop}}-
\li{\frac{1-\rop}{\lm-\rop}}-\li{\rom} \nn\\
&& -\li{\frac{\lp-\rom}{1-\rom}}  -\li{-\frac{\rom}{\lp-\rom}}+
\li{\frac{\lm}{\lm-\rom}}\nn\\
&& + \li{\frac{\rom-\lm}{1-\lm}}+
\frac{1}{2}\log^2\frac{\lp-\rop}{1-\rop}
-\frac{1}{2}\log^2\frac{\rop-\lm}{\rop}-\frac{1}{2}\log^2\rom \nn\\
&&-\log\rom\log\frac{1-\rom}{\rom}
-\frac{1}{2}\log^2\frac{\lp-\rom}{1-\rom}+\frac{1}{2}\log^2
\frac{\rom-\lm}{1-\lm} \nn\\
&&-\log\frac{\rom-1}{\lm-\rom}\log\frac{\lm-1}{\lm-\rom}
+\log\frac{-\rom}{\lm-\rom}\log\frac{-\lm}{\lm-\rom} + \frac{\pi^2}{6}
\nn\\
&&\left. +\,i\, \pi\left[
2\,\log\frac{\lp-\rom}{\lp-\rop} +\log\frac{1-\rop}{1-\rom}
\right] \right\} \;,
\eeqn
where we have factorized out the term $(m^2)^{-\e}$, that gives zero
contribution in the limit $\e\,\to\, 0$.

\section[Second triangle with two massive propagators:
        $T_{2m}$]{Second triangle with two massive propagators:
        {\mylarge $T_{2m}$}}
\label{app:t_2m}
%%%%%%%%%%%%%%%%%%%%%%%%%%%%%%%%%%%%%%%%%%%
\begin{figure}[htb]
\centerline{\epsfig{figure=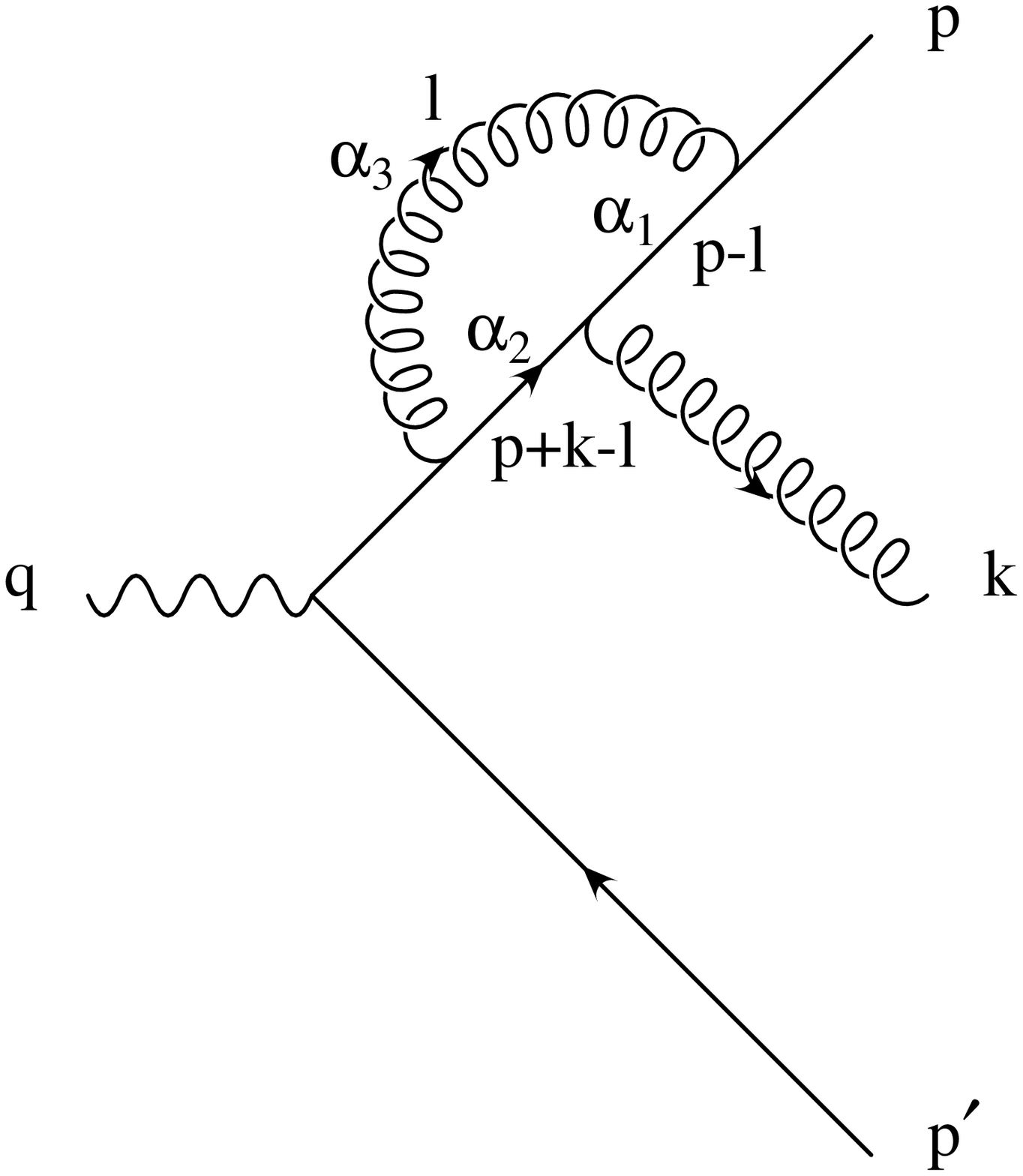,
width=0.36\textwidth,clip=}}\ccaption{}{ \label{fig:t_2m}
Diagram representing $T_{2m}$.}
\end{figure}
%%%%%%%%%%%%%%%%%%%%%%%%%%%%%%%%%%%%%%%%%%%
We are referring, as far as the denominator structure is concerned, to
a Feynman diagram of the type depicted in Fig.~\ref{fig:t_2m}
\beqn
T_{2m} &\equiv& \int \frac{d^dl}{(2\pi)^d}\ \frac{1}{l^2} \
\frac{1}{(l-p)^2-m^2} \
\frac{1}{(p+k-l)^2-m^2}  \nn\\
\label{eq:t_2m}
 &=& \frac{i}{16\pi^2}\ (4\pi)^\e \ \Gamma(1+\e) (-1)\int
\frac{[d\a]}{D^{1+\e}}\;, \quad\quad\quad  d=4 -2\e \;,
\eeqn
where we have performed the usual steps, already done for the previous
integrations, and where 
\beqn
D &=& -\lq \a_1\, \a_3 \,m^2 +\a_2\, \a_3 \,(p+k)^2 \rq+(\a_1+\a_2)\, m^2
 -i \eta \nn\\
 &=& -\lq-(\a_1+\a_2-\a_1\, \a_3)\,m^2 +\a_2\, \a_3\, s_1
+ i\eta\rq   \nn\\
 &=& -\lq-m^2(1-\a_3)^2  + \su \,\a_2\, \a_3  + i\eta\rq \;.
\eeqn
Continuing the integral into the unphysical region, according to
eq.~(\ref{eq:M2}), we have
\beq
 D = -\lq M^2 (1-\a_3)^2  + \su \, \a_2\, \a_3 \rq\;.
\eeq
The integral~(\ref{eq:t_2m}) becomes
\beq
  T_{2m} = \frac{i}{16\pi^2}\ (4\pi)^\e \ \Gamma(1+\e) \
{\rm e}^{i\pi\e} \int
  \frac{[d\a]}{\left[-D\right]^{1+\e}} = \Nep \, {\rm e}^{i\pi\e} \, I
\;,
\eeq
and, with the following change of variables
\[\begin{array}{l}
  \a_1 = x(1-y)  \\
  \a_2 = xy      \\
  \a_3 = 1-x \;, \\
\end{array}
\]
with Jacobian
\[  \frac{\partial(\a_1\,\a_2)}{\partial(x\,y)} = x \;,
\]
we get
\beq
I = \int_0^1 dx\, dy\,
\frac{x}{\left[M^2x^2 +\su x(1-x)y\right]^{1+\e}} = \int_0^1 dx\, dy\,
\frac{x^{-\e}}{\left[M^2x +\su (1-x)y\right]^{1+\e}}\;.
\eeq
This integral has no soft or collinear divergences, so we can put $\e=0$
\beq
I = \int_0^1 dx\, dy\,
\frac{1}{M^2x +\su (1-x)y}= \int_0^1 dx\, \frac{1}{\su(1-x)}
\log\frac{M^2x +\su (1-x)}{M^2x} \;.
\eeq
With the change of variable $z=1-x$ we obtain
\beqn
I &=& \frac{1}{\su}\int_0^1 \frac{dz}{z}
\left\{\log\left[1-\(1-\frac{\su}{M^2} \)z\right] -\log(1-z)
\right\} \nn\\
 &=&\frac{1}{\su} \left[ -\li{1-\frac{\su}{M^2}} +\frac{\pi^2}{6}
\right] \;.
\eeqn
The analytic continuation into the physical region $\(M^2 = -m^2 +
i\eta\) $ is straightforward. In fact we have
\dl{
1-\frac{\su}{M^2} \tol  1 +\frac{\su}{m^2}+i\eta\;,
\hfill}
which gives
\beq
T_{2m}= N(\e)\,\(m^2\)^{-\e}\frac{1}{\su}\left\{
\li{-\frac{\su}{m^2}} +\log\frac{\su}{m^2} \log\(1+\frac{\su}{m^2}
\) -i\, \pi \log\(1+\frac{\su}{m^2}\)\right\} \;,
\eeq
where we have factorized out the term $(m^2)^{-\e}$, that gives zero
contribution in the limit $\e\,\to\, 0$.

\section[Triangle with one massive propagator: $T_{1m}$]
{Triangle with one massive propagator: {\mylarge $T_{1m}$}}
\label{app:t_1m}
%%%%%%%%%%%%%%%%%%%%%%%%%%%%%%%%%%%%%%%%%%%
\begin{figure}[htb]
\centerline{\epsfig{figure=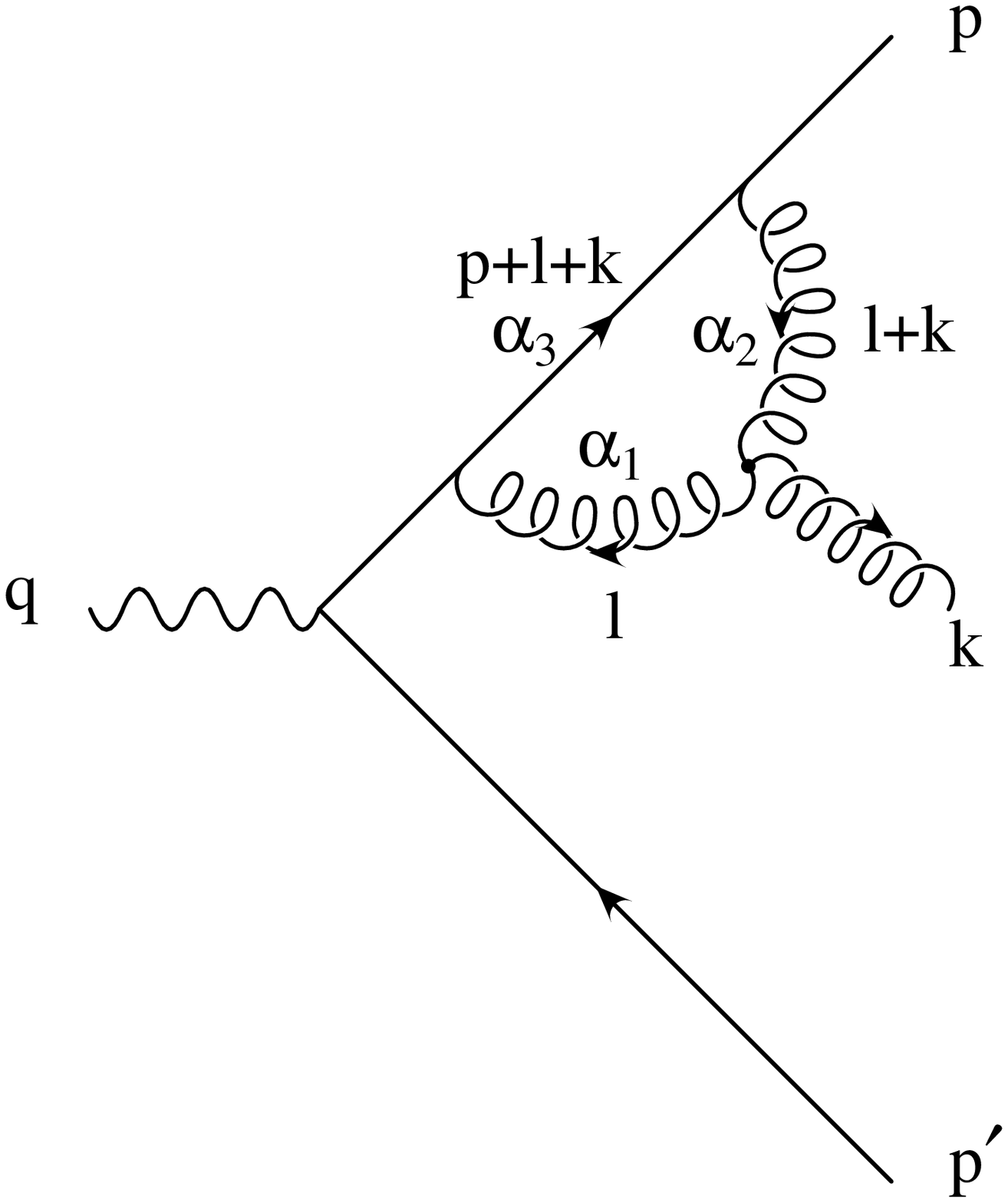,
width=0.36\textwidth,clip=}}\ccaption{}{ \label{fig:t_1m}
Diagram representing $T_{1m}$.}
\end{figure}
%%%%%%%%%%%%%%%%%%%%%%%%%%%%%%%%%%%%%%%%%%%
We are referring, as far as the denominator structure is concerned, to
a Feynman diagram of the type represented in Fig.~\ref{fig:t_1m}
\beqn
T_{1m} &\equiv& \int \frac{d^dl}{(2\pi)^d}\ \frac{1}{l^2} \
\frac{1}{(l+k)^2} \
\frac{1}{(l+p+k)^2-m^2}  \nn\\
\label{eq:t_1m}
&=&  \frac{i}{16\pi^2}\ (4\pi)^\e \ \Gamma(1+\e) (-1)\int
\frac{[d\a]}{D^{1+\e}}\;,\quad\quad\quad  d=4 -2\e \;,
\eeqn
where, with the definition~(\ref{eq:M2}), 
\beqn
 D &=& -\lq \a_1\,\a_3\,(p+k)^2 +\a_2\,\a_3\, m^2 \rq +\a_3\, m^2
 -i \eta \nn\\
 &=& -\lq\a_1\,\a_3\, s_1 + \a_2\,\a_3\, m^2 -\a_3\, m^2 +
 i\eta\rq \nn\\
 &=& -\lq-m^2 \a_3^2  + \su \,\a_1\,\a_3 + i\eta\rq  \nn\\
 &=& -\lq M^2 \a_3^2+ \su \, \a_1\,\a_3 \rq \;.
\eeqn
The integral~(\ref{eq:t_1m}) becomes
\beq
 T_{1m} = \frac{i}{16\pi^2}\ (4\pi)^\e \ \Gamma(1+\e) \
{\rm e}^{i\pi\e} \int
  \frac{[d\a]}{\left[-D\right]^{1+\e}} = \Nep \, {\rm e}^{i\pi\e} \, I
\;,
\eeq
so that, with the following change of variables
\[\begin{array}{l}
  \a_1 = x(1-y)  \\
  \a_2 = 1-x      \\
  \a_3 = x y \;, \\
\end{array}
\]
and the Jacobian
\[   \frac{\partial(\a_1\,\a_2)}{\partial(x\,y)} = x \;,
\]
we obtain
\beq
I = \int_0^1 dx\, dy\,
\frac{x}{\left[M^2 x^2 y^2 +\su x^2(1-y)y\right]^{1+\e}} =
\int_0^1 dx\, dy\,
\frac{x^{-1-2\e}}{\left[M^2 y^2 +\su (1-y)y\right]^{1+\e}} \;.
\eeq
Integrating in $x$
\beq
I = -\frac{1}{2\e} \int_0^1  dy\,  \frac{y^{-1-\e}} {\su^{1+\e}\left[
1+\(\frac{M^2}{\su}-1\)y\right]^{1+\e}}\;,
\eeq
and using the identity~(\ref{eq:projective}), we have
\beqn
I &=& -\frac{1}{2\e} \(M^2\)^{-\e} \frac{1}{\su}
\(\frac{M^2}{\su}\)^{2\e} \int_0^1  dy\, y^{-1-\e} \left[
1- y \(1-\frac{\su}{M^2} \)\right]^{2\e} \nn\\
 &=& -\frac{1}{2\e} \(M^2\)^{-\e} \frac{1}{\su}
\(\frac{M^2}{\su}\)^{2\e}
\left[ -\frac{1}{\e} -2\e\li{1-\frac{\su}{M^2}}\right] \;.
\eeqn
Expanding  $\(\frac{M^2}{\su}\)^{2\e}$ in powers of  $\e$
\beq
T_{1m} = N(\e)\, {\rm e}^{i\pi\e} \(M^2\)^{-\e} \frac{1}{\su}
\left[\frac{1}{2\e^2}+\frac{1}{\e}\log\frac{M^2}{\su}+\li{1-\frac{\su}{M^2}}
+\log^2\frac{M^2}{\su}\right]
\eeq
and continuing analytically the solution
\beqn
T_{1m} &=& N(\e)\,\(m^2\)^{-\e} \frac{1}{\su} \left\{
\frac{1}{2\e^2}+\frac{1}{\e}\log\frac{m^2}{\su} +
\log\frac{m^2}{\su}\log\(1+\frac{m^2}{\su}\) \right. \nn\\
 && \left.-\li{-\frac{\su}{m^2}}-\frac{5}{6}\pi^2 + 
 i\,\pi \left[ \frac{1}{\e}
+\log\(1+\frac{\su}{m^2}\) + 2\log\frac{m^2}{\su}\right]
\right\} \;. \phantom{aaa}
\eeqn

\section[First check triangle: $T_{2m}^{q-k}$]
{First check triangle: {\mylarge $T_{2m}^{q-k}$}}
\label{app:t_check1}
%%%%%%%%%%%%%%%%%%%%%%%%%%%%%%%%%%%%%%%%%%%
\begin{figure}[htb]
\centerline{\epsfig{figure=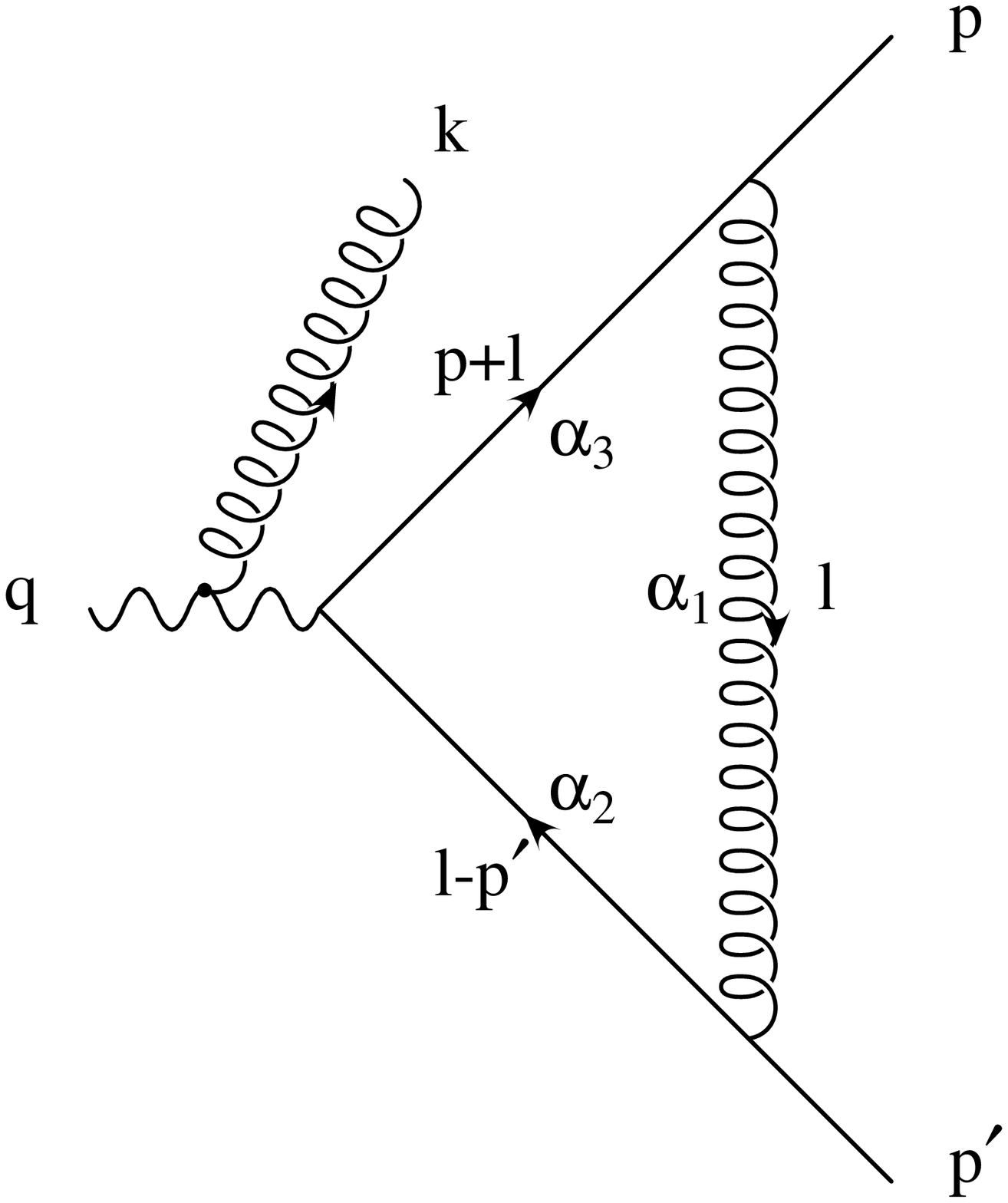,
width=0.36\textwidth,clip=}}\ccaption{}{ \label{fig:t_check1}
Diagram representing $T_{2m}^{q-k}$.}
\end{figure}
%%%%%%%%%%%%%%%%%%%%%%%%%%%%%%%%%%%%%%%%%%%
This appendix and the following one are deserved to compute two diagrams
useful for a partial check of the calculation.
The denominator structure is illustrated in Fig.~\ref{fig:t_check1}
\beqn
T_{2m}^{q-k} &\equiv& \int \frac{d^dl}{(2\pi)^d}\ \frac{1}{l^2} \
\frac{1}{(l-p')^2-m^2} \
\frac{1}{(l+p)^2-m^2} \nn\\
\label{eq:t_2m^q-k}
&=& \frac{i}{16\pi^2}\ (4\pi)^\e \ \Gamma(1+\e) (-1)\int
\frac{[d\a]}{D^{1+\e}}\;,\quad\quad\quad d=4 -2\e \;,
\eeqn
where
\beqn
D &=& -\lq \a_1\,\a_2 \,m^2 +\a_1\,\a_3\,m^2 +\a_2\,\a_3 \,(q-k)^2\rq
 +(\a_2+\a_3)\,m^2 -i \eta \nn\\
 &=& -\lq- m^2 (1-\a_1)^2 + s_3 \,\a_2\,\a_3 + i\eta\rq \nn\\
 &=& -\lq M^2 (1-\a_1)^2  + s_3 \,\a_2\,\a_3 \rq  \;.
\eeqn 
The integral~(\ref{eq:t_2m^q-k}) becomes
\beq
 T_{2m}^{q-k} = \frac{i}{16\pi^2}\ (4\pi)^\e \ \Gamma(1+\e) \
{\rm e}^{i\pi\e} \int
  \frac{[d\a]}{\left[-D\right]^{1+\e}}= \Nep \, {\rm e}^{i\pi\e} \, I \;,
\eeq
and, by changing the variables of integration,
\[\begin{array}{l}
  \a_1 = 1-x  \\
  \a_2 = x z      \\
  \a_3 = x (1-z)\;,  \\
\end{array}
\]
with Jacobian
\[    \frac{\partial(\a_1\,\a_2)}{\partial(x\,y)} = x \;,
\]
we have
\beq
I = \int_0^1 dx\,dz\, \frac{x}{\left[M^2 x^2+x^2 z(1-z) s_3
\right]^{1+\e}} = \int_0^1 dx\,dz\, \frac{x^{-1-2\e}} {\left[M^2+
z(1-z) s_3\right]^{1+\e}} \;.
\eeq
Performing the $x$ integration, we obtain
\beq
I = -\frac{1}{2\e}\(M^2\)^{-\e}\int_0^1 dz \frac{1}{M^2 \left[
1+z(1-z)\frac{s_3}{M^2}\right]^{1+\e}} \;,
\eeq
which is equal to the integral computed in eq.~(\ref{eq:I_R}), so that
we have
\beqn
T_{2m}^{q-k} &=& N(\e)\,{\rm e}^{i\pi\e}
\(M^2\)^{-\e}\frac{1}{s_3 \dpr} \left\{
\frac{1}{\e}\log\frac{-\cm}{\cp}+\log\frac{M^2}{s_3}\log\frac{-\cm}{\cp}
-\frac{1}{2}\log^2(-\cm) \right.\nn\\
&& \left.{}+\frac{1}{2}\log^2\cp - \log\dpr\log\frac{-\cm}{\cp}
+\li{-\frac{\cm}{\dpr}}-\li{\frac{\cp}{\dpr}} \right\} \;.
\eeqn
The analytic continuation of this solution, using
\dl{
\frac{\cp}{\dpr} > 1 \tol \frac{\cp}{\dpr} -i\eta\;,
\hfill}
is
\beqn
T_{2m}^{q-k} &=& N(\e)\,\(m^2\)^{-\e}\frac{1}{s_3 \dpr} \left\{
\frac{1}{\e}\log\frac{\cm}{\cp} +\log\frac{m^2}{s_3}\log\frac{\cm}{\cp}-
\frac{1}{2}\log^2\cm +\frac{1}{2}\log^2 \cp \right. \nn\\
&& -\log\dpr\log\frac{\cm}{\cp}
+\li{-\frac{\cm}{\dpr}}+\li{\frac{\dpr}{\cp}}
+\frac{1}{2}\log^2\frac{\cp}{\dpr}
-\frac{5}{6}\pi^2 \nn\\
&&\left.+\,i\,\pi \left[ \frac{1}{\e} +\log\frac{m^2}{s_3}-2\log\dpr
\right] \right\} \;.
\eeqn

\section[Second check triangle: $T_{3m}$]
{Second check triangle: {\mylarge $T_{3m}$}}
\label{app:t_check2}
%%%%%%%%%%%%%%%%%%%%%%%%%%%%%%%%%%%%%%%%%%%
\begin{figure}[htb]
\centerline{\epsfig{figure=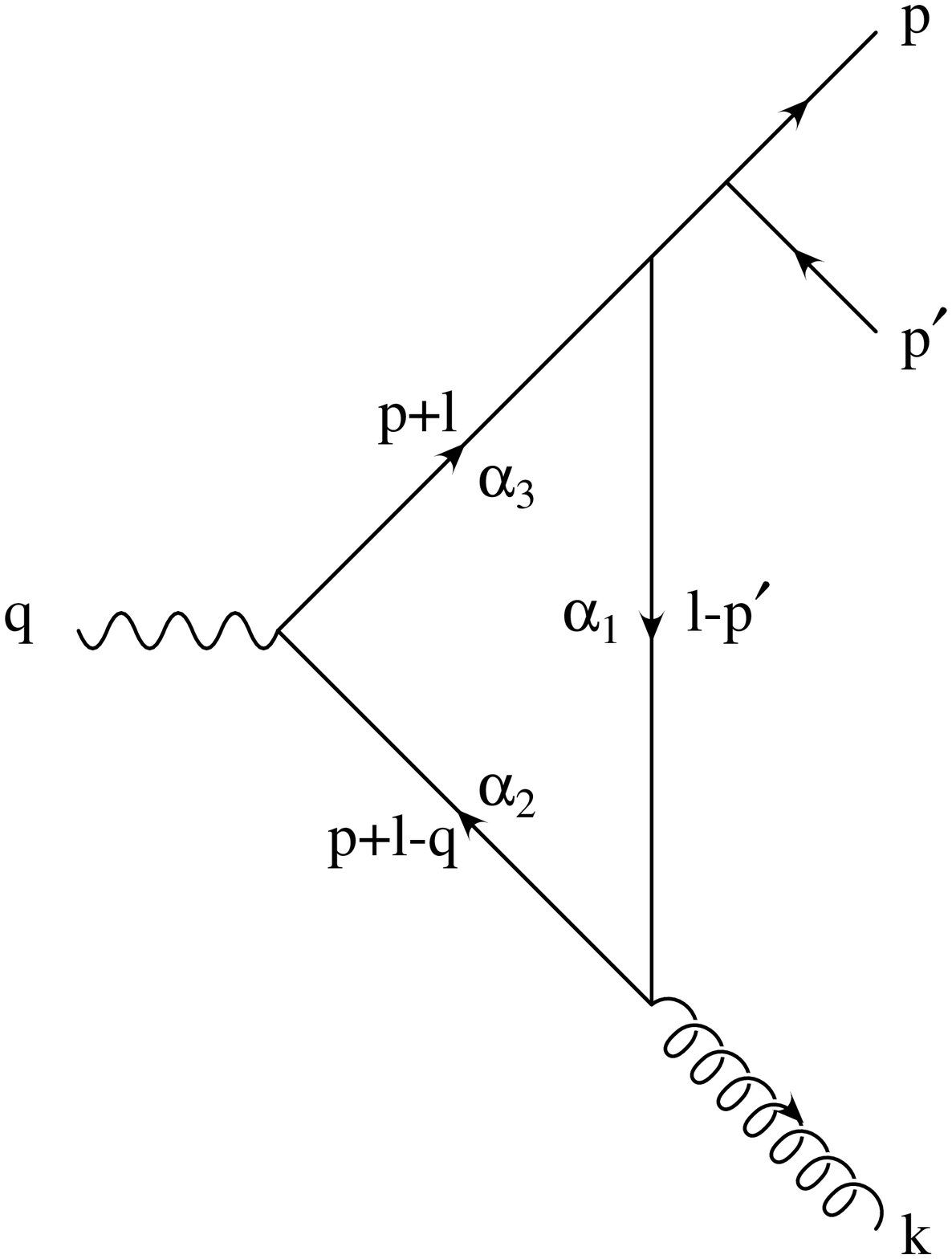,
width=\largfig,clip=}}\ccaption{}{ \label{fig:t_check2}
Diagram representing $T_{3m}$.}
\end{figure}
%%%%%%%%%%%%%%%%%%%%%%%%%%%%%%%%%%%%%%%%%%%
This appendix (as  the previous one) is  deserved to calculate a 
diagram used for a partial check of the calculation.
The denominator structure is depicted in Fig.~\ref{fig:t_check2}
\beqn
T_{3m} &\equiv& \int \frac{d^dl}{(2\pi)^d}\ \frac{1}{(l-p')^2-m^2} \
\frac{1}{(l+p-q)^2-m^2} \
\frac{1}{(l+p)^2-m^2}  \nn\\
\label{eq:t_3m}
&=& \frac{i}{16\pi^2}\ (4\pi)^\e \ \Gamma(1+\e) (-1)\int
\frac{[d\a]}{D^{1+\e}}\;,\quad\quad\quad d=4 -2\e \;,
\eeqn
where
\beqn
D &=& -\lq \a_1\,\a_2 \,(p+p')^2 +\a_2\,\a_3\, q^2 \rq +m^2
 -i \eta \nn\\
& =& -\lq -m^2 +\a_1\,\a_2\, s_3 +\a_2\,\a_3\, q^2 + i\eta\rq  \nn\\
 &=& -\lq M^2 + s_3 \,\a_1\,\a_3 +q^2 \a_2\,\a_3\rq   \;.
\eeqn
The integral~(\ref{eq:t_3m}) becomes
\beq
  T_{3m} = \frac{i}{16\pi^2}\ (4\pi)^\e \ \Gamma(1+\e) \
{\rm e}^{i\pi\e} \int
  \frac{[d\a]}{\left[-D\right]^{1+\e}} = \Nep \, {\rm e}^{i\pi\e} \, I\;,
\eeq
and, with the following change of variables,
\[\begin{array}{l}
  \a_1 = xy  \\
  \a_2 = x (1-y)      \\
  \a_3 = 1-x  \;,\\
\end{array}
\]
we have a  Jacobian
\[   \frac{\partial(\a_1\,\a_2)}{\partial(x\,y)} = x\;,
\]
so that
\beq
I = \int_0^1 dx\,dy\, \frac{x}{\left[  s_3\, x(1-x)y + q^2 x(1-x)(1-y) +M^2
\right]^{1+\e}} \;.
\eeq
This diagram has no soft or collinear divergences, and then we can put
$\e=0$, and we can integrate in $y$ to obtain
\beq
I = \frac{1}{s_3-q^2} \int_0^1 dx \ \frac{1}{1-x} \log\frac{M^2+x(1-x)
s_3}{M^2+x(1-x) q^2}\;.
\eeq
By splitting $\log\(a/b\)=\log a - \log b$ and using
\beqn
\int_0^1 \frac{dx}{x}\log\left[ 1+x(1-x)\frac{s_3}{M^2}\right] &=&
\int_0^1 \frac{dx}{x}\log\( 1-\frac{x}{\cp}\) +
\int_0^1 \frac{dx}{x}\log\( 1-\frac{x}{\cm}\) \nn\\
&=&-\li{\frac{1}{\cp}}-\li{\frac{1}{\cm}} = \frac{1}{2} \log^2 \(
1-\frac{1}{\cm}\) \;,\nn
\eeqn
we have
\beq
T_{3m}= N(\e) \,\frac{1}{2}\, \frac{1}{s_3-q^2} \left\{
\log^2 \(
1-\frac{1}{\cm}\) - \log^2 \( 1-\frac{1}{\lm}\) \right\} \;.
\eeq
The analytic continuation  gives
\beqn
T_{3m} &=& N(\e) \(m^2\)^{-\e} \frac{1}{2}\, \frac{1}{s_3-q^2}
\left\{ \log^2\(\frac{1}{\cm}-1\) -
\log^2\(\frac{1}{\lm}-1\)  \right.\nn\\
&&\left.{}-2\, i \,\pi \left[
\log\(\frac{1}{\cm}-1\)-
\log\(\frac{1}{\lm}-1\)
 \right] \right\} \;.
\eeqn

\section{Partial check of the virtual integrals}
\label{app:check}
\subsection[Check of $B_{2m}$]{Check of {\mylarge $B_{2m}$}}
\label{app:check_b_2m}
A partial check of the correctness of the above formulae can be
performed in the following way.
We consider first a check of $B_{2m}$. To this purpose, we define $I$ as
\beqn
\label{eq:I_check}
I  &\equiv& \int\frac{d^dl}{(2\pi)^d}\ \frac{1 + A(l-k)^2 + B \left[
(l+p-q)^2-m^2 \right] +C\left[ (l+p)^2-m^2\right]}{
 l^2 (l-k)^2 \left[(l+p-q)^2-m^2\right] \left[(l+p)^2-m^2\right]} \\
&=&\int \frac{d^dl}{(2\pi)^d}\  \frac{
1 + B [q^2-2p\cdot q] + 2 l\cdot \left[ -A k + B (p-q) + Cp\right]
+l^2(A+B+C)} { l^2 (l-k)^2 \left[(l+p-q)^2-m^2\right] 
\left[(l+p)^2-m^2\right]}\;.\nn
\eeqn
If we impose that $I$ has no soft and collinear divergences, we have
$$
\left\{
 \eqalign{
  1 + B [q^2-2p\cdot q] = 0 \cr
  k \cdot  \left[ -A k + B (p-q) + Cp\right] = 0 \;.
 }
\right.
$$
The solution of  this system is
$$
\left\{
 \eqalign{
   B = -\frac{1}{\sd} \cr
   C = -\frac{1}{\su}\;,
 }
\right.
$$
with the term $A$ undefined. We can then rewrite $I$ as
\beqn
I &=& \int \frac{d^dl}{(2\pi)^d}\ \frac{
2 l\cdot \left[ -A k -\frac{1}{\sd} (p-q) -\frac{1}{\su}p\right] +
l^2\(A-\frac{1}{\su}-\frac{1}{\sd}\)}{
 l^2 (l-k)^2 \left[(l+p-q)^2-m^2\right] \left[(l+p)^2-m^2\right]} 
\nn\\
\label{eq:first_check}
&=& B_{2m} + A T_{2m}^{'q} -\frac{1}{\sd} T_{1m} -\frac{1}{\su}
T_{1m}^{'} \;,
\eeqn
where the second line is obtained directly from eq.~(\ref{eq:I_check}), and
the primed quantities are obtained  with the
substitution $p\leftrightarrow p'$, that is $\su \leftrightarrow \sd$.
The integral $I$ is now convergent and the cancellation of the
divergent part in the right-hand side can be checked directly (both in
the real part and in the absorptive one).  In fact: \newline
{\bf cancellation of the terms proportional to ${\displaystyle
 \mathbf{\frac{1}{\e^2}}}$} 
\[
\underbrace{\frac{1}{\su\sd}}_{B_{2m}} -\frac{1}{\sd}
\underbrace{\frac{1}{2\su}}_{T_{1m}} -
\frac{1}{\su}\underbrace{\frac{1}{2\sd}}_{T_{1m}^{'}} = 0
\]
{\bf cancellation of the terms proportional to ${\displaystyle
  \mathbf{\frac{1}{\e}}}$} \mbox{}
\[
\underbrace{\frac{1}{\su\sd}\left\{ \log\frac{m^2}{\su} +
\log\frac{m^2}{\su} + 2\pi i \right\}}_{B_{2m}}
-\frac{1}{\sd}
\underbrace{\frac{1}{\su} \left\{ \log\frac{m^2}{\su}+ i\pi
\right\}}_{T_{1m}} - \frac{1}{\su}
\underbrace{\frac{1}{\sd} \left\{ \log\frac{m^2}{\sd}+ i\pi \right\} }_
{T_{1m}^{'}} = 0
\]
{\bf cancellation of the finite terms:} the finite integral $I$ of
eq.~(\ref{eq:first_check}) can be written as
\beqn
 I &=& \tilde{I} +  \(A-\frac{1}{\su}-\frac{1}{\sd}\) T_{2m}^{q} \nn \\
\tilde{I} &=&  \int \frac{d^dl}{(2\pi)^d}\ \frac{
2\, l\cdot \left[ -A k -\frac{1}{\sd} (p-q) -\frac{1}{\su}p\right] }{
 l^2 (l-k)^2 \left[(l+p-q)^2-m^2\right]
 \left[(l+p)^2-m^2\right]} \;. 
\eeqn
In order to compute $\tilde{I}$, we introduce  the
Feynman parameters (see eq.~(\ref{eq:Feynman_param}))  to rewrite the
denominator in the following form
\beqn
{\rm den} &=& \lg \a_1 l^2 + \a_4 (l-k)^2 + \a_3 \lq(l+p-q)^2-m^2\rq
+ \a_2 \lq(l+p)^2 -m^2\rq \rg^4 \nn\\
 &=& \underbrace{\(\a_1+\a_2+\a_3+\a_4\)}_{=1} l^2 + 2 l\cdot
\left[ -k\, \a_4 +(p-q)\, \a_3 +p \,\a_2\right] + \ldots\;,
\eeqn
where the $\a_i$ are illustrated in Fig.~\ref{fig:b_2m}.
With the following change of variable
\beq  l' = l + \left[ -k\, \a_4 +(p-q)\,\a_3 +p\, \a_2\right]\;,
\eeq
we have a denominator that contains only  $l'^2$, so that  $\tilde{I}$
becomes
\beqn
\tilde{I}\! &=& \!\!\Gamma(4) \int \frac{d^dl'}{(2\pi)^d}\,\int [d\a]\ \nn\\
&& \!\!
\times  \frac{ 2\,\Bigl[ l' - \left[ -k \a_4 +(p-q)\a_3 +p \a_2\right]
\Bigr] \cdot
 \left[ -A k -\frac{1}{\sd} (p-q) -\frac{1}{\su}p\right] }{\left[
 l'^2 +\ldots \right]^4}  \nn\\
&=& \!\! N(\e)\,(1+\e)\,{\rm e}^{i\pi \e} \!\! \int[d\a]
\frac{-2 \, \left[ -k \a_4 +(p-q)\a_3 +p\, \a_2\right]
\cdot \left[ -A k -\frac{1}{\sd} (p-q) -\frac{1}{\su}p\right]}
{\lq-m^2 (\a_2+a_3)^2 +\sd\,
\a_1\,\a_3+q^2\a_2\,\a_3+\su\,\a_2\,\a_4\rq^2}\;,\nn
\eeqn
where we have put $\e= 0$ in the integral because it is finite by
construction. Expanding the dot-product we get
\beqn
\tilde{I} &=& N(\e)\,(1+\e)\, {\rm e}^{i\pi \e} \int[d\a]
\frac{-2 \( \a_3 H +\a_2 K \)}
{[-(\a_2+a_3)^2 m^2+\sd \a_1\a_3+q^2\a_2\a_3+\su\a_2\a_4]^2}\nn\\
&\equiv& N(\e)\,(1+\e)\, {\rm e}^{i\pi \e} (-2)\left[
H I_3 + K I_2 \right] \;,
\eeqn
where
$$
 \eqalign{
  H = \frac{\sd}{2}A -1 -m^2\(\frac{1}{\su}+\frac{1}{\sd}\)+
    \frac{q^2-\sd}{2\su}  \cr
  K = -\frac{\su}{2}A-m^2\(\frac{1}{\su}+\frac{1}{\sd}\)+
    \frac{q^2-\sd}{2\sd} \;. \cr
 }
$$
In the Euclidean region used to compute the one-loop integrals and
with the change of variables of eq.~(\ref{eq:change_var_b_2m}), we have
\[
I_{\ga 2,3\gc} = \int_0^1 dx\,dy\,dz\, \frac{(1-y) \ga x,(1-x)\gc}{
\left[M^2y+\sd z(1-x)(1-y)+q^2
    xy(1-x)+\su x(1-y)(1-z)\right]^2} \;.
\]
Performing the $z$ and $y$ integration, we obtain
\beqn
I_{\ga 2,3\gc}&=&  \int_0^1 dx \, \ga \frac{1}{M^2+q^2x(1-x)-\su x}
 \log\frac{M^2+q^2x(1-x)}{\su x}  \right.\nn\\
&&\left. - \frac{1}{M^2+q^2x(1-x)-\sd(1- x)}
 \log\frac{M^2+q^2x(1-x)}{\sd(1- x)} \gc \frac{\ga x,(1-x)\gc}{\sd(1-x)
 -\su x} \;. \nn
\eeqn
This integral can be expressed in terms of dilogarithm and logarithm
functions; nevertheless we have performed a numerical integration in
order to check the identity~(\ref{eq:first_check}).

\subsection[Check of $B_{3m}$]{Check of {\mylarge $B_{3m}$}}

With a reasoning similar to the previous one, we can check
$B_{3m}$. We introduce the following integral
\beqn
I &\equiv& \int \frac{d^dl}{(2\pi)^d}\ \frac{1 + A \left[(l+p)^2-m^2\right]
+ B \left[(l+p-q)^2-m^2\right] + C \left[(l-p')^2-m^2\right] }{
 l^2 \left[(l+p)^2-m^2\right] \left[(l+p-q)^2-m^2\right]
 \left[(l-p')^2-m^2\right]  }\nn\\
&=&\int \frac{d^dl}{(2\pi)^d}\  \frac{
1 + B \left[q^2-2p\cdot q\right] + 2 l\cdot \left[ A p + B (p-q) - 
Cp'\right] +l^2(A+B+C) }{ l^2 \left[(l+p)^2-m^2\right] 
\left[(l+p-q)^2-m^2\right] \left[(l-p')^2-m^2\right]  } \;. \nn
\eeqn
This integral has only soft singularities, but no collinear ones, which
can be removed if we require that
\[
   1+ B\left[q^2-2p\cdot q\right] = 0 \;,
\]
that is
\[
   B = -\frac{1}{\sd} \;,
\]
with $A$ and $C$ undefined.
We can rewrite $I$ as
\beqn
I  &=& \int \frac{d^dl}{(2\pi)^d}\  \frac{
 2\, l\cdot \left[ A p + -\frac{1}{\sd}(p-q) - Cp'\right]
+l^2(A-\frac{1}{\sd}+C)
}{ l^2 \left[(l+p)^2-m^2\right] \left[(l+p-q)^2-m^2\right]
 \left[(l-p')^2-m^2\right]  }\nn\\
\label{eq:second_check}
&=& B_{3m} + A T_{2m}^{'} -\frac{1}{\sd} T_{2m}^{q-k} +C
T_{2m}^{'q}\;,
\eeqn
where the primed quantities are obtained  with the
substitution $p\leftrightarrow p'$, that is 
\mbox{$\su \leftrightarrow \sd$}.
The integral $I$ is now convergent and the cancellation of the
divergent part in the right-hand side can be checked directly (both in
the real part and in the absorptive one).  In fact: \newline
{\bf cancellation of the terms proportional to ${\displaystyle
 \mathbf{\frac{1}{\e}}}$} 
\[
\underbrace{\frac{1}{\sd s_3 \dpr} \(\log\frac{\cm}{\cp} + i\pi \)
}_{B_{3m}}
 -\frac{1}{\sd}
\underbrace{\frac{1}{s_3 \dpr} \(\log\frac{\cm}{\cp} + i\pi \) }
_{T_{2m}^{q-k}} = 0
\]
{\bf cancellation of the finite terms:} the finite integral $I$ of
eq.~(\ref{eq:second_check}) can be written as
\beqn
I &=& \tilde{I} + \(A-\frac{1}{\sd}+ C\) T_{3m} \nn\\
\tilde{I} &\equiv& \int \frac{d^dl}{(2\pi)^d}\,\frac{
2 \,l\cdot \left[ A p -\frac{1}{\sd} (p-q) - C p' \right] }{
  l^2 \left[(l+p)^2-m^2\right] \left[(l+p-q)^2-m^2\right]
 \left[(l-p')^2-m^2\right] } \;.
\eeqn
With the same procedure used to check $B_{2m}$ in the previous
appendix, we can write
\[
\tilde{I} =  N(\e)\,(1+\e)\  {\rm e}^{i\pi \e} \int[d\a]
\frac{-2\left[ \a_2 p+\a_3(p-q) -\a_4 p'\right] \cdot
\left[ A p -\frac{1}{\sd}(p-q) -Cp' \right] }
{\left[-m^2(1-\a_1)^2  +\sd\,\a_1\,\a_3+ q^2 \a_2\,
\a_3+s_3\,\a_2\,\a_4 \right]^2} \;,
\]
where we have put $\e= 0$ in the integral because it is now finite,
and the $\a_i$ are illustrated in Fig.~\ref{fig:b_3m}.
Expanding the dot-product, we have
\beqn
\tilde{I} &=&  N(\e)\,(1+\e)\ {\rm e}^{i\pi \e} \int[d\a]
\frac{-2 \( \a_2 J + \a_3 L + \a_4 R
\)}
{\left[-m^2(1-\a_1)^2  +\sd\,\a_1\,\a_3+ q^2 \a_2\, \a_3+s_3\,
\a_2\,\a_4 \right]^2} \nn\\
&\equiv&  N(\e)\,(1+\e)\  {\rm e}^{i\pi \e}
(-2) \left[ J I_2 + L I_3 + R I_4   \right] \;,
\eeqn
where
$$
 \eqalign{
  J = \( A-\frac{1}{\sd}+C \) m^2
  +\frac{1}{\sd}\frac{q^2-\sd}{2} -\frac{q^2-\su-\sd}{2}C \cr
  L = \( A-\frac{1}{\sd}+C \) m^2 - \frac{q^2-\sd}{2}A -1
  +\frac{\sd}{2} C  \cr
  R = \( A-\frac{1}{\sd}+C \) m^2 - \frac{q^2-\su-\sd}{2} A
  -\frac{1}{2} \;. \cr
 }
$$
In the Euclidean region used to compute the one-loop integrals and
with the change of variables of eq.~(\ref{eq:change_var_b_3m}), we have
\[
I_{\ga 2,3,4\gc} = \int_0^1 dx\,dy\,dz\, \frac{x z \ga (1-z),yz,(1-y)z \gc}
{[ M^2 x + \sd(1-x)yz
  +s_3 x(1-y)z(1-z) + q^2 xyz(1-z)]^2} \;.
\]
From the identity
\beq
\int_0^1 dx\, \frac{x}{[ a x +b]^2} = \frac{1}{a^2}
\left[\log\frac{a+b}{b}+\frac{b}{a+b}-1 \right]\;,
\eeq
we can integrate in $x$, where $a$ and $b$ are given by
$$
 \eqalign{
   a =  M^2  - \sd yz +s_3 (1-y)z(1-z) + q^2 yz(1-z)\cr
   b = \sd y z \;.  \cr
 }
$$
At this stage of the integration, $I_{\ga 2,3,4\gc}$ are 
two-variable integrals, which can be expressed in terms of
dilogarithm and logarithm functions; nevertheless we have performed a
numerical integration in order to check the
identity~(\ref{eq:second_check}).

\section[Self energy with one massive propagator: $S_{1m}$]
{Self energy with one massive propagator: {\mylarge $S_{1m}$}}
\label{app:s_1m}
%%%%%%%%%%%%%%%%%%%%%%%%%%%%%%%%%%%%%%%%%%%
\begin{figure}[htb]
\centerline{\epsfig{figure=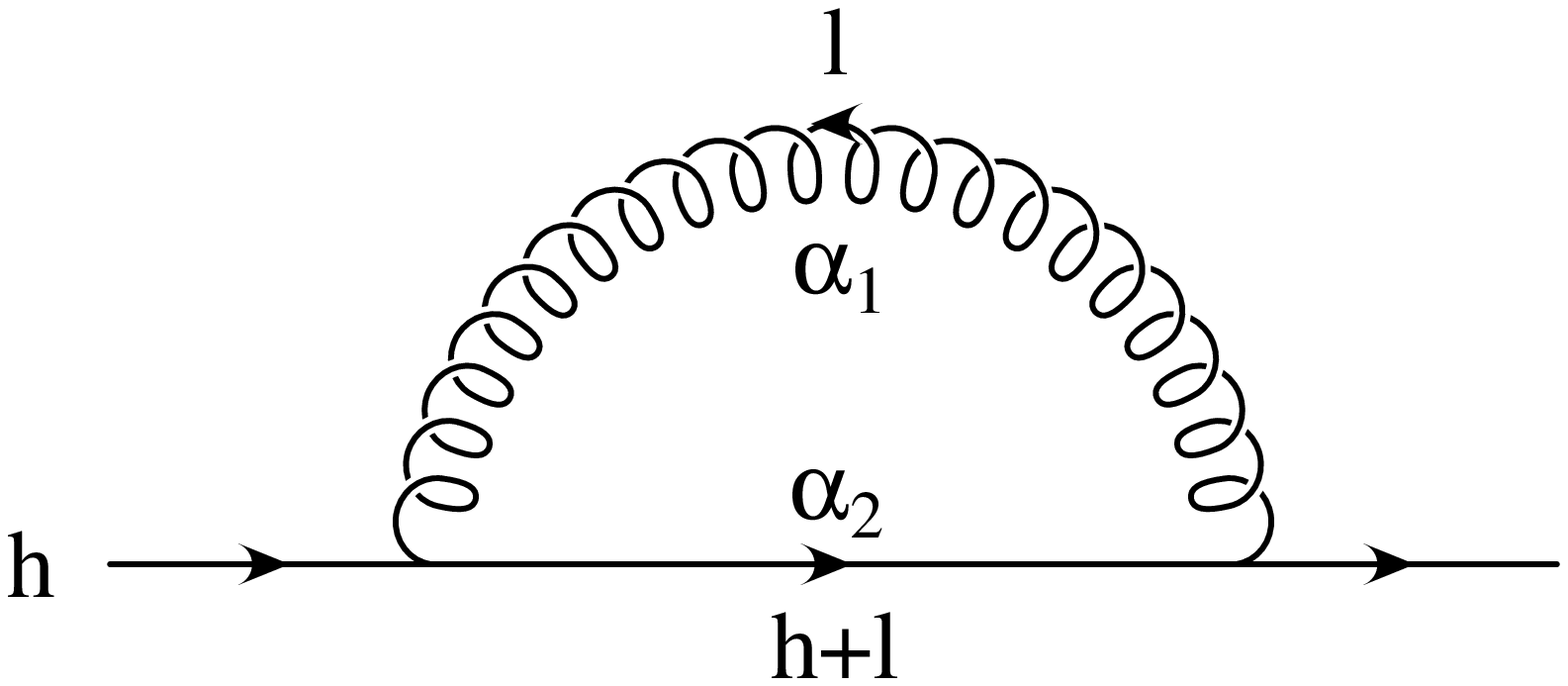,
width=\largfig,clip=}}\ccaption{}{ \label{fig:s_1m}
Diagram representing $S_{1m}$.}
\end{figure}
%%%%%%%%%%%%%%%%%%%%%%%%%%%%%%%%%%%%%%%%%%%
We are referring, as far as the denominator structure is concerned, to
a Feynman diagram of the type illustrated in Fig.~\ref{fig:s_1m}
\beq
S_{1m}\equiv\int \frac{d^dl}{(2\pi)^d}\ \frac{1}{l^2} \
\frac{1}{(h+l)^2-m^2}
\label{eq:s_1m}
= \frac{i}{16\pi^2}\ (4\pi)^\e \ \Gamma(\e) \int
\frac{[d\a]}{D^{\e}}\;,\quad\quad\quad d=4 -2\e \;,\phantom{aaa}
\eeq
where $h^2 \geq m^2$ and
\beq
D = -\a_1\,\a_2 \,h^2+  \a_2\, m^2 -i\eta  \;.
\eeq
We assume $h^2<0$ and, at the end, we return to the physical region
according to the $i\eta$ prescription. We then define
\[   H^2 = -(h^2+i\eta) > 0 \;,
\]
so that
\beq
  D =  H^2\a_1\,\a_2 + m^2 \a_2 \;,
\eeq
and, with the following change of variables,
\[\begin{array}{l}
  \a_1 = 1-x  \\
  \a_2 = x  \;,    \\
\end{array}
\]
we obtain (the Jacobian is equal to 1)
\beqn
S_{1m} &=& \frac{i}{16\pi^2}\ (4\pi)^\e \ \frac{\Gamma(1+\e)}{\e}
 \, \int_0^1dx\, \frac{1}{\Bigl[m^2 x + H^2 x (1-x)\Bigr]^\e} \nn\\
\label{eq:s_1m_intermedia}
& =& N(\e) \,\(m^2\)^{-\e} \frac{1}{\e}
 \int_0^1dx\, \frac{x^{-\ep}}{\left[1 + \frac{H^2}{m^2}
 (1-x)\right]^\e} \;.
\eeqn
The integral is finite, and then, expanding in $\e$, we get
\beqn
S_{1m} &=&
N(\e) \,\(m^2\)^{-\e} \frac{1}{\e} \ga 1 -\e \int_0^1 dx\, \log x -\e
 \int_0^1 dx\, \log\left[1 + \frac{H^2}{m^2} (1-x)\right]  \gc\nn\\
&=&
N(\e) \,\(m^2\)^{-\e}  \ga \frac{1}{\e} +2 - \(1+\frac{m^2}{H^2} \)
\log\(1+\frac{H^2}{m^2} \) \gc \;.
\eeqn
For $h^2 \geq m^2$, we have to continue our solution according to
\dl{
\log\(1+\frac{H^2}{m^2}\) \tol \log\(1-\frac{h^2}{m^2} -i\eta\) =
\log\(\frac{h^2}{m^2} -1\) -i\pi \;,
\hfill}
and we obtain
\beq
\label{eq:s_1m_result}
S_{1m} =
N(\e) \,\(m^2\)^{-\e}  \ga \frac{1}{\e} + 2 - \(1-\frac{m^2}{h^2}\)
\log\(\frac{h^2}{m^2} -1\)  + i\pi \(1-\frac{m^2}{h^2}\) \gc \;.
\eeq
Two particular expressions are interesting: $S_{1m}$ and
its derivate computed at  $h^2=m^2$. The first expression is easily done
\beq
\label{eq:s_1mm}
\left.S_{1m}\right|_{h^2=m^2} \equiv S_{1mm} =
N(\e) \,\(m^2\)^{-\e}  \( \frac{1}{\e} + 2 \) \;. \nn\\
\eeq
For the derivate, we have to start from eq.~(\ref{eq:s_1m_intermedia})
\beqn
\left. \frac{\partial S_{1m}}{\partial h^2} \right|_{h^2=m^2} &=& 
N(\e) \,\(m^2\)^{-\e} \frac{1}{\e}
 \int_0^1dx\, \left.\frac{\e \, \frac{1}{m^2}\, x^{-\e} (1-x)}
{\left[1 - \frac{h^2}{m^2} (1-x)\right]^{1+\e}} \right|_{h^2=m^2} \nn\\
&=&N(\e) \,\(m^2\)^{-\e} \frac{1}{m^2}
 \int_0^1dx\,  x^{-1-2\e} (1-x) \nn\\
&=&N(\e) \,\(m^2\)^{-\e} \frac{1}{m^2} \(-\frac{1}{2\e}-1\) \;.
\label{eq:s_1m_deriv}
\eeqn

\section[Self energy with two massive propagators: $S_{2m}$]
{Self energy with two massive propagators: {\mylarge $S_{2m}$}}
\label{app:s_2m}
%%%%%%%%%%%%%%%%%%%%%%%%%%%%%%%%%%%%%%%%%%%
\begin{figure}[htb]
\centerline{\epsfig{figure=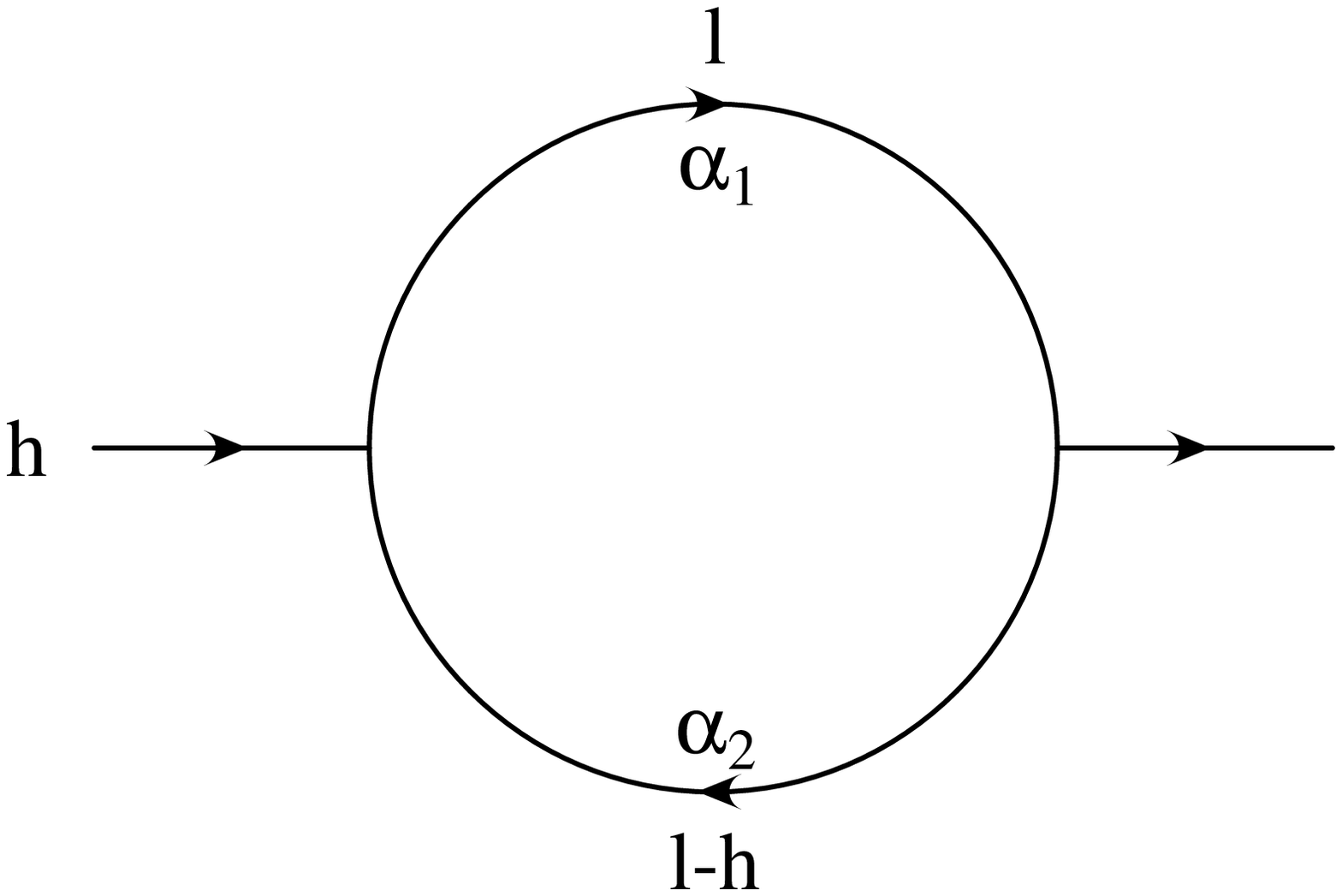,
width=\largfig,clip=}}\ccaption{}{ \label{fig:s_2m}
Diagram representing $S_{2m}$.}
\end{figure}
%%%%%%%%%%%%%%%%%%%%%%%%%%%%%%%%%%%%%%%%%%%
We are referring, as far as the denominator structure is concerned, to
a Feynman diagram of the type represented in Fig.~\ref{fig:s_2m}
\beq
\label{eq:s_2m}
S_{2m}\equiv\int \frac{d^dl}{(2\pi)^d}\ \frac{1}{l^2-m^2} \
\frac{1}{(l-h)^2-m^2}
= \frac{i}{16\pi^2}\ (4\pi)^\e \ \Gamma(\e) \int
\frac{[d\a]}{D^{\e}}\;,\quad\quad\quad d=4 -2\e \;,
\eeq
where $h^2 \geq 4 m^2$ and
\beq
D = - h^2 \a_1\,\a_2+ m^2-i\eta \;.
\eeq
Again we assume $h^2 < 0$ and define
\[
   H^2 = -(h^2+i\eta) > 0 \;.
\]
The Jacobian of the following change of variables is equal to 1,
\[\begin{array}{l}
  \a_1 = x  \\
  \a_2 = 1-x   \;,   \\
\end{array}
\]
so that
\beqn
\label{eq:s_2m_interm}
S_{2m} &=& \frac{i}{16\pi^2}\ (4\pi)^\e \ \frac{\Gamma(1+\e)}{\e} \,
 \int_0^1dx\, \frac{1}{\left[  m^2 + H^2 x(1-x)\right]^\e} 
\\
 &=& N(\e) \,\(m^2\)^{-\e} \frac{1}{\e}
 \int_0^1dx\, \frac{1}{\left[1 + \frac{H^2}{m^2} x(1-x)\right]^\e} \;.\nn
\eeqn
The integral is finite, and we can expand in $\e$
\beqn
S_{2m} &=&
N(\e) \,\(m^2\)^{-\e} \frac{1}{\e} \ga 1
-\e \int_0^1 dx\, \log\left[1 + \frac{H^2}{m^2} x (1-x)\right]  \gc\nn\\
&=& N(\e) \,\(m^2\)^{-\e}  \ga \frac{1}{\e} + 2 + \(\taup-\taum\)
\log\frac{-\taum}{\taup}  \gc \;,
\eeqn
where we have used the fact that
\[  1+\frac{H^2}{m^2}x(1-x) = \frac{H^2}{m^2}(\taup-x) (x-\taum)
\]
with
\[
\taupm = \frac{1}{2}\( 1 \pm \sqrt{1+\frac{4m^2}{H^2}}\) \;.
\]
If we are interested in the value of $h^2$ such that $h^2 \geq 4 m^2$,
we continue analytically our solution
\beq
\taupm = \frac{1}{2}\( 1 \pm \sqrt{1-\frac{4m^2}{h^2}}\)  \pm i\eta \;,
\eeq
and we obtain
\beq
S_{2m} =
N(\e) \,\(m^2\)^{-\e}  \ga \frac{1}{\e} + 2 + \(\taup-\taum\)
\log\frac{\taum}{\taup}   + i\,\pi \(\taup-\taum\) \gc \;.
\eeq
Two particular cases are interesting: the one with $m=0$ and the one
with $h^2 \approx 0 $.
For both cases, we start from eq.~(\ref{eq:s_2m_interm}). If $m=0$ we
have
\beq
\label{eq:s_2m_m=0}
\left. S_{2m}\right|_{m=0} = \Nep\, \(H^2\)^{-\e}\,\frac{1}{\e}
\int_0^1dx\, x^{-\e}(1-x)^{-\e}
= \Nep \(h^2\)^{-\e} {\rm e}^{i\pi \e}
\(\frac{1}{\e} + 2 \) \;.
\eeq
When $h^2 \approx 0 $, we can expand eq.~(\ref{eq:s_2m_interm}) in
$\e$ and $h^2$
\beqn
\left.S_{2m}\right|_{h^2 \approx 0} &=& 
N(\e) \,\(m^2\)^{-\e} \frac{1}{\e} \int_0^1dx\,  
\left[1 - \e \log \left( 1- \frac{h^2}{m^2}\, x\,(1-x)\right)
+ {\cal O}\(\e^2\) \right]
\nn\\
&=& N(\e) \,\(m^2\)^{-\e} \frac{1}{\e} \int_0^1dx\,  
\left[1 + \e  \left( \frac{h^2}{m^2}\, x\,(1-x)\right) + 
{\cal O}\(h^4\)\right] \nn\\
&=&  N(\e) \,\(m^2\)^{-\e} \left\{ \frac{1}{\e} +\frac{1}{6}\,
\frac{h^2}{m^2} \right\}\;.
\label{eq:s_2m_h^2=0}
\eeqn
In addition
\beq
\label{eq:s_2m_mu}
\(S_{2m}\)_\mu \equiv \int \frac{d^dl}{(2\pi)^d}\ \frac{l_\mu}{\(l^2-m^2\)
 \left[ (l-h)^2-m^2 \right] } = \frac{h_\mu}{2}\, S_{2m} \;,
\eeq
that can be demonstrated by contracting both sides with $h^\mu$, and
expanding
\beq
 h\cdot l = -\frac{1}{2} 
\left\{ \left[ (l-h)^2 -m^2 \right] - \left[ l^2 -m^2
 \right] -h^2 \right\} \;.
\eeq
In the same way, we can write
\beq
\label{eq:s_2m_munu}
\(S_{2m}\)_{\mu\nu} \equiv \int \frac{d^dl}{(2\pi)^d}\ 
\frac{l_\mu\,l_\nu}{\(l^2-m^2\)
 \left[ (l-h)^2-m^2 \right] } = A\, \frac{h_\mu\, h_\nu}{h^2} +
 B\,g_{\mu\nu} \;,
\eeq
with
\beqn
   A &=& \frac{1}{d-1} \left[ \(\frac{d}{2}-1\) L_{1m} + \(
   \frac{d}{4}\, h^2 -m^2\) S_{2m} \right] \nn\\
 B &=& \frac{1}{d-1}\left[ \frac{1}{2} \,L_{1m} + \(m^2 - \frac{h^2}{4}
   \) S_{2m} \right] \;,
\eeqn
that can be obtained by contracting  eq.~(\ref{eq:s_2m_munu}) with
$h^\nu$ and with $g^{\mu\nu}$. 
The expression of $L_{1m}$ is given in eq.~(\ref{eq:l_1m_ris}).

\section[Vacuum polarization with two massive propagators:
 $P_{2m}$]{Vacuum polarization with two massive propagators:
 {\mylarge $P_{2m}$}}
\label{app:p_2m}
%%%%%%%%%%%%%%%%%%%%%%%%%%%%%%%%%%%%%%%%%%%
\begin{figure}[htb]
\centerline{\epsfig{figure=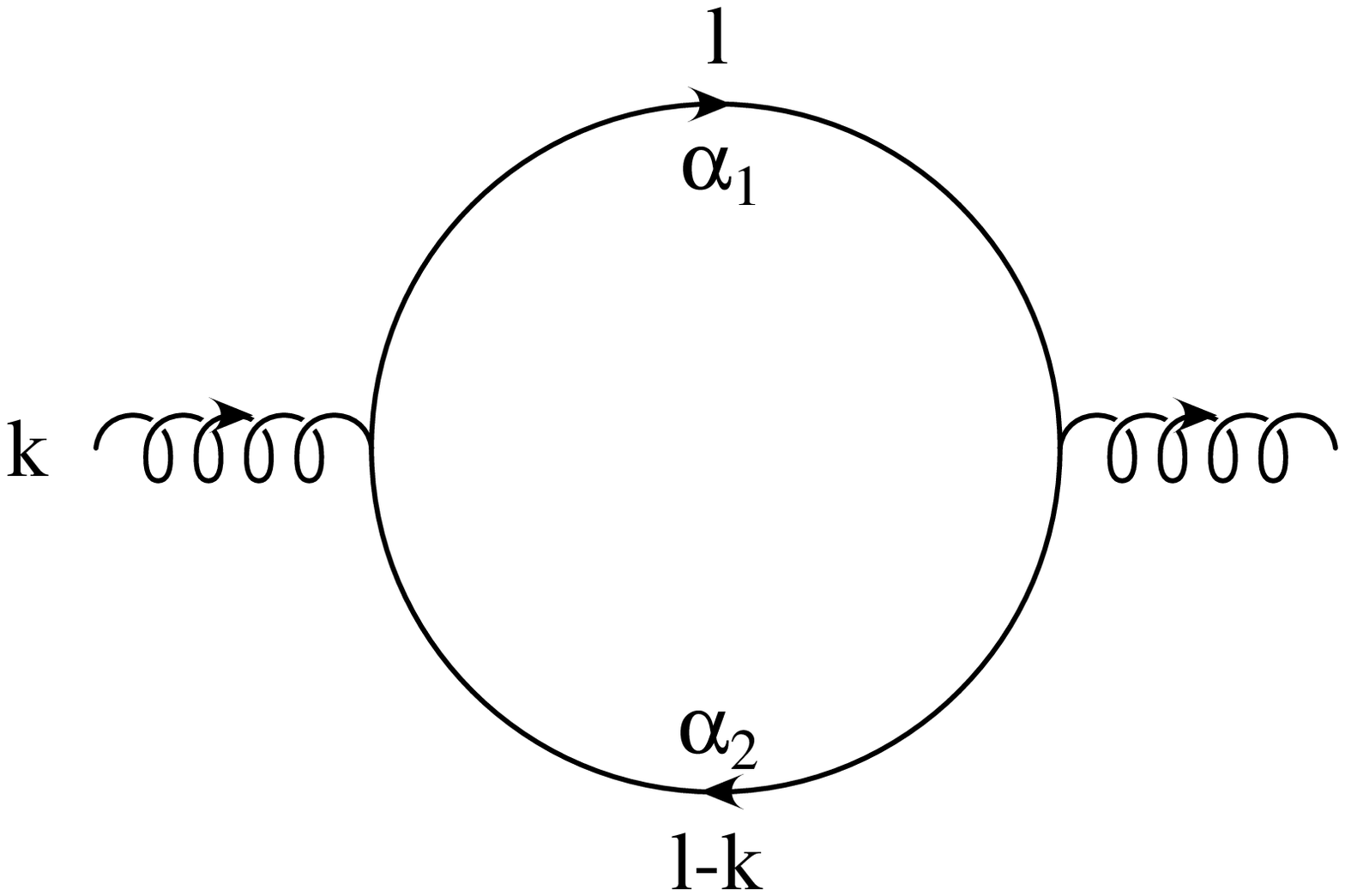,
width=\largfig,clip=}}\ccaption{}{ \label{fig:p_2m}
Diagram representing $P_{2m}$.}
\end{figure}
%%%%%%%%%%%%%%%%%%%%%%%%%%%%%%%%%%%%%%%%%%%
This appendix and the following one are to be regarded as particular
cases of Appendix~\ref{app:s_2m}. Nevertheless, we summarize here the
results. The structure of the denominators of this Feynman diagram is
depicted in Fig.~\ref{fig:p_2m}
\beq
\label{eq:p_2m}
P_{2m}\equiv\int \frac{d^dl}{(2\pi)^d}\ \frac{1}{l^2-m^2} \
\frac{1}{(l-k)^2-m^2} =
\frac{i}{16\pi^2}\ (4\pi)^\e \ \Gamma(\e) \int
\frac{[d\a]}{D^{\e}}\;,\quad\quad\quad d=4 -2\e \;,
\eeq
with
\beq
k^2 = 0\;,\quad\quad\quad  D = (\a_1 + \a_2)\,m^2-i\eta  = m^2 \;.
\eeq
The result is simply given by
\beq
 P_{2m} = \frac{i}{16\pi^2}\ (4\pi)^\e \ \frac{\Gamma(1+\e)}{\e} \
 \int_0^1 d\a\ \frac{1}{\(m^2\)^\e} = N(\e)\,\(m^2\)^{-\e} \frac{1}{\e}\;,
\eeq
that agrees with eq.~(\ref{eq:s_2m_h^2=0}) for $h^2 = 0$.
\section[Vacuum polarization with no massive propagator:
  $P_{0m}$]{Vacuum polarization with no massive propagator:
 {\mylarge $P_{0m}$}}
\label{app:p_m}
%%%%%%%%%%%%%%%%%%%%%%%%%%%%%%%%%%%%%%%%%%%
\begin{figure}[htb]
\centerline{\epsfig{figure=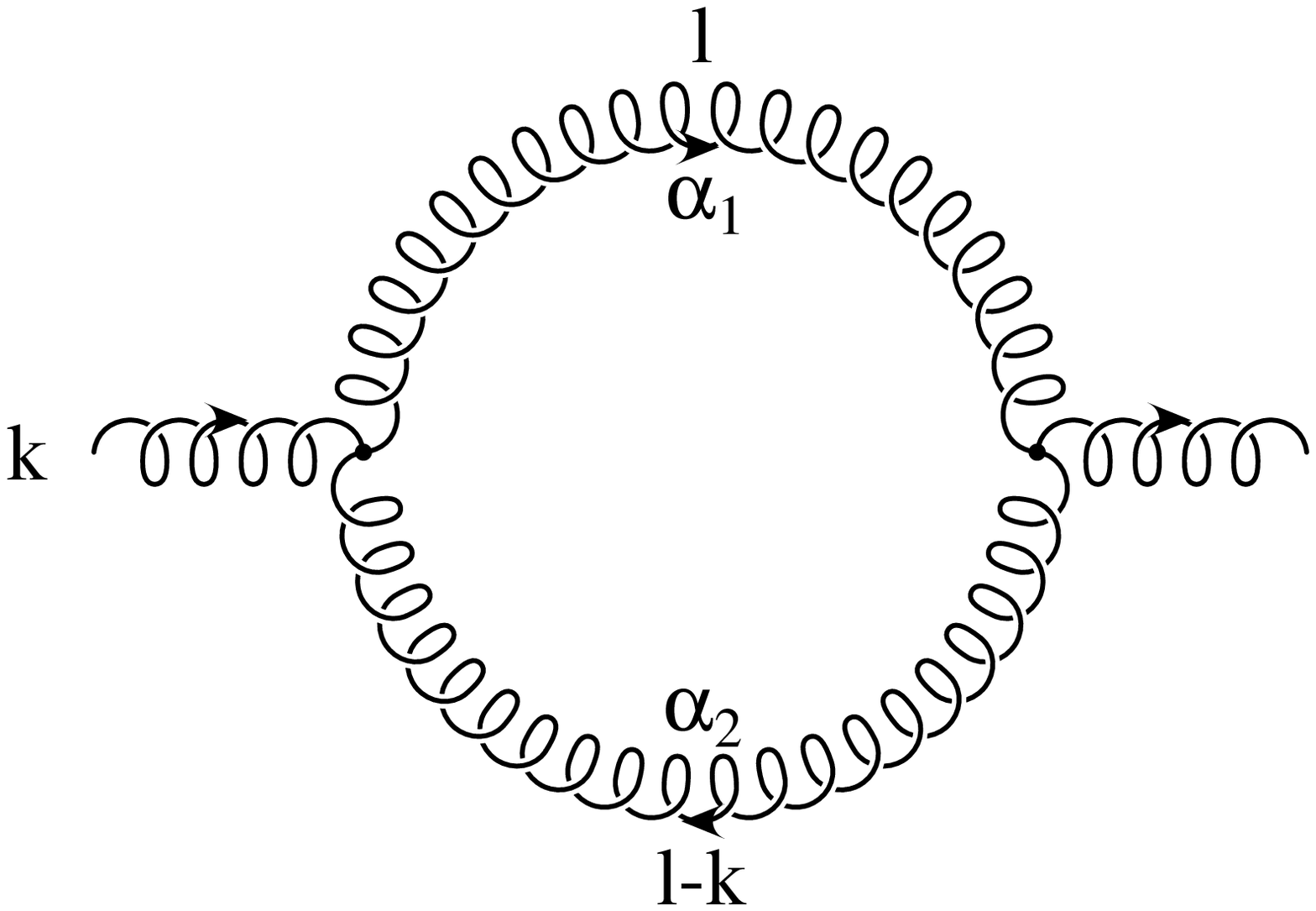,
width=\largfig,clip=}}\ccaption{}{ \label{fig:p_0m}
Diagram representing $P_{0m}$.}
\end{figure}
%%%%%%%%%%%%%%%%%%%%%%%%%%%%%%%%%%%%%%%%%%%
We are referring, as far as the denominator structure is concerned, to
a Feynman diagram of the type represented in Fig.~\ref{fig:p_0m}
\beq
P_{0m}\equiv\int \frac{d^dl}{(2\pi)^d}\ \frac{1}{l^2} \
\frac{1}{(l-k)^2}\;,\quad\quad\quad  d=4 -2\e \;.
\eeq
Since no invariant scalars are present, because $k^2=0$, we have
\beq
 P_{0m} = 0 \;.
\eeq

\section[Massive loop: $L_{1m}$]
{Massive loop: {\mylarge $L_{1m}$}}
\label{app:l_1m}
%%%%%%%%%%%%%%%%%%%%%%%%%%%%%%%%%%%%%%%%%%%
\begin{figure}[htb]
\centerline{\epsfig{figure=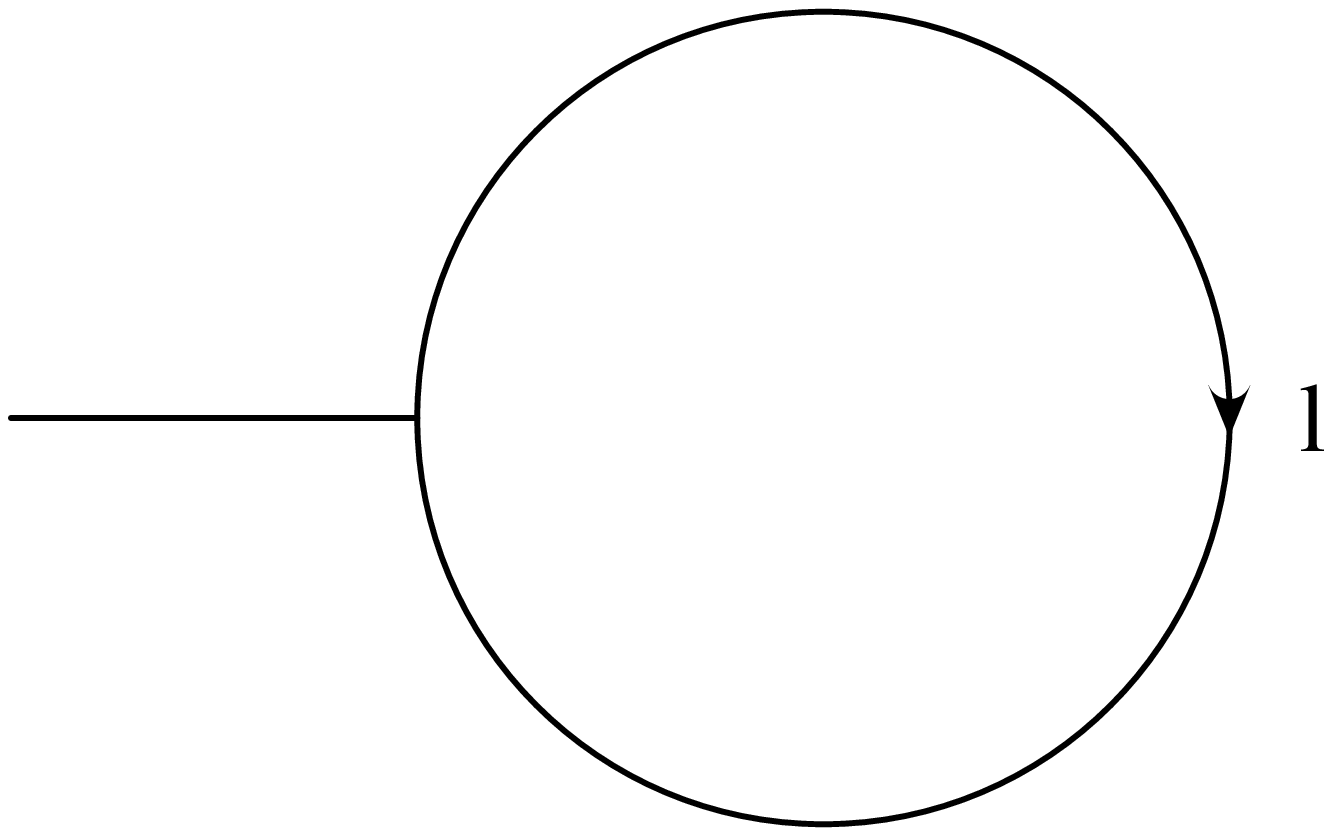,
width=0.34\textwidth,clip=}}\ccaption{}{ \label{fig:l_1m}
Diagram representing $L_{1m}$.}
\end{figure}
%%%%%%%%%%%%%%%%%%%%%%%%%%%%%%%%%%%%%%%%%%%
The single loop is illustrated in Fig.~\ref{fig:l_1m}
\beq
L_{1m}\equiv\int \frac{d^dl}{(2\pi)^d}\ \frac{1}{l^2-m^2}\;,
\quad\quad\quad   d=4 -2\e\;,
\eeq
and can be easily integarted
\beq
\label{eq:l_1m_ris}
L_{1m} = - \frac{i}{(4\pi)^2}\, (4\pi)^\e\,
\frac{\Gamma(-1+\e)}{\(m^2\)^{-1+\e}} = N(\e) \,\(m^2\)^{-\e} m^2
\(\frac{1}{\e}+1 \)\;.
\eeq
A particular case is the massless loop, where, again, we do not have
invariants to build the solution
\beq
\label{eq:l_0m}
L_{0m}\equiv\int \frac{d^dl}{(2\pi)^d}\ \frac{1}{l^2} = 0 \;.
\eeq

\chapter{Renormalization}
\thispagestyle{plain}
\section{Radiative corrections to external heavy-quark lines}
\label{app:mass_cnt}
%%%%%%%%%%%%%%%%%%%%%%%%%%%%%%%%%%%%%%%%%%%
\begin{figure}[htb]
\centerline{\epsfig{figure=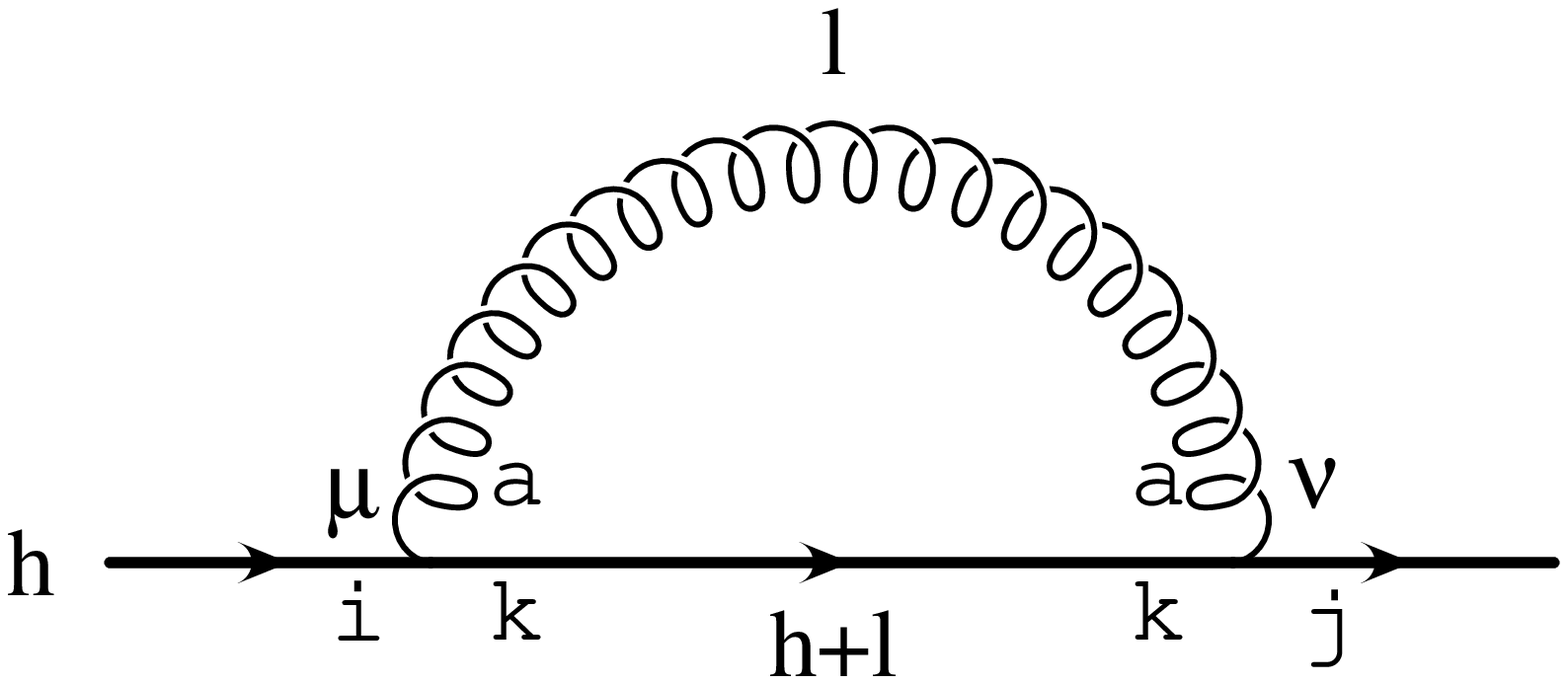,
width=\largfig,clip=}}
\ccaption{}{ \label{fig:quark_self_energy}
Diagram representing the quark self-energy $\Sigma_{ij}(h)$.}
\end{figure}
%%%%%%%%%%%%%%%%%%%%%%%%%%%%%%%%%%%%%%%%%%%
In this appendix we describe how to treat loop corrections to {\bf external}
heavy-quark lines and how to compute the mass counterterm.

The one loop correction to a quark propagator reads
\beq
\Sigma_{ij}(h) = \int\frac{d^dl}{(2\pi)^d} \(-ig_s\mu^\e
\gamma^\nu t_{jk}^a\)
\frac{i}{\dsh+\ds{l}-m} \( -ig_s\mu^\e \gamma^\mu t^a_{ki} \)
\frac{-ig_{\mu\nu}}{l^2} \;,
\eeq
where the sum over repeated indexes is meant and $\mu$ is the mass
parameter of the dimensional regularization, introduced in order to
keep $g_s$ dimensionless.  With a little algebra we have
\beqn
\Sigma_{ij}(h) &=& -g_s^2\mu^{2\e} C_F\, \delta_{ij}
\int\frac{d^dl}{(2\pi)^d} \frac{\gamma^\mu(\ds{l}+\dsh+m)\gamma_\mu}
{l^2 \left[(h+l)^2-m^2 \right]} \nn\\
&=& -g_s^2\mu^{2\e} C_F \,\delta_{ij}
\int\frac{d^dl}{(2\pi)^d} \frac{(2-d)(\dsh+\ds{l})+d\,m}
{l^2 \left[(h+l)^2-m^2 \right]} \nn\\
&=& -g_s^2 \mu^{2\e}C_F\, \delta_{ij} \left\{
\left[ (2-d) \dsh +d\,m\right] \int\frac{d^dl}{(2\pi)^d} \,\frac{1}
{l^2 \left[(h+l)^2-m^2 \right]}\right.\nn\\
&&+ \left.(2-d)  \int\frac{d^dl}{(2\pi)^d} \,\frac{\ds{l}}
{l^2 \left[(h+l)^2-m^2 \right]}
\right\}\;.
\eeqn
In the last integral we can replace
\beq
l_\mu \, \to\, l\cdot h \;\frac{h_\mu}{h^2}= \frac{1}{2}
\left\{ \left[ (h+l)^2 -m^2\right] -h^2-l^2+m^2 \right\}
\,\frac{h_\mu}{h^2} \;,
\eeq
and obtain
\beqn
\Sigma_{ij}(h)\!\! &=& \!\! -g_s^2 \mu^{2\e} C_F\, \delta_{ij} \left\{
\left[ (2-d) \dsh +d\,m\right] S_{1m} + \frac{\dsh}{2h^2}(2-d)\!
\left[\(m^2-h^2\)\, S_{1m}- L_{1m}   \right] \!\right\}\nn\\
&=& \!\! -g_s^2\mu^{2\e} C_F\, \delta_{ij} \left\{
\left[ \(-1+\e -(1-\e) \frac{m^2}{h^2} \) \dsh + (4-2\e)\,m \right]
S_{1m} \right.\nn\\
&& {} + \left.\frac{\dsh}{h^2}(1-\e)L_{1m}  \right\} \;,
\label{eq:quark_propagator}
\eeqn
where we have used the definitions given by
eqs.~(\ref{eq:s_1m}),~(\ref{eq:l_1m_ris}) and~(\ref{eq:l_0m}) and we
have specified $d=4-2\e$.  We are interested in the expansion of
$\Sigma_{ij}(h)$ around $\dsh=m$
\beq
\label{eq:sviluppo_Sigma}
\Sigma_{ij}(h) =
\left. \Sigma_{ij}(h)\right|_{\ds{h}=m} + \( \dsh-m\) \left.
\frac{\partial\Sigma_{ij}(h)}{\partial \dsh} \right|_{\ds{h}=m} +
{\cal O}\(\(\dsh-m\)^2\) \;.
\eeq
Using the identity
\beq
\dsh\, \dsh = h^2 \ \Rightarrow \ 2\dsh \,\partial\dsh =
\partial h^2 \
\Rightarrow\ \frac{\partial}{\partial\dsh} = 2 \dsh
\frac{\partial}{\partial  h^2}  \;,
\eeq
so that
\beq
\left.\frac{\partial\Sigma_{ij}(h)}{\partial \dsh}
\right|_{\ds{h}=m} =
2\,m \left.\frac{\partial\Sigma_{ij}(h)}{\partial h^2}
\right|_{\ds{h}=m} \;,
\eeq
and eqs.~(\ref{eq:s_1mm}) and~(\ref{eq:s_1m_deriv}), with simple
algebraic passages, we can rewrite~(\ref{eq:sviluppo_Sigma}) as
\beqn
\Sigma_{ij}(h) \!\!&=& \!\!
-g_s^2 C_F\,  N(\e) \(\frac{\mu^2}{m^2}\)^{\e} \!\delta_{ij}
\left\{ \left[ \frac{3}{\e}+4 \right]\,m + \( \dsh-m\) \left[
-\frac{3}{\e} -4\right]\right\}+ {\cal O}\(\(\dsh-m\)^2\) \nn\\
&\equiv& \!\!\delta_{ij}\,  \Bigl[ -i\,\delta m -i\, \( \dsh-m\)
z_Q\Bigr]+ {\cal O}\(\(\dsh-m\)^2\)  \;,
\eeqn
where
\beqn
\delta m &=& -i\, g_s^2 C_F\,  N(\e) \(\frac{\mu^2}{m^2}\)^{\e}
\left[ \frac{3}{\e}+4 \right]\,m \nn\\
z_Q &= &   i \,g_s^2 C_F\,  N(\e) \(\frac{\mu^2}{m^2}\)^{\e}
\left[ \frac{3}{\e}+4 \right] \;.
\eeqn
The full quark propagator at first order reads
\beqn
G_Q(h) &=&
\frac{i\, \delta_{ij}}{\dsh-m} + \frac{i\,\delta_{ik}}{\dsh-m}\
\Sigma_{kl}(h) \ \frac{i\,\delta_{lj}}{\dsh-m} +
{\cal O}\(\as^2\) \nn\\
&=& \frac{i\,\delta_{ij}}{\dsh-m} \(1 + z_Q\) +
\frac{i\,\delta_{ik}}{\dsh-m} \( -i\,\delta m\)
\frac{i\,\delta_{kj}}{\dsh-m} +  \frac{{\cal
O}\(h^2-m^2\)}{\dsh-m} \;.\phantom{aaaaa}
\eeqn
If we want that the pole of the propagator is not displaced by
radiative corrections, so that $m$ corresponds to the pole mass definition,
we have to add a mass counterterm to cancel the second term of the
above expression. For this reason, we define the Feynman rule for the
mass counterterm as the insertion, in the fermion propagator, of the
vertex $-i\,m_c$, where
\beq
\label{eq:mass_counterterm}
m_c = -\delta m = - g_s^2 C_F\,  N_\e \(\frac{\mu^2}{m^2}\)^{\e}
\left[ \frac{3}{\e}+4 \right]\,m\;,
\eeq
and $N_\e$ is defined in eq.~(\ref{eq:def_Nep}), that is
\[
N_\e = -i\,N(\e) = \frac{1}{16\pi^2} \(4\pi\)^\e \Gamma(1+\e)\;.
\]
In this way, {\bf slightly off-shell}, the quark propagator behaves like
\beq
G_Q(h) = \frac{i\, \delta_{ij}}{\dsh-m}\,Z_Q \;,
\eeq
with
\beq
\label{eq:Z_Q}
Z_Q = 1+z_Q = 1  - g_s^2 C_F\,  N_\e \(\frac{\mu^2}{m^2}\)^{\e}
\left[ \frac{3}{\e}+4 \right] \;.
\eeq

\section{Radiative corrections to  external gluon lines}
\label{app:gluon_renorm}
%%%%%%%%%%%%%%%%%%%%%%%%%%%%%%%%%%%%%%%%%%%
\begin{figure}[htb]
\centerline{\epsfig{figure=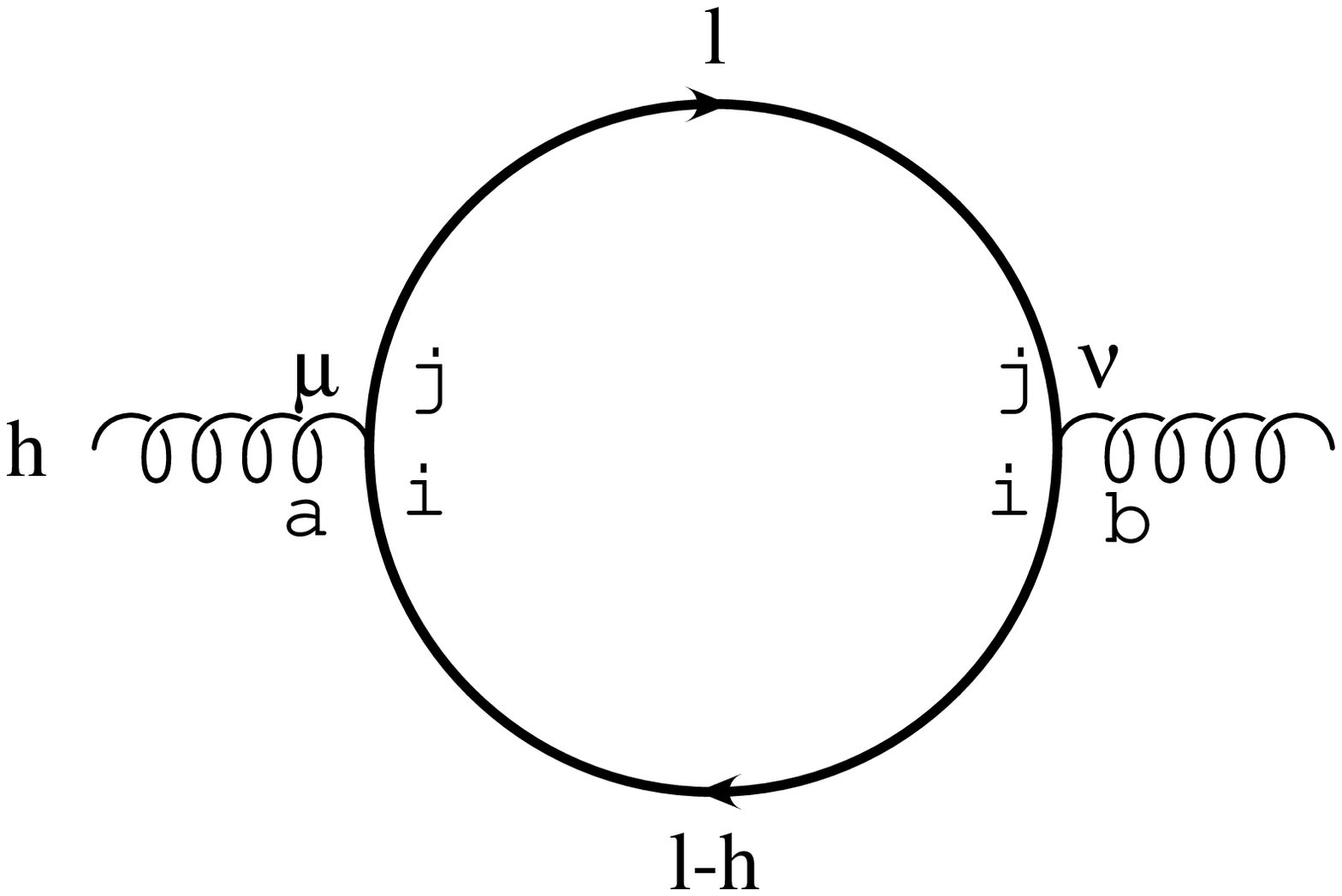,
width=\largfig,clip=}}
\ccaption{}{ \label{fig:gluon_self_energy}
Diagram representing the gluon self-energy contribution coming from a
massive-quark loop.}
\end{figure}
%%%%%%%%%%%%%%%%%%%%%%%%%%%%%%%%%%%%%%%%%%%
Gluon, light-fermion and ghost self-energy corrections to {\bf external}
gluon lines vanish in dimensional regularization, so  we need to
consider only the correction coming from the heavy-flavour loop,
represented in Fig.~\ref{fig:gluon_self_energy}.
This  contribution is given by
\beq
\Pi_{\mu\nu}^{ab} = -
\int\frac{d^dl}{(2\pi)^d} \Tr\left[ \(-ig_s\mu^\e \gamma_\nu t_{ij}^b\)
\frac{i}{\ds{l}-m} \( -ig_s\mu^\e \gamma_\mu t^a_{ji} \)
\frac{i}{\ds{l}-\dsh-m} \right]\;,
\eeq
where the minus sign takes care of the fermion loop. With simple
algebra
\beqn
\Pi_{\mu\nu}^{ab} \!\!&=& \!\!- g_s^2 \mu^{2\e}\, T_F\, \delta^{ab}
\int\frac{d^dl}{(2\pi)^d} \frac{\Tr\Bigl[
\gamma_\mu \( \ds{l}+m\) \gamma_\nu
\left[\(\ds{l}-\dsh\)+m\right]\Bigr]}
{\(l^2-m^2\) \left[\(l-h\)^2-m^2\right]} \nn\\
&=& \!\!
4 \,g_s^2 \mu^{2\e}\, T_F\, \delta^{ab}
\int\frac{d^dl}{(2\pi)^d} \frac{g_{\mu\nu}\(l^2-m^2\)
-g_{\mu\nu}\,l\cdot h - 2\,l_\mu l_\nu + l_\mu h_\nu + l_\nu h_\mu }
{\(l^2-m^2\) \left[\(l-h\)^2-m^2\right]}\;,\phantom{aaaa}
\eeqn
where $\Tr\(t^a t^b\)=T_F\,\delta^{ab}$,  $T_F = 1/2$.
Using eqs.~(\ref{eq:s_2m}), (\ref{eq:s_2m_mu})
and~(\ref{eq:s_2m_munu}), we can write
\beq
\Pi_{\mu\nu}^{ab} = 4 \,g_s^2 \mu^{2\e}\, T_F\, \delta^{ab} h^2
\left[g_{\mu\nu} -\frac{h_\mu h_\nu}{h^2} \right] \Pi(h^2) \;,
\eeq
where
\beq
\Pi(h^2) = \frac{d-2}{d-1}\, \frac{1}{h^2}\ L_{1m} -
\frac{1}{d-1}\left[\frac{d}{2} -1 + 2 \frac{m^2}{h^2}\right] S_{2m} \;.
\eeq
In $d=4-2\e$ dimensions, we get
\beqn
\Pi_{\mu\nu}^{ab} &=& \frac{4}{3} \,g_s^2 \mu^{2\e}\, T_F\, \delta^{ab} h^2
\(g_{\mu\nu} -\frac{h_\mu h_\nu}{h^2} \)
\Bigg\{ 2\(1-\frac{\e}{3}\)\frac{1}{h^2}\, L_{1m} \nn\\
&& {} -  \left[ 1-\frac{\e}{3}
+ 2 \,\frac{m^2}{h^2}\(1+\frac{2}{3}\,\e\)\right] S_{2m} \Bigg\} \;.
\label{eq:Pi}
\eeqn
We are interested in the form of the propagator for small gluon
virtuality: using eqs.~(\ref{eq:s_2m_h^2=0})
and~(\ref{eq:l_1m_ris}) we obtain
\beq
\Pi_{\mu\nu}^{ab} = -\frac{4}{3} \,\Nep \,g_s^2 \(\frac{\mu^2}{m^2}\)^\e
\, T_F\, \delta^{ab} h^2
\(g_{\mu\nu} -\frac{h_\mu h_\nu}{h^2} \) \( \frac{1}{\e}+
\frac{1}{6}\, \frac{h^2}{m^2} \)\;.
\eeq
It is worth noticing that, as previously stated, for small gluon
virtuality and for the massless-quark loop, $\Pi(h^2)$ is zero, because
$L_{1m}=0$ and $S_{2m} \,\to\, P_{0m}=0$.  The full gluon propagator
at first order reads
\beqn
G_g(h) \!\!&=&\!\! \frac{-i \,\delta^{ab} g_{\mu\nu}}{h^2} +
\frac{-i\,\delta^{ac}  g_{\mu}^{\alpha}}{h^2} \  \Pi_{\alpha\beta}^{cd}\
\frac{-i \,\delta^{db}g^{\beta}_{\nu}}{h^2} + {\cal O}\(\as^2\) \nn\\
&=&\!\! \frac{-i \,\delta^{ab} g_{\mu\nu}}{h^2} -N_\e T_F\,g_s^2
\(\frac{\mu^2}{m^2}\)^\e \frac{4}{3\,\e} \frac{-i \,\delta^{ab}
\(g_{\mu\nu} -h_\mu h_\nu/h^2 \)}{h^2} + \frac{{\cal
O}\(h^2\)}{h^2} \;.\phantom{aaaaa}
\eeqn
The part of the full propagator proportional to $h_\mu h_\nu$ gives
no contribution, since it is contracted with a conserved current. For
this reason we can write, for the {\bf slightly off-shell} gluon propagator
\beq
G_g(h) = \frac{-i \,\delta^{ab} g_{\mu\nu}}{h^2}  Z_g\;,
\eeq
with
\beq
\label{eq:Z_g}
Z_g = 1+z_g = 1 - N_\e T_F\,g_s^2 \(\frac{\mu^2}{m^2}\)^\e
\frac{4}{3\,\e}\;.
\eeq

\section{Charge renormalization}
\label{app:charge_renorm}
We carry out the charge renormalization in the mixed scheme of
Ref.~\cite{CWZ}, in which the \nlf light flavours are subtracted in the
\MSB\ scheme, while the heavy-flavour loop is subtracted at zero
momentum.

In this scheme, the heavy flavour decouples at low energies.
The prescription for charge renormalization is
\beq
\as\; \tol \;  Z_{\as} \, \as \equiv  Z_{\rm vertex}^2\,
Z_{\rm quark}^{-2} \, Z_{\rm gluon}^{-1} \, \as
\label{eq:Z_as}
\eeq
\beq
\label{eq:charge_ren}
Z_{\as} =  1+z_{\as} = 1 + g_s^2 N_\e
\left[ \( \frac{4}{3\e}T_F\,n_{\rm lf} -\frac{11}{3\e}C_A \)
+ \(\frac{\mu^2}{m^2}\)^\e
\frac{4}{3\e}T_F \right]\;,
\eeq
where $C_A = N_c = 3$ for an SU(3) gauge theory.

We try to justify the $T_F$ structure of this expression without
entering in the details of the calculation (albeit straightforward).
Three renormalization constants determine
the charge renormalization  $Z_{\as}$:
$Z_{\rm vertex}$, that takes account  of the two Feynman
diagrams describing quark-quark-gluon interaction; $Z_{\rm quark}$,
which contains the contribution coming from the quark self-energy and
$Z_{\rm gluon}$, that includes the effects of vacuum polarization.

The contributions proportional to $T_F$ come only from the gluon
self-energy part that takes account of the corrections due to quark
loops. In fact:
\begin{itemize}
\item[-] the vertex correction contains the colour
factors $C_A$ and $C_F$, and, in the \MSB\ scheme, gives
\beq
\label{eq:Z_vertex}
Z_{\rm vertex} = 1 - g_s^2 N_\e \frac{1}{\e}\lq
\(C_F-\frac{1}{2}\,C_A\)  + \frac{3}{2}\, C_A \rq =  1 - g_s^2 N_\e
\frac{1}{\e}\( C_F+C_A\)
\eeq
\item[-] the  radiative corrections to quark
propagator brings (see eq.~(\ref{eq:quark_propagator}))
\beq
\label{eq:Z_quark}
Z_{\rm quark} = 1 - g_s^2 N_\e \frac{1}{\e}\, C_F
\eeq
\item[-] the gluon- and ghost-loop corrections to  gluon
propagator contribute to the renormalization constant with
\beq
\label{eq:Z_g^gh}
Z_{\rm gluon}^{g,\,gh} = 1 +  g_s^2 N_\e \frac{1}{\e}\(
\frac{19}{12}\, C_A + \frac{1}{12}\,C_A\) =  1 +  g_s^2 N_\e
\frac{5}{3\e}\, C_A
\eeq
\item[-]
the contribution to the gluon self-energy coming from a massless fermion
loop is given by (see eq.~(\ref{eq:Pi}))
\beq
\Pi_{\mu\nu}^{ab} = -\frac{4}{3} \,\Nep \,g_s^2
\, T_F\, \delta^{ab} h^2
\(g_{\mu\nu} -\frac{h_\mu h_\nu}{h^2} \) \(\frac{\mu^2}{h^2}\)^\e
{\rm e}^{i\pi\e} \( \frac{1}{\e}+\frac{5}{3} \) \;,
\eeq
where we have used eq.~(\ref{eq:l_1m_ris}) and
the value of $S_{2m}$ given by eq.~(\ref{eq:s_2m_m=0}).
In this way, in the \MSB\ scheme, the massless loop contributes to the
renormalization constant of the  gluon propagator by an amount
\beq
\label{eq:Z_g^q}
Z_{\rm gluon}^{q} = 1 -  N_\e \,T_F\, \nlf \,g_s^2 \frac{4}{3\e}\;,
\eeq
where we have summed up the contributions of \nlf light quarks.
\item[-]
if we subtract the heavy-flavour loop at zero momentum, we perform the
same steps  we have done to obtain the expression of $Z_g$ (see
eq.~(\ref{eq:Z_g})) and we get
\beq
\label{eq:Z_g^Q}
Z_{\rm gluon}^Q = Z_g =  1 - N_\e T_F\,g_s^2 \(\frac{\mu^2}{m^2}\)^\e
\frac{4}{3\e} \;.
\eeq
\end{itemize}
Combining eqs.~(\ref{eq:Z_g^gh}), (\ref{eq:Z_g^q})
and~(\ref{eq:Z_g^Q}), we have
\beq
\label{eq:Z_gluon}
Z_{\rm gluon} =  Z_{\rm gluon}^{g,\,gh} \times  Z_{\rm gluon}^q
\times  Z_{\rm gluon}^Q =  1 - g_s^2 N_\e
\left[  \frac{4}{3\e}\,T_F\,n_{\rm lf} -\frac{5}{3\e}\,C_A
+ \(\frac{\mu^2}{m^2}\)^\e \frac{4}{3\e}\,T_F \right]\;.
\eeq
From the definition~(\ref{eq:Z_as}) and the eqs.~(\ref{eq:Z_vertex}),
(\ref{eq:Z_quark}) and~(\ref{eq:Z_gluon}), we obtain the expression of
$Z_{\as}$ given in eq.~(\ref{eq:charge_ren}).

Observe that, in this scheme, the term corresponding to the
heavy-flavour loop (eq.~(\ref{eq:Z_g^Q})) compensates exactly the
self-energy correction to the external gluon line coming from the
heavy-flavour loop (eq.~(\ref{eq:Z_g})): in fact, the final-state
gluon is on the mass shell, so it is effectively renormalized at zero
momentum by the heavy-quark loop, and thus decoupling applies.

\chapter{Soft and colliner amplitudes}
\thispagestyle{plain}
\section[Collinear limit for $g\,\to\, gg$ splitting]
{Collinear limit for {\mylarge $g\,\to\, gg$} splitting}
\label{app:gg_splitting}
%%%%%%%%%%%%%%%%%%%%%%%%%%%%%%%%%%%%%%%%%%%
\begin{figure}[htb]
\centerline{\epsfig{figure=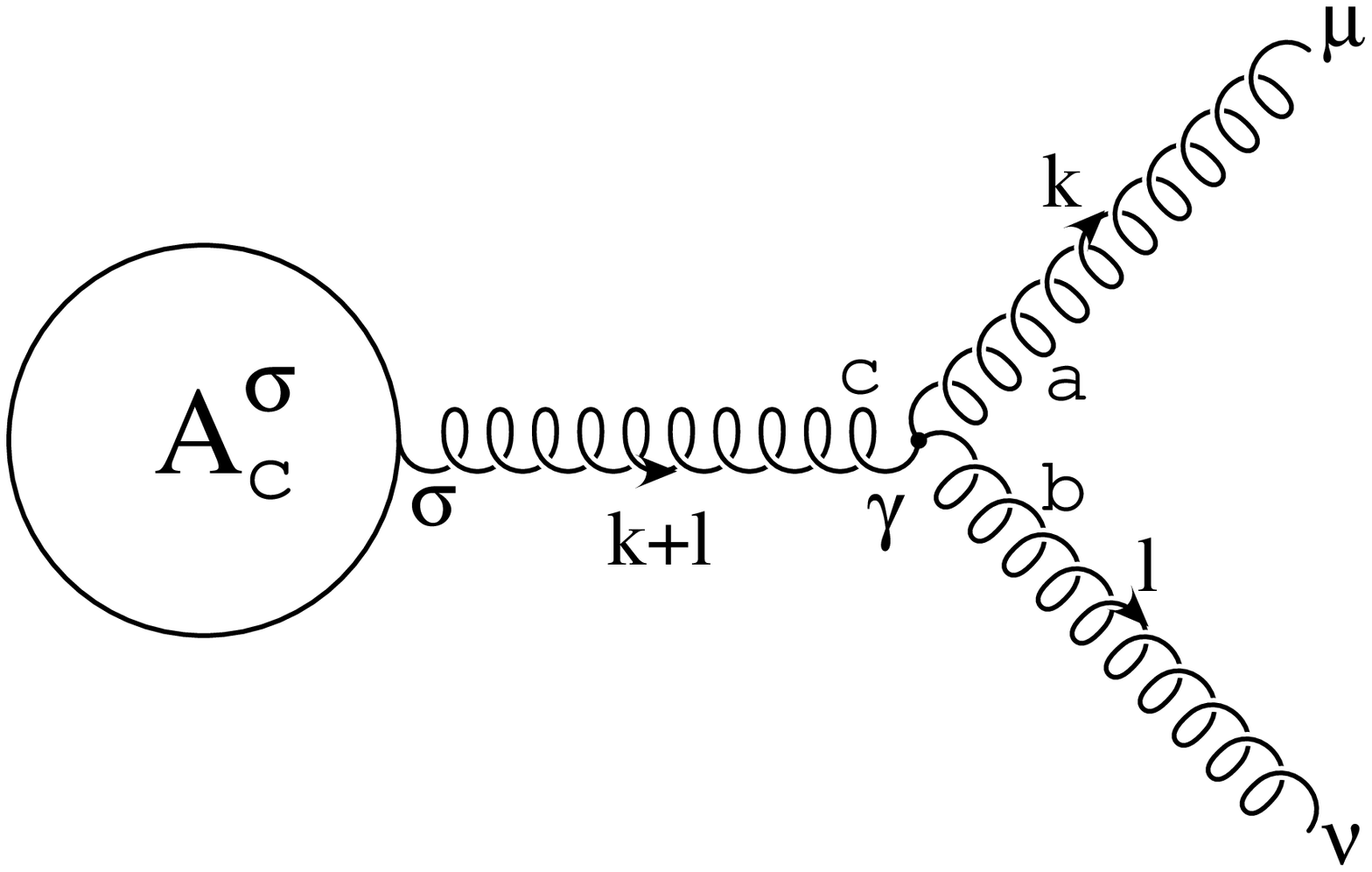,
width=\Largfig,clip=}}\ccaption{}{ \label{ggcoll}
Gluon splitting into a $gg$ couple.}
\end{figure}
%%%%%%%%%%%%%%%%%%%%%%%%%%%%%%%%%%%%%%%%%%%
In this appendix we  derive the singular part of the square of the
invariant amplitude when two collinear gluons are produced.
In the collinear limit,
the amplitude for the emission of two gluons in the final state can be
decomposed into two parts: the first one contains the graphs where the two
gluons are emitted by a single virtual one (see
Fig.~\ref{ggcoll}), and the other one contains all the other graphs
\beq
\label{eq:ampcoll}
{\cal A}^{ab} = \ga {\cal A}^{\s}_c(l+k)\, \frac{i
P_{\s\gamma}(k+l)}{(k+l)^2} \, (-g_s)\,
f^{abc} \, \Gamma^{\mu\nu\gamma}(-k,-l,k+l)
+{\cal R}^{\mu\nu}_{ab} \gc \e_{\mu}(k)\, \tilde{\e}_{\nu}(l)\;,
\eeq
where $a$ and $b$ are the colour indexes of the final gluons,
$P$ is the spin projector of the gluon propagator, $g_s$ is the strong
coupling constant, $f^{abc}$ are
the structure constants of the SU(3) gauge group, $\e$ and $\tilde{\e}$
are the polarization vectors of the final gluons, and
$\Gamma^{\mu\nu\gamma}$ is the Lorentz part of the three-gluon vertex
\beq
\label{eq:3gvertex}
\Gamma^{\mu\nu\gamma}(-k,-l,k+l) = (-k+l)^{\gamma}g^{\mu\nu} +
(-2l-k)^{\mu} g^{\nu\gamma} + (2k+l)^{\nu} g^{\mu\gamma}\;.
\eeq
Only the first term of~(\ref{eq:ampcoll}) is singular in the collinear
limit. We want to stress the fact that this term is singular in the
soft limit too. Therefore one has to be careful, when considering the
soft and collinear limit of the square amplitude, not to include this
contribution twice.

We introduce two light-like vectors $t$ and $\eta$, that, in the
centre-of-mass system, have components
\beq
\label{eq:t_eta_CMS}
\eqalign{
t=\(\left|\vet{k}+\vet{l}\,\right|,\,\vet{k}+\vet{l}\,\,\) \cr
\eta = c \times\( \frac{1}{\left|\vet{k}+\vet{l}\,\right|}\;,\, -
\frac{\vet{k}+\vet{l}}{\left|\vet{k}+\vet{l}\,\right|^2}\)}
\eeq
and choose $c=1/4$, so that $2\,t\cdot\eta=1$.
We then decompose, in the collinear limit,
\beq
\label{eq:def_t}
l^{\mu}+k^{\mu} = t^{\mu} + \xi \, \eta^{\mu}\;,
\eeq
where
\beq
\xi = (l+k)^2 = q^2 y\;.
\eeq
We will work in the light-cone gauge, characterized by the light-like
vector $\eta$, because, in this gauge, as we will see, the
interference of the divergent term of~(\ref{eq:ampcoll}) and of the finite
term ${\cal R}$ does not contribute to the singular part.

In this gauge, characterized by
\beq
\label{eq:eta_transv}
\eta^2 = 0\;, \quad\quad\quad\quad \eta\cdot\e = 0 \;, \quad\quad\quad\quad
\eta\cdot\tilde{\e} = 0 \;,
\eeq
the gluon spin projector becomes
\beq
\label{eq:projector}
P^{\s\gamma}(p) = -g^{\s\gamma} + \frac{\eta^{\s}p^{\gamma} +
\eta^{\gamma}p^{\s}}{\eta\cdot p}  \;,
\eeq
and the transversality of gluon polarization gives
\beq
\label{eq:transver}
k \cdot \e\,(k) = 0\;,  \quad\quad\quad\quad
l \cdot \tilde{\e}\,(l) = 0\;.
\eeq
In addition, for every momentum $p$
\beq
\label{eq:eta_contract}
\eta_\s P^{\s\gamma}(p) = \eta_\gamma P^{\s\gamma}(p) = 0 \;.
\eeq
We write $l$ and $k$ as (Sudakov decomposition)
\beq
\label{eq:def_tperp}
\eqalign{
k^{\mu} = v\, t^{\mu} + \xi'\eta^{\mu} + \kperp^{\mu}
\cr
l^{\mu} = (1-v)\,t^{\mu} + \xi''\eta^{\mu} -\kperp^{\mu}\;,
}
\eeq
with $\kperp$ such that
\beq
\label{eq:kperp_conds}
t\cdot\kperp=0 \;, \quad\quad\quad\quad \eta\cdot\kperp=0 \;.
\eeq
By imposing that $k^2=l^2=0$ and that $(l+k)^2=q^2 y$,  we have
\beq
\label{eq:kperp_norm}
\kperp^2 = -v(1-v)\,q^2 y\;,   \quad\quad\quad\quad
\xi' = (1-v)\, q^2 y\;, \quad\quad\quad\quad  \xi'' = v\, q^2 y\;.
\eeq
We can explicitly build the components of $\kperp$: in fact, in the
centre-of-mass system, the space-like four-vector $\kperp$ can be
written
\beq
\kperp = \(0, \vet{k}_\perp\)\;,
\eeq
with $\vet{k}_\perp$ belonging to the plane spanned by $\vet{k}$ and
$\vet{l}$, and perpendicular to $\vet{k}+\vet{l}$, so that
\beq
\label{eq:further_conditions}
\vet{k}_\perp = a\,\vet{k} + b\, \vet{l}\;, \quad\quad\quad\quad
\vet{k}_\perp \cdot (\vet{k}+\vet{l}) = 0\;, \quad\quad\quad\quad
\kperp \cdot q = 0\;.
\eeq
In this way, the conditions~(\ref{eq:kperp_conds}) are fulfilled, by
virtue of eqs.~(\ref{eq:t_eta_CMS}).
It can be easily shown that eqs.~(\ref{eq:further_conditions})
are satisfied, in the collinear limit, if
\beq
a\, (q\cdot k) + b\, (q\cdot l) =0\;,
\eeq
so that
\beq
a = \frac{N}{q\cdot k}\;,\quad\quad\quad\quad
b = -\frac{N}{q \cdot l}\;,
\eeq
with $N$ a normalization constant, determined by the first formula of
eqs.~(\ref{eq:kperp_norm}), but whose value is irrelevant to us,
because we are interested in normalized quantities, like
\beq
\frac{\kperp^{\mu}}{\sqrt{\kperp^2}} = \frac{1}{\sqrt{-2k\cdot l}}
\left[ \sqrt{\frac{q\cdot l}{q\cdot k}}\; k^{\mu} -
\sqrt{\frac{q\cdot k}{q\cdot l}}\; l^{\mu} \right] \;.
\eeq
From eqs.~(\ref{eq:def_t}) and~(\ref{eq:def_tperp}) we obtain
\beqn
k^{\mu} &=& \frac{1}{1-v} \left[ v\,l^{\mu} +(1-2\,v) q^2 y\,\eta^{\mu}
+\kperp^{\mu} \right] \label{eq:k_mu} \\
l^{\mu} &=& \frac{1}{v}\left[(1-v)\,k^{\mu}-(1-2\,v) q^2 y\,\eta^{\mu}
-\kperp^{\mu} \right]\label{eq:l_mu} \\
l^\mu-k^\mu &=& (1-2\,v)\(l^\mu+k^\mu-\xi\,\eta^\mu\) +
\(\xi''-\xi'\)\eta^\mu -2\,\kperp^\mu  \label{eq:l_minus_k_mu}\;.
\eeqn
We are now in a position to simplify the expression of the invariant
amplitude~(\ref{eq:ampcoll}). In fact, the first term of
$\Gamma^{\mu\nu\gamma}$ of eq.~(\ref{eq:3gvertex}) becomes, with the
help of eq.~(\ref{eq:l_minus_k_mu}),
\beqn
(-k+l)^{\gamma} P_{\s\gamma}(k+l) \e_{\mu}(k)\,
\tilde{\e}_{\nu}(l) &=& \lq(1-2\,v) \(l+k\)^\gamma
-2\,\kperp^\gamma  \rq   P_{\s\gamma}(k+l) \e_{\mu}(k)\,
\tilde{\e}_{\nu}(l)  \nn\\
&=&  \lq (1-2\,v) 2\,q^2 y\, \eta_\s   -2\,\kperp^\gamma
P_{\s\gamma}(k+l) \rq \e_{\mu}(k)\, \tilde{\e}_{\nu}(l)\;, \nn
\eeqn
where we have used eq.~(\ref{eq:eta_contract}) and, in the last line,
\beq
(k+l)^{\gamma} P_{\s\gamma}(k+l) = 2\,q^2 y\, \eta_{\s} \;.
\eeq
Using eqs.~(\ref{eq:eta_transv}), (\ref{eq:transver}), (\ref{eq:k_mu})
and~(\ref{eq:l_mu}) we can write
the second and third term of eq.~(\ref{eq:3gvertex}) in the following
way
\beqn
(-2l-k)^{\mu} \e_{\mu}(k)\, \tilde{\e}_{\nu}(l) &=& -2\,l^\mu
\e_{\mu}(k)\, \tilde{\e}_{\nu}(l)
=  \frac{2}{v} \, \kperp^\mu \e_{\mu}(k)\, \tilde{\e}_{\nu}(l)\nn\\
(2k+l)^{\nu} \e_{\mu}(k)\, \tilde{\e}_{\nu}(l)&=&  2\, k^\nu
\e_{\mu}(k)\, \tilde{\e}_{\nu}(l)
= \frac{2}{1-v}\, \kperp^\nu  \e_{\mu}(k)\, \tilde{\e}_{\nu}(l) \;.
\eeqn
The amplitude~(\ref{eq:ampcoll}) becomes
\beqn
{\cal A}^{ab} \!\!&=&\!\!\Biggl\{ {\cal A}^{\s}_c(l+k)\, \frac{i
P_{\s\gamma}(k+l)}{q^2 y} \, (-g_s)\,  f^{abc}  \nn\\
&& \times \left[
-2\,\kperp^{\gamma}\, g^{\mu\nu} + \frac{2}{v} \,\kperp^{\mu}\,
g^{\nu\gamma} + \frac{2}{1-v}\,\kperp^{\nu}\, g^{\mu\gamma} +{\cal O}\(y\)
\right] +{\cal R}^{\mu\nu}_{ab} \Biggr\} \e_{\mu}(k)\,
\tilde{\e}_{\nu}(l)\;. \phantom{aaaaa}
\eeqn
Observe that the first term is of order $1/\sqrt{y}$ , so that a
singularity with strength $1/y$ can arise only from the square of the
first term, and the interference term does not contribute.
For this reason  we can now substitute
\beq
{\cal A}^{\s}_{c}(l+k) \ \tol\  {\cal A}^{\s}_{c}(t) \,\equiv\,
\mbox{\rm tree-level\ amplitude}
\eeq
\beq
\label{eq:g_perp}
P^{\s\gamma}(k+l) \ \tol\  P^{\s\gamma}(t) =
-g^{\s\gamma} + \frac{\eta^{\s}t^{\gamma} +
\eta^{\gamma}t^{\s}}{\eta\cdot t} \equiv -\gperp^{\s\gamma}\;,
\eeq
where we have neglected terms of order $\kperp$, which give no
contributions to the divergent part of the amplitude.
Keeping only terms that contribute to the collinear singular part, we
can write, with the help of eq.~(\ref{eq:eta_transv}),
\beq
{\cal A}^{ab} =  {\cal A}_{c\s}(t) \,\frac{g_s}{q^2 y} \, i
f^{abc} \left[
-2\,\kperp^{\s}\, \gperp^{\mu\nu} + \frac{2}{v} \,\kperp^{\mu}\,
\gperp^{\nu\s} +
\frac{2}{1-v}\,\kperp^{\nu}\, \gperp^{\mu\s} \right]
\e_{\mu}(k)\, \tilde{\e}_{\nu}(l)\;.
\eeq

By squaring the amplitude and summing over the colours and spins of
the final gluons, we obtain, for the collinear singular part,
\beqn
{\cal M}_{\rm gg}^{\rm col} &=& \frac{g_s^2}{q^2}\,\frac{4\,C_A}{y} \ga
-\left[ -2+\frac{1}{v} +\frac{1}{1-v}+v\,(1-v) \right] g_{\s\s'}\right.\nn\\
\label{eq:ggcoll}
&& \left. {}- 2\, v\, (1-v) (1-\e)
\left[\frac{k_{\perp\s}k_{\perp\s'}}{\kperp^2}
-\frac{g_{\perp\s\s'}}{2-2\,\e}\right] \gc  {\cal A}_{c}^{\s}(t)
{\cal A}_{c}^{*\s'}(t)\;,
\eeqn
where we have used the gauge invariance $ t_{\s}\,{\cal
A}_{c}^{\s}(t)=0$ to write the following identity
\[
{\cal A}_{c}^{\s}(t) {\cal A}_{c}^{*\s'}(t) \,g_{\perp\s\s'}=
{\cal A}_{c}^{\s}(t) {\cal A}_{c}^{*\s'}(t)\, g_{\s\s'}\;,
\]
and
\beq
\kperp^{\mu\nu} k_{\perp\mu\nu} = d-2 = 2-2\,\e\;.
\eeq
The first term of eq.~(\ref{eq:ggcoll}) is recognized to be
the Altarelli-Parisi splitting function for the gluon-gluon process,
in $d=4-2\e$ dimensions. The second term vanishes after azimuthal
average in $4-2\e$ dimensions, as can be seen from
eq.~(\ref{eq:azimuth_integration}).

Coming now to our problem, we can further specify the structure of
${\cal A}_{c}^{\s}(t) {\cal A}_{c}^{*\s'}(t)$.
In fact, by using eq.~(\ref{eq:Mcalssp}), we can write eq.~(\ref{eq:ggcoll})
in the following form
\beqn
{\cal M}_{\rm gg}^{\rm col} &=& g_s^2 \,\mu^{2\e} \,
\frac{4\,C_A}{q^2\,y}\, \Biggl\{ -\left[- 2+\frac{1}{v}
+\frac{1}{1-v}+v\,(1-v) \right] g_{\s\s'}\nn\\
\label{eq:ggcollfin}
&&  {}- 2\, v\, (1-v) (1-\e)
\left[\frac{k_{\perp\s}k_{\perp\s'}}{\kperp^2}
-\frac{g_{\perp\s\s'}}{2-2\,\e}\right] \Biggr\}
\times {\cal M}_b^{\s\s'} \;.
\eeqn

\section[Collinear limit for $g\,\to \, q \bar{q}$ splitting]
{Collinear limit for {\mylarge $g\,\to \, q \bar{q}$} splitting}
%%%%%%%%%%%%%%%%%%%%%%%%%%%%%%%%%%%%%%%%%%%
\begin{figure}[htb]
\centerline{\epsfig{figure=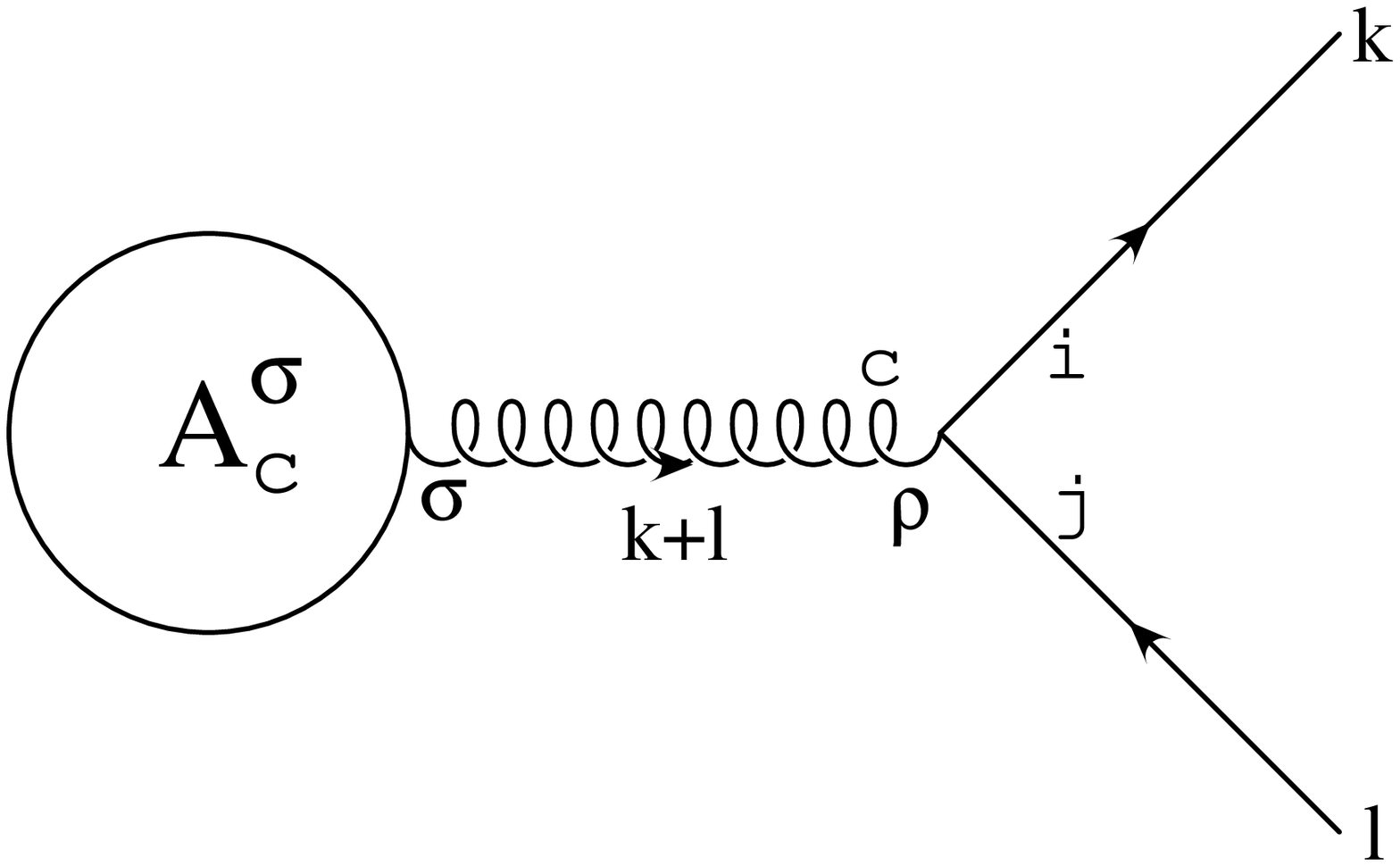,
width=\Largfig,clip=}}\ccaption{}{ \label{qqcoll}
Gluon splitting into a $q\bar{q}$ couple.}
\end{figure}
%%%%%%%%%%%%%%%%%%%%%%%%%%%%%%%%%%%%%%%%%%%
In this appendix, we derive the singular part of the square of the
amplitude for the emission of a collinear couple of massless
quark-antiquark.  The invariant amplitude of the process, represented in
Fig.~\ref{qqcoll}, is
\beq
{\cal A}_{ij} = {\cal A}_{c}^{\s}(k+l)\, \frac{i P_{\rho\s}(k+l)}
{(k+l)^2}\,
\bar{u}(k)\(-ig_s\gamma^{\rho}t^{c}_{ij}\) v(l) \;,
\eeq
where $P$ is given by eq.~(\ref{eq:projector}) and $t^c$ are the generators
of SU(3) gauge symmetry.
By squaring this amplitude and summing over the spins and colours
of the final quarks, we obtain
\beq
{\cal M}_{\rm q\bar{q}}^{\rm col} = \frac{T_F\,
g_s^2}{q^4 y^2} {\cal A}_{c}^{\s}(k+l) {\cal
A}_{c}^{*\s'}(k+l) P_{\rho\s}(k+l)  P_{\rho'\s'}(k+l)
\tr\(\ds{k}\gamma^{\rho}\ds{l}\gamma^{\rho'}\) \;,
\eeq
where $t^c$ are normalized such that $\tr\(t^a t^b\) = T_F\,\delta^{ab}$.

Considering now eqs.~(\ref{eq:def_tperp}), we see that, in the collinear
limit, the trace is of the order of $y$, so that the singular part can be
obtained by putting $y=0$ in the rest of the numerator
\beq
{\cal M}_{\rm q\bar{q}}^{\rm col} =
\frac{T_F\,g_s^2}{q^4 y^2} {\cal A}_{c}^{\s}(t) {\cal
A}_{c}^{*\s'}(t)\, g_{\perp\rho\s}  g_{\perp\rho'\s'}
\tr\(\ds{k}\gamma^{\rho}\ds{l}\gamma^{\rho'}\) \;,
\eeq
where we have used the definition of $t$ given in eq.~(\ref{eq:def_t}).
Evaluating the trace and keeping in the numerator only the terms
proportional to $y$, we obtain
\beq
{\cal M}_{\rm q\bar{q}}^{\rm col} = \frac{T_F\,g_s^2}{q^4 y^2}\,
4 \left[ -2 k_{\perp\s} k_{\perp\s'} -\frac{q^2 y}{2} g_{\s\s'}
\right] {\cal A}_{c}^{\s}(t) {\cal A}_{c}^{*\s'}(t) \;,
\eeq
that is
\beqn
{\cal M}_{\rm q\bar{q}}^{\rm col} &=& \frac{g_s^2}{q^2} \, \frac{4\, T_F}{y}
\Biggl\{-\frac{1}{2-2\e}\left[v^2+(1-v)^2-\e\right]g_{\s\s'} \nn\\
\label{eq:qqcoll}
&& {} + 2\,v\,(1-v)
\left[\frac{k_{\perp\s}k_{\perp\s'}}{\kperp^2}
-\frac{g_{\perp\s\s'}}{2-2\e}\right]  \Biggr\}
{\cal A}_{c}^{\s}(t) {\cal A}_{c}^{*\s'}(t) \;.
\eeqn
Here again we can recognize the Altarelli-Parisi kernel for
$g\,\to\,q \bar{q}$ splitting.

As previously done for eq.~(\ref{eq:ggcoll}) , we can specify this
formula to the problem we are studying: with the help of
eq.~(\ref{eq:Mcalssp}), we can write
\beqn
{\cal M}_{\rm q\bar{q}}^{\rm col} &=&  g_s^2 \, \mu^{2\e}  \,
\frac{4\, T_F}{q^2\,y} \,\Bigg\{ -\frac{1}{2-2\e}\left[v^2+(1-v)^2 -
\e\right]g_{\s\s'} \nn\\
\label{eq:qqcollfin}
&& {} + 2\,v\,(1-v) \left[\frac{k_{\perp\s}k_{\perp\s'}}{\kperp^2}
-\frac{g_{\perp\s\s'}}{2-2\e}\right] \Biggr\}
\times {\cal M}_b^{\s\s'} \;.
\eeqn

\section[Soft limit for the  invariant amplitude
$Q\overline{Q} g g$]{Soft limit for the  invariant amplitude
{\mylarge $Q\overline{Q} g g$}}
\label{app:soft}
In this appendix we derive the divergent part of the invariant
amplitude for the process
\beq
Z/\gamma(q) \;\to\; Q(p) + \overline{Q}(p') + g(k) + g(l)\;,
\eeq
in the limit when the momentum $l$ of the gluon is soft.
A soft singularity appears only if the soft gluon is emitted from one of
the external legs.

If the emitting external particle is the gluon, the amplitude of the
process, in the Feynman gauge, is
\beq
{\cal A}^{ab{\rm (g)}}_{ij}
= {\cal A}^{c\s}_{ij}(l+k)\, \frac{-i}{(k+l)^2} \, (-g_s)\,
f^{abc} \, \Gamma^{\mu\nu}_{\s}(-k,-l,k+l)
\, \e_{\mu}(k)\, \tilde{\e}_{\nu}(l)\;,
\eeq
where we have added to
eq.~(\ref{eq:ampcoll}) the colour indexes $i,j$ of the produced quarks. \\
As $l$ goes to zero, this term develops a singularity: in fact, by using the
gauge condition $k^{\s}\,{\cal A}_{c\s}^{ij}(k) =0$ and
eq.~(\ref{eq:transver}), we can write this amplitude as
\beq
\label{eq:ggsoft}
{\cal A}^{ab{\rm (g)}}_{ij}= g_s f^{abc}\, \frac{k^{\nu}}{k\cdot l}\,
{\cal A}^{c\s}_{ij}(k)\, \e_{\s}(k) \,\tilde{\e}_{\nu}(l) + \mbox{\rm
non-singular terms.}
\eeq

Similarly, if we  consider the emission of a soft gluon of colour index
$b$ from an external
quark leg with momentum $p$ and colour index $i$, that is
\[
Q_n(p+l)\, \to \, Q_i(p) + g_b(l)\;,
\]
we can write the invariant amplitude
\beq
{\cal A}_{ij}^{ab{\rm(Q)}} = \bar{u}(p) (-ig_s\gamma^{\nu} t^{b}_{in})
\,\frac{i}{\ds{p}+\ds{l}-m}\, \tilde{\cal {A}}_{nj}^{a \mu}(p+l)
\, \e_{\mu}(k) \,\tilde{\e}_{\nu}(l)\;,
\eeq
where $\tilde{\cal {A}}$ refers to the rest
of the process from which the quark external line takes origin.
In the limit of $l$ going to zero, we can rewrite this amplitude as
\beq
\label{eq:qgsoff}
{\cal A}_{ij}^{ab{\rm(Q)}} = g_s
\,\frac{p^{\nu}}{p\cdot l}\,t^{b}_{in} \, {\cal {A}}_{nj}^{a \mu}(p)
\, \e_{\mu}(k) \,\tilde{\e}_{\nu}(l) + \mbox{\rm
non-singular terms,}
\eeq
where we have defined
\beq
{\cal {A}}_{nj}^{a \mu}(p) = \bar{u}(p) \tilde{\cal
{A}}_{nj}^{a \mu}(p) \;.
\eeq

In the same way, we can obtain the limit of the amplitude for the
soft emission from an antiquark with momentum $p'$ and colour index $j$
\beq
\label{eq:qbargsoff}
{\cal A}_{ij}^{ab{\rm(\overline{Q})}} = -g_s
\,\frac{p'^{\nu}}{p'\cdot l} \, {\cal {A}}_{in}^{a \mu}(p')\, t^{b}_{nj}
\, \e_{\mu}(k) \,\tilde{\e}_{\nu}(l) + \mbox{\rm
non-singular terms.}
\eeq

Considering that ${\cal {A}}_{ij}^{c \s}$ of eq.~(\ref{eq:ggsoft}) can
be written as
\beq
{\cal {A}}_{ij}^{c \s} = t^c_{ij} {\cal {A}}^{\s}\;,
\eeq
where $ {\cal {A}}^{\s}$ does not contain any colour element, and the
similar ones for eqs.~(\ref{eq:qgsoff}) and~(\ref{eq:qbargsoff}),
we can sum the three amplitudes to obtain
\beq
{\cal A}_{ij}^{ab} = g_s \ga
i f^{abc}\, \frac{k^{\nu}}{k\cdot l}\,t^c_{ij}
+ \frac{p^{\nu}}{p\cdot l}\,t^{b}_{in} \, t^{a}_{nj} -
\frac{p'^{\nu}}{p'\cdot l} \,t^{a}_{in}\,t^{b}_{nj} \gc
{\cal {A}}^{\mu}
\, \e_{\mu}(k) \,\tilde{\e}_{\nu}(l)  \;,
\eeq
where we have disregarded the non-singular terms.

By squaring the amplitude and summing over the spins and colours of
the final gluons and quarks, we have
\beqn
{\cal M}_{\rm gg}^{\rm soft}(l) \!\!&=& \!\! g_s^2\,\mu^{2\e}\ga
- C_A \left[\frac{p\cdot k} {(p\cdot l) \, (k\cdot l)} +
\frac{p'\cdot k} {(p'\cdot l) \, (k\cdot l)} \right] \right.\nn\\
\label{eq:lsoft}
&&  \!\! \left.{}-2 \(C_F-\frac{C_A}{2}\) \frac{p\cdot p'}{(p\cdot l) \,
(p'\cdot l)} + C_F \left[ \frac{m^2}{(p\cdot l)^2} +
\frac{m^2}{(p'\cdot l)^2} \right]  \gc \times {\cal M}^\s_\s \phantom{aaaaa}
\eeqn
where we have made use of eq.~({\ref{eq:Mcalssp}).

The same result applies in the case of $k$ soft, once the interchange $l
\leftrightarrow k $ is made.

\chapter{List of integrals for the four-jet singular soft contributions}
\thispagestyle{plain}
%\input{phd_4jet_singular}
%\section{List of integrals for the four-jet singular soft contributions}
\label{app:soft_integrals}
In this appendix, we summarize the values of the integrals required to
isolate the singular terms of the four-jet cross section,
in the soft-gluon limit. The form of the integrals we are going to
compute is the following
\beq
I =  \frac{1}{N_{\phi}}\int_0^\pi d\phi\,(\sin\phi)^{-2\e}
\int_0^x d y \int_0^1 dv\, f(y,v,\phi)\;.
\eeq
We fix our attention to the $y$ and $v$ variables and we
split the integration range into two triangles
\beq
\int_0^x d y \int_0^1 dv\, f(y,v,\phi) = \int_0^x d y \int_0^{y/x} dv\,
f(y,v,\phi) +\int_0^x d y \int_{y/x}^1 dv\, f(y,v,\phi)\;.
\eeq
We perform the following change of variables: $v=y\,t$ in the
first integral and $v=y/t$ in the last one. In this way, we obtain
\beqn
\int_0^x d y \int_0^1 dv\,f(y,v,\phi) &=&
\int_0^x d y \int_0^{1/x} dt \,y\, f(y,y\,t,\phi) +
\int_0^x d y \int_{y}^{x} dt \, \frac{y}{t^2} \,
f\(y,\frac{y}{t},\phi\) \nn\\
&=& \int_0^{1/x}dt \int_0^x dy \, y \, f(y,y\,t,\phi)  +
\int_{0}^x dt \int_0^{t} dy \, \frac{y}{t^2} \,
f\(y,\frac{y}{t},\phi\)\;,\nn
\eeqn
where, in the second line, we have exchanged the order of integration.

In this way we succeed in writing $I$ as the sum of two contributions
\beqn
\label{eq:change_of_variables}
I &\equiv&  I^{(1)} + I^{(2)}\\
I^{(1)} &\equiv&  \frac{1}{N_{\phi}}\int_0^\pi
d\phi\,(\sin\phi)^{-2\e}\int_0^{1/x}dt \int_0^x dy \, y \,
f(y,y\,t,\phi)
\label{eq:I^1}\\
I^{(2)} &\equiv& \frac{1}{N_{\phi}}\int_0^\pi d\phi\,(\sin\phi)^{-2\e}
\int_{0}^x dt \int_0^{t} dy \, \frac{y}{t^2} \,
f\(y,\frac{y}{t},\phi\) \;.
\label{eq:I^2}
\eeqn

\subsubsection{${\mathbf I_1}$-type integrals }
The first type of integrals to compute is
\beq
\label{eq:I_1}
I_1(x,h) \equiv \frac{1}{N_{\phi}}\int_0^\pi d\phi\,(\sin\phi)^{-2\e}
\int_0^x d y \int_0^1 dv\,[v(1-v)]^{-\e} y^{-\e}  \frac{1}
{y\,[y+h\,v]} \;.
\eeq
As $f(y,v,\phi)=1/y\,[y+h\,v]$ is $\phi$-independent, the $\phi$
integration is straightforward and, according to eq.~(\ref{eq:I^1}),
we have
\beqn
I_1^{(1)} &=& \int_0^{1/x}dt \int_0^x dy \, y^{-1-2\e}\(1-y\,t\)^{-\e}
\frac{t^{-\e}}{1+h\,t} \nn \\
&=& \int_0^{1/x}dt \,\frac{t^{-\e}}{1+h\,t} \int_0^x dy \,
y^{-1-2\e} \left[ 1-\e\log(1-y\,t) + {\cal O}\(\e^2\) \right]\nn\\
&=& \int_0^{1/x}dt \,\frac{t^{-\e}}{1+h\,t} \left[
-\frac{1}{2\e} x^{-2\e} - \e
\int_0^x dy \, y^{-2\,\e} \frac{\log(1-y\,t)}{y} + {\cal
O}\(\e^2\)\right] \;. \phantom{aaaaa}
\eeqn
The integral in $y$ is finite  so it gives contribution of order
$\e$ that  can be neglected. In this way we have
\beqn
I_1^{(1)} &=& -\frac{1}{2\,\e}\, x^{-2\e} \int_0^{1/x}dt
\(1-\e\log t + {\cal O}\(\e^2\)\) \frac{1}{1+h\,t} \nn\\
& =& -\frac{1}{2\,h \,\e}\, x^{-2\e}
\left\{ \log\(1+\frac{h}{x}\)
+ \e  \lq \log x \log\(1+\frac{h}{x}\) -\li{-\frac{h}{x}} \rq\rg \;,
\phantom{aaaa}  \label{eq:I_1^1}
\eeqn
where terms of order $\e$ are not computed.
The integral of eq.~(\ref{eq:I^2}) gives rise to
\beqn
I_1^{(2)} &=& \int_{0}^x dt \int_0^{t} dy \, y^{-1-2\e}
\(1-\frac{y}{t}\)^{-\e} \frac{t^{-1+\e}}{t+h}\nn\\
&=& \int_{0}^x dt \,\frac{t^{-1+\e}}{t+h} \int_0^{t} dy \, y^{-1-2\e}
\left[ 1-\e\log\(1-\frac{y}{t} \) + {\cal O}\(\e^2\)  \right] \nn\\
&=& I_{1a}^{(2)} + I_{1b}^{(2)}  \;,
\label{eq:I_1^2}
\eeqn
where, for simplicity, we have separated out the integration into two
terms. Performing the $y$ integration in $I_{1a}^{(2)}$, we get
\beqn
I_{1a}^{(2)} &=& -\frac{1}{2\,\e} \int_{0}^x dt \,\frac{t^{-1-\e}}{t+h}
=-\frac{1}{2\, h\, \e} \int_{0}^x dt \, \( \frac{1}{t} -
\frac{1}{t+h} \) t^{-\e} \nn\\
&=&   -\frac{1}{2\, h\, \e}  \left\{ -\frac{1}{\e}x^{-\e} -\int_{0}^x
dt \, \frac{1-\e\log t + {\cal O}\(\e^2\)  }{t+h}\right\} \nn\\
&=&  -\frac{1}{2\, h\, \e}  \left\{ -\frac{1}{\e}x^{-\e}
-\log\(1+\frac{x}{h}\) + \e \left[\log x \log\((1+\frac{x}{h}\) +
\li{-\frac{x}{h}} \right]   \right\} \phantom{aaaaa}
\label{eq:I_1a^2}
\eeqn
where we have neglected terms of order $\e$. To compute
$I_{1b}^{(2)}$, we perform the change of variable $z=y/t$ to obtain
\beqn
I_{1b}^{(2)} &=& -\e \int_{0}^x dt \, \frac{t^{-1-\e}}{t+h} \int_0^1
dz \, \frac{\log(1-z)}{z^{1+2\e}}
= \e \,\frac{\pi^2}{6}\frac{1}{h} \lq -\frac{1}{\e}x^{-\e} + {\cal O}\(1\)
\rq \;. \label{eq:I_1b^2}
\eeqn
Collecting all terms together, according to
eqs.~(\ref{eq:change_of_variables}), (\ref{eq:I_1^1})--(\ref{eq:I_1b^2}),
we have
\beq
\label{eq:I_1_finale}
I_1(x,h) =  \frac{1}{2\,h} \ga
\frac{1}{\e^2}-\frac{1}{\e} \log h - \log^2\frac{x}{h}
+\frac{1}{2}\log^2 h -\frac{\pi^2}{2}-2 \li{-\frac{x}{h}} \gc
+ {\cal O}\(\e\) \;.
\eeq

%%%%%%%%%%%%%%%%%%%%%%%%%%%%%%%%%%%%%%%%%%%%%%%%%%%%%%%%%%%%%%%%%%%%%%%%
\subsubsection{${\mathbf I_2}$-type integrals }
The second type of integrals has the following expression
\beq
I_2(x,h) \equiv \frac{1}{N_{\phi}}\int_0^\pi d\phi\,(\sin\phi)^{-2\e}
\int_0^x d y \int_0^1  dv\,[v(1-v)]^{-\e} y^{-\e}  \frac{1} {[y+a
v]^2}\;.
\eeq
Here again, the $\phi$ integration is straightforward, and we have,
according to eq.~(\ref{eq:I^1}),
\beqn
I_2^{(1)} &=& \int_0^{1/x}dt \int_0^x dy \, y^{-1-2\e}\(1-y\,t\)^{-\e}
\frac{t^{-\e}}{\(1+h\,t\)^2} \nn \\
&=& \int_0^{1/x}dt \,\frac{t^{-\e}}{\(1+h\,t\)^2} \int_0^x dy \,
y^{-1-2\e} \left[ 1-\e\log(1-y\,t) + {\cal O}\(\e^2\) \right]\nn\\
&=&  -\frac{1}{2\,\e}\, x^{-2\e}\int_0^{1/x}dt \, \(1-\e\log t\)
\frac{1}{\(1+h\,t\)^2}+ {\cal O}\(\e\)\nn\\
&=&-\frac{1}{2\,h\,\e} \,x^{-2\e} \lg 1-\frac{x}{x+h} +\e\lq
\log\(1+\frac{h}{x}\) + \frac{h}{x+h}\log x\rq \rg + {\cal
O}\(\e\)\;. \phantom{aaa}
\eeqn
From eq.~(\ref{eq:I^2}) we obtain
\beqn
I_2^{(2)} &=& \int_{0}^x dt \int_0^{t} dy \, y^{-1-2\e}
\(1-\frac{y}{t}\)^{-\e} \frac{t^{\e}}{\(t+h\)^2}\nn\\
&=& \int_{0}^x dt \,\frac{t^{\e}}{\(t+h\)^2} \int_0^{t} dy \, y^{-1-2\e}
\left[ 1-\e\log\(1-\frac{y}{t} \) + {\cal O}\(\e^2\)  \right] \nn\\
&=& -\frac{1}{2\,\e}\int_0^x dt\, \frac{t^{-\e}}{\(t+h\)^2} =
-\frac{1}{2\,\e}\int_0^x dt\, \( 1-\e\log t+ {\cal O}\(\e^2\)\)
\frac{1}{\(t+h\)^2} \nn\\
&=&  -\frac{1}{2\,h\,\e}\lg \frac{x}{x+h} -\e \lq \log \frac{h}{x+h} +
\frac{x}{x+h}\log x \rq \rg + {\cal O}\(\e\)\;.\phantom{aaa}
\eeqn
Adding together these two contributions, we have
\beq
\label{eq:I_2_finale}
I_2(x,h) = I_2^{(1)}+I_2^{(2)}
= \frac{1}{2\,h} \ga -\frac{1}{\e} -2\log\(1+\frac{h}{x}\)
+\log h \gc + {\cal O}\(\e\)\;.
\eeq

%%%%%%%%%%%%%%%%%%%%%%%%%%%%%%%%%%%%%%%%%%%%%%%%%%%%%%%%%%%%%%%%%%%%%%%%
\subsubsection{${\mathbf I_3}$-type integrals }
The last type of integrals has the following form
\beq
\label{eq:I_3}
I_3  \equiv  \frac{1}{N_{\phi}} \int_0^\pi d\phi\,(\sin\phi)^{-2\e}
I_3(x) \;,
\eeq
where
\beq
I_3(x) \equiv \int_0^x d y \int_0^1 dv\,[v(1-v)]^{-\e} y^{-\e}
\frac{1} {y + h\,v} \,
\frac{1}{y-c\cos\phi\sqrt{y}\sqrt{v} + g\,v} \;.
\label{eq:I_3(x)}
\eeq
According to eq.~(\ref{eq:I^1}) and ~(\ref{eq:I^2}), we have
\beqn
%%%%%%%%%%%%%%%%%%%%%%%%%%%%%%%%%%%%%%%%%%%%%%
I_3^{(1)}(x) &=& \int_0^{1/x}dt \int_0^x dy \, y^{-1-2\e}\(1-y\,t\)^{-\e}
\frac{t^{-\e}}{(1+h\,t)\(1-c\cos\phi\sqrt{t} + g\,t\)} \nn \\
&=& -\frac{1}{2\,\e}\,x^{-2\e} \int_0^{1/x}dt
\frac{t^{-\e}}{(1+h\,t)\(1-c\cos\phi\sqrt{t} + g\,t\)}
+{\cal O}\(\e\) \nn \\
&=& -\frac{1}{2\,\e}\(1-2\e\log x\)  \int_0^{1/x}dt
\frac{1-\e\log t}{(1+h\,t)\(1-c\cos\phi\sqrt{t} + g\,t\)} +{\cal
O}\(\e\) \nn \\
&=&  -\frac{1}{2\,\e} \int_0^{1/x}dt
\frac{1}{(1+h\,t)\(1-c\cos\phi\sqrt{t} + g\,t\)} \nn\\
&& + \int_0^{1/x}dt
\frac{\frac{1}{2}\log t +\log x}{(1+h\,t)\(1-c\cos\phi\sqrt{t} + g\,t\)}
+{\cal O}\(\e\)\\
%\eeqn
%\beqn
%%%%%%%%%%%%%%%%%%%%%%%%%%%%%%%%%%%%%%%%%%%%%%%%%%%%%%%%%
I_3^{(2)}(x) &=& \int_0^{x}dt \int_0^t dy \, y^{-1-2\e}\(1-
\frac{y}{t}\)^{-\e}
\frac{t^{\e}}{(t+h)\(t-c\cos\phi\sqrt{t} + g\)} \nn \\
&=& -\frac{1}{2\,\e} \int_0^{x}dt
\frac{t^{-\e}}{(t+h)\(t-c\cos\phi\sqrt{t} + g\)} +{\cal O}\(\e\)\nn \\
&=& -\frac{1}{2\,\e} \int_0^{x}dt
\frac{1-\e\log t}{(t+h)\(t-c\cos\phi\sqrt{t} + g\)} +{\cal O}\(\e\)\nn \\
&=&  -\frac{1}{2\,\e} \int_0^{x}dt
\frac{1}{(t+h)\(t-c\cos\phi\sqrt{t} + g\)} \nn\\
&& + \int_0^{x}dt
\frac{\frac{1}{2}\log t}{(t+h)\(t-c\cos\phi\sqrt{t} + g\)} +{\cal
O}\(\e\) \;.
\eeqn
From eqs.~(\ref{eq:I_3}) and~(\ref{eq:I_3(x)}) we have
\beqn
I_3 &=& \frac{1}{N_{\phi}} \int_0^\pi d\phi\,(\sin\phi)^{-2\e}
\lq I_3^{(1)}(x)+ I_3^{(2)}(x) \rq \nn\\
&=&  \frac{1}{N_{\phi}} \int_0^\pi d\phi\,\lq1-2\e\log\(\sin\phi\)\rq
\lq I_3^{(1)}(x)+ I_3^{(2)}(x) \rq \nn\\
&=& \frac{1}{N_{\phi}}
\lg -\frac{1}{2\,\e} \,I_\e + \frac{1}{2}\lq
I_{3a} +  I_{3b} \rq   + I_{\phi} + {\cal O}\(\e\) \rg \;,
\label{eq:I_3_finale}
\eeqn
with
\beqn
\label{eq:I_e}
I_\e &=& \int_0^{\infty} dt\,\int_0^{\pi} d\phi\,
\frac{1}{\(1+h\,t\) \(1 -c\cos\phi \sqrt{t} + g \,t\)} \\
\label{eq:I_3a}
I_{3a} &=& \int_0^{\frac{1}{x}} dt\, \int_0^{\pi}d\phi\,
\frac{2\log x +\log t} {\(1+h\,t\) \(1 -c\cos\phi\sqrt{t} + g\, t\)}\\
\label{eq:I_3b}
I_{3b} &=& \int_{0}^{x} dt\, \int_0^{\pi} d\phi\,
\frac{\log t}{\(h+t\) \(t -c\cos\phi \sqrt{t} + g \)}\\
\label{eq:I_phi}
I_{\phi}&=& \int_0^{\infty} dt\,\int_0^{\pi} d\phi\,
\frac{\log\(\sin\phi\)}{\(1+h\,t\) \(1 -c\cos\phi \sqrt{t} + g\, t\)} \;.
\eeqn
We can now integrate in $\phi$  the first three expressions, using the
identity
\[
\int_0^{\pi} d\phi \,\frac{1}{a-b\cos\phi}= \frac{\pi}{\sqrt{a^2-b^2}}\;,
\]
and in $t$ the last one, to obtain
\beqn
I_\e &=& \int_0^{\infty} dt\,
\frac{1}{\(1+h\,t\)} \frac{\pi}{\sqrt{\(1+g\,t\)^2-c^2t}} \nn\\
&=& \frac{\pi}{\sqrt{(g-h)^2+c^2h}} \ \log\frac{h\(c^2-2g+2h +
2\sqrt{(g-h)^2+c^2h}\)}{h^2-\(g-\sqrt{(g-h)^2+c^2h}\)^2}\\
%%%%%%%%%%%%%%%%%%%%%%%%%%%%%%%%%%%
I_{3a} &=&  \int_0^{\frac{1}{x}} dt\,
\frac{2\log x +\log t} {\(1+h\,t\)} \frac{\pi}{\sqrt{\(1+g\,t\)^2-c^2t}}\\
%%%%%%%%%%%%%%%%%%%%%%%%%%%%%%%%%%%
I_{3b} &=& \int_{0}^{x} dt\,
\frac{\log t}{\(h+t\)}  \frac{\pi}{\sqrt{\(t+g\)^2-c^2t}}\\
%%%%%%%%%%%%%%%%%%%%%%%%%%%%%%%%%%%
I_{\phi}&=& \int_0^{\frac{\pi}{2}} d\phi\, \log\(\sin\phi\) \,
\frac{2}{(g-h)^2+h\(c\cos\phi\)^2} \nn\\
&&{}\times \lg \frac{2\,c\,(g+h)\cos\phi} {\sqrt{4g-c^2\cos^2\phi}}\
\arctan\frac{c\cos\phi}{\sqrt{4g-c^2\cos^2\phi}} +
(g-h)\log\frac{g}{h}\rg\nn\\
%%%%%%%%%%%%%%%%%%%%%%%%%%%%%%%%%%%
&=&\int_0^{1} dt\, \frac{\log\sqrt{1-t^2}}{\sqrt{1-t^2}}\,
\frac{2}{(g-h)^2+h\,c^2\, t^2} \nn\\
&&{}\times \lg \frac{2\,c\,(g+h)\,t} {\sqrt{4g-c^2\, t^2}}\
\arctan\frac{c\,t}{\sqrt{4g-c^2\, t^2}} +
(g-h)\log\frac{g}{h}\rg \;,
\eeqn
where we have used the symmetry of the integral $I_\phi$ around the point
$\pi/2$ and made the change of variable $t=\cos \phi$.

We are now in a position of further reduce the expression of
$I_\e$, to account for the actual values of the parameters: in fact,
with the help of eqs.~(\ref{eq:parameter1})--(\ref{eq:parameter2}),
(\ref{eq:def_sig3}) and~(\ref{eq:csipm}), we have
\beq
I_{\e} = \frac{\pi}{K}\, \frac{\(1+\dpr^2\)\(1-\dpr^2\)}{4\dpr}
\(1-x_g\) \log\(\frac{\zp}{\zm}\)^2   \;.
\eeq

%\input{phd_biblio}
%\newpage
%{\bf Acknowledgements}\newline\noindent
%We wish to thank G.~Altarelli, A.~Blondel, G.~Cowen, L.~L\"onnblad,
%M.~Mangano, C.~Mariotti, K.~M\"onig, M.~Seymour, J.~Steinberger and B. Webber
%for useful discussions.

%\input{phd_thanking}
\newpage
\thispagestyle{plain}
\mbox{}
\newpage
\thispagestyle{empty}
\markboth{\dunhxii \underline{}}
{\dunhxii \underline{}}
\mbox{}
\vfill
\mbox{}
\hfill
\begin{minipage}[t]{11cm}
{\it
Ringrazio innanzi tutto il Prof.~Paolo Nason, per la costante e
fondamentale presenza durante il mio dottorato e per gli utili
suggerimenti che mi ha dato nello svolgimento di questo calcolo.

Un grazie va anche al Prof.~Stefano Catani, per l'attenta lettura di
questa tesi.

Infine, vorrei ringraziare i miei genitori ed i miei amici di sempre,
per essermi stati accanto in questi ultimi anni.

}
\end{minipage}
\vspace{3cm}
\mbox{}
\begin{thebibliography}{99}
%refdef.tex
\addcontentsline{toc}{chapter}{\bibname}%
\markboth{\dunhxii \underline{\bibname}}
{\dunhxii \underline{\bibname}}

%\genericchapter{\bibname}{}
\relax
\def\pl#1#2#3{{\it Phys. Lett. }{\bf #1}\ (19#2)\ #3}
\def\zp#1#2#3{{\it Z. Phys. }{\bf #1}\ (19#2)\ #3}
\def\prl#1#2#3{{\it Phys. Rev. Lett. }{\bf #1}\ (19#2)\ #3}
\def\rmp#1#2#3{{\it Rev. Mod. Phys. }{\bf#1}\ (19#2)\ #3}
\def\prep#1#2#3{{\it Phys. Rep. }{\bf #1}\ (19#2)\ #3}
\def\pr#1#2#3{{\it Phys. Rev. }{\bf #1}\ (19#2)\ #3}
\def\np#1#2#3{{\it Nucl. Phys. }{\bf #1}\ (19#2)\ #3}
\def\sjnp#1#2#3{{\it Sov. J. Nucl. Phys. }{\bf #1}\ (19#2)\ #3}
\def\app#1#2#3{{\it Acta Phys. Polon. }{\bf #1}\ (19#2)\ #3}
\def\jmp#1#2#3{{\it J. Math. Phys. }{\bf #1}\ (19#2)\ #3}
\def\nc#1#2#3{{\it Nuovo Cim. }{\bf #1}\ (19#2)\ #3}
\relax
%\def    \nuke   #1#2#3{{\sl Nucl. Phys.} {\bf B#1}(#2)#3}
%\def    \pl     #1#2#3{{\sl Phys. Lett.} {\bf #1B}(#2)#3}
%\def    \prl    #1#2#3{{\sl Phys. Rev. Lett.} {\bf #1}(#2)#3}
%\def    \pr     #1#2#3{{\sl Phys. Rev.} {\bf #1}(#2)#3}
%\def    \prd    #1#2#3{{\sl Phys. Rev.} {\bf D#1}(#2)#3}
%\def    \prep   #1#2#3{{\sl Phys. Rep.} {\bf #1}(#2)#3}
%\def    \zeit   #1#2#3{{\sl Z. Phys.} {\bf #1}(#2)#3}
%\def    \cpc   #1#2#3{{\sl Comput. Phys. Commun.} {\bf #1}(#2)#3}

\bibitem{ERT}
R.K.~Ellis, D.A.~Ross and A.E.~Terrano, \np{B178}{81}{421}.
\bibitem{VGO}
J.A.M.~Vermaseren, K.J.F.~Gaemers and S.J.~Oldham, \np{B187}{81}{301}.
\bibitem{FKSS}
K.~Fabricius, G.~Kramer, G.~Schierholz and I.~Schmitt, \zp{C11}{81}{315}.
\bibitem{NO}
P.~Nason and C.~Oleari, \pl{B387}{96}{623}, \hepph{9607347}.
\bibitem{NO1}
P.~Nason and C.~Oleari, \pl{B407}{97}{57}, \hepph{9705295}.
\bibitem{Rodrigo}
G.~Rodrigo, {\it Nucl. Phys. Proc. Suppl.} {\bf 54A}\ (1997)\ 60,
\hepph{9609213}; \newline
G.~Rodrigo, Ph.~D.~Thesis,
Univ.~of Val\`encia, 1996, \hepph{9703359};\newline
G.~Rodrigo, A.~Santamaria and M.~Bilenkii, \prl{79}{97}{193},
\hepph{9703358}.
\bibitem{Bil_Rod_Santa}
M.~Bilenkii, G.~Rodrigo and A.~Santamaria, \np{B439}{95}{505}.
\bibitem{Bernreuther}
W.~Bernreuther, A.~Brandenburg and P.~Uwer, \prl{79}{97}{189},
\hepph{9703305}.
\bibitem{Brandenburg}
A.~Brandenburg and P.~Uwer, preprint PITHA-97-29, \hepph{9708350}.
\bibitem{YellowBook}
Z.~Kunszt and P.~Nason, ``QCD'', in ``Z Physics at LEP 1'',
eds. G.~Altarelli, R.~Kleiss and C.~Verzegnassi
(Report CERN 89-08, Geneva, 1989).
\bibitem{Ballestrero}
A.~Ballestrero, E.~Maina and S.~Moretti, \pl{B294}{92}{425};
\np{B415}{94}{265}.
\bibitem{Magnea}
L. Magnea and E. Maina, \pl{B385}{96}{395}, \hepph{9604385}.
\bibitem{Hagiwara}
K.~Hagiwara, T.~Kuruma and Y.~Yamada, \np{B358}{91}{80}.
\bibitem{NasonWebber}
P.~Nason and B.R.~Webber, \np{B421}{94}{473}.
\bibitem{PDG}
Particle Data Group, R.M.~Barnett et al., \pr{D54}{96}{1}.
\bibitem{NasonWebber_PL}
P.~Nason and B.R.~Webber, \pl{B395}{97}{355}, \hepph{9612353}.
\bibitem{MeleNason}
B.~Mele and P.~Nason, \np{B361}{91}{626}.
\bibitem{Harlander}
R.~Harlander and M.~Steinhauser, \hepph{9710413}.
\bibitem{AltarelliParisi}
G.~Altarelli and G.~Parisi, \np{B126}{77}{298}.
\bibitem{CurciFurmanskiPetronzio}
G.~Curci, W.~Furmanski and R.~Petronzio, \np{B175}{80}{27}.
\bibitem{Floratos}
E.G.~Floratos, R.~Lacaze and C.~Kounnas, \np{B192}{81}{417};
\pl{B98}{81}{89}.
\bibitem{Konishi}
J.~Kalinowski, K.~Konishi, P.N.~Scharbach and T.R.~Taylor,
\np{B181}{81}{253};\newline
J.~Kalinowski, K.~Konishi and T.R.~Taylor,
\np{B181}{81}{221}.
\bibitem{Spira}
B.A.~Kniehl, G.~Kramer and M.~Spira, \zp{C76}{97}{689},\newline
\hepph{9610267} $2^{\rm nd}$ version;\newline
J.~Binnewies, B.A.~Kniehl and  G.~Kramer, \zp{C76}{97}{677},\newline
\hepph{9702408}.
\bibitem{FurmanskiPetronzio}
W.~Furmanski and R.~Petronzio, \pl{97B}{80}{437}.
%\bibitem{Spira}
%B.~A.~Kniehl, G.~Kramer and M.~Spira, preprint DESY-96-210,
%\hepph{9610267}.
\bibitem{CWZ} J.~Collins, F.~Wilczek and A.~Zee, \pr{D18}{78}{242}.
\bibitem{BN} F.~Bloch and A.~Nordsiek, \pr{52}{37}{54}.
\bibitem{KLN} T.~Kinoshita, \jmp{3}{62}{650},\\
T.~D.~Lee and M.~Nauenberg, \pr{B133}{64}{1549}.
\bibitem{Reinders} L.J.~Reinders, H.~Rubinstein and S.~Yazaki,
\prep{127}{85}{1}.
\bibitem{NDE} P.~Nason, S.~Dawson and R.~K.~Ellis,
\np{B327}{89}{49}.
\bibitem{no_fragmentation} P.~Nason and C.~Oleari, preprint CERN-TH/97-209,
\hepph{9709358}, to appear in {\it Phys. Lett.}
\bibitem{LEP2}
P.~Nason and B.W.~Webber, ``QCD'', in ``Physics at LEP2'',
eds. G.~Altarelli, T.~Sj\"ostrand and F.~Zwirner (Report CERN96-01,
Geneva 1996).
\bibitem{NasonOleari}
P.~Nason and C.~Oleari, preprint CERN-TH/97-219, \hepph{9709360},
to appear in  {\it Nucl. Phys. }{\bf B}.

%\bibitem{Rbpapers}
%\bibitem{ALEPH3}
%  ALEPH Collab., R.~Barate \etal, preprint CERN-PPE-97-018, Feb 1997;\newline
%\bibitem{ALEPH4}
%  ALEPH Collab., R.~Barate \etal, preprint CERN-PPE-97-017, Feb 1997;\newline
%\bibitem{ALEPH5}
%ALEPH Collab., O.~Bazarko, \hepex{9609005}; \newline
%\bibitem{ALEPH1}
%  ALEPH Collab., D.~Buskulic et al., \pl{B313}{93}{535};\newline
%\bibitem{ALEPH2}
%  ALEPH Collab., D.~Buskulic et al., \pl{B313}{93}{549};\newline
%\bibitem{DELPHI1}
%  DELPHI Collab., P.~Abreu et al., \zp{C70}{96}{531};\newline
%\bibitem{DELPHI2}
%  DELPHI Collab., P.~Abreu et al., \zp{C66}{95}{323};\newline
%\bibitem{DELPHI3}
%  DELPHI Collab., P.~Abreu et al., \zp{C65}{95}{555};\newline
%\bibitem{L3}
%  L3 Collab., O.~Adriani et al., \pl{B307}{93}{237};\newline
%\bibitem{OPAL1}
%  OPAL Collab., K.~Ackerstaff et al., \zp{C74}{97}{1};\newline
%\bibitem{OPAL2}
%  OPAL Collab., R.~Akers et al., \zp{C65}{95}{17};\newline
%\bibitem{SLD}
%  SLD Collab., K.~Abe et al., \pr{D53}{96}{1023}.
%\bibitem{GluonSplitting}
%H.~Przysiezniak, ``Gluon Splitting into Heavy Quarks in $e^+e^-$
% Annihilations'', talk given at the
%``QCD and High Energy Hadronic Interactions'',
%XXXI Rencontres de Moriond, March 1996, OPAL CR248.
%\bibitem{MCmodels}
%L.~L\"onnblad, \cpc{71}{1992}{15};\newline
%G.~Marchesini \etal, \cpc{67}{1992}{465};\newline
%T.~Sj''ostrand, \cpc{82}{1994}{74} and Lund Univ. report LU~TP~95-20.
%\bibitem{MontecarloCalc}
%Private communications from M.~Seymour, L.~L\"onnblad, G.~Cowen,
%and K.~M\"onig.
%\bibitem{Jersak}
%J.~Jers\'ak, E.~Laermann and P.~Zerwas, \prd{25}{1982}{1218}.
%\bibitem{NasonWebber}
%P.~Nason and B.R.~Webber, \nuke{421}{1994}{473}.
%\bibitem{PDG}
%  Particle Data Group, R.M.~Barnett et al., \pr{D54}{96}{1}.
%\bibitem{NasonWebber}
%  P.~Nason and B.R.~Webber, \pl{B395}{97}{355}, \hepph{9612353}.
%\bibitem{OPALbfrag}
%G.~Alexander \etal, OPAL Col. \pl{B364}{1995}{93}.
%\bibitem{ALEPHbfrag}
%D.~Buskulic \etal, ALEPH Col., \pl{B357}{1995}{699}.
\end{thebibliography}
\end{document}